\documentclass[12pt,a4paper,twoside,sort&compress]{book}
\usepackage{epsfig}
\usepackage{amsfonts}
\usepackage{amssymb}
\usepackage{amsmath}
\usepackage{setspace}
\usepackage{slashed}
\usepackage{lscape}
\usepackage{pdfpages}
\RequirePackage[a-1b]{pdfx}
\RequirePackage[latin1,utf8]{inputenc}
\usepackage{csquotes}
\usepackage{amsthm}
\usepackage{latexsym}
\usepackage{psfrag}
\usepackage{graphicx}
\usepackage{rotating}
\usepackage{graphics}
\usepackage{subfigure}
\usepackage{mathenv}
\usepackage{amsthm}
\usepackage{xcolor}
\usepackage[thinlines]{easytable}
\usepackage{multirow}
\usepackage{booktabs}
\usepackage{soul}
\usepackage{lineno,hyperref}
\usepackage{lipsum}
\usepackage{mathtools}
\usepackage{fancyhdr}

\newcommand{\ba}{\begin{eqnarray}}
\newcommand{\ea}{\end{eqnarray}}

\newcommand{\ignore}[1]{}
\usepackage[latin1]{inputenc}
\usepackage{cite}
\usepackage{subfigure}
\usepackage[small,bf]{caption}

\usepackage[Conny]{fncychap} 
\ChNameVar{\LARGE\sffamily\bfseries}
\ChNumVar{\Huge} \ChTitleVar{\Huge\sffamily\bfseries}
\ChRuleWidth{0.0pt} \ChNameUpperCase \ChTitleUpperCase 
\usepackage{fancyhdr}

   {\setlength\paperheight {297mm}%
   \setlength\paperwidth  {210mm}}
\catcode`@=11

\catcode`@=11
\begin{document}

    \pagestyle{fancy}                       
    \fancyfoot{}                            
    \renewcommand{\chaptermark}[1]{         
    \markboth{\chaptername\ \thechapter.\ #1}{}} %
    \renewcommand{\sectionmark}[1]{         
    \markright{\thesection.\ #1}}         %
    \fancyhead[LE,RO]{\bfseries\thepage}    
   \fancyhead[RE]{\bfseries\leftmark}      
   \fancyhead[LO]{\bfseries\rightmark}     
   \renewcommand{\headrulewidth}{0.3pt}    
    \makeatletter
    \def\cleardoublepage{\clearpage\if@twoside \ifodd\c@page\else%
        \hbox{}%
        \thispagestyle{empty}
        \newpage%
        \if@twocolumn\hbox{}\newpage\fi\fi\fi}
    \makeatother

\includepdf[page={1-4}]{First_three}


\thispagestyle{empty}
\begin{center}
\vspace*{6cm}
{\Large{\emph{Dedicated to my parents, Mr. Yogesh Rathi and Mrs. Reeta Rathi, family members \& my first teachers Bulbul, ``Mr. Vikas Kapoor Sir''.}}} \\
\vspace*{0.5cm}

\end{center} 
\frontmatter
\clearpage
\phantomsection
\addcontentsline{toc}{chapter}{Abstract}

\begin{center}
{\Huge{\textbf{Abstract}}}
\end{center}
In the present thesis, we study various models of 2D dilaton gravity known as JT gravity coupled with non-trivial gauge interactions. In particular, we investigate the effects of the non-trivial gauge couplings on the thermal properties of black holes and wormholes in two dimensions and compute various physical observables like entropy, free energy, etc. Further, we examine the possibilities of the Hawking-Page transition and wormhole to black hole phase transition in two dimensions. In addition, we also study the transformation properties of boundary stress-energy tensor under the diffeomorphism and $U(1)$ gauge transformation and hence compute the central charge associated with the 1D boundary theory.

To begin with, we first construct the model of JT gravity coupled with Abelian ($ U(1) $) and $ SU(2) $ Yang-Mills (YM) fields using the compactification of 5D gravitational theory. We obtain the vacuum and black hole solutions associated with the 2D model perturbatively, treating the gauge couplings as the expansion parameters. We find that the \emph{interpolating} vacuum solution asymptotes to AdS$_2$ in the IR and Lifshitz$_2$ in the UV, which we identify as the thermal radiation background for our analysis. On the other hand, the charged 2D black hole solution asymptotes to Lifshitz$_2$ geometry. Our analysis on thermal stability reveals the existence of a first order phase transition at $ T \sim T_0 $ such that for $ T < T_0 $ the thermal radiation dominates. On the other hand, as we increase the temperature of the system $ T\sim T_2 (>T_0 )$, a \emph{globally} stable black hole emerges, which clearly indicates the onset of the Hawking-Page (HP) transition in 2D gravity models.

Next, we construct a model of JT gravity that uncovers the phase transition between the Euclidean wormhole and the black hole phase at finite charge density and/or chemical potential. In the low temperature regime, the charged wormhole solution dominates the system, while, on the other hand, as we increase the temperature, the wormhole phase undergoes a first order phase transition into a pair of black holes at finite chemical potential. Moreover, the Free energy (density) and the total charge of the system suffer a discontinuous change at the critical point ($T = T_0$). We finally conjectured that the boundary theory dual to the aforementioned model of JT gravity could be a two-site (uncoupled) complex SYK model with finite chemical potential.

In the third example, we propose a diffeomorphism invariant action of the JT gravity coupled with $U(1)$ gauge fields that contain all the possible 2-derivative as well as 4-derivative interactions. We compute the vacuum and black hole solutions of 2D gravity in the perturbative regime, where we consider the couplings associated with the higher derivative interactions as the expansion parameters. We find that the vacuum solution in the UV limit is dominated by the Lifshitz$_2$ geometry with dynamical exponent ($z=\frac{7}{3}$). On the other hand, the Ricci scalar blows up in the deep IR limit due to the presence of higher derivative interactions in the theory. We compute the holographic stress-energy tensor and obtain the central charge associated with the 1D boundary theory. Our analysis reveals that the central charge depends on the negative power of the coupling associated with the 4-derivative interactions.

We further explore the thermal properties of 2D black holes in the presence of quartic interactions. To be specific, we compute the Wald entropy associated with the 2D black holes and interpret its near-horizon divergence in terms of the density of states. Then, we investigate the properties of the 2D black hole in the near-horizon limit and calculate the associated central charge. We find that the near-horizon CFT may be recast as a 2D Liouville theory with higher derivative corrections using suitable coordinate transformations. We study the vacuum structure and invariant properties of the generalized Liouville theory and obtain the Weyl anomaly associated with it. 

In the final example, we construct a model of 2D dilaton gravity by compactifying the four dimensional Einstein gravity coupled with the ModMax Lagrangian. We perform the perturbative analysis to find out the vacuum solutions of the theory, which reduce to $AdS_2$ in the absence of ModMax interactions. We further compute the holographic central charge associated with the 1D boundary theory up to quadratic order in the gauge couplings. We finally obtain the ModMax corrected 2D black hole solutions and explore their extremal limits.

\clearpage 
\thispagestyle{empty}
\phantomsection\addcontentsline{toc}{chapter}{Acknowledgments}
\pagestyle{fancy}
\fancyhead[LO,RE]{{\bf Acknowledgments}}
\fancyhead[RO,LE]{{\bf \thepage}}


\begin{center}
{\Huge{\textbf{\emph{Acknowledgments}}}}
\end{center}
I would like to extend my heartfelt gratitude to my supervisor, Dr. Dibakar Roychowdhury, who taught me this subject from scratch. He was always available in the office for discussions, even on holidays and after official hours. He consistently supported me in academic activities and inspired me to pursue higher research. I am also deeply thankful to my supervisor's family for their continuous blessings and support.

I would like to express my gratitude to all the members of my ``Student Research Committee'', namely Prof. Aalok Misra (Chairperson), Prof. Rajdeep Chatterjee (Member), and Associate Prof. P.C. Srivastava (Member). I would also like to extend my thanks to Prof. Binoy Patra and the other professors of the Physics Department at IIT Roorkee.

I would like to extend special thanks to my collaborators, Dr. Arindam Lala, Dr. Aditya Mehra, Jitendra Pal, Gopal Yadav, Salman Khan, Debarshi, Sumit, Pushpa, Ashwani and Abhishek for their valuable discussions and comments. I also wish to express my gratitude to Dr. K. P. Yogendran (IISER Mohali), Prof. Patrick Das Gupta (University of Delhi), Prof. Sanjay Jain (University of Delhi), Prof. Debajyoti Choudhury (University of Delhi), Prof. Sunil Mukhi (ICTS, Bengaluru), Prof. Suvrat Raju (ICTS, Bengaluru), Prof. Shiraz Minwalla (TIFR, Mumbai), Prof. Gautam Mandal (TIFR, Mumbai), Prof. Justin David (IISc, Bengaluru), Prof. Chethan Krishnan (IISc, Bengaluru), Prof. Ashoke Sen (ICTS, Bengaluru), Dr. Sarthak Parikh (IIT Delhi) and Prof. Rajesh Gopakumar (ICTS, Bengaluru) for their valuable discussions and questions.

The journey of my PhD at IIT Roorkee would be incomplete without the support and companionship of Brijmohan, Shivam, Ankur, Sonu, Shubham Sharma (a true friend who helped me during my class 7 final exams), Avi (a strong man who always stands by me), Nikhil, Md. Anas, Vikram, Vinay, Shilpi, Vartika, Dibyendu, Nidhi, Ravina, Shruti, Pooja, Neha, and all our batchmates from 2019. I would like to extend my thanks to all the members of Research Scholar Room 1, including Shikha, Nitish, Ayushi, and Aalok. I would like to thank all my cousins, namely Somya, Anushka, Abhiraj, and Kismis for their consistent support and for playing with me all the time when I visit their home. Finally, I would like to express my gratitude to all the teachers, especially Mr. Hemo (Silver Bells, Shamli), and many others who have guided me at various stages of my life.

I also acknowledge the authorities at the Indian Institute of Technology, Roorkee, for their unconditional support towards researches in basic sciences.

\cleardoublepage
\thispagestyle{empty}
\phantomsection
\addcontentsline{toc}{chapter}{Table of Contents}
\pagestyle{fancy}
\fancyhead[LO,RE]{{\bf Contents}}
\fancyhead[RO,LE]{{\bf \thepage}}
\tableofcontents

\clearpage
\thispagestyle{empty} 
\phantomsection
\addcontentsline{toc}{chapter}{List of Publications}

\pagestyle{fancy}
\fancyhead[LO,RE]{{\bf Publications}}
\fancyhead[RO,LE]{{\bf \thepage}}

\begin{center}
{\Large\textbf{List of Publications}}
\vspace{0.6cm}
\end{center}
\textbf{A. Refereed Journal Publications (included in the 
thesis)}
\begin{enumerate}
\item

A.~Lala, \textbf{H.~Rathi} and D.~Roychowdhury,
``Jackiw-Teitelboim gravity and the models of a Hawking-Page transition for 2D black holes,''
\textbf{Phys. Rev. D} \textbf{102} (2020) no.10, 104024
doi:10.1103/PhysRevD.102.104024
[arXiv:2005.08018 [hep-th]].
\item
\textbf{H.~Rathi} and D.~Roychowdhury,
``Phases of Euclidean wormholes in JT gravity,''
\textbf{Nucl. Phys. B} \textbf{994} (2023), 116315
doi:10.1016/j.nuclphysb.2023.116315
[arXiv:2111.11279 [hep-th]].
\item
\textbf{H.~Rathi} and D.~Roychowdhury,
``Holographic JT gravity with quartic couplings,''
\textbf{JHEP} \textbf{10} (2021), 209
doi:10.1007/JHEP10(2021)209
[arXiv:2107.11632 [hep-th]].
\item
\textbf{H.~Rathi} and D.~Roychowdhury,
``AdS$_{2}$ holography and ModMax,''
\textbf{JHEP} \textbf{07} (2023), 026
doi:10.1007/JHEP07(2023)026
[arXiv:2303.14379 [hep-th]].


\end{enumerate}
\vspace{0.6cm}
\textbf{B. Under review}

\begin{enumerate}
\vspace{0.5cm}	
\item 
J.~Pal, \textbf{H.~Rathi}, A.~Lala and D.~Roychowdhury,
``Non-chaotic dynamics for Yang-Baxter deformed $\text{AdS}_{4}\times \text{CP}^{3}$ superstrings,''
[arXiv:2208.09599 [hep-th]] [Under review in EPJC].

\item
G.~Yadav and \textbf{H.~Rathi},
``Yang-Baxter Deformed Wedge Holography,''

[arXiv:2307.01263 [hep-th]] [Under review in PRD]. 

\item 
A.~Mehra, \textbf{H.~Rathi} and D.~Roychowdhury,
``Carrollian Born-Infeld Electrodynamics,''
[arXiv:2401.06958 [hep-th]] [Under review in JHEP].

\end{enumerate}
\vspace{0.6cm}
\newpage
\textbf{C. Participation in Conferences/Schools}
\vspace{0.6cm}
\begin{enumerate}
\item
 I attended a \textbf{``National Strings Meeting 2019''} at IISER Bhopal, India.
\item I attended the online conference \textbf{``Virtual Strings 2020''} at Cape Town South Africa.
\item 
 I presented a poster (online) at \textbf{``DAE-BRNS High Energy Physics Symposium 2020''} at the NISER Bhubaneswar, India. The title of the poster was \textbf{``Hawking-Page transition in JT gravity''}.
\item 
I gave an oral (online) presentation on \textbf{``JT gravity and Hawking-Page transition'' } at the \textbf{``Shivalik HEPCTAS Meeting- Summer 2020''} at IISER Mohali, India.
\item 
I attended the online seminar  \textbf{``Virtual Meeting on Gravity, Strings and Fields''} at IIT Roorkee India.
\item I attended the online seminar \textbf{``Mysteries of Universe Lecture series-I''} at IIT Roorkee, India.
\item I attended the online seminar \textbf{``Mysteries of Universe Lecture series-II''} at IIT Roorkee, India.
\item  I presented an online poster titled \textbf{``Hawking-Page transition in 2D Gravity''} at \textbf{``Strings 2021''} conference in São Paulo Brazil.
\item I attended the online conference \textbf{``School on Superstring Theory and Related Topics''} in October-2021 at ICTP, Italy.

\item I attended the online meeting \textbf{``Indian Strings Meeting-2021''} at IIT Roorkee, India.
\item I gave an oral presentation on \textbf{``Holographic JT gravity with quartic couplings'' } at the \textbf{``Shivalik HEPCTAS Meeting''} in December-2021 at IISER Mohali, India.
\item  I attended the online seminar \textbf{``$1^{st}$ school on Non-Relativistic Quantum Field theory, Gravity and Geometry''} at Nordita, Stockholm, Sweden.
\item  I attended the online seminar \textbf{``Quantum Information in QFT and AdS/CFT-I''}, jointly organized by  IIT Gandhinagar, IISc Bengaluru and IIT Hyderabad, India. 
\item  I attended the online seminar \textbf{``Quantum Information in QFT and AdS/CFT- II''}, jointly organized by  IIT Gandhinagar, IISc Bengaluru and IIT Hyderabad, India. 
\item I have attended the program, \textbf{``Nonperturbative and Numerical Approaches to Quantum Gravity, String theory and Holography''} in August-2022 at ICTS Bengaluru, India.

\item I presented a poster  titled \textbf{``Phases of complex SYK from Euclidean wormholes'' } at \textbf{``XXV DAE-BRNS High Energy Physics symposium 2022''} held at IISER Mohali, India.
\item I gave an oral presentation, \textbf{``$AdS_2/CFT_1$ holography at finite charge density''}, at the \textbf{``CHEP Seminar''} on 30-August-2023 at IISc Bengaluru, India.

\item I attended the online seminar at \textbf{``ICTS Strings''} and gave an oral talk \textbf{``$AdS_2/CFT_1$ holography at finite charge density''}, at the \textbf{``ICTS Strings''} on 31-August-2023 at ICTS Bengaluru, India.

\item I attended the \textbf{``Indian Strings Meeting-2023''} in \textbf{IIT Mumbai, India} during 10-16 December 2023 and gave an \textbf{oral talk} titled \textbf{``$AdS_{2}/CFT_{1}$ at finite density and Holographic
aspects of 2D Black Holes.''}
\item I visited  \textbf{Prof. Gautam Mandal} at \textbf{TIFR, India} during 18-20 December 2023 and presented our work on \textbf{``$AdS_{2}$ holography and ModMax.''}

\item I visited  \textbf{Dr. Sarthak Parikh} at \textbf{IIT Delhi, India} during 03-06 January 2024 and discussed our work on \textbf{``$AdS_{2}$ holography and ModMax.''}

\item{I visited  \textbf{Prof. Nabamita Banerjee} at \textbf{ IISER, Bhopal, India} during 15-16 January 2024 and gave an \textbf{oral talk} titled \textbf{``$AdS_{2}/CFT_{1}$ at finite density and Holographic
aspects of 2D Black Holes.''}}

\end{enumerate}

\clearpage
\phantomsection
\addcontentsline{toc}{chapter}{List of Figures}
\pagestyle{fancy}
\fancyhead[LO,RE]{{\bf List of Figures}}
\fancyhead[RO,LE]{{\bf \thepage}}
\listoffigures
\cleardoublepage
\phantomsection
\mainmatter \pagestyle{fancy} \setcounter{chapter}{0}
\setcounter{section}{0} \setcounter{subsection}{0}
\setcounter{tocdepth}{3}

\chapter{Introduction and Overview}
\allowdisplaybreaks
\pagestyle{fancy}
\fancyhead[LE]{\emph{\leftmark}}
\fancyhead[LO]{\emph{\rightmark}}
\rhead{\thepage} 


Over the past several decades, the $AdS_{d+1}/CFT_d$ correspondence (also known as the gauge/gravity duality \cite{Maldacena:1997re}-\cite{Gubser:1998bc}), has emerged as the most radical concept in modern theoretical physics. This remarkable duality conjecture establishes a profound connection between classical gravitational theories in $d+1$-dimensional Anti-de Sitter ($AdS_{d+1}$) space-time and a class of Large $N$ gauge theories that live in $d$-dimensional Minkowski space-time. These Large $N$ gauge theories possess a remarkable property called the conformal invariance and are therefore typically referred as Conformal Field Theory ($CFT_d$).

In the original proposal, the duality was conjectured between the gravitational theory in $AdS_5\times S^5$ and $\mathcal{N} =4$ Yang Mills theory (SYM) living in $d=4$ space-time dimension \cite{Maldacena:1997re}. $\mathcal{N}=4$ SYM is a theory that encompasses non-Abelian gauge fields, six scalar fields, four Weyl fermions, and four bosons, all of which are due to the principles of supersymmetry and their interactions. In this context, $\mathcal{N}=4$ signifies the number of supercharges within the theory. Notably, in four-dimensional spacetime, the $\beta$-function for $\mathcal{N}=4$ SYM precisely vanishes for all couplings, indicating the theory's invariance under the conformal group \cite{Sohnius:1985qm}. Furthermore, the correlation functions of the theory remain finite under suitable regularization schemes, rendering that the theory is free from divergences \cite{Sohnius:1985qm}.

The discovery of $AdS_5/CFT_4$ duality took place while studying $D3$ branes in type IIB superstring theory \cite{Maldacena:1997re}. In the closed string description, the spacetime geometry of a stack of $N$ $D3$ branes in the near-horizon limit is characterized by the $AdS_5\times S^5$ geometry. Conversely, the low-energy description of $N$ $D3$ branes is characterised by the $\mathcal{N}=4$ Super Yang Mills theory living in four space-time dimensions. Remarkably, both these theories share the same global symmetry group namely $SO(2,4)\times SO(6)$ which provides additional evidence supporting the duality.

In a nutshell, $AdS_5/CFT_4$ duality establishes a relationship between the generating functions of the gravitational theory in $AdS_5\times S^5$ and $\mathcal{N} =4$ SYM in four space-time dimensions\footnote{See \cite{Ramallo:2013bua}-\cite{Natsuume:2014sfa} for recent reviews on $AdS/CFT$ duality. }. It is found that in the large $N$ limit, the partition functions of both these theories are equivalent, $Z_{CFT} = Z_{AdS}$ \cite{Gubser:1998bc}. Although this duality is primarily tested in the large $N$ limit, in principle, this equality should hold for all values of $N$.

$AdS/CFT$ duality has emerged as an immensely valuable tool for solving various strongly coupled phenomena in Large $N$ gauge theories \cite{Policastro:2002se}-\cite{Sadeghi:2013zma}. In the Large $N$ limit, one can attempt to solve these theories perturbatively, treating $1/N$ as an expansion parameter. However, this approach is considerably challenging within the framework of Quantum Field Theory. Alternatively, $AdS/CFT$ duality suggests that Large $N$ gauge theories are effectively described by classical gravitational theories in some space-time that is a solution of Einstein's equation of general relativity. Therefore, one could utilize the classical gravitational theory as a dual description to study strongly coupled gauge theories in the Large $N$ limit.

In general, the $AdS_{d+1}/CFT_d$ duality holds for any space-time dimension ($d$). However, for the purpose of the present thesis, we are interested in exploring a particular class of duality known as $AdS_2/CFT_1$
correspondence. Below, we delve into the details of the $AdS_2$ holography and motivate the correspondence in the first place.

\section{$AdS_2$ holography : An overview }

 The $AdS_2$ space-time can be obtained as a solution of Einstein's gravity in two dimensions \cite{Maldacena:2016upp},
\begin{align}\label{exactads2}
    S=-\frac{\Phi_0}{16\pi G_2}\left(\int_M d^2x\sqrt{-g}R^{(2)}+2\int_{\partial M} dt\sqrt{-\gamma}K\right)+S_M,
\end{align}
where $\Phi_0$ is a constant, $R^{(2)}$ and $G_2$ denote the respective Ricci scalar and Newton's constant in two dimensions. The second term in the action (\ref{exactads2}) denotes the boundary action known as the Gibbons-Hawking-York (GHY) term \cite{Gibbons:1976ue} which is required for a successful execution of the variational principle. In this context, $\gamma$ represents the induced boundary metric and $K$ denotes the trace of the extrinsic curvature. Finally, $S_M$ stands for the action describing the matter content of the theory ($\Phi,A_{\mu}...$).

In the Poincaré coordinates\footnote{Here, we set the $AdS$ length $L=1.$}, the metric of the  $AdS_2$ space-time can be expressed as \cite{Natsuume:2014sfa}
\begin{align}\label{pcads2}
ds^2=\frac{1}{z^2}\left(-dt^2+dz^2\right),
\end{align}
where the boundary is located at $z=0$. 

The $AdS_2$ space-time (\ref{pcads2}) exhibits maximal symmetry and remains invariant under the $SL(2,\mathbb{R})$ group. Additionally, it is interesting to notice that the $AdS_2$ space-time has two boundaries. While this property is less apparent in the Poincaré coordinates (\ref{pcads2}), it becomes quite transparent in the conformal coordinates \cite{Zee:2013dea}, where the line element is given by
\begin{align}
    ds^2=\frac{1}{\cos^2{\psi}}\left(-dt^2+d\psi^2\right).
\end{align}
In this coordinate system, the variable $\psi$ ranges from $-\frac{\pi}{2}$ to $\frac{\pi}{2}$, indicating that $AdS_2$ is represented by a rectangular strip in the $(t - \psi)$ plane, having two boundaries located at $\psi = \pm \frac{\pi}{2}$.

Unfortunately, the pure $AdS_2$ geometry does not support any finite energy excitation above the vacuum \cite{Maldacena:2016upp, Maldacena:1998uz}. Therefore, pure $AdS_2$ gravity is suitable only for describing the ground state of the system. Additionally, one can easily check that the stress-energy tensor obtained by varying (\ref{exactads2}) in the absence of matter field ($S_M$) with respect to the space-time metric ($g_{\mu\nu}$) becomes identically zero. In other words, the action (\ref{exactads2}) is purely topological and does not play a role in determining the dynamics of the space-time metric.

However, it is possible to construct a meaningful theory of gravity in two dimensions by introducing a coupling between the space-time metric and a scalar field known as the dilaton ($\Phi$).  Such theories are commonly referred to as 2D Einstein-dilaton (ED) gravity that supports a nearly $AdS_2$ ($nAdS_2$) geometry \cite{Maldacena:2016upp}. It is noteworthy to mention that one can derive the 2D Einstein-dilaton gravity following a compactification of Einstein’s gravity in higher dimensions \cite{Strominger:1998yg}.

The action of the ED gravity in two dimensions is given by \cite{Maldacena:2016upp} 
\begin{align}\label{jtads2}
    S=&\hspace{1mm}-\frac{\Phi_0}{16\pi G_2}\left(\int_M d^2x\sqrt{-g}R^{(2)}+2\int_{\partial M} dt\sqrt{-\gamma}K\right)-\frac{1}{16\pi G_2}\Bigg(\int_M d^2x\sqrt{-g}\Phi\big(R^{(2)}\nonumber\\
    &+2\big)+\hspace{2mm}2\int_{\partial M} dt\sqrt{-\gamma}\Phi_b K\Bigg)+S_M,
\end{align}
where $\Phi_b$ is the value of dilaton at the boundary. 

In the above action (\ref{jtads2}), the first part is purely topological, while the second part is known as the Jackiw-Teitelboim (JT) gravity \cite{Jackiw:1984je}-\cite{Teitelboim:1983ux}. It is important to mention that both the space-time metric ($g_{\mu\nu}$) and the dilaton ($\Phi$) diverge near the boundary. Therefore, one needs to implement the following boundary conditions \cite{Maldacena:2016upp},
\begin{align}\label{introbc}
    \Phi_{\text{boundary}}=\Phi_b\sim\frac{\Phi_r}{\epsilon}\hspace{1mm},\hspace{2mm}
    g_{\tau\tau}\big |_{\text{(boundary)}}\sim\frac{1}{\epsilon^2},
\end{align}
where $\tau$ denotes the time in boundary theory, $\Phi_r$ denotes the ``renormalised'' value of $\Phi$ at the boundary and $\epsilon$ is the UV cut-off parameter. Furthermore, one can parameterize the $AdS_2$ metric\footnote{Here, we consider the Euclidean version of the metric (\ref{pcads2}). } (\ref{pcads2}) in terms of the boundary time ($\tau$) which yields 
\begin{align}\label{introadsrepara}
    ds^2= \frac{1}{z^2}\left(t'^2+z'^2\right)d\tau^2,
\end{align}
where prime $(')$ represents the derivative with respect to the boundary time ($\tau$). One can further express the coordinate $z$ in terms of $\tau$ as $z(\tau) = \epsilon t(\tau)$, where $t(\tau)$ is any arbitrary function which also satisfied the boundary condition (\ref{introbc}).

To gain deeper insights into the $AdS_2/CFT_1$ duality, it is quite useful to construct the effective action \cite{Maldacena:2016upp} or the boundary action for JT gravity using the aforementioned data (\ref{jtads2})-(\ref{introadsrepara}). The effective action can be obtained by integrating out the dilaton $(\Phi)$ in the on-shell action which yields
\begin{align}\label{reparactint}
    S_{\text{eff}}=-\frac{1}{8\pi G_2}\int \frac{d\tau}{\epsilon} \frac{\Phi_r(\tau)}{\epsilon}K,
\end{align}
where $\Phi_r(\tau)$ denotes the coupling associated with the ($0+1$)$d$ boundary theory. 

On the other hand, the trace of extrinsic curvature ($K$) can be expressed in terms of the parametrized coordinates  
\begin{align}\label{introktrace}
    K=\frac{t'\left(t'^2+z'^2+zz''\right)-zz't''}{\left(t'^2+z'^2\right)^{\frac{3}{2}}}\hspace{1mm}=\hspace{1mm}1+\epsilon^2\text{Sch}(t, \tau),
\end{align}
where we have used the solution $z(\tau)=\epsilon t(\tau)$ and $\text{Sch}(t, \tau)$ represents the Schwarzian 
\begin{align}\label{introschjt}
    \text{Sch}(t, \tau)=-\frac{1}{2}\frac{t''^2}{t'^2}+\left(\frac{t''}{t'}\right)'.
\end{align}

Finally, plugging (\ref{introktrace})  back into (\ref{reparactint}), we obtain the renormalised effective action for JT gravity
\begin{align}\label{jtintrosch}
    S_{\text{eff}}=-\frac{1}{8\pi G_2}\int d\tau \Phi_r(\tau)\text{Sch}(t, \tau).
\end{align}

Notice that, (\ref{jtintrosch}) is invariant under the under the $SL(2,\mathbb{R})$ transformation
\begin{align}
    \tau\rightarrow\frac{a \tau+b}{c\tau+d},
\end{align}
where $ad-bc=1$. In other words, the effective action (\ref{jtintrosch}) is invariant under the isometry group of $AdS_2$ mentioned earlier. 

We now present a quantum mechanical model known as the Sachdev-Ye-Kitaev (SYK) model whose deep infrared (IR) dynamics is conjectured to be dual to the  JT gravity set up presented above. In particular, we are interested in a quantum many  body system whose dynamics in the deep IR  can be expressed in terms of the Schwarzian (\ref{introschjt}) and which looks similar to the effective action for JT gravity (\ref{jtintrosch}). To summarise, in the soft/Schwarzian limit, the SYK model \cite{Sachdev:1992fk}-\cite{Maldacena:2016hyu} is conjectured to be dual to  JT gravity in ($1+1$) dimension, which forms the basis for the rest of the discussion. 

The SYK model represents a one-dimensional quantum mechanical system that comprises of $N$ Majorana fermions. Specifically, it accounts for the interactions among $q=4$ fermions out of the total $N$ at the same time. The Hamiltonian of the SYK model can be expressed as \cite{Maldacena:2016hyu} 
\begin{align}\label{sykhamiltonian}
    H=\frac{1}{4!}\sum_{ijkl}J_{ijkl}\psi_i\psi_j\psi_k\psi_l,
\end{align}
where each $\psi_i$ corresponds to the Majorana fermion, while $J_{ijkl}$ is the dimensionful coupling constant with energy dimensions. Additionally, the Majorana fermions satisfy the commutation relations $\{\psi_i,\psi_j\}=\delta_{ij}$, indicating that the coupling constant ($J_{ijkl}$) must be fully antisymmetric in all indices to maintain the hermicity of the Hamiltonian (\ref{sykhamiltonian}).

The values of the couplings ($J_{ijkl}$) are randomly chosen from the following Gaussian distribution function \cite{Maldacena:2016hyu}
\begin{align}\label{distrifunc}
    P\left(J_{ijkl}\right)=\sqrt{\frac{N^3}{12\pi J^2}}\text{exp}\left(-\frac{N^3J^2_{ijkl}}{12J^2}\right),
\end{align}
which implies that the system has quenched disorder and $J$ is the dimensionful parameter that describes the distribution function. By employing the distribution function (\ref{distrifunc}), the average and root mean square values of the coupling $J_{ijkl}$ turns out to be
\begin{align}
    <J_{ijkl}>=0\hspace{1mm},\hspace{2mm}<J^2_{ijkl}>=\sqrt{\frac{3!J^2}{N^3}}.
\end{align}

Let us now explore some key features of the SYK model that makes it worthy for further investigations.  Firstly, in the strong coupling and Large $N$ limit, the $2$-point function can be solved exactly \cite{Polchinski:2016xgd}-\cite{EMarcus thesis}. To explore the 2-point function, we begin by writing the Lagrangian of the SYK model as follows \cite{Polchinski:2016xgd}
\begin{align}\label{lagSYK}
    L=\frac{1}{2}\psi_i\frac{d}{d\tau}\psi_i-H,
\end{align}
where the Hamiltonian ($H$) is given in (\ref{sykhamiltonian}). It is important to notice that the first term in (\ref{lagSYK}) represents the Lagrangian of free Majorana fermions, while the second term represents the interaction between them.

The 2-point function (or the Green's function) in the Euclidean space is defined as \cite{Maldacena:2016hyu}
\begin{align}\label{gfsyk}
    G_{ij}(\tau)\hspace{1mm}\equiv\hspace{1mm}<T\psi_i(\tau)\psi_j(0)>\hspace{1mm}=\hspace{1mm}<\psi_i(\tau)\psi_j(0)>\theta(\tau)-<\psi_i(0)\psi_j(\tau)>\theta(-\tau),
\end{align}
where $\theta(\tau)$ denotes the step function and the expectation value $<...>$ can be obtained by taking the derivatives of the partition function ($Z[\mathcal{J}]$) with respect to the source field $\mathcal{J}_i$. The partition function is given by  
\begin{align}
    Z\left[\mathcal{J}\right]=\int D\psi_1D\psi_2...\psi_N\exp^{-\int d\tau L+\int d\tau\psi_i\mathcal{J}_i},
\end{align}
where the 2-point function is defined as 
\begin{align}\label{gfsykf}
    G_{ij}(\tau)=\frac{\delta}{\delta \mathcal{J}_i(\tau)}\frac{\delta}{\delta \mathcal{J}_j(0)}\ln\left(Z\left[\mathcal{J}\right]\right)\Big|_{\mathcal{J}=0}.
\end{align}

The 2-point function for free fermions is straightforward to obtain, and can be obtained using (\ref{lagSYK}) and (\ref{gfsykf})
\begin{align}\label{f2psykintro}
    G_{ij}^{(0)}(\tau)=\frac{1}{2}\text{sgn}(\tau)\delta_{ij}\hspace{1mm},\hspace{2mm}G_{ij}^{(0)}(\omega)=-\frac{1}{i\omega}\delta_{ij},
\end{align}
where `0' denotes the Green's function in the absence of any interactions, and $\text{sgn}(\tau)$ represents the signum function. Notice that, the second expression in (\ref{f2psykintro}) denotes the 2-point function in the frequency ($\omega$) space.

\begin{figure}[htp]
\centering
\includegraphics[scale=.7]{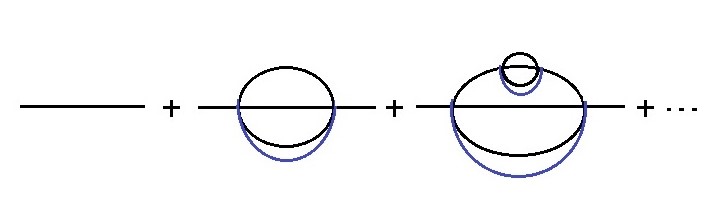}
\caption{Pictorial representation of the corrections in the 2-point function where blue line denotes the disorder averaging.} 
\label{introloop}
\end{figure}

Now, we compute the Green's function in the presence of quartic interactions. After performing the ``disorder average'', one can schematically represent the leading-order diagrams for the 2-point function as shown in Figure (\ref{introloop}). 

Summing over all the diagrams as in Figure (\ref{introloop}), we obtain the full 2-point function (G$(\omega)$) in the frequency space
\begin{align}\label{intgull2p}
   \frac{1}{ G(\omega)}=-i\omega-\Sigma(\omega),
\end{align}
where $\Sigma(\omega)$ represents the self-energy containing all the one-particle-irreducible (1PI) diagrams and can be expressed as \cite{Maldacena:2016hyu}
\begin{align}\label{int2g0}
    \Sigma(\tau_1,\tau_2)=J^2G(\tau_1,\tau_2)^3.
\end{align}

In the strong coupling limit (or low energy regime), one can drop the first term in (\ref{intgull2p}) which leads to the following expression
\begin{align}\label{int2ga}
    \int d\tau'G(\tau,\tau')\Sigma(\tau',\tau'')=-\delta(\tau-\tau'').
\end{align}

Having the expression for the Green's function in the above form (\ref{int2ga}), it becomes quite evident that the system exhibits an emergent conformal symmetry in the deep infrared (IR). In other words, the expression (\ref{int2ga}) remains invariant under the transformation $\tau\rightarrow f(\tau)$, provided that 
\begin{align}\label{gtranssykint}
    G(\tau,\tau')=\Big|f'(\tau)f'(\tau')\Big|^\frac{1}{4} G(f(\tau),f(\tau')).
\end{align}

Using (\ref{int2g0}) and (\ref{int2ga}), one can write down the final form of the full 2-point function as \cite{Maldacena:2016hyu}
\begin{align}\label{gfullsykint}
    G(\tau)=\left(\frac{1}{4\pi J^2}\right)^{\frac{1}{4}}\frac{\text{sgn}(\tau)}{\sqrt{|\tau|}}.
\end{align}

At the first place, it might appear that the model exhibits a complete Virasoro symmetry group.  However, the explicit form of the full Green's function (\ref{gfullsykint}) is found to be invariant only under the $SL(2,\mathbb{R})$ group \cite{Maldacena:2016hyu}. Therefore, the system is considered to be nearly conformal and is referred to as $nCFT_1$.

Now, we shift gears and present an alternative derivation that leads to an effective action for the SYK model (\ref{lagSYK}), which in turn provides a crucial hint for the $nAdS_2/nCFT_1$ correspondence. To derive an effective action for the model, we consider $M$ copies of fermion path integral and perform the disorder average \cite{Maldacena:2016hyu}

\begin{align}\label{replicapartion}
    \overline{Z^M}=&\hspace{1mm}\int D\psi_i^{\alpha}DJ_{ijkl}\exp\left(-\frac{N^3}{12J^2}J^2_{ijkl}\right)\times\exp\Bigg[-\sum_{\alpha=1}^{M}\int d\tau\Bigg(\frac{1}{2}\sum_i\psi_i^{\alpha}\frac{d}{d\tau}\psi_i^{\alpha}\nonumber\\
    &\hspace{1mm}-\frac{1}{4!}\sum_{ijkl}J_{ijkl}\psi_i^{\alpha}\psi_j^{\alpha}\psi_k^{\alpha}\psi_l^{\alpha}\Bigg)\Bigg],
\end{align}
where the index ``$\alpha$" represents the copies of the path integral. Notice that one could easily integrate out the coupling $J_{ijkl}$ to obtain 
\begin{align}\label{replicapartion1}
    \overline{Z^M}=&\int D\psi_i^{\alpha}\exp\Bigg[-\frac{1}{2}\Bigg(\sum_{\alpha=1}^M\sum_{i=1}^N\int d\tau \psi_i^{\alpha}\frac{d}{d\tau}\psi_i^{\alpha}-\nonumber\\
    &\frac{J^2N}{4}\sum_{\alpha\beta}\int d\tau_1d\tau_2\left(\sum_{i=1}^N\frac{1}{N}\psi_i^{\alpha}(\tau_1)\psi_i^{\beta}(\tau_2)\right)^4\Bigg)\Bigg],
\end{align}
where the last term in the integral (\ref{replicapartion1}) is a bilocal function of the variables $\tau_1$ and $\tau_2$.

The next step involves inserting the bilocal fields \cite{Maldacena:2016hyu}
\begin{align}
   \Tilde{G}^{\alpha\beta}(\tau_1,\tau_2)=\frac{1}{N}\sum_{i=1}^N\psi_{i}^{\alpha}(\tau_1)\psi_{i}^{\beta}(\tau_2), 
\end{align}
and the Lagrange multiplier ($\Tilde{\Sigma}^{\alpha\beta}(\tau_1,\tau_2)$) into the disorder average partition function (\ref{replicapartion1}).

Integrating out the fermionic degrees of freedom, one finds
\begin{align}
     \overline{Z^M}=\int D\Tilde{G}D\Tilde{\Sigma}\exp\left(-M S_{\text{eff}}\right),
\end{align}
where we have used the symmetric condition (i.e. $\Tilde{G}^{\alpha\beta}=\delta^{\alpha\beta}\Tilde{G}$) and the effective action ($S_{\text{eff}}$) is given by 
\begin{align}\label{sykinteff}
   S_{\text{eff}}=-\frac{N}{2}\log\text{det}\left(\partial_{\tau}-\Tilde{\Sigma}\right) +\frac{1}{2}\int d\tau_1d\tau_2\Bigg(N\Tilde{\Sigma}(\tau_1,\tau_2)\Tilde{G}(\tau_1,\tau_2)-\frac{J^2N}{4}\Tilde{G}(\tau_1,\tau_2)^4\Bigg).
\end{align}

It is important to notice that the previous results for the self-energy (\ref{int2g0}) and the full 2-point function (\ref{intgull2p}) can be obtained by varying the effective action (\ref{sykinteff}) with respect to $\Tilde{G}(\tau_1,\tau_2)$ and $\Tilde{\Sigma}(\tau_1,\tau_2)$, respectively. Furthermore, $S_{\text{eff}}$ (\ref{sykinteff}) remains invariant under the reparametrization $\tau_1\rightarrow f(\tau_1)$ and $\tau_2\rightarrow f(\tau_2)$ provided the Green's function transforms as in (\ref{gtranssykint}). 

Our next task is to expand the effective action (\ref{sykinteff}) around its saddle points ($G$ (\ref{intgull2p}) and $\Sigma$ (\ref{int2g0})) and obtain the leading order corrections due to the fluctuations. We propose following expansions
\begin{align}\label{introflut}
\Tilde{G}=G+|G|^{-1}g\hspace{1mm},\hspace{2mm}\Tilde{\Sigma}=\Sigma+|G|\sigma,
\end{align}
where $g$ and $\sigma$ denote the fluctuations.

Using (\ref{introflut}), the effective action (\ref{sykinteff}) can be schematically expressed as 
\begin{align}\label{sfluctintra}
   S^*_{\text{eff}}=&\hspace{1mm}-\frac{N}{12J^2}\int d\tau_1d\tau_2d\tau_3d\tau_4\sigma(\tau_1,\tau_2)\Tilde{K}(\tau_1,\tau_2,\tau_3,\tau_4)\sigma(\tau_3,\tau_4)+\nonumber\\
    &\hspace{5mm}\frac{N}{2}\int d\tau_1d\tau_2\left(g(\tau_1,\tau_2)\sigma(\tau_1,\tau_2)-\frac{1}{2}J^2g(\tau_1,\tau_2)^2\right),
\end{align}
where $*$ denotes the leading order corrections to the effective action (\ref{sykinteff}) and $\Tilde{K}(\tau_1,\tau_2,\tau_3,\tau_4)$ represents the symmetric kernel \cite{Maldacena:2016hyu}
\begin{align}\label{kerneldefcori}
    \Tilde{K}(\tau_1,\tau_2,\tau_3,\tau_4)=-3J^2|G(\tau_1,\tau_2)|G(\tau_1,\tau_3)G(\tau_2,\tau_4)|G(\tau_3,\tau_4)|
\end{align}
which appears in the computation of 4-point function. 

After integrating out $\sigma$, we obtain
\begin{align}\label{ises1}
 \frac{S^*_{\text{eff}}}{N}=\frac{3J^2}{4}g.\left(\Tilde{K}^{-1}-1\right)g,   
\end{align}
where we use the shorthand notation 
\begin{align}
    g.\mathcal{F}(\Tilde{K})g=\int d\tau_1d\tau_2d\tau_3d\tau_4g(\tau_1,\tau_2)\mathcal{F}(\Tilde{K}(\tau_1,\tau_2,\tau_3,\tau_4))g(\tau_3,\tau_4),
\end{align}
together with the fact $\mathcal{F}(\Tilde{K})=\left(\Tilde{K}^{-1}-1\right)$.

It is worth mentioning that the above expression for the effective action (\ref{ises1}) is valid at low energy. However, in the deep IR limit, when we use the 2-point function (\ref{gfullsykint}) to compute the kernel ($\Tilde{K}$) and hence the effective action (\ref{ises1}), we find that the effective action turns out to be zero. In other words, the fluctuation ($g$) acts as the eigenfunction of the kernel $\Tilde{K}$ with a unit eigenvalue. Therefore, one should include the first order non-conformal contributions \cite{Maldacena:2016hyu} in the eigen values of the kernel to obtain a finite value for the effective action. 

In order to evaluate a finite value for the effective action, we perform a small reparametrization, i.e., $\tau\rightarrow \tau+\epsilon(\tau)$, and compute the corrections (see \cite{Maldacena:2016hyu} for the detailed calculations) due to the reparametrization in the 2-point function (\ref{gfullsykint}) and the kernel $\Tilde{K}$ (\ref{kerneldefcori}). After incorporating these corrections, the effective action (\ref{ises1}) in the frequency space for $\epsilon$ turns out to be proportional to $n^2(n-1)$, where $n$ stands for the Matsubara frequency. In the position space, the effective action can be expressed as 
\begin{align}\label{iactlocal}
    \frac{S^*_{\text{eff}}}{N}\sim\int d\tau\left(\epsilon''^2-\left(\frac{2\pi}{\beta}\right)^2\epsilon'^2\right).
\end{align}

Finally, one can further generalize the result (\ref{iactlocal}) by considering the finite reparametrization i.e. $\tau\rightarrow f(\tau)$, which yields \cite{Maldacena:2016hyu}
\begin{align}\label{sykintrosch}
    S^*_{\text{eff}}\sim\int d\tau\text{Sch}(f, \tau),
\end{align}
where $\text{Sch}(f, \tau)$ denotes the Schwarzian derivative (\ref{introschjt}) which is invariant under $SL(2,\mathbb{R})$ transformation. Remarkably, the effective action of the SYK model (\ref{sykintrosch}) in the near IR limit is similar to that of JT gravity (\ref{jtintrosch}), providing a strong evidence in favour of the JT/SYK correspondence.

Another intriguing feature of the SYK model is that it is maximally chaotic \cite{Maldacena:2015waa}-\cite{Jensen:2016cah}. The chaos of the system is quantified in terms of the  Lyapunov exponent ($\lambda_L$) which is derived using the four-point out of time ordered correlation functions. More precisely, the measure of chaos can be estimated using a suitable combination of Hermitian operators $V(0)$ and $W(t)$, which are separated in time by an amount $t$. This can be schematically expressed as \cite{Maldacena:2015waa}
\begin{align}\label{introf1}
    F(t)=\text{tr}[yV(0)yW(t)yV(0)yW(t)],
\end{align}
where the quantity $y$ is given by 
\begin{align}
    y=\left(\frac{1}{Z}\exp\left(-\beta H\right)\right)^{\frac{1}{4}},
\end{align}
where $Z$ denotes the partition function, $H$ stands for the Hamiltonian and $\beta$ represents the inverse temperature. 

Large $N$ Conformal field theories that are holographically described by Einstein's theory of gravity, allow us to compute the above function (\ref{introf1}) using the methods discussed in \cite{Shenker:2013pqa}-\cite{Shenker:2014cwa}, which yields
\begin{align}\label{introf2}
    F(t)= F_0-\frac{F_1}{N^2}\exp\left(\frac{2\pi}{\beta}t\right)+O(N^{-4}),
\end{align}
where $F_0$ and $F_1$ are the constants which are determined by the specific choices of the operators $V(0)$ and $W(t)$. 

Using (\ref{introf2}), one can readily calculate the dissipation time\footnote{The dissipation time is defined as the time at which the expectation values of the 2-point function (for example $<V(0)V(t)>$) decays to the exponential of its initial value \cite{Maldacena:2015waa}. } ($t_d$) and the scrambling time\footnote{The scrambling time is defined as the time at which the function $F(t)$ attains a significant value.} ($t_*$) for the system which is given by
\begin{align}
    t_*=\frac{\beta}{2\pi}\log(N^2)\hspace{2mm}\text{and}\hspace{2mm}t_d\sim\beta.  
\end{align}

Furthermore, the authors in \cite{Maldacena:2015waa} conjectured that thermal quantum systems with a large number of degrees of freedom cannot be more chaotic than the result obtained in (\ref{introf2}). In such cases, the function ($F(t)$) in the time interval $t_d < t < t_*$ can be approximated as follows
\begin{align}\label{corad1}
    F_d\sim tr[y^2Vy^2V]tr[y^2Wy^2W]
\end{align}
where the above expression (\ref{corad1}) represents the product of two disconnected correlation functions. 

Additionally, the authors further argue that the rate at which $F(t)-F_d$ can grow is bounded by the following expression
\begin{align}\label{choint1}
    \frac{d}{dt}\left(F_d-F(t)\right)\leq\frac{2\pi}{\beta}\left(F_d-F(t)\right).
\end{align}

According to Kitaev \cite{ktt1}, the correlator (\ref{corad1}) of a maximally chaotic system (for example the SYK model) is expected to grow as 
\begin{align}   \label{choint2}
F_d-F(t)\sim\exp(\lambda_Lt),
\end{align}
where $\lambda_L$ is the  Lyapunov exponent.

On comparing the above results (\ref{choint1}) and (\ref{choint2}), we obtain the following bound on the Lyapunov exponent
\begin{align}
    \lambda_L\leq\frac{2\pi}{\beta}=2\pi T,
\end{align}
where $T$ denotes the temperature of the system.

This shows that the Lyapunov exponent estimated in the context of the SYK model bears a remarkable similarity to the Lyapunov exponent for the black hole phase in Einstein gravity \cite{Shenker:2013pqa}-\cite{Shenker:2014cwa}. To summarise, combining all the above facts together, we now have a growing evidence behind the $nAdS_2/nCFT_1$ correspondence where the JT gravity in 2D acts as a dual counterpart of the SYK model in the large $N\rightarrow \infty$ and strong coupling $ J>>1$ limit.

\section{JT/SYK duality: A literature survey}\label{sectionlr}

In this section, we briefly explore the various interesting generalizations of JT/SYK duality. An immediate extension of JT gravity was investigated by the authors in \cite{Almheiri:2014cka}, known as the Almheiri-Polchinski (AP) model. In particular, the authors in \cite{Almheiri:2014cka} examined the asymptotic structure, holographic properties, and back-reaction problem in the generalized 2D dilaton gravity, whose Lagrangian was defined as follows
\begin{align}\label{apls1}
    L=\frac{1}{16\pi G}\sqrt{-g}\left(\Phi^2R+\lambda(\nabla \Phi)^2-U(\Phi)\right),
\end{align}
where $\lambda$ was the parameter, $\Phi$ denoted the dilaton, $R$ denoted the Ricci scalar, and $U(\Phi)$ represented the potential. These authors considered a special case ($\lambda=0$ and $U(\Phi)=2(1-\Phi^2$)), where the system could be solved exactly in the classical regime.

Their analysis revealed that the solutions of (\ref{apls1}) interpolated to $AdS_2$ space-time in the deep IR limit and approached conformal Lifshitz geometry in the asymptotic UV limit, which regulated the back-reaction. Furthermore, they investigated the response of the geometry to the infalling matter pulse of energy $E$. Finally, they computed the holographic correlation function of the model and found that the back-reaction broke the conformal invariance of the 4-point function.

In the literature, the AP model (\ref{apls1}) had been extensively explored by the authors in \cite{Kyono:2017pxs}-\cite{Lala:2019inz}. In particular, the authors in \cite{Kyono:2017jtc} explored the Yang-Baxter (YB) deformation \cite{Klimcik:2002zj}-\cite{Matsumoto:2015jja} of the AP model (\ref{apls1}). It was found that the YB deformed $AdS_2$ metric was supported as a consistent solution of the AP model if one deformed the dilaton potential $U(\Phi)$ from a simple polynomial to a hyperbolic function, analogous to the integrable deformations. Furthermore, they computed the YB deformed black hole solutions and investigated their thermal properties.

Their calculations suggested that the Hawking temperature of the YB deformed black hole remained unaffected however, Bekenstein-Hawking (BH) entropy changed significantly. In addition, they found that the BH entropy of the YB deformed black hole exactly matched with the entropy computed using the renormalized boundary stress-energy tensor. Finally, they showed that all the results of the YB deformed case were consistent with the standard AP model (\ref{apls1}) in the absence of deformations.

Moreover, the effects of YB deformations on the spectrum of the SYK model in the strong coupling regime were also explored by the authors in \cite{Lala:2018yib}, using the notion of $AdS/SYK$ holography. In particular, these authors uplifted the YB deformed AP model into 3D and computed the Kaluza-Klein spectrum of a scalar field in the YB deformed background.

Their findings indicated that the effects of YB deformation on the spectrum of the SYK model appeared at the order $\sim\frac{1}{J^2}$ in the expansion. Additionally, they interpreted the YB deformation in the context of collective field excitations pertaining to the SYK model. Finally, they calculated the effective action of the theory up to the quadratic order in the fluctuations around the infrared points and obtained the associated correlation function up to the order $\sim\frac{1}{J^2}$.

The charged version of the AP model (\ref{apls1}) was constructed by the authors in \cite{Lala:2019inz}, where they included the interaction of the dilaton $(\Phi)$ with the $U(1)$ gauge fields ($L_{int}$) of the following form,
\begin{align}\label{adls1}
    L_{int}\sim Z(\Phi)F_{\mu\nu}F^{\mu\nu}.
\end{align}
In particular, they explored the asymptotic structure and phase stability of the vacuum and black hole solutions for two special choices of the coupling, namely, $Z(\Phi)\sim(\Phi^2)^2$ and $Z(\Phi)\sim e^{-\Phi^2}$.

Their findings revealed that the $U(1)$ gauge fields substantially changed the asymptotic geometry of both vacuum and black hole solutions in a similar way for both choices of couplings. More precisely, the vacuum solution approached $AdS_2$ in the deep IR limit and Lifshitz$_2$ with a dynamical exponent $z=\frac{3}{2}$ in the UV limit. On the other hand, the black hole solution asymptoted to Lifshitz$_2$ geometry with a dynamical exponent $z=\frac{3}{2}$. Finally, they computed the thermodynamic observables, i.e., free energy and entropy, associated with the 2D black holes and ruled out the possibility of any phase transition between vacuum and black hole solutions in both cases.

The boundary theory dual to the $U(1)$ gauge-deformed AP model \cite{Lala:2019inz} has not been completely understood so far. However, in \cite{Davison:2016ngz}, the authors constructed a SYK model\footnote{For the supersymmetric SYK model, refer to the reference \cite{Fu:2016vas}. } that exhibited the global $U(1)$ gauge symmetry, corresponding to the conserved charge. To be precise, these authors studied a SYK model composed of $N$ complex (or Dirac) fermions, also known as the complex SYK (cSYK) model. They constructed the effective Schwarzian action corresponding to the cSYK model and computed its various thermodynamic properties.  

The authors in \cite{Gross:2016kjj} further generalized the SYK model by adding flavors to the fermions, denoted as $\chi_i^f$, where $f$ denoted the flavor and $i=1,..,N_a$ denoted the site, with $a=1,..,f$. The Hamiltonian of the system could be schematically expressed as follows

\begin{align}\label{flals1}
H = \sum_{I} J_I \left(\chi_{i_1}^1...\chi_{i_{q1}}^1\right)...\left(\chi_{j_1}^f...\chi_{j_{qf}}^f\right),
\end{align}

where $I$ and $J_I$ respectively denoted the collective site index and the couplings.

They computed the 2-point function for the generalized SYK (gSYK) model (\ref{flals1}) using the Schwinger-Dyson equation and obtained the IR fixed point. Furthermore, they computed the spectrum of the gSYK and showed that it had the global $O(N_1) \times O(N_2) \times .. \times O(N_f)$ symmetry that appeared after performing the disorder average. Finally, they demonstrated that the conformal symmetry was broken in the 4-point function, and the system (gSYK) exhibited maximal chaotic behavior.

The gravitational theory dual to the complex SYK model \cite{Davison:2016ngz} in the low-energy sector was developed by the authors in \cite{Gaikwad:2018dfc}. Particularly, they considered 3D Einstein gravity with a negative cosmological constant, along with the presence of a $U(1)$ Chern-Simons (CS) term and a coupling between the CS term and bulk gravity. They obtained a 2D theory of gravity using suitable dimensional reduction (Kaluza-Klein reduction) and demonstrated that the effective action for 2D gravity was identical to the effective Schwarzian action corresponding to the cSYK model in the low-energy sector \cite{Davison:2016ngz}.

The holographic properties of $U(1)$ gauged charged JT gravity were elegantly investigated by the authors in \cite{Castro:2008ms}. In particular, they studied a model of 2D gravity coupled with $U(1)$ gauge fields, the action of which was defined as
\begin{align}\label{cls1}
    I_{bulk}\sim\int_{\mathcal{M}} d^2x\sqrt{-g}\left(e^{-2\phi}\left(R+\frac{8}{L^2}\right)-\frac{L^2}{4}F^2\right),
\end{align}
where $R$ represented the Ricci scalar, $F^2$ represented the field strength of $U(1)$ gauge fields, $\phi$ denoted the dilaton and $L$ denoted the $AdS$ length scale.
They systematically renormalized the action given in (\ref{cls1}) at the boundary and found that a mass term for a $U(1)$ gauge field
\begin{align}\label{cls2}
    I_{mass}\sim \int_{\partial\mathcal{M}}dx\sqrt{-\gamma}A^aA_a
\end{align}
was required to cancel the boundary divergences. However, they also demonstrated that the unusual mass term (\ref{cls2}) remained invariant under the $U(1)$ gauge transformations, where $A_{\mu}\rightarrow A_{\mu}+\partial_{\mu}\lambda$ that preserved the boundary conditions.

These authors computed the boundary stress-energy tensor and examined its transformation properties under the combined action of diffeomorphism and $U(1)$ gauge transformations. Consequently, they obtained the central charge associated with the 1D boundary theory, which agreed with the result of Hartman and Strominger \cite{Hartman:2008dq}.

Furthermore, they showed that the central charge obtained from the transformation properties of the boundary stress-energy tensor for 2D gravity (\ref{cls1}) was consistent with the dimensional reduction of the Brown-Henneaux central charge \cite{Brown:1986nw} for 3D gravity. Finally, these authors examined the thermal properties of 2D black holes and, in particular, computed the Bekenstein-Hawking entropy. Interestingly, they found that this entropy precisely matched with the ground state entropy for the 2D CFT obtained from the Cardy formula \cite{Cardy:1986ie}.

A similar calculation was also performed by the authors in \cite{Alishahiha:2008tv}, where they considered the action (\ref{cls1}) and found that the trace of the stress-energy near the black hole horizon vanished, indicating the presence of another CFT near the horizon. They computed the central charge associated with the near horizon CFT and showed that it precisely matched with the central charge of the CFT living at the boundary of space-time.

As another example, these authors \cite{Alishahiha:2008tv} also studied a case in which the dilaton ($e^{\phi}$) was directly coupled with the $U(1)$ gauge field ($A_{\mu}$), and its action was given by 
\begin{align}\label{alishals1}
    I_{bulk}\sim\int_{\mathcal{M}} d^2x\sqrt{-g}e^{\phi}\left(R+\frac{2}{L^2}-\frac{L^2}{4}e^{2\phi}F^2\right).
\end{align}

The above action (\ref{alishals1}) could be obtained using a suitable dimensional reduction of Einstein's gravity in three dimensions coupled with a negative cosmological constant. Finally, these authors pointed out that the black hole entropy, computed using the central charge and the Cardy formula \cite{Cardy:1986ie}, exactly matched with the Bekenstein-Hawking entropy for both cases (\ref{cls1}) and (\ref{alishals1}), providing direct evidence of the $AdS_2/CFT_1$ duality.

A detailed holographic dictionary for the action (\ref{alishals1}) was developed by the authors in \cite{Cvetic:2016eiv}. In particular, they computed the vacuum and black hole solutions of (\ref{alishals1}) for both running and constant dilaton. For both cases, they holographically renormalized the action and obtained the exact renormalized 1-point function. Furthermore, they uplifted the 2D solutions to the 3D, which turned out to be the general solutions of 3D Einstein-Hilbert action coupled with negative cosmological constant. Interestingly, these authors found that the 2D holographic dictionary obtained above exactly matched the holographic dictionary for $AdS_3/CFT_2$. Finally, they computed the asymptotic symmetries and obtained the associated conserved charges. Their analysis revealed that, in the case of a running dilaton, the conserved charges were both electric charge and mass. On the other hand, for a constant dilaton, the non-vanishing conserved entity was only the electric charge.

In addition to the vacuum and black hole solutions, the authors in \cite{Maldacena:2018lmt}-\cite{Zhang:2020szi} extensively studied the wormhole solutions and the associated phase stability within the framework of the JT/SYK duality. More precisely, the authors in \cite{Maldacena:2018lmt} examined a nearly $AdS_2$ space-time geometry that depicted the static (or eternal) traversable wormholes. The word ``traversable" meant that both boundaries of $AdS_2$ space-time were connected to each other. This interaction between the two boundaries was obtained by adding the following boundary interaction
\begin{align}
S_{int} = g\sum_{i=1}^N \int du O^i_L(u)O^i_R(u),
\end{align}
where $g$ denoted the coupling constant, $u$ denoted the boundary time, $O^i$ represented the bulk field operators having the conformal dimension $\Delta$, which was computed at the boundary, and $L/R$ denoted the left/right boundaries. Therefore, the bulk contained $N$ matter fields, which were required to generate the negative null energy in the bulk.

In addition, these authors studied a two-site coupled SYK model composed of Majorana fermions, and its Hamiltonian ($H$) was defined as \cite{Maldacena:2018lmt} 
\begin{align}
    H=H_L+H_R+H_{int},
\end{align}
where $H_{L}$ and $H_R$ represented two copies of the usual Majorana SYK Hamiltonian. Furthermore, $H_{int}$ denoted the interaction between the two sites, and was given by
\begin{align}
    H_{int}=i\mu\sum_{i}\psi^i_L\psi^i_R,
\end{align}
where $\mu$ represented the coupling constant. For a sufficiently small coupling ($\mu \ll 1$) and in the low-energy regime, they observed that the system possessed a constant gapped phase, which was identified with the traversable wormhole solution of JT gravity coupled with matter fields. Interestingly, these authors found that the two-site coupled SYK model exhibited a Hawking-Page transition from thermal radiation at low temperatures to the black hole phase at high temperatures.

On the other hand, the authors in \cite{Garcia-Garcia:2020ttf} studied a two-site uncoupled SYK model composed of Majorana fermions with complex couplings. They calculated the spectrum of the two-site SYK model and investigated its thermodynamic properties. Interestingly, they found that the ensemble average of the free energy over the complex couplings showed a constant gapped phase at sufficiently low temperatures, indicating the presence of a wormhole phase.

Next, these authors proposed a gravitational theory dual to a two-site uncoupled SYK model, which was a JT gravity coupled with a massless scalar field, and its action was given by
\begin{align}\label{garcials1}
    S=S_{JT}+\frac{1}{2}\int d^2x\sqrt{-g}\left(\partial\chi\right)^2,
\end{align}
where $S_{JT}$ represented the action for JT gravity (\ref{jtads2}), and $\chi$ denoted the massless scalar field. They found that the solution to the action (\ref{garcials1}) turned out to be a Euclidean wormhole, characterized by the double trumpet geometry without coupling between the boundaries. Furthermore, this geometry was consistent only when one switched on an imaginary source for the scalar field $\chi$. Finally, they computed the free energy of the system and showed that it underwent a phase transition from the two disconnected black hole phase at high temperature to a connected wormhole phase at low temperature, which was consistent with the observations on the SYK side.

Finally, the authors in \cite{Garcia-Garcia:2020vyr} investigated the thermal properties of a two-site coupled SYK model composed of Dirac fermions, and its Hamiltonian $(H^{c})$ was defined as
\begin{align}\label{garcompls1}
 H^{c}=H^{c}_L+H^{c}_R+H^{c}_{int},
\end{align}
where $H^{c}_{L}$ and $H^{c}_R$ denoted the two copies of the complex SYK Hamiltonian. Furthermore, $H^{c}_{int}$ represented the interaction between the two sites, which was given by
\begin{align}
    H^c_{int}=-\frac{1}{2}\sum_{i}\left(\eta\psi^{i\dagger}_L\psi^{i}_R+\eta^*\psi^{i\dagger}_R\psi^{i}_L\right),
\end{align}
where $\eta$ denoted the complex coupling, and its modulus was defined as $|\eta|=\kappa$. Using the above Hamiltonian (\ref{garcompls1}), they computed the Schwinger-Dyson equation for Green function and self-energy in the large $N$ limit and obtained the grand potential of the system.

In the low-temperature regime, they found that the grand potential of the system was almost constant for sufficiently small values of coupling ($\kappa$) and chemical potential ($\mu$), indicating the existence of a wormhole phase. In this limit, the charge of the system was also found to be nearly zero. As the system's temperature increased, it was likely to undergo a first-order phase transition into a charged wormhole phase for a particular range of $\kappa$ and $\mu$. As the temperature continued to increase, the charged wormhole further transitioned into a system of two charged black holes through another first-order phase transition. In addition, a discontinuous change in the charge of the system was observed at the critical points where both first-order phase transitions occurred. Finally, these authors derived the low-energy effective action (or Schwarzian action) of the theory and showed that the results obtained from the Schwarzian action were consistent with those computed from the Schwinger-Dyson equation in the large $N$ limit.

At present, we have enough evidence to support the idea that the SYK model could be a holographic dual to JT gravity in the IR regime. However, the authors in \cite{Saad:2019lba} proposed a radical idea and demonstrated that JT gravity is dual to a random ensemble of quantum mechanical systems, also known as the matrix model\footnote{See \cite{Stanford:2019vob}-\cite{Johnson:2021rsh} and references therein for supersymmetric generalizations of JT gravity and the matrix model. }, rather than a specific quantum system. 
In particular, these authors computed the (Euclidean) partition function of JT gravity on 2D surfaces that could have any number of Schwarzian boundaries and an arbitrary number of genus. Their calculations suggested that these partition functions could be directly mapped to the genus expansion of a matrix integral,
\begin{align}
    \mathcal{Z}=\int dH e^{-N TrV(H)},
\end{align}
where $H$ was an $N\times N$ Hermitian matrix that represented the Hamiltonian corresponding to the boundary theory, and $V(H)$ denoted the potential function. Furthermore, to compare the JT gravity results with the matrix model, one needed to take the limit as $N$ goes to infinity. Such matrix integrals were also referred to as double-scaled matrix integrals \cite{Brezin:1990rb}-\cite{Gross:1989vs}.

This matrix model approach provided a non-perturbative (and non-unique) completion of the genus expansion, and it was found to be useful in understanding the ``plateau" region in the spectral form factor. However, this model encountered significant instabilities in the low-energy regime, the resolution of which led to non-unique physics. To rectify this and achieve a stable, unique, and non-perturbative completion of physics in the low-energy domain, an alternative approach was formulated by the authors in \cite{Johnson:2019eik}. They achieved this by constructing a different model using a complex matrix integral \cite{Morris:1990cq}-\cite{Dalley:1991vr}.

The author in \cite{Witten:2020wvy} further investigated the duality between JT gravity and the matrix integral model by introducing a deformation to the JT gravity theory
\begin{align}\label{witls1}
    I_{deformed}\sim\int d^2x\sqrt{-g}\left(\phi(R+2)+U(\phi)\right),
\end{align}
where the function $U(\phi)$ was treated as the perturbation over the pure JT gravity. In particular, the authors considered various forms of the function $U(\phi)$, for instance
\begin{align}
U(\phi)=2\sum_{i=1}^{r}\epsilon_ie^{-\alpha_i\phi},
\end{align}
where $\pi<\alpha_i<2\pi$ represented the deficit angle, $\epsilon_i$ denoted the amplitude, and they computed the density of eigenvalues associated with the dual matrix model. The phase transition and non-perturbative effects in the context of deformed JT gravity (\ref{witls1}) were investigated by the authors in \cite{Johnson:2020lns}.

Finally, the models of 2D dilaton gravity have also proven valuable in investigating the Information Paradox \cite{Hawking:1975vcx}. In particular, the authors in \cite{Almheiri:2019hni}-\cite{Almheiri:2019psf} studied the JT gravity model coupled with a 2D $CFT$ (that is holographically dual to $AdS_3$), and the action could be schematically expressed as follows
\begin{align}\label{lsip1}
    I(g_{\mu\nu}^{(2)},\phi,\chi)= I_{JT}(g_{\mu\nu}^{(2)},\phi)+I_{CFT_2}(g_{\mu\nu}^{(2)},\chi),
\end{align}
where the first term in (\ref{lsip1}) denoted the action for standard JT gravity, and the second term represented the action for $CFT_2$. Furthermore, here $g_{\mu\nu}^{(2)}$ denoted the 2D space-time metric, $\phi$ represented the dilaton, and $\chi$ collectively represented all the matter fields pertaining to the $CFT_2$.

It is important to notice that the $CFT_2$ was defined on the fixed space-time background $g_{\mu\nu}^{(2)}$, which was related to the three-dimensional space-time metric ($g_{\mu\nu}^{(3)}$) as follows
\begin{align}
    g_{\mu\nu}^{(3)}\Big|_{boundary}=\frac{1}{\epsilon^2}g_{\mu\nu}^{(2)},
\end{align}
where $\epsilon$ represented the UV cut-off.

The authors in \cite{Almheiri:2019hni} considered a two-dimensional evaporating black hole. In particular, they computed the entropy associated with the Hawking radiation and the black hole using the Ryu-Takayanagi (RT) prescription \cite{Ryu:2006bv}. Their analysis revealed that the entropy of the radiation initially increased, and after crossing the Page time \cite{Page:1993wv}-\cite{Page:2013dx}, it started decreasing. This observation was consistent with the unitary evolution of a quantum mechanical system.

The computation of the entropy of Hawking radiation from the RT formula \cite{Ryu:2006bv} was further justified by the authors in \cite{Almheiri:2019qdq}-\cite{Penington:2019kki}. They achieved this by introducing the concept of replica wormholes within the framework of JT gravity coupled with conformal matter. Moreover, the authors in \cite{Penington:2019kki} pointed out that the geometries of wormholes could also be interpreted as some sort of ensemble average in the gravitational theory.

Recently, the authors in \cite{Geng:2022slq} found that JT gravity could also be realized in the framework of the Karch-Randall (KR) braneworld \cite{Karch:2000ct}-\cite{Karch:2000gx}. To be specific, the authors in \cite{Geng:2022slq} studied a wedge holography\footnote{See the recent work \cite{Yadav:2023sdg} for the Yang-Baxter deformed version of Wedge holography.} \cite{Akal:2020wfl}-\cite{Miao:2020oey} setup that involved two KR $AdS_2$ branes within the bulk of $AdS_3$. These KR branes intersected each other at the conformal boundary of the $AdS_3$ space-time. They considered the KR branes to be fluctuating and showed that the effective theory of fluctuations in the low energy limit was precisely JT gravity. Additionally, the authors in \cite{Bhattacharya:2023drv} calculated the holographic complexity of JT gravity within the context of KR braneworld using the ``complexity equals volume" proposal \cite{Susskind:2014moa}-\cite{Alishahiha:2015rta}. They found that complexity exhibited a linear dependence on the boundary time during late times, which was consistent with the computation in \cite{Brown:2018bms}.

\section{Theme of the thesis}\label{introthm}
Given a brief literature survey of the JT/SYK duality, this thesis serves a dual purpose. First, we explore the novel Hawking-Page transition \cite{Hawking:1982dh}-\cite{Witten:1998zw} and the wormhole to  black hole phase transition \cite{Maldacena:2018lmt}-\cite{Zhang:2020szi} within the framework of JT gravity coupled with non-trivial gauge couplings. Second, we investigate the transformation properties of the (boundary) stress-energy tensor under the combined action of the diffeomorphism and the $U(1)$ gauge transformation, aiming to calculate the central charge associated with the 1D boundary theory. In the forth-coming subsections, we will delve into greater detail regarding the Hawking-Page transition and the central charge of the 1D boundary theory, which holds a pivotal role in the rest of this thesis.   

\subsection{Hawking-Page transition}

After the groundbreaking discovery that the dynamics of black holes had a strong resemblance to the thermodynamics of ordinary matter \cite{Bardeen:1973gs}, the authors in \cite{Hawking:1982dh}-\cite{Witten:1998zw} immediately investigated the thermal properties of black holes in $AdS$ space-time. In particular, the authors in \cite{Hawking:1982dh} constructed the Euclidean action that possessed a Schwarzschild $AdS$ solution in four dimensions\footnote{For higher dimensions, see the reference \cite{Belhaj:2015hha}} and studied these solutions in the canonical ensemble.

Their analysis revealed the existence of a critical temperature, $T_0=\frac{1}{2\pi}\sqrt{-\Lambda}$, where $\Lambda$ represented a negative cosmological constant below which the free energy of the system was negative without a black hole, and the system was dominated by thermal radiation. As the temperature of the system increased, two types of black holes (with lower and higher masses) could exist in equilibrium with the thermal radiation. The black hole with lower mass had a negative heat capacity, $\mathcal{C}=\frac{dE}{dT}<0$, therefore it was unstable and would transition either into pure thermal radiation or a larger mass black hole. However, the lower mass black hole also had a positive free energy, which suggested that the chance of this state occurring was less than that of pure thermal radiation.

On the other hand, the heat capacity and free energy of a black hole with higher mass turned out to be positive in the temperature range $T_0<T<T_1$, where $T_1=\frac{1}{\pi}\sqrt{-\frac{\Lambda}{3}}$. Moreover, its free energy changed sign and became less than that of pure thermal radiation if the temperature of the system was greater than or equal to $T_1$. Therefore, the black hole with higher mass was locally stable. On further increasing the temperature of the system, i.e., $T>T_2\sim \left(-m_p^2\Lambda\right)^{\frac{1}{4}}$, where $m_p$ was the Planck mass, it was found that the presence of a black hole was necessary to achieve global stable equilibrium. Collectively, all these features were called the Hawking-Page (HP) transition in the literature. In the boundary gauge theory, this novel HP transition was interpreted as the confinement/deconfinement transition \cite{Witten:1998zw}.

Just after the proposal of the novel Hawking-Page transition from $AdS$ thermal radiation to the black hole phase, the authors in \cite{Chamblin:1999tk}-\cite{Dehghani:2020jcw} readily explored the thermal properties of charged and rotating black holes in various space-time dimensions. In particular, the authors in \cite{Chamblin:1999tk} obtained the Einstein-Hilbert action in 4D and 5D coupled with $U(1)$ gauge fields and a Chern-Simons term using suitable dimensional reduction from the higher-dimensional gauge super-gravity theory. They further computed the associated charged black hole solutions and studied their phase stability. They found that the thermodynamic phase structures associated with these models were analogous to the van der Waals Maxwell liquid-gas system. Moreover, they further showed that these results were consistent with the dual boundary theory.

Similar calculations were also performed by the authors in \cite{Banerjee:2011raa}, where they studied charged (Reissner-Nordström) and rotating (Kerr) $AdS$ black holes in the grand canonical ensemble. They found that the second-order derivatives of the Gibbs potential exhibited discontinuity at critical points, and the system obeyed the Ehrenfest relations, indicating the onset of a second-order phase transition from smaller mass to larger mass black holes. 

The phase stability of 5D Schwarzschild AdS black holes in the grand canonical ensemble was investigated by the authors in \cite{Zhang:2014uoa}. In this case, the chemical potential of the system was defined using the number of colors associated with the dual boundary gauge theory. Their calculations suggested that the chemical potential for the stable black hole phase was always negative, but it changed sign for the unstable phase. Similar results were also reported by the authors in \cite{Maity:2015ida} for Kerr-AdS black holes in four and five space-time dimensions.

The HP transitions in higher derivative theories were investigated by the authors in \cite{Su:2019gby}-\cite{Wang:2020pmb}. In particular, these authors explored the thermal properties of $AdS$ black holes in Gauss-Bonnet (GB) gravity and Einstein-Gauss-Bonnet (EGB) gravity \cite{Glavan:2019inb} in the extended phase space \cite{Kastor:2009wy}-\cite{Cvetic:2010jb}. In the extended phase space, the negative cosmological constant was considered as a new variable and treated as an effective thermodynamic pressure. EGB gravity was a generalization of the usual GB gravity, obtained by rescaling the GB coupling constant $\alpha\rightarrow\frac{\alpha}{d-4}$ in $d$ space-time dimensions and finally taking the limit $d\rightarrow 4$. In both cases, the higher derivative terms were found to reduce the HP transition temperature, and the HP transition in EGB gravity occurred only within a certain range of pressure.

Moreover, the Hawking-Page (HP) transition has been studied in massive gravity \cite{Cai:2014znn}-\cite{Li:2020khm}. In particular, the authors in \cite{Cai:2014znn} studied the phase transition of ($n+2$)-dimensional charged black holes in massive gravity \cite{Vegh:2013sk} in both the canonical and grand canonical ensembles. They found that in the grand canonical ensemble, a phase transition in four dimensions could occur only if the combination $k-\mu^2/4+c_2m^2$ is positive, where $\mu$ is the chemical potential, $c_2m^2$ is the coefficient that appeared in the potential in massive gravity, and the constant $k$ characterized the curvature. Furthermore, the first-order phase transition between smaller and larger mass black holes could occur even in the absence of charge if $n+2 \geq 5$. 

The authors in \cite{Zhang:2015wna} explored the phase stability of 3D black holes (Banados-Titelboim-Zanelli) in a new class of massive gravity\footnote{Also, see \cite{Detournay:2015ysa} for the HP phase transition in 3D topological massive gravity.} \cite{Bergshoeff:2009hq}-\cite{Bergshoeff:2009aq}, which notably incorporates higher curvature terms. Their investigation revealed the presence of a critical temperature, below which the thermal soliton with a mass of $M=-1$ became more probable. Conversely, above the critical temperature, the system was dominated by the black hole phase with a mass of $M\geq0$.

Recently, the authors in \cite{Witten:2020ert}-\cite{Sahoo:2020unu} explored the Hawking-Page transition in the context of JT gravity. In particular, the author in \cite{Witten:2020ert} considered JT gravity with an arbitrary dilaton potential $(W(\phi))$ such that $W(\phi)$ reduced to $2\phi$ at spatial infinity. It was discovered that there existed a set of black holes, one with a negative heat capacity and another with a positive heat capacity and lower free energy. As the temperature increased, the energy and entropy of the system encountered a discontinuity, indicating the onset of a first-order phase transition, which was identical to the HP phase transition. The authors in \cite{Forste:2021roo} further investigated the HP transition in deformed JT gravity and established its connection with the double-scaled matrix model.

The authors in \cite{Cao:2021upq} further examined the phase structure of the complex SYK model and a charged black hole in deformed JT gravity. Interestingly, they discovered that the phase diagram of a charged black hole was identical to the van der Waals Maxwell liquid-gas system. Furthermore, in the limit where the charge approached zero, this behavior reduced to the standard Hawking-Page (HP) phase transition.

Despite the extensive study of Hawking-Page (HP) transitions in various classes of gravitational theories, as discussed above, the HP transitions have been explored the least within the framework of 2D gravity. Therefore, in the first part of this thesis, we aim to address some of the remaining gaps in the literature.

In this regard, our initial focus is on investigating the asymptotic structure and phase stability of vacuum, which serves as the thermal radiation background, and 2D black holes in the context of JT gravity coupled with $U(1)$ gauge fields and $SU(2)$ Yang-Mills fields, as outlined in \cite{Lala:2020lge}. More precisely, we compute thermal observables such as free energy and Wald entropy related to thermal radiation and 2D black holes, and explore their dependencies on the Hawking temperature.

We found that at a certain critical temperature ($T=T_0$), the effective free energy and effective entropy of the system undergo a discontinuity, indicating the onset of the Hawking-Page (HP) transition between thermal radiation and 2D black holes. Our analysis on phase stability suggests that the presence of $SU(2)$ Yang-Mills fields plays a crucial role in obtaining the HP transitions in 2D gravity.

After investigating the Hawking-Page transition in the 2D gravity model, we proceed to explore another type of phase transition known as the wormhole to black hole phase transition within the framework of JT gravity coupled with $U(1)$ gauge fields and a Chern-Simons (CS) term, as discussed in \cite{Rathi:2021mla}. It is important to mention that the CS term is necessary to obtain a consistent wormhole solution at finite chemical potential.

We investigate the thermal properties of both the wormhole and black hole phases. To be specific, we calculate the total charge and the free energy density associated with both the wormhole and black hole solutions in two dimensions. Interestingly, we find that the free energy density and the total charge of the system exhibit a discontinuity at a critical point ($T=T_0$), indicating the phase transition from the wormhole phase at lower temperatures ($T<T_0$) to the black hole phase at higher temperatures ($T>T_0$).

\subsection{Central charge}
In this section, we present the systematic derivation of the central charge\footnote{The derivation of the central charge can be found in \cite{Kiritsis:2019npv}-\cite{Polchinski:1998rq}. } associated with the Conformal Field Theory living on the boundary, which will play a crucial role in the second part of the thesis.

To begin with, first, we define the conformal transformations and briefly review classical conformal group algebra.  Conformal transformations are the transformations that transform the space-time metric ($g_{\mu\nu}(x)$) up to an overall factor, $\Omega(x)$ under the general coordinate transformation \cite{Dbook1}
\begin{align}
    g_{\mu\nu}(x)\rightarrow g'_{\mu\nu}(x')=\Omega(x)g_{\mu\nu}(x).
\end{align}

Now, we study the generators associated with conformal transformation. Under infinitesimal coordinate transformation i.e. $x^{\mu}\rightarrow x'^{\mu}=x^{\mu}+\epsilon^{\mu}$, the space-time metric changes as 
\begin{align}\label{icca1}
    \delta g_{\mu\nu}=-(\partial^{\lambda}\epsilon_{\mu}g_{\lambda\nu}+\partial^{\lambda}\epsilon_{\nu}g_{\lambda\mu})-\epsilon^{\lambda}\partial_{\lambda}g_{\mu\nu}.
\end{align}

In order for these transformations (\ref{icca1}) to be conformal, the parameter $\epsilon$ must satisfy the following constraint
\begin{align}\label{icca2}
\partial_{\mu}\epsilon_{\nu}+\partial_{\nu}\epsilon_{\mu}=\frac{2}{d}(\partial\cdot\epsilon)\delta_{\mu\nu},
\end{align}
where $d$ is the space-time dimension, and we set the space-time metric ($g_{\mu\nu}$) to be flat.

We are interested in a particular case of $d=2$, and consider the space-time metric in Euclidean signature. For this choice, the above constraint (\ref{icca2}) simplifies to
\begin{align}\label{icca3}
\partial_1\epsilon_1=\partial_2\epsilon_2\hspace{1mm},\hspace{2mm}\partial_1\epsilon_2=-\partial_2\epsilon_1.
\end{align}

One can further simplify the analysis by working in complex coordinates, i.e., $z=x_1+ix_2$ and $\overline{z}=x_1-ix_2$. In this coordinate system, equations (\ref{icca3}) takes the following form
\begin{align} \label{icca4}
\partial\overline{\epsilon}=0\hspace{1mm},\hspace{2mm}\overline{\partial}\epsilon=0,
\end{align}
where we introduce parameters as $\epsilon=\epsilon_1+i\epsilon_2$, $\overline{\epsilon}=\epsilon_1-i\epsilon_2$, and denote the derivatives as $\overline{\partial}=\partial_{\overline{z}}$.

Notice that the constraint (\ref{icca4}) suggests that the parameter ($\epsilon$) can depend only on $z$ and not on its complex conjugate $\overline{z}$ and vice versa. In other words, the constraint still holds under the following coordinate transformations
\begin{align}
z\rightarrow f(z)\hspace{1mm},\hspace{2mm}\overline{z}\rightarrow\overline{f}(\overline{z}).
\end{align}

Next, we expand the parameter $\epsilon$ in terms of $z$ as follows
\begin{align}\label{icalg}
\epsilon(z)=-\sum a_nz^{n+1},
\end{align}
where $l_n=-z^{n+1}\partial_z$ are the generators associated with these transformations.

One can easily check that the above generators ($l_n$) satisfy the following commutation relations
\begin{align}\label{introclassl}
[l_m,l_n]=(m-n)l_{m+n}\hspace{1mm}, \hspace{2mm}[\overline{l}_m,\overline{l}n]=(m-n)\overline{l}{m+n},
\end{align}
and $[\overline{l}_m,l_n]=0$. This algebra suggests that the 2D conformal group is infinite-dimensional \cite{Kiritsis:2019npv}. However, the only sub-algebra formed by the generators $l_{0,\pm1}$ and $\overline{l}_{0,\pm1}$ is well defined on the Riemann sphere. Therefore, it is sufficient to consider only this sub-algebra and it has the representation $SL(2,\mathbb{C})$. Moreover,  these generators $l_{-1}$, $l_0$, $l_{1}$ respectively generate the translation, scaling, and special conformal transformation and the transformation generated by these generators $l_{0,\pm1}$ can be expressed as
\begin{align}\label{introcontr}
    z\rightarrow f(z)=\frac{az+b}{cz+d},
\end{align}
where $a,b,c,d$ $\epsilon $ $\mathbb{C}$ and satisfy the relation, $ad-bc=1$.

It is important to note that under the transformations (\ref{introcontr}), the field $\Phi(z,\overline{z})$ of a conformally invariant field theory transforms as
\begin{align}
\Phi(z,\overline{z})=\left(\frac{\partial f}{\partial z}\right)^{\Delta}\left(\frac{\partial \overline{f}}{\partial \overline{z}}\right)^{\overline{\Delta}} \Phi'(f(z),\overline{f}(\overline{z})),
\end{align}
where ($\Delta$, $\overline{\Delta}$) are the conformal weights associated with the field $\Phi$.

Our next task is to investigate the Hilbert space of the 2D conformal field theory whose coordinates are denoted by $\tau$ and $\sigma$. We begin by compactifying the spatial coordinate $\sigma=\sigma+2\pi$ to avoid IR problems. This compactification converts $\tau$ and $\sigma$ plane into a 2D cylinder. Next, we map the cylinder into the complex plane using the following conformal transformations \cite{Kiritsis:2019npv}
\begin{align}
z=e^{\tau+i\sigma}\hspace{1mm},\hspace{2mm} \overline{z}=e^{\tau-i\sigma}.
\end{align}

Recall that generator $l_0$ generates the scaling $z\rightarrow \lambda z$, therefore the generator $l_0+\overline{l}_0$ translates the time coordinate on the cylinder. In other words, the Hamiltonian ($H$) of the system can be expressed as 
\begin{align}
    H=l_0+\overline{l}_0.
\end{align}

Furthermore, the trace of the stress-energy tensor of a CFT is zero in flat space-time. In complex coordinates, the trace is denoted by $T_{z\overline{z}}=0$, where the other components of the stress-energy tensor $T_{zz}$ and $T_{\overline{z}\overline{z}}$ are non-vanishing. However, the conservation law, i.e., $\partial^{\mu}T_{\mu\nu}=0$, along with the condition $T_{z\overline{z}}=0$, gives
\begin{align}\label{intrccd1}
\partial_z T_{\overline{z}\overline{z}}=\partial_{\overline{z}}T_{zz}=0.
\end{align}

The above condition (\ref{intrccd1}) on the components of the stress-energy tensor implies that
\begin{align}
T(z)\equiv T_{zz}\hspace{1mm},\hspace{2mm} \overline{T}(\overline{z})\equiv T_{\overline{z}\overline{z}},
\end{align}
indicating the existence of conserved charges. If $T(z)$ is conserved, then the product $\epsilon(z) T(z)$ is also conserved. Therefore, the current associated with conserved charges is given by
\begin{align}\label{inccd1}
    Q_\epsilon=\frac{1}{2\pi i}\oint dz \epsilon(z) T(z)\hspace{1mm},\hspace{2mm}Q_{\overline{\epsilon}}=\frac{1}{2\pi i}\oint d\overline{z} \overline{\epsilon}(\overline{z}) \overline{T}(\overline{z}).
\end{align}

These conserved charges (\ref{inccd1}) could be interpreted as the generators associated with the infinitesimally small conformal transformations
\begin{align}\label{inccd2}
z\rightarrow z+\epsilon(z)\hspace{1mm},\hspace{2mm}\overline{z}\rightarrow \overline{z}+\overline{\epsilon}(\overline{z}).
\end{align}
The variation of the field $\Phi(z,\overline{z})$ under the above transformations (\ref{inccd2}) is usually expressed in terms of the commutator as follows
\begin{align}\label{intvarcd1}
\delta_{\epsilon,\overline{\epsilon}}=[ Q_\epsilon+Q_{\overline{\epsilon}},\Phi(z,\overline{z})].
\end{align}

Now we define another useful quantity in the CFT known as Operator Product Expansion (OPE). The OPE is the expansion of two radially product operators in terms of the local orthonormal operators as follows \cite{Kiritsis:2019npv}
\begin{align}
    \Phi_i(z,\overline{z})\Phi_j(w,\overline{w})=\sum_{k}C_{ijk}(z-w)^{\Delta_k-\Delta_i-\Delta_j}(\overline{z}-\overline{w})^{\overline{\Delta}_k-\overline{\Delta}_i-\overline{\Delta}_j}\Phi_k(w,\overline{w}),
\end{align}
where the $C_{ijk}$ are the constants which is related to the 3-point function  $<\Phi_i\Phi_j\Phi_k>$.

Next, we define the mode expansion of the stress-energy tensor,
\begin{align}\label{itinv}
    T(z)=\sum_{n\epsilon \mathbb{Z}} z^{-n-2}L_n\hspace{1mm},\hspace{2mm}  \overline{T}(\overline{z})=\sum_{n\epsilon \mathbb{Z}} \overline{z}^{-n-2}\overline{L}_n,
\end{align}
where $L_n$ are the generators. One can obtain $L_n$ by inverting above the expressions (\ref{itinv}),
\begin{align}\label{intln}
    L_n=\frac{1}{2\pi i}\oint dz z^{n+1}T(z)\hspace{1mm},\hspace{2mm} \overline{L}_n=\frac{1}{2\pi i}\oint d\overline{z} \overline{z}^{n+1}\overline{T}(\overline{z}).
\end{align}

With all the important definitions in hand, we can now compute the OPE of two stress-energy tensors ($T(z)$), which can be expressed as \cite{Kiritsis:2019npv}
\begin{align}\label{iccd1}
T(z)T(w)=\frac{c}{2(z-w)^4}+2\frac{T(w)}{(z-w)^2}+\frac{\partial T(w)}{z-w}+...
\end{align}
Here, we have utilized the fact that the conformal weights of $T(z)$ and $\overline{T}(\overline{z})$ are $(2,0)$ and $(0,2)$, respectively. A similar OPE can also be computed for the stress-tensor $\overline{T}(\overline{z})$ by replacing $z$ with $\overline{z}$ and $c$ with $\overline{c}$. The constant `$c$' that appears in (\ref{iccd1}) is called the left central charge, and `$\overline{c}$' is called the right central charge of the CFT. Furthermore, $c=\overline{c}$ for the diffeomorphism-invariant theory. 

Now, one can work out the commutation relation between the two generators $L_n$ using the OPE between the two stress-energy tensors and the following identity
\begin{align}\label{introidenti}
\left[\int d\sigma B,A\right]=\oint dz B(z)A(w).
\end{align}
Here, $A(w)$ and $B(z)$ are the two operators.

Using (\ref{intln}), (\ref{iccd1}), and (\ref{introidenti}), we find
\begin{align}
[L_n,L_m]=(n-m)L_{n+m}+\frac{c}{12}(n^3-n)\delta_{n+m,0},
\end{align}
where $c$ is the usual central charge associated with the CFT. In literature, this algebra is known as the Virasoro algebra and it further reduces to the classical algebra (\ref{introclassl}) in the classical limit $c=0$.

Next, we consider the conformal transformation $z\rightarrow z+\epsilon(z)$ and compute the variation of the stress-energy tensor under these transformations on an arbitrary Riemann surface. Furthermore, we tag the tensor indices to the stress-energy symbol ($T_{zz}$) for clarity.

Using (\ref{inccd1}), (\ref{intvarcd1}), and (\ref{introidenti}), the variation of the stress-energy tensor can be expressed as:
\begin{align}\label{intexp1}
\delta_{\epsilon}T_{zz}(z)=\frac{1}{2\pi i}\oint dw \epsilon(w)T_{ww}(w)T_{zz}(z).
\end{align}
Now, by substituting the OPE for two stress-energy tensors (\ref{iccd1}) into (\ref{intexp1}) and solving the integrals, the variation can be elegantly written as
\begin{align}\label{centralderint}
\delta_{\epsilon}T_{zz}(z)=\frac{c}{12}\partial^3_z\epsilon(z)+2\partial_z\epsilon(z)T_{zz}(z)+\epsilon(z)\partial_zT_{zz}(z),
\end{align}
where the constant `$c$' denotes the central charge.

Recall that the above derivation for the central charge (\ref{centralderint}) is valid only for 2D CFT. However, one can obtain the result for a 1D CFT by simply switching off one coordinate ($x_2$) in $z$ and identifying $x_1$ as $\tau$ or, more simply, by setting $z = \tau$.

Furthermore, it is important to notice that the last two terms in the expression (\ref{centralderint}) arise due to the fact that $T_{zz}$ transforms as a tensor of weight $(2,0)$. On the other hand, the first term turns out to be proportional to the central charge, which indicates that conformal symmetry is breaking in curved space-time. We identify this as the conformal anomaly associated with the CFT in a non-flat space-time. Moreover, one can check that the trace of the stress-energy tensor in curved space-time is non-vanishing i.e. $T^{\mu}_{\mu}\sim c R^{(2)}$, where $c$ is the central charge, and $R^{(2)}$ is the Ricci scalar in 2D.

The transformation properties of the boundary stress-energy tensor under the combined action of diffeomorphism and a $U(1)$ gauge transformation, as well as the computation of the central charge in the context of Maxwell-dilaton gravity, were extensively explored by the authors in \cite{Castro:2008ms, Hartman:2008dq}. 

Following a similar approach, in the second part of the thesis, we aim to further investigate the effects of non-trivial gauge interactions on the transformation properties of the boundary stress-energy tensor and, consequently, calculate the central charge associated with the 1D boundary theory. In particular, we utilize the expression (\ref{centralderint}) to calculate the central charge of the boundary theory in the presence of quartic couplings \cite{Rathi:2021aaw} and ModMax interactions \cite{Rathi:2023vhw}.

In this regard, we initially compute the holographic stress-energy tensor and central charge associated with the boundary theory whose bulk part contains all possible 2-derivative as well as 4-derivative interactions between gravity and the $U(1)$ gauge fields. We find that the central charge is largely dominated by the 4-derivative interactions in the theory. Moreover, it grows as a negative power of the coupling associated with the 4-derivative interactions. Our analysis suggests that these results are highly non-perturbative, which means that one cannot simply switch off the higher derivative interactions to obtain the results of \cite{Castro:2008ms, Hartman:2008dq}.

Next, we investigate the effects of 2D projected ModMax interactions \cite{Rathi:2023vhw} on the boundary theory observables, particularly the central charge. Our analysis reveals that ModMax corrections ($\sim O(\gamma\kappa)$) appear with a negative sign, and $U(1)$ gauge corrections ($\sim O(\kappa^2$)) appear with a positive sign in the central charge. However, we treat the ModMax parameter ($\gamma$) as sufficiently small, such that the overall central charge turns out to be a positive quantity.

\section{Organisation of the thesis } 
In this section, we describe the organization of the thesis. Broadly, this thesis contains two parts, which are based on the papers \cite{Lala:2020lge, Rathi:2021mla,Rathi:2021aaw,Rathi:2023vhw}. The first part comprises two chapters, namely Chapters 2 and 3. In Chapter 2, we investigate the phase stability of 2D black holes in the presence of SU(2) Yang-Mills fields \cite{Lala:2020lge}, and in Chapter 3, we explore the wormhole to black hole phase transition at finite charge density (or chemical potential) \cite{Rathi:2021mla}. The second part of the thesis also contains two chapters, i.e., Chapters 4 and 5. Chapter 4 delves into the study of holographic properties of 2D Einstein-dilaton gravity in the presence of quartic couplings \cite{Rathi:2021aaw}, and in Chapter 5, we examine the holographic properties of $AdS_2$ gravity coupled with `ModMax' interactions \cite{Rathi:2023vhw}. Finally, we conclude the thesis with some intriguing future directions in Chapter 6.

Below, we briefly discuss the details of each chapters.

\subsection{Chapter 2: Jackiw-Teitelboim gravity and the models of a Hawking-Page transition for 2D black holes}\label{HP}
In this chapter, we study the Jackiw-Teitelboim (JT) gravity coupled with $U(1)$ gauge fields and $SU(2)$ Yang-Mills fields that reveals an analogue of the Hawking-Page \cite{Hawking:1982dh} transition in 2D theories of gravity \cite{Lala:2020lge}. We construct the 2D theory following a compactification of Einstein's gravity in 5D \cite{Fan:2015aia} accompanied by Abelian gauge fields as well as $SU(2)$ Yang-Mills fields. We obtain the vacuum and black hole solutions associated with the 2D gravity using the perturbative method, where we treat the coupling constants of the $U(1)$ gauge fields ($\xi$) and $SU(2)$ Yang-Mills fields ($\kappa$) as  expansion parameters.

We show that the vacuum structure of the 2D theory is dominated by the Lifshitz$_2$ in the UV and $AdS_2$ in the deep IR limit. On the other hand, the 2D charged black hole solution asymptotes to Lifshitz$_2$ geometry. Our analysis on thermal stability reveals the existence of first order phase transition at the critical point $(T\sim T_0)$. At the low temperatures ($T<T_0$), the system is dominated by the thermal radiation. On the other hand, as we increase the temperature of the system and cross the threshold value ($T\sim T_2>T_0$), the system is dominated by the stable black hole phase. We interpret all these features as the 2D analog of the Hawking-Page transition for higher dimensional black holes.

\subsection{Chapter 3: Phases of Euclidean wormholes in JT gravity}\label{WH}
In this chapter, we cook up a theory of Einstein-dilaton gravity in two dimensions \cite{Rathi:2021mla} that exhibits the Euclidean wormhole to black hole phase transition in the presence of $U(1)$ gauge fields \cite{VanMechelen:2019ebr}. The low temperature phase of the system is identified as the ``charged wormhole''. On the other hand, as we increase the temperature of the system, it transits into two ``charged black hole'' system via first order phase transition. At the transition temperature $(T=T_0)$, we identify that the Free energy (density) and the total charge of the system change discontinuously. Furthermore, we conjecture that the boundary theory dual to this gravitational set up could be identified as a two-site uncoupled complex Sachdev-Ye-Kitaev (cSYK) model with non-vanishing chemcial potential \cite{Garcia-Garcia:2020vyr}-\cite{Zhang:2020szi}.

\subsection{Chapter 4: Holographic JT gravity with quartic couplings}
In this chapter, we formulate a most general theory of gravity coupled with  $U(1)$ gauge fields that includes all possible 2-derivative as well as 4-derivative interactions allowed by the diffeomorphism invariance in two dimensions \cite{Rathi:2021aaw}. We obtain the vacuum and black hole solutions for the 2D theory pertubatively upto linear order in the couplings associated with the 2-derivative $(\xi)$ and 4-derivative $(\kappa)$ interactions.

The vacuum solution asymptotes to Lifshitz$_2$ geometry with dynamical exponent \big($z=\frac{7}{3}$\big) in the UV limit. On the other hand, it diverges in the deep IR limit due to the presence of non-trivial 4-derivative interactions in the theory. We further compute the holographic stress-energy tensor and obtain the central charge for the 1D boundary theory \cite{Castro:2008ms}. To our surprise, the central charge grows as the ``inverse'' power of the coupling associated with the 4-derivative interactions $(\kappa)$, which does not have smooth $\kappa\rightarrow0$ limit. 

We further compute the thermodynamical variables pertaining to the 2D black holes. In particular, we compute the Hawking temperature \cite{Hawking:1975vcx} for 2D black holes and the Wald entropy \cite{Wald:1993nt}-\cite{Brustein:2007jj} associated with it. Interestingly, the Wald entropy for 2D black holes diverges near horizon due to the presence of novel 4-derivative interactions $(\kappa)$. We interpret the near horizon divergences in terms of the density of states \cite{tHooft:1984kcu}-\cite{Solodukhin:2011gn}.

Finally, we study the near horizon dynamics of 2D black holes and in particular compute the central charge associated with the near horizon Conformal Field Theory (nCFT) \cite{Alishahiha:2008tv}. Using the proper field redefinition, we further demonstrate that the nCFT can be recast into a generalised 2D Liouville gravity (gLG) \cite{Grumiller:2007wb}-\cite{Jackiw:2005su} accompanied by the higher derivative interactions. We examine the Weyl invariance of gLG and pin down the associated Weyl anomaly. We also discuss the vacuum structure of the generalised Liouville gravity at classical level.

\subsection{Chapter 5: $AdS_2$ holography and ModMax}
In this chapter, we study the 2D theory of Einstein's gravity coupled with the $U(1)$ gauge fields and the ``projected ModMax'' interactions \cite{Rathi:2023vhw}. We obtain the 2D theory of gravity using a suitable dimensional reduction of 4D Einstein's gravity accompanied by the ModMax interactions \cite{Bandos:2020jsw}-\cite{Kosyakov:2020wxv}. We obtain the vacuum structure  and 2D black hole solutions using perturbative techniques \cite{Rathi:2023vhw}. We explore the transformation properties of the 1D boundary stress-energy tensor under both diffeomorphism and $U(1)$ gauge transformations and compute the associated holographic central charge . On top of it, we explore the thermal properties of ModMax corrected black holes in two dimensions and comment on the their extremal limits.

\subsection{Chapter 6: Conclusion and Outlook}
In this chapter, we provide a summary of chapters 2 to 5 and discuss various interesting possible extensions of the work \cite{Lala:2020lge}-\cite{Rathi:2023vhw}.

\includepdf[page=1]{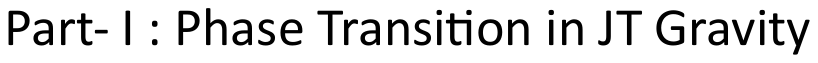}

\chapter{JT gravity and the models of Hawking-Page transition for 2D black holes}
\allowdisplaybreaks
\pagestyle{fancy}
\fancyhead[LE]{\emph{\leftmark}}
\fancyhead[LO]{\emph{\rightmark}}
\rhead{\thepage} 

In this chapter, we discuss a top down construction of Jackiw-Teitelboim (JT) gravity using the compactification of 5D gravitational theory in the presence of an Abelian ($ U(1) $) as well as $ SU(2) $ Yang-Mills (YM) fields. We compute the background solutions associated with the 2D model in the perturbative regime of the theory where the corrections have been estimated over the uncharged JT solutions while treating the gauge couplings as the parameters of the expansion. 

Our analysis reveals the existence of two classes of solutions namely, (i) the \emph{interpolating} vacuum solution with AdS$_2$ in the IR and Lifshitz$_2$ in the UV, which serves as the thermal radiation background for our analysis and (ii) the charged 2D black hole solution exhibiting Lifshitz$_2$ asymptotics. The analysis on thermal stability reveals the onset of a first order phase transition at $ T \sim T_0 $, such that for $ T < T_0 $ the only possible state is the thermal radiation background without any black hole. On the other hand, as the temperature is gradually increased beyond certain \emph{crossover} value $ T\sim T_2 (>T_0 )$, \emph{globally} stable black hole phase emerges indicating the onset of Hawking-Page (HP) transition in 2D gravity models.

\section{Overview and Motivation}

After the discovery of celebrated AdS/CFT correspondence \cite{Maldacena:1997re,Witten:1998qj}, several interesting examples of duality have been discovered as well as tested within the realm of strong/weak conjecture among which the very recent discovery of an interesting toy model of quantum holography \cite{ktt2,ktt3} is something worthy of praise. This proposal is based on the original work of \cite{Sachdev:1992fk} and therefore goes under the name of Sachdev-Ye-Kitaev (SYK) models \cite{Polchinski:2016xgd,Maldacena:2016hyu}. 

Ever since this proposal has been put forward, there have been growing interests in order to explore the Large N near infra-red (IR) dynamics in SYK models using its dual gravitational description(s) that lives in one higher dimension. The collective analyses unveil $(1+1)$D Jackiw-Teitelboim (JT) dilaton gravity as the dual model \cite{Teitelboim:1983ux,Jackiw:1984je} that has been conjectured. Putting all these pieces together, finally leads towards an emergent SYK/AdS$ _2 $ correspondence \cite{Jackiw:1984je}-\cite{Lala:2018yib}. These analyses have been subsequently extended for charged SYK models and their corresponding dual gravitational counterparts \cite{Lala:2019inz}, \cite{Davison:2016ngz}-\cite{Castro:2008ms}.

The purpose of the present chapter is to elaborate an example of dimensional compactification which finally leads towards $(1+1)$D Jackiw-Teitelboim (JT) dilaton gravity in the presence of non-trivial matter (gauge) couplings. The parent 5D model \cite{Fan:2015aia,Fan:2015yza}, that we choose to start with, was actually proposed in order to construct electrically charged space-time solutions with anisotropic Lifshitz scaling \cite{Kachru:2008yh,Taylor:2008tg} and its Lagrangian is given by
\begin{align}\label{newaddedexpchapt21}
    \mathcal{L}= \sqrt{-g}\left(\mathcal{R}-3\Lambda-\frac{1}{4g_{s}^{2}}
F^{a}_{MN}F^{aMN}-\frac{1}{4}F_{MN}F^{MN}\right)-\frac{\sigma}{2g_{s}^{2}}
\epsilon^{MNPQR}F^{a}_{MN}F^{a}_{PQ}A_{R},
\end{align}
where $\mathcal{R}$ is the Ricci scalar in five dimensions, $\Lambda$ is the negative cosmological constant, $\epsilon^{MNPQR}$ is the completely antisymmetric tensor. Furthermore, the Maxwell ($F_{MN}$) and the Yang-Mills ($F^a_{MN}$) field strengths are given by  
\begin{align}
    F_{MN}=\partial_MA_N-\partial_NA_M\hspace{1mm},\hspace{2mm} F^{a}_{MN}=\partial_MA^{a}_N-\partial_NA^{a}_M+\epsilon^{abc}A^{b}_MA^{c}_N.
\end{align}

The last term in the above expression (\ref{newaddedexpchapt21}) is known as the Chern-Simons (CS) term which is actually a supergravity-inspired ``$FFA$" type term \cite{Fan:2015aia, Romans:1985ps}. However, the dimensional reduction, on the other hand, results in some form of 2D gravity model where both the gravity as well as the dilation are found to be non minimally coupled with the abelian (U(1)) and the $SU(2)$ Yang-Mills (YM) sectors of the theory. \\\\
Below we enumerate some notable facts about the 2D model that is constructed in this chapter using a suitable dimensional reduction ansatz.\\ \\ 
$ \bullet $ The space-time geometry of the vacuum solution of the theory has been found to \emph{interpolate} between a Lifshitz$_2$ state in the UV and an AdS$_2$ state in the IR. Moreover, the vacuum solution also contains matter fields, namely Abelian ($U(1)$) as well as $SU(2)$ Yang-Mills (YM) fields, which contribute to the thermal radiation \cite{Hawking:1982dh}. Therefore, we identify the vacuum solution as the thermal radiation background for our subsequent analysis in the Euclidean formalism.\\\\
$ \bullet $ Thermal excitations have been identified as \emph{charged} black holes with Lifshitz$_2$ asymptotics. In the Euclidean framework, these black holes serve as the basis for what we call the analogue of Hawking-Page (HP) transition \cite{Hawking:1982dh}-\cite{Witten:1998zw}, \cite{Carlip:2003ne}-\cite{Myung:2007ti} in 2D gravity models.\\\\
$ \bullet $ Considering the Euclidean framework, we perform our analysis of thermal stability in black holes using a canonical ensemble. Our analysis reveals the existence of certain critical temperature, $ T \sim T_0 \sim \sqrt{\mu_0} $ such that for $ T> T_0 $ there exists two possible phases of black holes - one with lower mass and a negative heat capacity, and the other with higher mass and a positive heat capacity. As the temperature of the system is reduced below $ T_0 $, the black hole phase with negative heat capacity decays into pure thermal radiation via a \emph{first} order phase transition at $ T=T_0 $. On the other hand, as the temperature is increased above $ T_0 $, the lower mass black hole gradually transits into a globally stable phase of larger mass black hole that is in thermal equilibrium with its surrounding radiation background. During this course of transition, we observe several transition temperatures ($T_0< T_1< T_2$) and finally \emph{crossover} to a globally stable phase of black hole for $ T> T_2 $. We collectively identify all these features as being the analogue of HP transitions \cite{Hawking:1982dh} in ``charged" JT model whose holographic interpretation is yet to be unfolded.\\ \\
The organisation for the rest of the chapter is as follows. We construct our 2D model in Section \ref{model}, where we give a brief account for the dimensional reduction procedure and obtain the corresponding background solutions following a perturbative approach. Adopting to a Euclidean framework in Section \ref{phase}, we explore the thermal stability in black holes using a canonical ensemble. Finally, we conclude in Section \ref{conclusion}. 

\section{$D=2$ model}\label{model}
\subsection{A top down construction}
We consider $ D=5 $ gravity model minimally coupled to both abelian ($A_{M}$) as well as $SU(2)$ Yang-Mills (YM) fields ($A_{M}^{a}$ ($a=1,2,3$)) \cite{Fan:2015aia},
\begin{align}\label{act:5D}
S_{5D}=&\int d^{5}x \sqrt{-g}\Bigg(\mathcal{R}-3\Lambda-\frac{\kappa}{4g_{s}^{2}}
F^{a}_{MN}F^{aMN}-\frac{\xi}{4}F_{MN}F^{MN}-\nonumber\\
&\frac{\sigma}{2g_{2}^{2}}
\frac{\epsilon^{MNPQR}}{\sqrt{-g}}F^{a}_{MN}F^{a}_{PQ}A_{R}\Bigg)
\end{align}
where $ \Lambda (<0)$ is the cosmological constant\footnote{We set the AdS radius $L=1$ and $16\pi G=1$ in the subsequent analysis.}. Here  $\kappa$ and $\xi$ are the gauge coupling constants which would be treated as an expansion parameter in the subsequent analysis. The CS piece ($ \sim F^a \wedge F^a \wedge A $), on the other hand, is quite ubiquitous to $ D=5 $ theories and does not seem to have left with any of its imprints on the reduced $ D=2 $ model. 

The $ D=2 $ theory is obtained using the reduction ansatz \cite{Davison:2016ngz}
\begin{equation}\label{anz:red}
ds^{2}=\Phi^{\alpha}\;d\tilde{s}^{2}+\Phi^{\beta}dx_{i}^{2},\quad A_{M}^{a}dx^{M}
=A_{\mu}^{a}dx^{\mu}, \quad A_{M}dx^{M}=A_{\mu}dx^{\mu}~;~\alpha,\beta  \in \mathbb{R},
\end{equation}
together with the metric of the reduced spacetime,
\begin{equation}\label{met:2D}
d\tilde{s}^{2}=\tilde{g}_{\mu\nu}dx^{\mu}dx^{\nu}~; ~
g_{\mu\nu}=\Phi^{\alpha}\tilde{g}_{\mu\nu}~;~\mu , \nu= t, z.
\end{equation}

The $ D=2 $ action (modulo a total derivative) could be formally expressed as\footnote{The methodology of dimensional
reduction from $5D$ to $2D$ has been briefly discussed in Appendix \ref{dimred}.
}
\begin{eqnarray}\label{actsim:2D}
S_{2D}&=&\mathcal{V}_{3}\int d^{2}x\sqrt{-\tilde{g}}\;\Bigg(\tilde{\mathcal{R}}\Phi -V(\Phi)
-\Phi \mathcal{L}(\tilde{A}_{\mu},\tilde{A}^a_{\mu})\Bigg)+S_{\text{GH}}+S_{ct},       \nonumber\\
\mathcal{L}(\tilde{A}_{\mu},\tilde{A}^a_{\mu})&=&\frac{\kappa}{4g_{s}^{2}}
\tilde{F}^{a}_{\mu\nu}\tilde{F}^{a\mu\nu}+\frac{\xi}{4}\tilde{F}_{\mu\nu}\tilde{F}^{\mu\nu}~;
\quad V(\Phi)=3 \Lambda \Phi,
\end{eqnarray}
where, we set $\alpha=0$ and $\beta=2/3$ without any loss of generality. Notice that our model (\ref{actsim:2D}) is a special
case of \cite{Almheiri:2014cka} with $ C=0 $ and $ A=-3\Lambda $. Moreover, here $ S_{\text{GH}}=-\int dt \sqrt{-\gamma} 
\mathcal{K}\Phi $ is the standard Gibbons-Hawking (GH) term ($ \mathcal{K} $ being the extrinsic curvature associated with
the boundary hypersurface \cite{Gibbons:1976ue}) and $ S_{\text{ct}} $ is the so called counter term\footnote{In this chapter, we conduct our analysis in the bulk, so the boundary counter term would not affect the observables significantly. However, it plays a crucial role in Chapters 3-5, where we perform the boundary analysis and compute the observables pertaining to the 1D boundary theory.}  which cures the divergences of the on-shell
action near its asymptotic limits.
\subsection{Equations of motion}
The equations of motion that readily follow from the variation of the parent action (\ref{actsim:2D}) are listed below,
\begin{subequations}\label{eom:col}
\setlength{\jot}{8pt}
\begin{eqnarray}
\label{5a}
\left(\nabla_{\mu}\nabla_{\nu}-\tilde{g}_{\mu\nu}\Box\right)\Phi+\frac{\xi\Phi}{2}\left(
\tilde{F}_{\mu\rho}\tilde{F}_{\nu}^{\;\rho}-\frac{1}{4}\tilde{F}^{2}\tilde{g}_{\mu\nu}\right) \\\notag
\qquad\qquad\qquad\quad+\frac{\kappa\Phi}{2g_{s}^{2}}\left(\tilde{F}^{a}_{\mu\rho}
\tilde{F}_{\nu}^{a\rho}-\frac{1}{4}\tilde{F}^{a^2}\tilde{g}_{\mu\nu}\right)-\frac{3\Lambda}{2}
\Phi\tilde{g}_{\mu\nu}&=&0 \label{eom:guge}\\
\tilde{\mathcal{R}}-3\Lambda-\frac{\kappa}{4g_{s}^{2}}\tilde{F}^{a}_{\mu\nu}
\tilde{F}^{a\mu\nu}-\frac{\xi}{4}\tilde{F}_{\mu\nu}\tilde{F}^{\mu\nu}&=&0\label{eom:sclr}\\
\frac{1}{\sqrt{-\tilde{g}}}\partial_{\mu}\left(\sqrt{-\tilde{g}}\Phi\tilde{F}^{a\mu\nu}\right)
+\Phi\epsilon^{abc}\tilde{A}_{\mu}^{b}\tilde{F}^{c\mu\nu}&=&0\label{eom:nab}\\
\partial_{\mu}\left(\sqrt{-\tilde{g}}\Phi\tilde{F}^{\mu\nu}\right)&=&0. \label{eom:ab}
\end{eqnarray}
\end{subequations}

In order to proceed further, we choose to work with the static metric ansatz
\begin{equation}\label{met:ansz}
ds^{2}=e^{2\omega(z)}\left(-dt^{2}+dz^{2}\right)
\end{equation}
together with the ansatz for the gauge fields,
\begin{subequations}
\begin{align}
\tilde{A}_{\mu}&=\left(\tilde{A}_{t}(z),0\right)\label{ansz:ab}\\[10pt]
\tilde{A}_{\mu}^{a}&=\tilde{A}_{t}^{3}(z)\tau^{3}dt+\tilde{A}_{z}^{1}(z)\tau^{1}dz
\label{ansz:nab}
\end{align}
\end{subequations}
where $\tau^{a}=\sigma^{a}/2i$ are the Pauli matrices of $ SU(2) $ YM theory.

Using  (\ref{met:ansz}), (\ref{ansz:ab}) and (\ref{ansz:nab}) we finally note down the set of dynamical equations\footnote{In order to simplify our notations we set, $\tilde{A}_{t}^{3}=\chi(z)$ and $\tilde{A}_{z}^{1}=\eta(z)$.}, 
\begin{subequations}
\setlength{\jot}{8pt}
\begin{align}
2\Phi''+6\Lambda\Phi e^{2\omega}+\Phi e^{-2\omega}\left[\xi\left(\tilde{A}_{t}'\right)^{2}
+\frac{\kappa}{g_{s}^{2}}\left(\left(\chi'\right)^{2}+\chi^{2}\eta^{2}\right)\right]&=0 \label{nor:sclr1}\\
\Phi''-2\omega'\Phi'&=0 \label{nor:sclr2}\\
4\omega''+6\Lambda e^{2\omega}-e^{-2\omega}\left[\xi\left(\tilde{A}_{t}'\right)^{2}
+\frac{\kappa}{g_{s}^{2}}\left(\left(\chi'\right)^{2}+\chi^{2}\eta^{2}\right)\right]&=0\label{nor:guge}\\
\Phi \tilde{A}_{t}''+\Phi'\tilde{A}_{t}'-2\Phi\omega'\tilde{A}_{t}'&=0 \label{nor:ab}\\
\left(\Phi e^{-2\omega}\chi\eta\right)'+e^{-2\omega}\Phi\eta\chi'&=0 \label{nor:nab1}\\
\left(\Phi e^{-2\omega}\chi' \right)'-e^{-2\omega}\Phi\eta^{2}\chi&=0\label{nor:nab2}
\end{align}
\end{subequations}
together with the following constraint,
\begin{equation}\label{cnst:nab}
\Phi e^{-2\omega}\chi^{2}\eta =0.
\end{equation}
\subsection{Solving the dynamics}
We propose the following perturbative method of solving the dynamics\footnote{Here `$ab$' and `$na$' stand for respective perturbative corrections (to the uncharged JT solutions \cite{Almheiri:2014cka}) due to abelian and non-abelian sectors of the $ D=2 $ model (\ref{actsim:2D}).} (\ref{nor:sclr1})-(\ref{nor:nab2})
\begin{equation}\label{pert:exp}
\mathcal{A}(z)=\mathcal{A}_{(0)}(z)+\xi\mathcal{A}_{(1)}^{ab}(z)
+\kappa\mathcal{A}_{(1)}^{na}(z)+\cdots
\end{equation}
which is a perturbation in the gauge coupling parameters over the uncharged background solutions \cite{Almheiri:2014cka}. In the present analysis, we retain ourselves only upto leading order in the perturbative expansion (\ref{pert:exp}). The general strategy would be to substitute (\ref{pert:exp}) into (\ref{nor:sclr1})-(\ref{nor:nab2}) and obtain equations of motion at different order in the perturbative expansion\footnote{See Appendix \ref{apen:eom} for details.}. Here, $\mathcal{A}(z)$ stands for either of the dynamical variables $\Phi$, $\omega$, $\chi$ and $\eta$.\\ \\ 
Before we proceed further, the following observations are noteworthy to mention:\\ \\
$ \bullet $ Since we are interested in expanding the action (\ref{actsim:2D}) upto leading order in the gauge coupling, therefore it is sufficient to estimate leading/zeroth order solutions for both abelian and non abelian gauge fields.\\\\
$ \bullet $ The abelian sector (\ref{nor:ab}) could be further simplified in terms of other variables,
\begin{equation}
F_{zt}\equiv A_{t}'=\frac{Qe^{2\omega}}{\Phi}  \label{max:gen}
\end{equation}
where $Q$ is the corresponding $ U(1) $ charge. \\\\
$ \bullet $ The dilaton equation of motion (\ref{nor:sclr2}) could be recast as,
\begin{equation}\label{dilaton}
\Phi'(z)=-\frac{e^{2\omega}}{2}
\end{equation} 
whose solution may be expressed as,
\begin{equation}\label{dila:gen}
\Phi(z)\simeq -\int \frac{dz}{2}~  e^{2\omega}.
\end{equation}
\subsection{Interpolating vacuum}\label{sol:vac}
In order to find vacuum solutions, the first step would be to note down zeroth order solutions\footnote{See Appendix \ref{sol:zero} 
for the details of the derivation.} for both the dilaton and the metric from (\ref{dil:zero}) and (\ref{omga:zero}), 
 \begin{subequations}
\setlength{\jot}{5pt}
\begin{align}
e^{2\omega_{(0)}^{\text{vac}}}&=-\frac{2}{3\Lambda z^{2}}\label{met:vac}\\
\Phi_{(0)}^{\text{vac}}&=-\frac{1}{3\Lambda z}. \label{diln:vac}
\end{align}
\end{subequations}
Combining (\ref{met:vac}) and (\ref{met:ansz}) it is by now quite evident that the vacuum solution corresponds to AdS$_2$
spacetime in standard Poincare coordinates
\begin{align}
ds^{2}\sim z^{-2}(-dt^2+dz^2),
\end{align}
where we identify $z\to 0$ as the asymptotic UV limit of the bulk space-time. On the other hand, $z\to\infty$ stands for the IR limit 
\cite{Almheiri:2014cka}.

Next, we note down solutions corresponding to (\ref{zero:nab1}) and (\ref{zero:nab2}) 
\begin{align}
\chi_{(0)}^{\text{vac}}&\simeq \log z \label{zero:chi}\\
\eta_{(0)}^{\text{vac}}&\simeq \frac{1}{z\left(1+\log 
z\right)^{2}} \label{zero:eta}
\end{align}
where we have multiplied (\ref{zero:nab2}) by $\chi_{(0)}/\eta_{(0)}$ and used (\ref{cnst:0}). 

On the other hand, using (\ref{omga:xi}) and (\ref{omga:kapa}), the solutions corresponding to $\omega_{(1)}^{ab}$ and $\omega_{(1)}^{na}$ can
be found as,
\begin{subequations}
\setlength{\jot}{5pt}
\begin{align}
\left(\omega_{(1)}^{ab}\right)^{\text{vac}}&=10z^{2}+\frac{\mathsf{C}}{z}
+\frac{Q^{2}\Lambda}{6}z^{2}\left(1-3\log z\right),\label{sol:w1ab}\\
\left(\omega_{(1)}^{na}\right)^{\text{vac}}&=-\frac{3\Lambda}
{z}.\label{sol:w1na}
\end{align}
\end{subequations}

Finally, the leading order solutions for dilaton may be obtained from (\ref{dil:xi}) and (\ref{dil:kapa}) 
\begin{subequations}
\setlength{\jot}{5pt}
\begin{align}
\left(\Phi_{(1)}^{ab}\right)^{\text{vac}}&=\frac{-3\left(\mathsf{C} -20z^{3}
\right)+4Q^{2}\Lambda z^{3}\left(1-\frac{3}{4}\cdot\log z\right)}{9\Lambda
z^{2}}, \label{sol:p1ab}\\
\left(\Phi_{(1)}^{na}\right)^{\text{vac}}&\simeq\frac{1}{z^{2}}.
\label{sol:p1na}
\end{align}
\end{subequations}

Combining all these pieces together we find,
\begin{align}\label{met:vacu}
\begin{split}
ds^{2}_{\text{vac}}&=e^{2\omega^{\text{vac}}}\left(-dt^{2}+dz^{2}\right) \\[6pt]
&\approx e^{2\omega_{(0)}^{\text{vac}}}\left(1+2\xi\left(\omega_{(1)}^{ab}
\right)^{\text{vac}}+2\kappa\left(\omega_{(1)}^{na}\right)^{\text{vac}}\right)
\left(-dt^{2}+dz^{2}\right).
\end{split}
\end{align}

Given the above metric structure (\ref{met:vacu}), it is customary to explore various asymptotic limits associated to it. For example, near the IR ($z\rightarrow\infty$) region one finds,
\begin{align}\label{met:IR}
\begin{split}
e^{2\omega^{\text{vac}}}_{IR}
&\simeq -\frac{2}{3\Lambda z^{2}}-\frac{2}{9\Lambda}\left(60\xi+Q^{2}\xi
\Lambda\left(1-3\log z\right)\right)+\mathcal{O}\left(z^{-3}\right)
\end{split}
\end{align}
which clearly reveals an emerging AdS$_2$ geometry.

On the other hand, the UV ($z\rightarrow0$) limit of the metric reveals,
\begin{align}\label{met:UV}
\begin{split}
e^{2\omega^{\text{vac}}}_{UV}
&\simeq \frac{1}{z^{3}}\left(-\frac{4\xi\mathsf{C}}{3\Lambda}+4
\kappa\right)-\frac{2}{3\Lambda z^{2}}  \\
&\quad -\frac{2}{9\Lambda}\left(60\xi+Q^{2}\xi\Lambda\left(1-3\log z\right)
\right)+\mathcal{O}\left(z^{2}\right)
\end{split}
\end{align}
an emerging Lifshitz$ _2 $ geometry with dynamical exponent $z_{\text{dyn}}=\frac{3}{2}$. A careful analysis further reveals that for $\xi=\kappa=0$, the resulting geometry becomes AdS$_2$ in both the asymptotic limits. This confirms that gauge fields in the theory are actually responsible for the change in asymptotic (UV) structure as found earlier in \cite{Lala:2019inz,Taylor:2008tg}.
\subsection{ 2D black holes}
The zeroth-order/ uncharged background solutions $\omega_{(0)}$ and $\Phi_{(0)}$ may be expressed as,
\begin{subequations}
\setlength{\jot}{5pt}
\begin{align}
e^{2\omega_{(0)}^{\text{BH}}}&=-\frac{8\mu}{3\Lambda\sinh^{2}
\left(2\sqrt{\mu}z\right)}  \label{met:bh}\\
\Phi_{(0)}^{\text{BH}}&=-\frac{2\sqrt{\mu}}{3\Lambda}\coth\left(
2\sqrt{\mu}z\right). \label{diln:bh}
\end{align}
\end{subequations}

Below we note down first-order solutions corresponding to the metric and the dilaton. In order to simplify our analysis, we consider the following change of coordinate,
\begin{equation}\label{coor:new}
z=\frac{1}{2\sqrt{\mu}}\coth^{-1}\left(\frac{\rho}{\sqrt{\mu}}\right).
\end{equation}

Using (\ref{coor:new}), the solution corresponding to (\ref{omga:xi}) can be expressed as,
\begin{align}\label{omga1a:BH}
\begin{split}
\left(\omega_{(1)}^{ab}\right)^{\text{BH}}&\simeq\frac{\rho}{\sqrt{\mu}}
+\left[\frac{\rho}{2\sqrt{\mu}}\cdot\log\left(\frac{\sqrt{\mu}+\rho}{\sqrt{\mu}
-\rho}\right)-1\right]  \\
&\quad -\frac{3Q^{2}\Lambda}{16\mu^{3/2}}\left\{2\sqrt{\mu}\left(1+\log\rho\right)+\rho\log
\rho\cdot\log\left(\frac{\sqrt{\mu}-\rho}{\sqrt{\mu}+\rho}\right)\right.  \\
&\qquad\qquad\qquad\quad \left. +\rho\left[\text{Li}\left(2,\frac{\rho}{\sqrt{\mu}}\right)-
\text{Li}\left(2,\frac{-\rho}{\sqrt{\mu}}\right)\right] \right\}.
\end{split}
\end{align}

On the other hand, substituting (\ref{omga:kapa}) in (\ref{dil:kapa}) and using (\ref{coor:new}) we find,
\begin{align}\label{omga1na:BH}
\begin{split}
\left(\omega_{(1)}^{na}\right)^{\text{BH}}\simeq\frac{\sqrt{\mu}\rho
+\left(\mu-\rho^{2}\right)\tanh^{-1}\left(\frac{\rho}{\sqrt{\mu}}\right)}
{2(\mu-1)\mu^{3/2}}.
\end{split}
\end{align}

Finally, we note down zeroth order solutions for gauge fields that readily follow from (\ref{zero:nab1}) and (\ref{zero:nab2}),
\begin{subequations}
\setlength{\jot}{5pt}
\begin{align}
\chi_{(0)}^{\text{BH}}&\simeq1-\frac{1}{2\sqrt{\mu}}
\log\left(\frac{\rho}{\sqrt{\mu}}\right)  \label{zero:chibh}  \\
\eta_{(0)}^{\text{BH}}&\simeq\frac{\left(\rho^{2}-\mu\right)}{4\rho
\mu^{3/2}}\left(1-\frac{1}{\sqrt{2\mu}}\log\left(\frac{\rho}{\sqrt{\mu}}
\right)\right)^{-2}.    \label{zero:etabh}
\end{align}
\end{subequations}

Collecting all the pieces together, we note down the black hole metric as,
\begin{align}\label{met:BH}
\begin{split}
ds^{2}_{\text{BH}}\simeq -\frac{8}{3\Lambda}\left(\rho^{2}-\mu\right)\left(1+2\xi\left(\omega_{(1)}^{ab}\right)^{\text{bh}}
+2\kappa\left(\omega_{(1)}^{na}\right)^{\text{bh}}\right)\left(-dt^{2}+\frac{d\rho^{2}}{4\left(
\rho^{2}-\mu\right)^{2}}\right).
\end{split}
\end{align}
The black hole solution (\ref{met:BH}) exhibits a horizon at $ \rho = \sqrt{\mu} $. On the other hand, substituting, $ \rho \sim \frac{1}{\delta} $ and thereby taking $ \delta \rightarrow 0 $ limit reveals, $ ds^{2}_{\text{BH}}\Big|_{\rho \rightarrow \infty} \sim  \frac{1}{\delta^3}$. This suggests that the UV asymptotic of the space time approaches Lifshitz$_2$ geometry with dynamical critical exponent $z_{\text{dyn}}=\tfrac{3}{2}$.

The Hawking temperature of the black hole (\ref{met:BH}) can be found as,
\begin{equation}\label{temp:BH}
T_{H}=\frac{\sqrt{\mu}}{\pi}.
\end{equation}

Finally, we note down the Wald entropy \cite{Wald:1993nt} of the black hole,\footnote{Here by entropy we mean entropy
per unit volume of the transverse space namely, $S_{W}/\mathcal{V}_{3}$.}
\begin{align}\label{ent:wald}
\begin{split}
S_{W}&=4\pi\Phi\big|_{\rho=\sqrt{\mu}}\\
&=-\frac{1}{96\Lambda\sqrt{\mu}(|\mu-1|)}\Big[(|\mu-1|)\Big\{-33Q^{2}\Lambda
\xi+64\mu\left(1+0.89\xi\right)   \\
&\qquad\qquad\qquad\qquad\quad-15Q^{2}\xi\left(\log\mu\right)\Big\}-240
\kappa\Big]
\end{split}
\end{align}
which clearly shows a discontinuity at $ \mu = \mu_0 =1 $. As we probe into details on thermal stability, we identify the above discontinuity as a signature of first order transition from an unstable black hole phase to pure radiation.

In order to compare the $2D$ black hole entropy (\ref{ent:wald}) with the black hole entropy in $5D$ we uplift the $2D$
solution into $5D$,
\begin{equation}
ds_{5D}^{2}=ds_{2D}^{2}+\Phi^{2/3}dx_{i}^{2}, \qquad i=1,2,3.
\end{equation}

In a manner similar to the $2D$ theory, the Wald entropy corresponding to the $5D$ black hole can be written as 
\cite{Jacobson:1993vj,Brustein:2007jj}
\begin{align}\label{wald:5D}
\begin{split}
\hat{S}_{W}&=-2\pi \int_{\Sigma}\Bigg(\frac{\delta L_{5D}}{\delta\mathcal{R}_{abcd}}\Bigg)
\hat{\epsilon}_{ab}\hat{\epsilon}_{cd}\bar{\epsilon}    \\
&=-2\pi \int_{\rho = \sqrt{\mu},t=\text{const.}}\Bigg(\frac{\delta L_{5D}}{\delta
\mathcal{R}_{\rho t \rho t}}\Bigg)\hat{\epsilon}_{\rho t}\hat{\epsilon}_{\rho t}\Phi
\prod_{i=1}^{3}dx_{i}\\
&=\mathcal{V}_{3}4\pi\Phi \big|_{\rho=\sqrt{\mu}},
\end{split}
\end{align}
where $\Sigma$ is the codimension-$2$ space-like bifurcation surface with binormal $\hat{\epsilon}_{ab}$ which is symmetric
in $a \leftrightarrow b$ and normalized as $\hat{\epsilon}^{ab}\hat{\epsilon}_{ab}=-2$, $\bar{\epsilon}$ is the induced volume
on $\Sigma$, and $\mathcal{R}_{abcd}$ is the Riemann curvature tensor in the theory. Also the bifurcation surface is at the
horizon $\rho = \sqrt{\mu}$, $t=$ constant. Notice that, in deriving the final form of the Wald entropy in (\ref{wald:5D}) we have
used the fact that the Lagrangian density $L_{5D}$ corresponding to the $5D$ theory (\ref{act:5D}) contains term linear 
in $\mathcal{R}$ only. In addition, we have used the form of the metric given in the ansatz (\ref{anz:red}) explicitly. Hence from
(\ref{ent:wald}) and (\ref{wald:5D}) we observe that the Wald entropy (per unit volume of the transverse space ($\mathcal{V}_{3}$))
for the two theories indeed match.

\section{Thermal stability and HP transition}\label{phase}
We now move on to the Euclidean formalism \cite{Witten:1998zw} and explore thermal stability in black holes. To start with, the full Euclidean \textit{onshell} action is schematically expressed as,
\begin{equation}\label{act:sep}
-S^{\text{os}}_{2D}=S_{\text{grav}}^{\text{os}}+S^{\text{os}}_{\text{GH}}+S^{\text{os}}_{ct}
\end{equation}
where each of the above entities on the R.H.S. of (\ref{act:sep}) are estimated using background solutions found in the previous Section. This is accompanied by an analytic continuation of the Lorentzian time, $ t\rightarrow i \tau $ with a periodicity $ \beta $. In the present analysis, there exists two distinct periodicities corresponding to thermal radiation bath ($ \beta_{\text{TH}} $) and the black hole phase ($ \beta_{\text{BH}} $) respectively. The black hole periodicity $ \beta_{\text{BH}} $ is uniquely fixed by (\ref{temp:BH}) and the former ($ \beta_{\text{TH}} $) is arbitrary.

Notice that both the vacuum and the black hole geometry are the same in the UV asymptotic limit $\rho \rightarrow \infty$, i.e., Lifshitz$_2$ with a dynamical critical exponent $z_{\text{dyn}} = \frac{3}{2}$. In other words, the effects of black holes die out near the boundary. Therefore, one can use this fact to fix the $\beta_{\text{TH}}$, which yields 
\begin{align}\label{fix:beta}
\begin{split}
\frac{\beta_{\text{TH}}}{\beta_{\text{BH}}}\approx 2+\mathcal{O}\left(\frac{1}{\rho_{c}}\right),
\end{split}
\end{align}
where $\rho=\rho_{c}(\rightarrow\infty)$ is the radial cutoff \cite{Witten:1998zw}.

Using (\ref{fix:beta}), the \emph{effective} free energy of the configuration could be obtained using the background subtraction method \cite{Witten:1998zw}, 
\begin{align}\label{dif:sos}
\begin{split}
\Delta \mathcal{F}
&=\mathcal{F}_{\text{BH}}-\mathcal{F}_{\text{TH}}   \\
&=\frac{1}{18\Lambda(\mu-1)}\Big[(\mu-1)\left\{-7Q^{2}\Lambda
\xi+16\xi\left(-30+0.22\mu\right)\right\}+120\Lambda\kappa\Big].
\end{split}
\end{align}

Notice that this free energy\footnote{It should be noted that the free energy associated with the vacuum solution (thermal radiation) grows as $\frac{1}{z^2}$ in the UV limit, which is quite different from that of the gauge fields and the space-time metric.} is quite unique to our original choices of parameters $\alpha$ and $\beta$ in (\ref{anz:red}) and
therefore to the JT gravity model proposed in (\ref{actsim:2D}). In other words, the background subtraction method used in this
paper is also quite pertinent to the choices of parameters $\alpha$ and $\beta$.

Finally, using a canonical ensemble ($ \mathcal{Z}\sim e^{-\Delta S^{\text{os}}_{2D}} $), the energy of the black hole configuration may be estimated as,
\begin{align}\label{avg:eng}
\mathcal{M}_{\text{BH}}\equiv \langle\mathcal{E}\rangle=\frac{1}{450\Lambda(\mu-1)^{2}}\left( 3000\kappa\Lambda(3\mu-1)-\xi(\mu-1)^{2}\left(12000
+175Q^{2}\Lambda+88\mu\right)\right).
\end{align}

\begin{figure}[htp]
\begin{subfigure}
  \centering
  \includegraphics[width=0.7\linewidth]{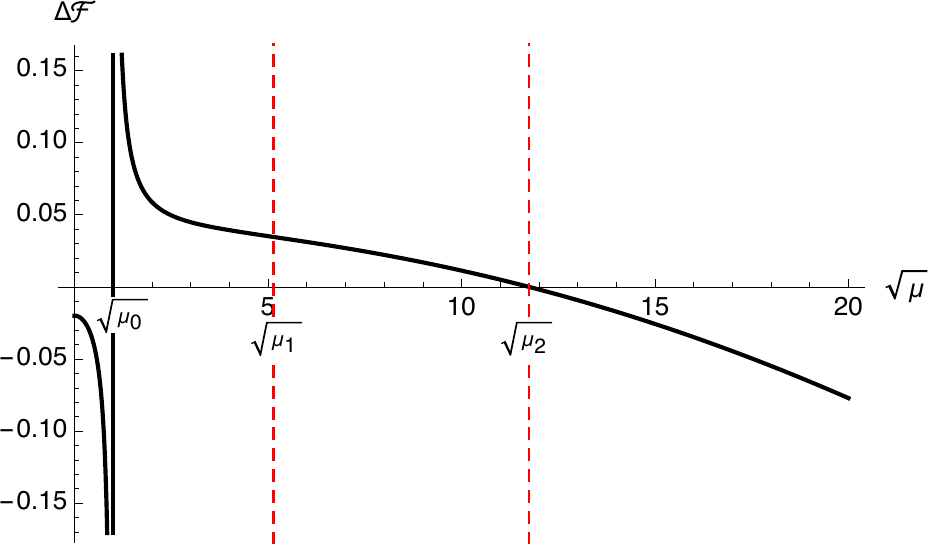} 
  \caption{Free energy \textit{vs.} temperature plot.} 
  \label{fig:1a}
\end{subfigure}
~~
\begin{subfigure}
  \centering
  \includegraphics[width=0.7\linewidth]{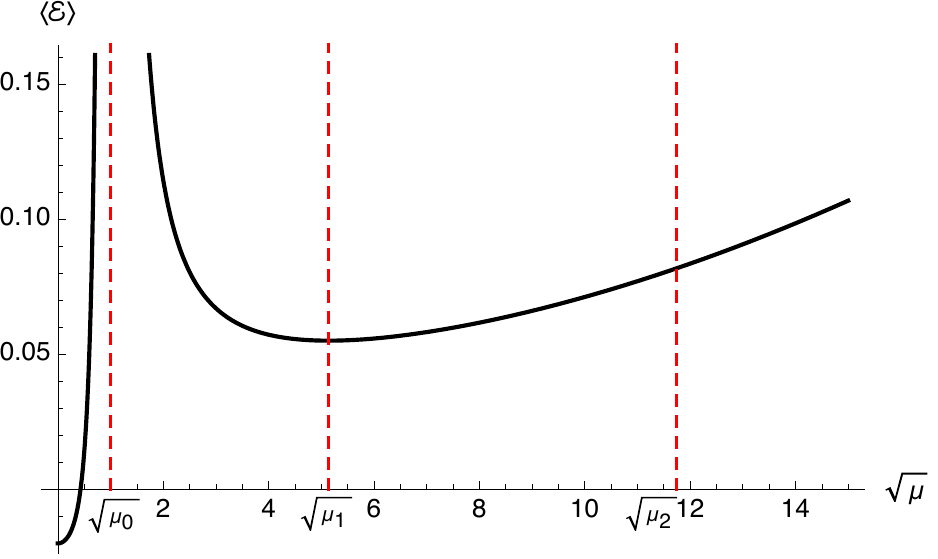}  
  \caption{Energy ($ \sim $Mass) \textit{vs.} temperature plot.}
  \label{fig:1b}
\end{subfigure}
\caption{Behaviour of effective free energy ($\Delta\mathcal{F}$) and
energy ($\mathcal{E} \sim \mathcal{M}_{\text{BH}}$) of the configuration with temperature
$T~ (\sim \sqrt{\mu})$.}
\label{fig:1}
\end{figure}

\begin{figure}[t!]
\begin{subfigure}
  \centering
  \includegraphics[width=0.7\linewidth]{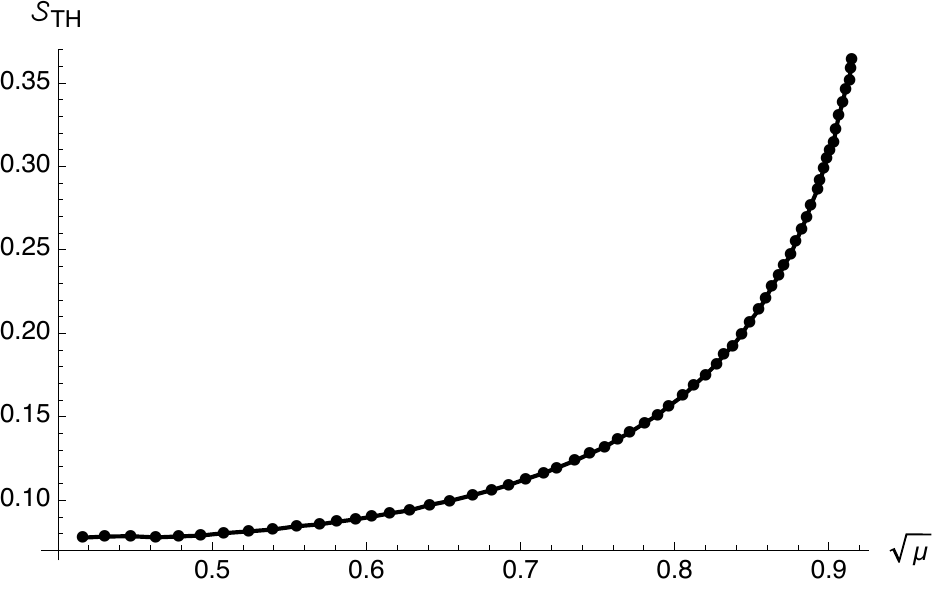}  
  \caption{Thermal entropy ($ \mathcal{S}_{\text{TH}} $) plot for $ \sqrt{\mu} <1 $ branch.}
  \label{fig:2a}
\end{subfigure}
~~
\begin{subfigure}
  \centering
  \includegraphics[width=0.7\linewidth]{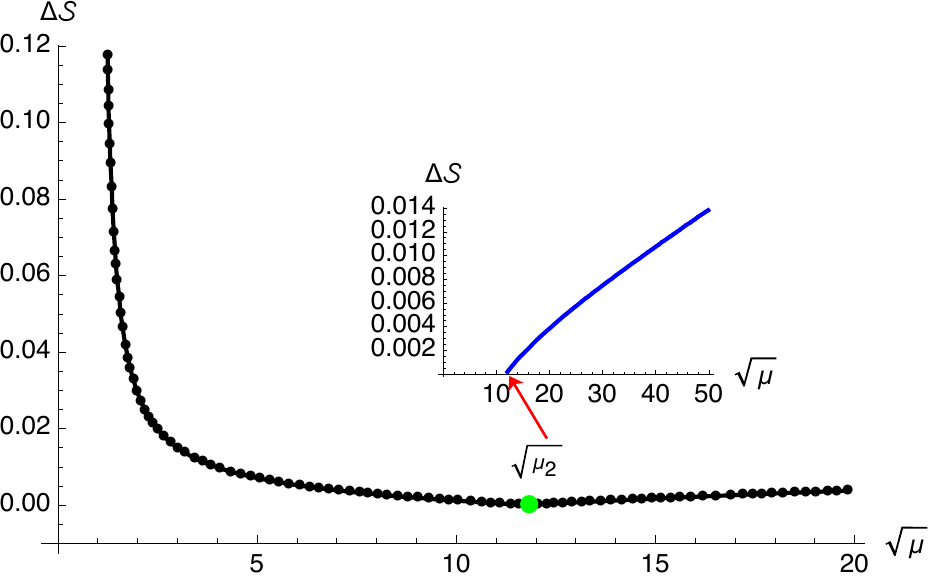}  
  \caption{Effective entropy ($ \Delta \mathcal{S} $) plot for the $\sqrt{\mu}>1$ branch.}
  \label{fig:2b}
\end{subfigure}
\caption{Behavior of the entropy ($ \mathcal{S} $) as a function of temperature $T(\sim
\sqrt{\mu})$.}
\label{fig:2}
\end{figure}

Figure \ref{fig:1} displays the behavior of both the free energy ($ \Delta \mathcal{F} $) and the energy ($ \sim $ mass) of the (black hole) configuration with temperature ($ T \sim \sqrt{\mu} $). In all the subsequent plots, we set $Q=0.01$, $\xi=0.009$ and $\kappa=0.009$. Below we enumerate the key observations regarding Figure \ref{fig:1} and Figure \ref{fig:2}.\\\\
$ \bullet $ To start with, we notice that the thermal radiation collapses to form an \emph{unstable} black hole in the region $ T_0<T<T_1 $ which can either completely decay into pure radiation without a black hole (for $ T<T_0  $) or transit to a larger mass black hole for $ T>T_0 $. Clearly in this phase, the smaller mass black hole has larger free energy than the thermal radiation bath ($ \mathcal{F}_{\text{BH}}>\mathcal{F}_{\text{TH}} $) and is also characterized by a negative heat capacity, $ C_{\text{BH}} \sim \frac{\partial \langle\mathcal{E}\rangle}{\partial \sqrt{\mu}}<0 $ (Fig.\ref{fig:1b}). Therefore, we identify $ T_0<T<T_1 $ as a radiation dominated phase with $ \Delta \mathcal{F}>0 $ and $ T<T_0 $ as a pure radiation phase with, $ \Delta \mathcal{F}\sim \mathcal{F}_{\text{TH}}<0 $ (Fig.\ref{fig:1a}). The corresponding entropy\footnote{The entropy ($ \mathcal{S} $) that we measure should be understood in two ways. For $ \sqrt{\mu} <1$ branch, one is left with pure radiation, therefore we identify the corresponding entropy simply as the entropy of the thermal bath ($ \mathcal{S}_{\text{TH}} $). On the other hand, for $ \sqrt{\mu} >1$ branch, we introduce the notion of  \emph{effective} (or difference) entropy, $\Delta \mathcal{S} \sim -\frac{\Delta \mathcal{F}}{\Delta \sqrt{\mu}}\sim  |\mathcal{S}_{\text{BH}}-\mathcal{S}_{\text{TH}} |$ which is basically the difference in entropy between two distinct configurations namely, the black hole and the heat bath surrounding it (see (\ref{dif:sos})). However, in both cases, the change in total entropy must be positive following the second law of thermodynamics.} plot (Fig.\ref{fig:2}) reveals the onset of a first order phase transition at $T_0 \sim \sqrt{\mu_0} =1 $. The $ \sqrt{\mu}<1 $ branch (Fig.\ref{fig:2a}) corresponds to entropy of the configuration that ultimately boils down into pure radiation. This part of the phase diagram represents a rapidly evaporating black hole (Fig.\ref{fig:1b}) placed in a heat bath which eventually results in a thermal ensemble with \emph{unique} entropy ($ \mathcal{S}_{\text{TH}} $). The $\sqrt{\mu}>1$ branch (Fig.\ref{fig:2b}), on the other hand, depicts a phase with decreasing difference entropy ($\Delta \mathcal{S} $) for $ T\gtrsim T_0 $ which corresponds to the fact that the entropy of the unstable black hole approaches the entropy of the thermal bath as it gradually shrinks in size (Fig.\ref{fig:1b}) with the increase in temperature. This in turn is related to the negative heat capacity of the black hole as mentioned earlier.\\\\


$ \bullet $ As the temperature of the configuration approaches $ T \sim T_1 ~\sim \sqrt{\mu_1}$, the corresponding mass of the black hole reaches a minima (Fig.\ref{fig:1b} ) and thereafter starts increasing slowly for $ T\gtrsim T_1 $. We identify this as a \emph{locally}\footnote{By local we mean that this phase is existing only in a small temperature window $ T_1<T<T_2 $.} stable phase of black hole ($ \mathcal{S}_{\text{BH}} \sim \mathcal{S}_{\text{TH}} $) with positive heat capacity $ C_{\text{BH}} \sim \frac{\partial \langle\mathcal{E}\rangle}{\partial \sqrt{\mu}}>0 $ (Fig.\ref{fig:1b}). However, the effective free energy ($ \Delta \mathcal{F} $) of the configuration reveals that this phase is still dominated by the thermal radiation background with $ \mathcal{F}_{\text{TH}}<\mathcal{F}_{\text{BH}} $ (Fig.\ref{fig:1a}). The total entropy of the configuration is therefore $\sim \mathcal{S}_{\text{TH}}  $ and hence the difference $ \Delta \mathcal{S}\sim 0$ over the range $ T_1 <T<T_2 $ (Fig.\ref{fig:2b}).\\\\
$ \bullet $ Finally, we reach the point of \emph{crossover} $ T=T_2 $ beyond which the black hole mass ($ \mathcal{M}_{\text{BH}} $) starts becoming ever increasing with the increase in temperature and thereby yields a positive heat capacity ($ C_{\text{BH}} >0 $) (Fig.\ref{fig:1b}). A careful analysis further reveals that this is the point of inflection where the difference entropy vanishes ($ \Delta \mathcal{S}=0 $) and thereafter starts increasing (Fig.\ref{fig:2b}) as one approaches the region $ T>T_2 $. This corresponds to a \emph{globally}\footnote{By global we mean that this phase persists for all temperatures $ T>T_2 $.} stable phase of black hole with $ \mathcal{F}_{\text{BH}}<\mathcal{F}_{\text{TH}} $ (Fig.\ref{fig:1a}). Needless to say that for $ T>T_2 $, the entropy content of the configuration is mostly dominated due to the presence of the larger mass black hole which largely contributes to both in the increase in total entropy as well as the difference entropy ($ \Delta \mathcal{S} \gtrsim 0$) of the system.  

In a nutshell, the black hole solutions in the presence of $SU(2)$ Yang-Mills interactions are unstable\footnote{See \cite{Bizon:1991nt} for the stability/instability of colored (i.e., non-Abelian) black holes.} in the region $T_0 < T < T_1$, where the free energy is positive, and the heat capacity is negative. These solutions become locally stable in the range $T_1 < T < T_2$, where the free energy is positive and the heat capacity of black holes is also positive. On the other hand, colored black hole solutions become globally stable beyond the temperature $T_2$. 
\section{Summary and final remarks}\label{conclusion}
Now, we conclude with a brief summary of the main results of the chapter. We propose a possible formulation of the Jackiw-Teitelboim (JT) gravity using dimensional reduction of parent 5D gravitational theory \cite{Fan:2015aia} in the presence of abelian as well $ SU(2) $ Yang-Mills fields. To our surprise, we notice an early (thermal) instability associated to black hole micro-states using a canonical ensemble for the 2D model. We identify that these early black hole states might either decay into pure radiation or they might switch over to a bigger black hole micro-state with higher entropy and lesser free energy. We summarise all these features collectively as being the analogue of Hawking-Page transition \cite{Hawking:1982dh} in 2D gravity models. However, the holographic interpretation of this analysis in terms of the (dual) quantum mechanical model living in one dimension is yet to be explored. 

In a nutshell, in this chapter, we investigate the phase transition between the thermal $AdS_2$ phase and the black hole phase, which we identify as the Hawking-Page (HP) transition in two dimensions. In the next Chapter, we extend our analysis further where we explore the phase stability of the (Euclidean) wormhole solutions \cite{Maldacena:2018lmt}-\cite{Garcia-Garcia:2020vyr}  in the context of the JT/SYK duality.



\chapter{Phases of Euclidean wormholes in JT gravity}
\allowdisplaybreaks
\pagestyle{fancy}
\fancyhead[LE]{\emph{\leftmark}}
\fancyhead[LO]{\emph{\rightmark}}
\rhead{\thepage} 

After investigating the thermal properties of the vacuum and the black hole solutions of the JT gravity that is coupled with $U(1)$ gauge fields and $ SU(2)$ Yang-Mills interaction, we now move on towards studying a model of JT gravity that reveals evidence of (Euclidean) wormhole to black hole phase transition at finite charge density and/or chemical potential. We identify the low temperature phase of the system as the charged wormhole solution.

As the temperature of the system is increased, it undergoes a first order phase transition to a two black hole system at finite charge density. At the critical point ($T = T_0$) of the phase transition, both the Free energy (density) and the charge undergoes a discontinuous change. Finally, we conjecture that the field theory dual to this gravitational set up is a two-site (uncoupled) complex SYK model at a finite chemical potential ($\mu$).

\section{Introduction and overview}
Wormholes are the geometrical bridges that connect the asymptotic regions of space-time \cite{Hawking:1988ae,Hawking:1990in}. These are the solutions of Einstein's equation in the classical limit. 

Wormhole solutions \cite{Maldacena:2018lmt}-\cite{Zhang:2020szi}, \cite{Sahoo:2020unu}, \cite{Garcia-Garcia:2021squ}-\cite{Garcia-Garcia:2019poj} are studied extensively in the context of Jackiw-Teitelboim (JT) gravity models \cite{Teitelboim:1983ux,Jackiw:1984je} those are conjectured to be dual to Sachdev-Ye-Kitaev (SYK) like models \cite{Sachdev:1992fk}-\cite{Lala:2019inz}, \cite{Davison:2016ngz}-\cite{Castro:2008ms}, \cite{Bulycheva:2017uqj} in one dimension. In particular, the authors in \cite{Maldacena:2018lmt} investigate a two site coupled SYK model, which for small values of the coupling, is found to exhibit a \emph{gapped} phase at sufficiently low energies. 

This \emph{gapped} phase in the coupled SYK model is identified with the traversable wormhole solution of the nearly-$AdS_2$ \cite{Jensen:2016cah, Almheiri:2014cka} gravity interacting with the matter fields. However, as the temperature is increased, the SYK model exhibits a phase transition which in the language of the dual gravity picture, corresponds to a Hawking-Page transition into a black hole phase at high temperatures.

The authors in \cite{Garcia-Garcia:2020ttf} further extend these results to explore a two site uncoupled Majorana SYK model which also exhibits a \emph{gapped} phase at low temperatures. The gravitational analogue of this phenomenon is proposed to be an Euclidean wormhole solution of JT gravity in the presence of matter couplings. These results were further generalised in \cite{Garcia-Garcia:2020vyr}, where the authors consider a weak coupling between the two site complex SYK whose gravitational dual corresponds to a traversable wormhole solution \cite{Maldacena:2018lmt} at zero charge density. Using the Schwinger-Dyson equation, they further establish the onset of a first order phase transition in which the wormhole phase transits into a two black hole system at high temperatures\footnote{For higher dimensional wormholes  and the associated phase structure see \cite{Maldacena:2004rf}-\cite{Kundu:2021nwp}.}.

Recently, the complex SYK model has been further investigated in the presence of different chemical potentials \cite{Zhang:2020szi}. At low energies, the authors in \cite{Zhang:2020szi} identify the ground state of the system as an eternal traversable wormhole that connects the two sides at low (averaged) chemical potential. Interestingly, these wormhole solutions transit into a two black hole system at high chemical potential.  

Given the above review on the literature, the purpose of the present chapter is to study the Euclidean wormhole solutions at finite (charge) density and in particular, to explore the associated phase stability of the solution at low temperatures. 

Below, we outline the key findings of our analysis.

In this chapter, we cook up a theory of Einstein-Maxwell-dilaton (EMD) gravity within the JT gravity framework that exhibits a first order phase transition between the charged wormhole solutions at low temperatures and black hole solutions at high temperatures and fixed chemical potential. In particular, we explore the thermal properties of both of these solutions. We observe that the regularised Free energy density $(\mathcal{F}_{(wh)}^{reg})$ of the wormhole configuration is almost constant at sufficiently low temperatures ($T<T_0$) indicating the presence of ``gapped'' phase in the dual (conjectured) two site complex SYK model \cite{Garcia-Garcia:2020ttf, Garcia-Garcia:2020vyr}.

The organization for the rest of the chapter is as follows.

$\bullet$ In Section \ref{grsetup}, we emphasize on the role of the Maxwell-Chern-Simons (MCS) term \cite{VanMechelen:2019ebr, Deser:1981wh} that appears in topological gauge theories. When coupled to $ AdS_3 $ gravity, following a suitable dimensional reduction (see Appendix \ref{dimreduction}), this results into a dynamical JT gravity model which exhibits a wormhole to black hole phase transition at fixed chemical potential. 

We also carry out a first principle derivation of the quantum stress-energy tensor for the $U(1)$ gauge fields ($A_{\mu}$) that takes into account the double trumpet background. Since gauge fields in 2D are non conformal, therefore the present derivation is significantly different from those of the earlier results reported in \cite{Garcia-Garcia:2020ttf}. 

$\bullet$ In Section \ref{thermoref2}, we carry out a detailed analysis on various thermodynamic entities pertaining to the wormhole as well as the black hole phase. These include computing entities like the ``boundary'' Free energy density ($\mathcal{F}$), total charge ($Q$), temperature ($T$) and the chemical potential ($\mu$). In particular, we express these quantities as, $\mathcal{F}=\mathcal{F}(T,\mu)$ and $Q=Q(T,\mu)$, where both $T$ and $\mu$ are treated as independent parameters of the system.

$\bullet$ In section \ref{secphasetrans}, We explore the variations of $\mathcal{F}$ and $Q$ with temperature $(T)$ while keeping the chemical potential $(\mu=\mu_0)$ fixed. Our analysis reveals that the wormhole phase at low temperature ($T<T_0$) undergoes a first order phase transition (at $T=T_0$) into a two black hole system at finite charge $(Q)$.

$\bullet$ In Section \ref{npbnewsecconject}, we qualitatively  discuss the structure of dual field theory (complex SYK model) associated with the 2D Einstein-Maxwell-dilaton gravity. 

Finally, we conclude the chapter in Section \ref{secconcnpb}. 
\section{JT gravity set up in 2D }\label{grsetup}
We begin by considering the following Einstein-Maxwell-dilation (EMD) gravity\footnote{In the Appendix \ref{dimreduction}, we show how the first integral on the R.H.S. of (\ref{action}) appears as a result of dimensional reduction. The second term ($ \sim \int (\partial \chi)^2 $), on the other hand, has been added by hand following the same spirit as that of \cite{Garcia-Garcia:2020ttf}.} in 2D
\begin{align}\label{action}
    S_{JT}=&\int_{\mathcal{M}} d^2x\sqrt{-g}\Big[\Phi(R+2)+a_1\Phi^{2}F^2+a_2\Phi\varepsilon^{\mu\nu}F_{\mu\nu}\Big]+\int_{\mathcal{M}} d^2 x\sqrt{-g}(\partial\chi)^2+\nonumber\\
    &\int_{\partial \mathcal{M}} d \tau \sqrt{-\gamma}\Phi2K+S_{ct},
\end{align}
where $a_1,a_2$ are some arbitrary coupling constants of the theory, $\Phi$ is the dilaton and $\chi$ is a scalar field. Here, $S_{ct}$ is the counter term that is introduced in order to keep the on-shell action finite. Finally, here $\varepsilon^{\mu\nu}=\frac{1}{\sqrt{-g}}\epsilon^{\mu\nu}$ and $K$ is the trace of the extrinsic curvature.

Here, we introduce $L_{CS}= \Phi\varepsilon^{\mu\nu}F_{\mu\nu} $ as the Chern-Simons density (CSd) term associated with the 2D gravity model (\ref{action}).  As our analysis reveals, the CSd term for the wormhole phase is non-zero as one approaches the boundary of the space time. On the other hand, it vanishes asymptotically for the black hole phase. 

This altogether makes a crucial difference between the boundary Free energy densities of these two phases. It is noteworthy to mention that the above is an ``on-shell'' result and does not depend on the choice of the coupling constant $a_2$.

The variation of the action (\ref{action}) yields the following set of equations 
\begin{align}
\Phi &: (R+2)+2a_1\Phi F^2+a_2\varepsilon^{\mu\nu}F_{\mu\nu}=0, \label{eom1}\\
\chi &:\Box\chi=0,\label{eom2}\\
A_{\mu} &: \bigtriangledown _{\mu}\big[2a_1\Phi^{2} F^{\mu\nu}+a_2\Phi\varepsilon^{\mu\nu}\big]=0,\label{eom3}\\
g_{\mu\nu} &:\Box\Phi g_{\mu\nu}-\bigtriangledown_{\mu}\bigtriangledown_{\nu}\Phi-g_{\mu\nu}\Phi+<T_{\mu\nu}>=0,\label{eom4}
\end{align} 
where $<T_{\mu\nu}>$ is the full stress-energy tensor\footnote{We define the stress tensor as $T_{\mu\nu}=\frac{1}{\sqrt{-g}}\frac{\delta S}{\delta g^{\mu\nu}}$.} combining gauge fields and the scalar field
\begin{align}
    <T_{\mu\nu}>\hspace{1mm}&=\hspace{1mm}<T^{(gauge)}_{\mu\nu}>+<T^{(\chi)}_{\mu\nu}>,\label{eom4a}\\
    <T^{(gauge)}_{\mu\nu}>\hspace{1mm}&=a_1\Phi^{2}\Big[2g^{\alpha\beta}F_{\mu\alpha}F_{\nu\beta}-\frac{1}{2}F^2g_{\mu\nu}\Big],\label{eom4b}\\
    <T^{(\chi)}_{\mu\nu}>\hspace{1mm}&=(\partial_{\mu}\chi)(\partial_{\nu}\chi)-\frac{1}{2}g_{\mu\nu}(\partial \chi)^2.\label{eom4c}
    \end{align}
\subsection{Euclidean wormholes}  \label{secwh}
We begin by considering the possibilities for a charged Euclidean wormhole solution of ($\ref{action}$) whose geometry is described by a double trumpet \cite{Garcia-Garcia:2020ttf} having two asymptotics. The corresponding space-time metric is expressed as
\begin{align}\label{gauge}
    &ds^2=\frac{1}{\cos^2\rho}(d\tau^2+d\rho^2)\hspace{1mm};\hspace{2mm}-\frac{\pi}{2}\leq\rho\leq\frac{\pi}{2}\hspace{1.5mm},\hspace{1.5mm} \tau\sim\tau+b,
\end{align}
where $b$ is the periodicity of the Euclidean time $(\tau)$.

Given (\ref{gauge}), we solve the equations of the motion (\ref{eom1})-(\ref{eom4}) using the static gauge
\begin{align}\label{gauge1}
A_{\tau}=\xi(\rho),\hspace{1mm}A_{\rho}=0,\hspace{1mm}\Phi=\Phi(\rho)\hspace{1mm},\hspace{1mm}\chi=\chi(\rho).
\end{align}

Using (\ref{gauge}) and (\ref{gauge1}), it is now trivial to find the solution for the scalar field
\begin{align}\label{chisol}
    \chi=C_1\rho+C_2,
\end{align}
where $C_1$ and $C_2$ are the integration constants. 

On the other hand, the equations of motion for the dilaton (\ref{eom1}) and the gauge field (\ref{eom3}) can be reduced down to a single equation of the form
\begin{align}\label{constraint}
    \Phi\cos^2\rho(\partial_{\rho}\xi)=\frac{a_2}{2a_1},
\end{align}
where we set the integration constant to zero for the consistency of the wormhole solution.
\vspace{2mm}


$\bullet$ {\bf{Stress-energy tensor}}

\vspace{2mm}
Our next task would be to compute the full stress-energy tensor $<T_{\mu\nu}>$ combining both the $U(1)$ gauge fields $(A_{\mu})$ as well as the scalar field $(\chi)$
\begin{align}\label{stfull}
    <T_{\mu\nu}>\hspace{1mm}&=\hspace{1mm}<T^{(gauge)}_{\mu\nu}>+<T^{(\chi)}_{\mu\nu}>,
\end{align}
where $<T^{(\chi)}_{\mu\nu}>$ denotes the stress-energy tensor for the scalar field $(\chi)$ and\\ $<T^{(gauge)}_{\mu\nu}>$ corresponds to the stress-energy tensor for the $U(1)$ gauge fields $(A_{\mu})$.

The general strategy would be to break the stress-energy tensor (\ref{stfull}) into classical and quantum pieces as follows
\begin{align}\label{breakref2}
    <T_{\mu\nu}>_{cl}\hspace{1mm}&=\hspace{1mm}<T_{\mu\nu}^{(\chi)}>_{cl}\hspace{1mm}+\hspace{1mm}<T_{\mu\nu}^{(gauge)}>_{cl},\nonumber\\
     <T_{\mu\nu}>_{qm}\hspace{1mm}&=\hspace{1mm}<T_{\mu\nu}^{(\chi)}>_{qm}\hspace{1mm}+\hspace{1mm}<T_{\mu\nu}^{(gauge)}>_{qm},
\end{align}
where $`cl$'stands for the classical and $`qm$' stands for the quantum stress-energy tensor.

The classical part of the stress-energy tensor can be computed directly substituting (\ref{gauge})-(\ref{chisol}) into (\ref{eom4b})-(\ref{eom4c}). On the other hand, we use the point-splitting method of \cite{book} in order to fix the quantum stress-energy tensor ($ <T_{\mu\nu}>_{qm} $).

 The classical and the quantum part of the scalar stress-energy tensor \\ ($ <T^{(\chi)}_{\mu\nu}>_{qm} $) is quite straightforward to obtain. These results can be summarised as follows \cite{Garcia-Garcia:2020ttf}
 \begin{align}
     <T^{(\chi)}_{\rho\rho}>_{cl}\hspace{1mm}&=- <T^{(\chi)}_{\tau\tau}>_{cl}\hspace{1mm}=\frac{C_1^2}{2},\label{sst1}\\
      <T_{\rho\rho}^{(\chi)}>_{qm}&=\frac{1}{24\pi}-\frac{1}{24\pi\cos^2\rho}+X_{\rho\rho}(b),\nonumber\\
      X_{\rho\rho}(b)&=-\sum_{p\in \mathbb{Z}}\frac{p\pi}{b^2\tanh(2\pi^2p/b)},\label{sst2}\\
      <T_{\tau\tau }^{(\chi)}>_{qm}&=-\frac{1}{24\pi}-\frac{1}{24\pi\cos^2\rho}+X_{\tau\tau}(b),\nonumber\\
      X_{\tau\tau}(b)&=\sum_{p\in \mathbb{Z}}\frac{p\pi}{b^2\tanh(2\pi^2p/b)}.\label{sst3}
 \end{align}

The major part of the present computation therefore involves estimating the quantum stress-energy tensor for the $U(1)$ gauge fields. The classical part of the stress-energy tensor can be obtained using (\ref{eom4b}) and (\ref{gauge})
\begin{align}\label{classst}
    <T_{\rho\rho}^{(gauge)}>_{cl}\hspace{1mm}=\hspace{1mm}<T_{\tau\tau}^{(gauge)}>_{cl}\hspace{1mm}=\hspace{1mm}a_1\Phi^{2}\cos^2\rho(\partial_{\rho}\xi)^2.
\end{align}

The derivation of the quantum stress-energy tensor ($ <T_{\mu\nu}^{(gauge)}>_{qm} $) for the $U(1)$ gauge fields (\ref{breakref2}) is discussed in detail in the Appendix \ref{QSTref2}. These results are obtained in a straightforward way by considering a double trumpet geometry (\ref{gauge}). 

Below, we summarise these results as
 \begin{align}
      <T_{\tau\tau +}^{(gauge)}>_{qm}\hspace{1mm}=\hspace{1mm}<T_{\rho\rho +}^{(gauge)}>_{qm}\hspace{1mm}=&\hspace{1mm}\frac{1}{\mathcal{A}_4(\rho)}\Bigg[a_1\exp\big(p_0(\pi-2\rho)/2\big)\cos^2\rho\Phi^2\Big(4\mathcal{A}_1(\rho)+\nonumber\\
      &(\pi-2\rho)\big(\mathcal{A}_2(\rho)-\mathcal{A}_3(\rho)\big)\Big)\Bigg],\label{stgaugepref2}\\
 <T_{\tau\tau -}^{(gauge)}>_{qm}\hspace{1mm}=\hspace{1mm}<T_{\rho\rho -}^{(gauge)}>_{qm}\hspace{1mm}=&\hspace{1mm}\frac{1}{\mathcal{B}_3(\rho)}\Bigg[a_1\exp(p_0\rho)\cos^2\rho\Phi^2\Big(\mathcal{B}_1(\rho)+\nonumber\\&(\pi+2\rho)\mathcal{B}_2(\rho)\Big)\Bigg],\label{stgaugemref2}
 \end{align}
 where the details of the functions $\mathcal{A}_1(\rho),.. \mathcal{A}_4(\rho), \mathcal{B}_1(\rho),.. \mathcal{B}_3(\rho)$ are given in the Appendix \ref{QSTref2} and the subscripts $`\pm$' denote the  expectation values near the boundary limits, $\rho=\pm\frac{\pi}{2}$.

Substituting (\ref{sst1})-(\ref{sst3}), (\ref{classst}), (\ref{stgaugepref2}) and (\ref{stgaugemref2}) into (\ref{stfull}) one finally obtains the full stress-energy tensor combining the $U(1)$ gauge field and the scalar field $(\chi)$. The complete stress-energy tensor (\ref{stfull}) is then used in the next section to calculate the dilaton $(\Phi)$ profile in the asymptotic limits $(\rho\rightarrow\pm\frac{\pi}{2})$ of the wormhole space time.

\vspace{2mm}
$\bullet$ {\bf{Solving for $\Phi$ and $\xi$}}\label{secsolphi}

  \vspace{2mm}
  Before we proceed further, let us first calculate the boundary stress-energy tensor\footnote{A similar calculation is also discussed in \cite{Rathi:2021aaw}. } using the  Gibbons-Hawking-York term (\ref{action})
  \begin{align}\label{boundary}
   T_{\tau\tau}\Big|_{boundary}=\frac{1}{\sqrt{-\gamma}}\frac{\delta S_{GHY}}{\delta \gamma ^{\tau\tau}}\hspace{1mm},\hspace{1mm}\text{where}\hspace{2mm}S_{GHY}=\int d \tau \sqrt{-\gamma}\Phi2K,
  \end{align}
  which plays a crucial role in obtaining the near boundary profile for the dilaton ($\Phi$).
  
  On solving (\ref{boundary}) using the double trumpet geometry (\ref{gauge}) one finds
  \begin{align}
      T_{\tau\tau}\Big|_{+}=\alpha\Phi\sec{\rho}\tan{\rho}\hspace{2mm}\text{and}\hspace{2mm}T_{\tau\tau}\Big|_{-}=\beta\Phi\sec{\rho}\tan{\rho},
  \end{align}
  where the subscripts $\pm$ denote the stress-energy tensors near the asymptotic limits $\rho_{L,R}\sim\pm\frac{\pi}{2}$ with $\alpha,\beta$ being constants.
  
 Given the double trumpet geometry (\ref{gauge}), the equation of motion (\ref{eom4}) for the metric turns out to be
  \begin{align}\label{phi1}
      \partial_{\rho}^2\Phi-\tan{\rho} \partial_{\rho}\Phi-\frac{\Phi}{\cos^2{\rho}}+<T_{\tau\tau}>=0,
  \end{align}
  where $<T_{\tau\tau}>\hspace{1mm}=\hspace{1mm}<T_{\tau\tau}^{(gauge)}>+<T_{\tau\tau}^{(scalar)}>+<T_{\tau\tau}^{(boundary)}>$. 
  
  Upon solving (\ref{phi1}), one finds the following ``asymptotic'' profiles for the dilaton\footnote{The functions $ F_{\pm} $ play crucial role while determining the Free energy (density) near the boundary of the wormhole space time. This is due to the fact these functions appear explicitly in the asymptotic profiles for the dilaton ($ \Phi_{\pm} $) which in turn carry information about the Free energy (density) near the asymptotic boundary of the wormhole spacetime. These constants in $ F_{\pm} $ are further constrained by the fact that the left and the right temperatures ($ T^{(wh)}_{\pm} $) of the wormhole solution must be identified. This further reduces the number of independent constants to $ r_0 $ and $ \alpha $ which finally appear in the expression for the Free energy (\ref{fedp1}). These constants can be further replaced in terms of the chemical potential ($ \mu $) (see (\ref{finalmu}) ) and the coupling constant(s) which eventually removes all the ambiguities in the expression of the Free energy (\ref{freenergy}).}
  \begin{align}\label{phisol}
    \Phi\Big|_{\rho\rightarrow\frac{\pi}{2}} = \Phi_+=\frac{-F_+}{\big(\frac{\pi}{2}-\rho\big)}-\frac{1}{24\pi}+\alpha r_0\hspace{2mm}\text{and}\hspace{2mm}\Phi\Big|_{\rho\rightarrow-\frac{\pi}{2}}=\Phi_-=\frac{F_-}{\big(\rho+\frac{\pi}{2}\big)}-\frac{1}{24\pi}-\beta s_0,
  \end{align}
  where the details are given in the Appendix \ref{expref2} where the subscripts $`\pm$' denote the leading order terms in $\Phi$.

Using (\ref{constraint}) and (\ref{phisol}), the asymptotic profiles for the gauge field turn out to be
 \begin{align}\label{gaugesol1}
           \xi_{\pm}=\frac{a_2}{2a_1 F_{\pm}}\log\Big(\pi\mp 2\rho\Big)+\mu_{\pm},\hspace{1mm}
       \end{align}
       where $\mu_{\pm}$ denote the chemical potentials near the boundaries $\rho_{L,R}\sim\pm\frac{\pi}{2}$
       \begin{align}\label{gaugesol2}
            \mu_{+}=\frac{a_2}{2a_1}\frac{1}{F_{+}}\Bigg(-1-\frac{192F_{+}^2\pi^2}{(1-24\pi r_0\alpha)^2}\Bigg)\log(-48F_+\pi),\\
            \mu_{-}=\frac{a_2}{2a_1}\frac{1}{F_{-}}\Bigg(-1-\frac{192F_{-}^2\pi^2}{(1+24\pi s_0\beta)^2}\Bigg)\log(-48F_-\pi).
       \end{align}      
       
     At this stage, it is important to notice that the associated Chern-Simons density (CSd) takes a finite value in the asymptotic limits
       \begin{align}\label{ref2cswh}
           L_{CS}^{(wh)}\Big|_{\rho\rightarrow\pm\frac{\pi}{2}}=-2\cos^2\rho\Phi\partial_{\rho}\xi=-\frac{a_2}{a_1},
       \end{align}
  where the on-shell condition (\ref{constraint}) is imposed. Therefore, it plays a significant role while estimating the Free energy (density) for the boundary theory.
 \subsection{Euclidean black holes}\label{secebh}
 We now move on towards constructing the Euclidean black hole solution of (\ref{action}). We further use these solutions to discuss the thermal properties of the 2D black hole. 

We solve these equations (\ref{eom1})-(\ref{eom4}) ``perturbatively'' treating the couplings $a_1$ and $a_2$ as expansion parameters. We express these background fields using the static gauge 
\begin{align}\label{bhansatzref2}
    ds^2&=e^{2\omega(z)}(d\tau^2+dz^2)\hspace{1mm},\nonumber\\
    A_{\mu}&\equiv(A_{\tau}(z),0)\hspace{1mm},\hspace{1mm}\Phi=\Phi(z)\hspace{1mm},\hspace{1mm}\chi=\chi(z).
\end{align}

Next, we expand these background fields schematically as \cite{Lala:2020lge}
\begin{align}\label{fexp}
    \mathcal{H}= \mathcal{H}_0+a_1\mathcal{H}_1+a_2\mathcal{H}_2\hspace{1mm},\hspace{2mm}|a_1|<<1\hspace{1mm},\hspace{1mm}|a_2|<<1,
\end{align}
where $\mathcal{H}$ stands for any of the fields $\omega,\Phi,A_{\tau}$ and $\chi$. 

Here, the subscript $`0$' stands for the pure JT gravity solution while the other two subscripts $`1$'and $`2$' denote the associated corrections due to the $U(1)$ gauge fields \cite{Lala:2020lge}. We discuss all these in detail in the Appendix \ref{bhsolref2}. 

A straightforward computation the on-shell Chern-Simons density (CSd) for the black hole phase shows 
\begin{align}\label{ref2bhcsdm}
  L_{CS}^{bh}\Big|_{z\rightarrow 0}=-2 e^{-2\omega_0}\Phi_0\partial_zA_{\tau}\Big|_{z\rightarrow 0}=-4 b_3z\Big|_{z\rightarrow 0} =0,
\end{align}
where $r=\sqrt{r_H}\coth(2\sqrt{r_H}z)$.  Therefore, unlike the wormhole phase, its contribution can be ignored while obtaining various thermodynamic entities (like Free energy density for example) near the boundary of the black hole spacetime. 

Finally, the space-time metric for the Euclidean black hole (\ref{bhansatzref2}) turns out to be
\begin{align}\label{stm}
    ds^2\approx4(r^2-r_H)(1+2a_1\omega_1)\Bigg(d\tau^2+\frac{dr^2}{4(r_H-r^2)^2}\Bigg),
\end{align}
where $\omega_1$ is given in Appendix \ref{bhsolref2} and the black hole horizon is located at $r=\sqrt{r_H}$.
\section{Thermodynamics}\label{thermoref2}
 We now examine the thermal properties of the wormhole and the black hole solutions those were obtained previously in Section \ref{grsetup}. From the periodicity of the Euclidean time $(\tau)$, one can identify the temperature ($T$) associated with these solutions. 
 
Finally, we estimate the Free energy density ($\mathcal{F}$) and the total charge ($ Q $) and express them as a function of temperature ($T$) and the chemical potential ($\mu$). In other words, we treat both the temperature $(T)$ and the chemical potential $(\mu)$ as independent thermodynamic variables where one of them can be tuned while keeping the other fixed.
 \subsection{Wormholes}
 $\bullet$ {\bf{Temperature}}\label{tempsecref2}
 \vspace{1mm}
 
In order to determine the temperature of the wormhole, we impose the following boundary conditions\footnote{These boundary conditions simply follow from the asymptotic structures of the dilaton ($\Phi$) (\ref{phisol}) and the space-time metric ($g_{\mu\nu}$) (\ref{gauge}), where $\epsilon$ is the UV cutoff. } \cite{Maldacena:2018lmt, Garcia-Garcia:2020ttf} on $\Phi$ and $g_{\mu\nu}$
 \begin{align}\label{bc1}
     \Phi\sim\frac{\phi}{\epsilon}\hspace{2mm}\text{and}\hspace{2mm}ds^2\Big|_{\rho\rightarrow\pm\frac{\pi}{2}}\sim\frac{du_{\pm}^2}{\epsilon^2},
 \end{align}
 where $u_{\pm}$ are identical to $\tau$ and with the periodicity conditions $u_{\pm}\sim u_{\pm}+\phi\beta_{\pm}$. Here,  $\beta_{\pm}$ correspond to the inverse temperatures near the asymptotics $\rho_{L,R}=\pm\frac{\pi}{2}$.
 
Finally, using (\ref{gauge}), (\ref{phisol}) and (\ref{bc1}), we identify the temperature associated with the wormhole solution near the asymptotics as 
 \begin{align}\label{npbcond1}
     T^{(wh)}_{\pm}=\mp\hspace{1mm}\frac{F_{\pm}}{b}, 
 \end{align}
 where the functions $F_{\pm}$ are given in the Appendix \ref{expref2}.
 
  Notice that, the right temperature $(T^{(wh)}_+)$ of the wormhole near $\rho_R\sim\frac{\pi}{2}$ is different from that of its left temperature $(T^{(wh)}_-)$ near $\rho_L=-\frac{\pi}{2}$. However, setting $\beta=-\alpha,\hspace{1mm}C_3=C_4=\eta,\hspace{1mm}q_0=-p_0,\hspace{1mm}B(-\pi/2,m)=-B(\pi/2,m),\hspace{1mm}D(-\pi/2,m)=-D(\pi/2,m),\hspace{1mm}s_0=r_0,\hspace{1mm}s_1=-r_1,\hspace{1mm}s_2=r_2,\hspace{1mm}s_3=-r_3,$ and $\mu_+=-\mu_-=\mu$, we find that $T^{(wh)}_+= T^{(wh)}_-=T_{(wh)}$.

The identification of the chemical potentials ($ \mu_{\pm} $) reveals an useful identity of the form
\begin{align}\label{finalmu}
      \mu = \frac{a_2}{2a_1bT_{(wh)}}\Bigg(1+\frac{192b^2\pi^2T_{(wh)}^2}{(1-24\pi r_0\alpha)^2}\Bigg)\log(48\pi b T_{(wh)}),
  \end{align}
  which is further used in order to remove ambiguities in the Free energy density.
  
  \vspace{2mm}
  $\bullet$ {\bf{Free energy}}

  \vspace{2mm}
  
 The Free energy is defined using the Euclidean path integral\footnote{An exact computation of the Free energy would indeed require the computation of the full bulk integral first and thereby taking its asymptotic limits ($ \rho \rightarrow \pm \frac{\pi}{2} $) for some fixed radial coordinate that approaches the boundary. Taking the boundary limit is important because the dual field theory we conjecture about is supposed live on this boundary. Ideally, this should be conjectured as the Free energy density of the dual field theory. However, as far as the present computation is concerned, this turns out to be a quite non-trivial task due to the complicated profile of the dilaton ($ \Phi $) which appears to be an important element of the bulk integral. Therefore, to deal with the situation, one has to \emph{approximate} the integral by considering its limiting value near the boundary of the spacetime. In other words, the ``boundary'' Free energy that is estimated in this paper, is defined as the integral that is evaluated using the \emph{asymptotic} data of the bulk fields where we ignore some of the IR degrees of freedom those might come from the interior of the bulk.}
  \begin{align}\label{fwh}
      F_{(wh)}=-\beta^{-1}\log Z_E^{(wh)},\hspace{2mm}Z_E^{(wh)}=e^{-S_{(wh)}^{(os)}},
  \end{align}
  where $Z_E$ is the Euclidean partition function and $S_{(wh)}^{(os)}$ stands for the Euclidean on-shell action corresponding to the wormhole solutions (\ref{phisol}) and (\ref{gaugesol1}).
  
Recall that, in Section \ref{secsolphi}, we determine asymptotic profile (\ref{phisol}) for the dilaton ($\Phi$). These asymptotic data are used to calculate the boundary Free energy density (\ref{action}) associated with the wormhole phase.
  
The regularised Free energy density $(\mathcal{F}^{(reg)}_{(wh)})$ of the boundary theory is defined through the following integral 
  \begin{align}\label{fedwh}
      F_{(wh)}^{(reg)}\Big|_{\rho\rightarrow\pm\frac{\pi}{2}}=\int d\tau\sqrt{-\gamma}\mathcal{F}_{(wh)}^{(reg)},
  \end{align}
 where the ``regularised'' Free energy density is expressed as\footnote{Here the divergent piece is absorbed using the following counter term $S_{ct}=\int d\tau\sqrt{-\gamma}\big(2\frac{F_+}{\delta}\big)$, where $\delta$ being the UV cutoff.}
  
       \begin{align}\label{fedp1}
      \mathcal{F}_{(wh)}^{(reg)}=&\hspace{1mm}T_{(wh)}\Bigg(-\frac{2a_1b^2T_{(wh)}^2(1-24\pi r_0\alpha)^4\mu^2}{(192b^2\pi^2T_{(wh)}^2+(1-24\pi r_0\alpha)^2)^2\log(48b\pi T_{(wh)})^2}-\frac{1}{12\pi}+2r_0\alpha\Bigg).
      \end{align}
  
  Using (\ref{finalmu}), the above expression (\ref{fedp1}) further simplifies as
  \begin{eqnarray}
  \label{freenergy}
\mathcal{F}_{(wh)}^{(reg)}=  \hspace{1mm}T_{(wh)}\Bigg(-\frac{a_2^2}{2a_1}-\frac{1}{12\pi}\sqrt{\frac{192a_2\pi^2b^2T_{(wh)}^2\log(48\pi b T_{(wh)})}{2a_1b T_{(wh)} \mu-a_2\log(48\pi b T_{(wh)})}}\Bigg).
  \end{eqnarray}

  \vspace{2mm}
 $\bullet$ {\bf{Charge}}
 \vspace{2mm}
 
 The derivation of the $ U(1) $ charge ($ Q $) follows from the definition of the $ U(1) $ current\footnote{Our analysis follows closely the algorithm developed by authors in \cite{Castro:2008ms}. } \cite{Castro:2008ms} 
 \begin{align}\label{chargedef}
     J^{\mu}=\frac{1}{\sqrt{-g}}\frac{\delta S_{JT}}{\delta A_{\mu}}.
 \end{align}
 
 The variation of (\ref{action}) with respect to the gauge field ($A_{\mu}$) yields
 \begin{align}\label{chargevar}
     \delta S_{JT}=&\int d^2x\sqrt{-g}\nabla_{\mu}\Big(-4 a_1\Phi^2F^{\mu\nu}-2a_2\Phi \varepsilon^{\mu\nu}\Big)\delta A_{\nu} \nonumber\\
     &+ \int d\tau\sqrt{-\gamma} n_{\rho}\Big(4 a_1\Phi^2F^{\rho\tau}+2a_2\Phi \varepsilon^{\rho\tau}\Big)\delta A_{\tau},
 \end{align}
 where $n_{\rho}$ is the unit normal vector to the boundary ($\rho=\pm \frac{\pi}{2}$). Notice that, the first term on the R.H.S. in (\ref{chargevar}) vanishes on-shell. In other words, one is only left with an integral that is evaluated in the asymptotic limit(s) where the dual field theory is living.
 
Using the on-shell condition in (\ref{chargevar}), we finally obtain the boundary current as 
\begin{align}\label{chargectbdy}
    J^{\tau}_{bdy}=n_{\rho}\Big(4 a_1\Phi^2F^{\rho\tau}+2a_2\Phi \varepsilon^{\rho\tau}\Big),
\end{align}
where the subscript $``bdy"$ denotes the current evaluated near the boundary. 

Finally, the $ U(1) $ charge associated with the wormhole phase is define as\footnote{We conjecture this as the global $ U(1) $ charge associated with the dual field theory living on the boundary of the wormhole spacetime.}
\begin{align}\label{bdycharge}
    Q_{(wh)}=\int d\tau \sqrt{-\gamma}J^{\tau}_{bdy}.
\end{align}

Using the asymptotic data (\ref{phisol}) and (\ref{gaugesol1}) together with (\ref{finalmu}), we finally obtain
\begin{align}
\label{boundarycharge}
    Q_{(wh)}= \frac{a_2}{12\pi T_{(wh)}}\sqrt{\frac{192a_2\pi^2b^2T_{(wh)}^2\log(48\pi b T_{(wh)})}{2a_1b T_{(wh)} \mu-a_2\log(48\pi b T_{(wh)})}}.
\end{align}
 
Like Free energy density (\ref{freenergy}), the boundary $ U(1) $ charge (\ref{boundarycharge}) is also free from ambiguities and is fixed by the coupling constant ($ a_2 $) and the periodicity ($ b $) in the Euclidean time. Here, it is noteworthy to mention that both the regularised Free energy density $ (\mathcal{F}_{(wh)}^{(reg)})$ and the total charge ($ Q_{(wh)}$) of the wormhole solution are equal on both the boundaries $\rho_{L,R}$. This follows using  the relations between the constants as mentioned in Section \ref{tempsecref2}.  
\subsection{Black holes}
We now compute the Free energy density and the total charge associated with the black hole solution those are obtained previously in Section \ref{secebh}. The basic philosophy and the physical considerations behind these derivations are the same as those for the wormholes which we therefore prefer not to repeat here.

To begin with, we compute the Hawking temperature\footnote{One can obtain the same expression (\ref{htnewph}) using the periodicity arguments of the Euclidean time $(\tau)$ in an expansion near the horizon.} ($T_H$) of the 2D black hole \cite{Rathi:2021aaw} 
\begin{align}\label{htnewph}
    T_H=\frac{1}{2\pi}\sqrt{\frac{1}{4}g^{\tau\tau}g^{rr}(\partial_{r}g_{\tau\tau})^2}\Bigg|_{r\rightarrow\sqrt{r_H}}=\frac{\sqrt{r_H}}{\pi}.
\end{align}
  
  The regularised Free energy density $(\mathcal{F}^{(reg)}_{(bh)})$ for the boundary theory is defined through the following integral
  \begin{align}\label{fedwhbhnew}
      F_{(bh)}^{(reg)}\Big|_{r\rightarrow\infty}=\int d\tau\sqrt{-\gamma}\mathcal{F}_{(bh)}^{(reg)},
  \end{align}
   where the ``regularised'' Free energy density is expressed as\footnote{Here the divergences are absorbed using the counter term, $S_{ct}=-\int d\tau\sqrt{-\gamma}\big(2+4a_1b_5+2a_1b_9)r$. The origin of these integration constants are shown in detail in the Appendix \ref{bhsolref2}.}
\begin{align}
\label{bhfreeenery}
    \mathcal{F}_{(bh)}^{(reg)}=-\frac{a_1}{2}T_{H}\Bigg(T_{H}+\frac{d_1}{T_{H}^2}+d_2\Bigg).
\end{align}

Here, we rescale the Free energy density by a constant $d_0=-\pi^2b_5+2\pi(b_4+b_5)$ and the other constants can be expressed in terms of $d_0$ as $d_1=-\frac{b_1b_6}{\pi^2d_0}$ and $d_2=-\frac{4b_8}{d_0}$.

The (regularised) boundary charge for the black hole phase can be obtained in a similar way as in the case for wormholes. After a suitable rescaling by $ b_3 $, this yields 
\begin{align}
\label{c2}
  Q_{(bh)}\Big|_{z=0}=\frac{4a_1}{T_{H}},
\end{align}
which finally depends only on the coupling constant $ a_1 (>0)$ of the theory.
\section{Phase transition}\label{secphasetrans}
Finally, with all these solutions in hand, we are now in a position to explore the thermal stability of our solutions with respect to the temperature ($ T $). The key observables in this regard are the regularised Free energy densities (\ref{freenergy}), (\ref{bhfreeenery}) and the global charges (\ref{boundarycharge}), (\ref{c2}) those were obtained previously in Section \ref{thermoref2}. 

\begin{figure}
\includegraphics[scale=.30]{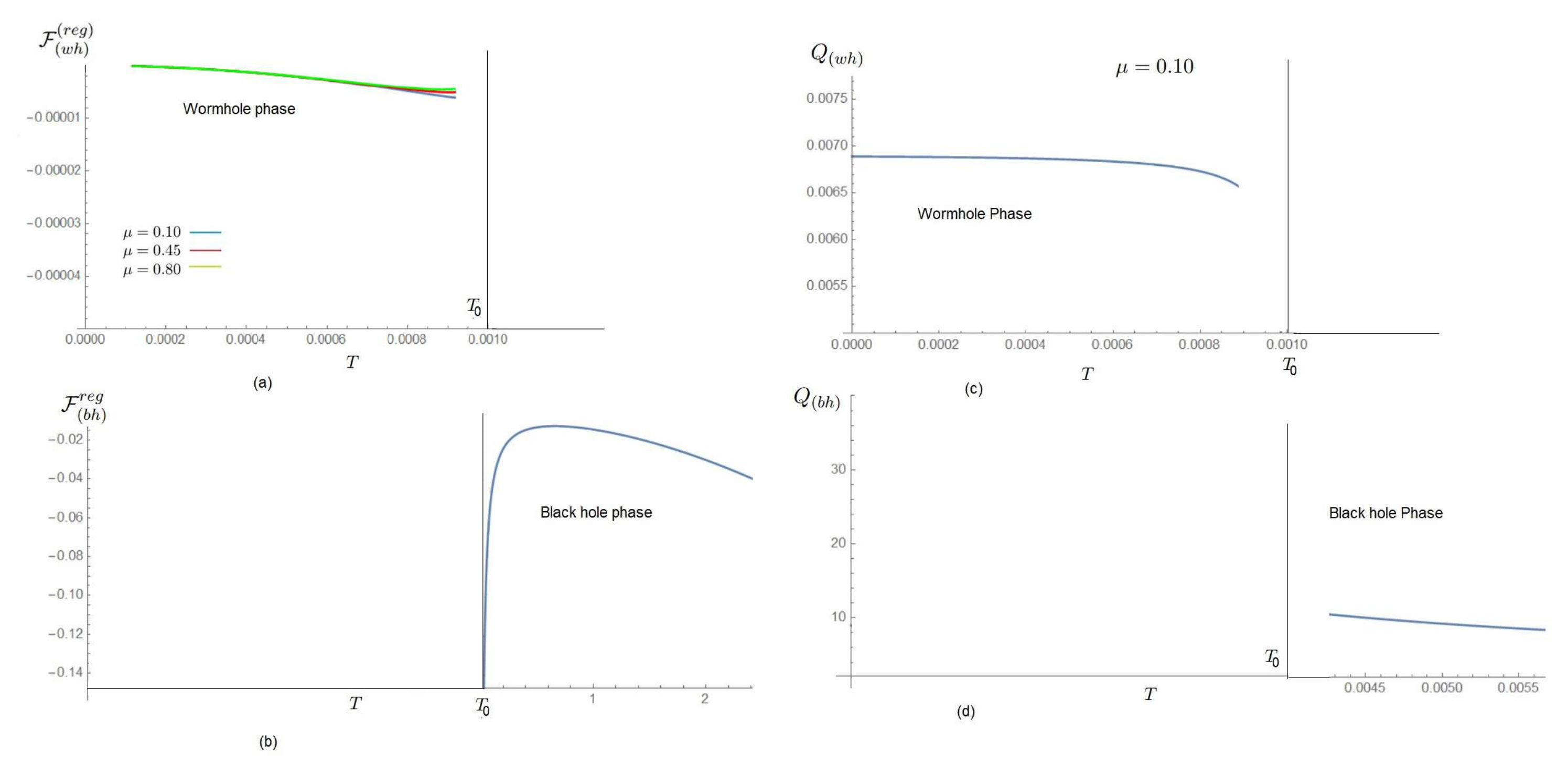}
\caption{Figures (a) and (b) represent the variations of the boundary Free energy densities $(\mathcal{F})$ with temperature ($T$) and at a fixed chemical potential $(\mu)$. On the other hand, the figures (c) and (d) illustrate the variation of the total charge $(Q)$ with temperature ($T$). Here we denote both the wormhole and the black hole temperatures as $T$. We identify the critical temperature as $T_0=\frac{1}{48b\pi}$, where $ b $ is the periodicity of the Euclidean time. Here, we set the basic parameters of the theory as $b=6.63145,a_1=0.009,a_2=0.0009$. These parameters are chosen in order to achieve a best fit for the plot. The choices for the coupling constants ($ a_1 $, $ a_2 \ll 1$) are always less than one in a perturbative expansion (see (\ref{fexp})). Finally, we set the remaining constants as, $d_1=1$ and $d_2=1$.} 
\label{figref2four}
\end{figure}

The  variations of the ``boundary'' Free energy densities ((\ref{freenergy}) and (\ref{bhfreeenery})) are shown in the figures \ref{figref2four}(a) and \ref{figref2four}(b). As these figures reveal, for sufficiently low temperatures $(T<<T_0)$, the regularised Free energy $(\mathcal{F}_{(wh)}^{reg})$ of the charged wormhole  solution remains as a constant indicating the presence of a ``gapped'' phase in the dual (conjectured) two-site complex SYK model at finite chemical potential $(\mu=\mu_0)$.

As the temperature of the system is further increased, we observe a ``discontinuous'' change (see figure \ref{figref2four}(a)) at $T= T_0$, characterising the onset of a \emph{first} order phase transition. On increasing the temperature beyond $T=T_0$, the wormhole phase becomes unstable and passes over to an Euclidean (two) black hole system as shown in figure \ref{figref2four}(b). The dual counterpart of this phase is conjectured to be a ``hot'' complex SYK model at finite chemical potential ($ \mu $). 

We also plot the total charge ($Q$) of the system as a function of temperature ($T$) (see figures \ref{figref2four}(c) and \ref{figref2four}(d)). Notice that, as we move towards the critical temperature ($T= T_0$), we observe a discontinuous jump in the total charge ($Q$) of the system which is quite reminiscent to that of \cite{Garcia-Garcia:2020vyr}. However, unlike in \cite{Garcia-Garcia:2020vyr}, here one should conjecture about a two-site ``uncoupled'' complex SYK rather than a coupled one. 

\section{A qualitative discussion on the conjectured SYK dual }\label{npbnewsecconject}
In this Section, we qualitatively argue the structure of the dual field theory corresponding to the JT gravity set up (\ref{action}). The notable features of this set up is two fold (i) the presence of $U(1)$ gauge field (which sources a chemical potential ($\mu$) for the dual SYK) and (ii) the boundary contributions coming from the bulk Chern-Simons term. The Chern-Simons term plays a significant role while obtaining the wormhole phase in the dual gravity description which is therefore expected to play a vital role while constructing the low temperature phase of the dual SYK model.

Notice that, the present analysis corresponds to the phase stability of (charged) Euclidean wormhole solutions and not  ``traversable wormhole'' solutions of \cite{Maldacena:2018lmt , Garcia-Garcia:2020vyr}. Therefore, it is expected that the field theory (that is dual to the gravitational set up (\ref{action})) should represent a two-site \emph{uncoupled} complex SYK model in the presence of a global $U(1)$ symmetry \cite{Davison:2016ngz , Gaikwad:2018dfc,  Garcia-Garcia:2020vyr, Zhang:2020szi, Bulycheva:2017uqj}.

However, unlike the previous examples  \cite{Davison:2016ngz , Gaikwad:2018dfc,  Garcia-Garcia:2020vyr, Zhang:2020szi, Bulycheva:2017uqj}, the conjectured dual Hamiltonian must contain an additional contribution ($H_{CS}$) due to the bulk Chern-Simons term in (\ref{action}). The dual Hamiltonian could be schematically expressed as
\begin{align}\label{neweomsyk}
    \tilde{H}_{SYK}=\sum_{m=1,2}H_{SYK}^{(m)}+H_{CS},
\end{align}
where $H_{SYK}^{(m)}$ , {($m=1,2$)} is the usual complex (uncoupled) Hamiltonian where the superscript $m$ denotes the number of copies of the SYK model \cite{Davison:2016ngz , Gaikwad:2018dfc,  Garcia-Garcia:2020vyr, Zhang:2020szi, Bulycheva:2017uqj}.

Given the above facts, we conjecture that there exists an Euclidean action corresponding to the dual Hamiltonian (\ref{neweomsyk}) which can be schematically expressed as
\begin{align}\label{npbactioncsyk1}
    S=\int d\tau\Bigg[\frac{1}{2}\psi^{\dag}_i(\partial_{\tau}-\mu)\psi_i-\tilde{H}_{SYK}\Bigg],
\end{align}
and should be thought of as a straightforward generalization of \cite{Gaikwad:2018dfc} in the presence of Chern-Simons contributions. 

The boundary contribution due to the bulk Chern-Simons term can be estimated by expanding the bulk action (\ref{action}) in the near boundary limit which turns out to be 

\begin{align}\label{npbactioncsyk2}
    S_{CS}\sim a_2\int d\tau\mu\Phi_b,
\end{align}
where $\Phi_b$ stands for the boundary value of the bulk dilaton ($\Phi$) that acts as a coupling constant for the dual SYK model under consideration. 

Notice that, here we perform the  bulk calculations using a static gauge (\ref{gauge1}) and therefore $\Phi_b$ does not explicitly depend on the Euclidean time ($\tau$). This suggests that, for the present model, the Chern-Simons contribution (\ref{npbactioncsyk2}) acts as a constant shift to the boundary Hamiltonian (\ref{npbactioncsyk1}).

Now, one can further investigate various thermodynamic properties and in particular the Free energy associated with the (two-site) complex SYK model (\ref{neweomsyk}) and compare it with the Free energy calculations in the dual gravitational description. For this purpose, one requires to define the statistical average of the path integral and the Green's function in the Large N limit associated with the complex SYK model (\ref{neweomsyk}). The next step would be to solve the corresponding Schwinger-Dyson\footnote{Alternatively, one can also compute the Free energy of the system by exact diagonalization technique of the Hamiltonian \cite{Garcia-Garcia:2020ttf}. This method is quite useful over the Schwinger-Dyson approach whenever there is no inter-site coupling between the two copies of the SYK model.} (SD) equations for Green's function and calculate the grand canonical potential of the system \cite{Garcia-Garcia:2020vyr},\cite{Garcia-Garcia:2022xsh}-\cite{Garcia-Garcia:2022adg}. Using the grand potential, one can further estimate the Free energy pertinent to the dual model (\ref{neweomsyk}).  

It is expected that the above Free energy calculation in its infrared limit (or low energy regime) would match with the Free energy of the 2D Einstein-Maxwell-dilaton gravity system (\ref{action}) in the regime of small temperature ($T<<1$) and charge ($Q<<1$) where the chemical potential ($\mu$) is held fixed and set to be small (see Figures \ref{figref2four}(a) and \ref{figref2four}(c)). In other words, the bulk wormhole solutions at low temperature ($T<<1$) and low charge densities ($Q<<1$) should represent the low energy phase of the dual SYK model (\ref{neweomsyk}).

In connection to the above, it is also expected that the Free energy computed on the dual SYK side should exhibit a zero slope (flat region) in its infrared which is similar to that of the JT gravity set up in the regime of small temperature ($T$) and chemical potential ($\mu$) (see Figure \ref{figref2four}(a)). This would indicate the possibilities for finding a ``wormhole phase'' on the SYK counterpart of the duality. It would be indeed an interesting project to explore all the above directions in the near future.

\section{Concluding remarks}\label{secconcnpb}
Now, we conclude the chapter with a brief summary of the key results. In this chapter, we investigate the phases of Euclidean wormhole solutions in the presence of an abelian one form. In other words, the current analysis is a generalisation of the gravitational sector of \cite{Garcia-Garcia:2020ttf} in the presence of a finite charge density and/or chemical potential ($\mu$). The presence of the chemical potential ($\mu$) eventually reveals a richer phase structure which we summarise below.

At low temperatures, the gravitational sector of the system turns out to be a charged wormhole solution. As the temperature is increased beyond $T=T_0$, the wormhole phase undergoes a first order phase transition and transits into a two black hole system at finite charge density. Like boundary Free energy (density), the boundary $ U(1) $ charge $(Q)$ also undergoes a discontinuous change \cite{Garcia-Garcia:2020vyr} at $ T=T_0 $. 

The above discussion on the wormhole to black hole phase transition at finite charge density and/or chemical potential ($\mu$), along with the Hawking-Page (HP) transition discussed in Chapter 1, complete the first part of the thesis.

In the second part of the thesis, we focus more on the QFT observables in the context of the JT/SYK correspondence, where in particular, we focus on estimating the holographic central charge at finite chemical potential ($\mu$).

\includepdf[page=1]{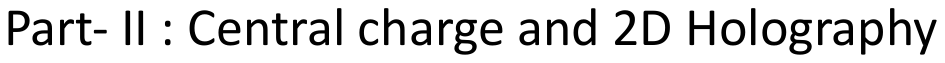}


\chapter{Holographic JT gravity with quartic couplings}
\allowdisplaybreaks
\pagestyle{fancy}
\fancyhead[LE]{\emph{\leftmark}}
\fancyhead[LO]{\emph{\rightmark}}
\rhead{\thepage} 

In this chapter, we construct the most general theory of 2D Einstein-dilaton gravity coupled with $U(1)$ gauge fields that contains all possible 2-derivative and 4-derivative interactions allowed by the diffeomorphism invariance. We renormalise the 2D action and obtain the vacuum solution as well as the black hole solution. The vacuum solution in the UV is dominated due to Lifshitz$_2$ with dynamical exponent $\left(z=\frac{7}{3}\right)$ while on the other hand, the space-time curvature diverges as we move towards the deep IR in the bulk. 
 
 We calculate the holographic stress-energy tensor and the central charge for the  boundary theory. Our analysis shows that the central charge goes as the inverse power of the coupling associated to 4-derivative interactions. We also compute the Wald entropy for 2D black holes and interpret its near horizon divergence in terms of the density of states.
 
 Finally, we explore the near horizon structure of 2D black holes and calculate the central charge corresponding to the CFT near horizon. We further show that the near horizon CFT could be recast
as a 2D Liouville theory with higher derivative corrections.  We study the Weyl invariance of this generalised Liouville theory and identify the Weyl anomaly associated to it.  We also comment on the classical vacuum structure of the theory. 

\section{Introduction and overview}
It has been more than two decades now since the discovery of the celebrated $AdS_{d+1}/CFT_d$ correspondence (also known as gauge/gravity duality) \cite{Maldacena:1997re}-\cite{ Gubser:1998bc}. However, the study of this duality in $d=1$ dimension provides a remarkable insight about models of quantum gravity living in two dimensions \cite{Strominger:1998yg,Hartman:2008dq},\cite{Navarro-Salas:1999zer}-\cite{Azeyanagi:2007bj}. In the present chapter, we focuses on a particular model of classical gravity in two dimensions known as the Jackiw-Teitelboim (JT) gravity \cite{Jackiw:1984je,Teitelboim:1983ux} whose holographic dual is conjectured to be the Sachdev-Ye-Kitaev (SYK) model \cite{Sachdev:1992fk}-\cite{Lala:2019inz}, \cite{Davison:2016ngz}-\cite{Castro:2008ms}, \cite{Bulycheva:2017uqj} which describes the quartic interactions among the Majorana fermions. 
 
 In literature, the duality between the SYK model and JT gravity has been explored in the presence of U(1) gauge fields \cite{Lala:2019inz, Davison:2016ngz, Gaikwad:2018dfc,Castro:2008ms} as well as the SU(2) Yang-Mills fields \cite{Lala:2020lge}. In particular, the authors in \cite{Castro:2008ms} consider the JT gravity model minimally coupled to the U(1) gauge fields and study the holographic stress tensor and the central charge \cite{Hartman:2008dq,Cadoni:2000gm,Castro:2008ms}, \cite{deHaro:2000wj}-\cite{Hohm:2010jc} for the boundary theory.

In this chapter, we look for a further generalisation on the gravitational side of the correspondence by incorporating higher derivative (quartic) interactions in the same spirit as that of its cousins living in higher dimensions\footnote{It was observed in \cite{Utiyama:1962sn} that the 4-derivative interactions are crucial to obtain the finite result of average stress tensor of quantum fields coupled with classical gravity. Higher derivative corrections are also found to be useful in cosmology in order to describe the inflationary models, see \cite{Castellanos:2018dub}-for a recent review.}\cite{Lovelock:1971yv}-\cite{Kraus:2005vz}. We start with the most generic higher derivative theories of gravity in five dimensions \cite{Myers:2009ij} and search for its imprints in the lower dimensional models in the context of SYK/JT gravity correspondence. We follow the standard procedure of dimensional reduction \cite{Davison:2016ngz} which results in the most generic higher derivative theories of gravity (including quartic interactions) in 2D. 

Following the standard AdS/CFT prescription, we compute various physical observables associated  with the dual quantum mechanical model in 1D and explore upon the effects of incorporating the higher derivative corrections on the dual field theory observables. In particular, we compute the holographic stress energy tensor and estimate the central charge associated with the 1D boundary theory.

We also construct the corresponding 2D black hole solutions and explore its thermal properties. Finally, we show that the model of JT gravity with quartic coupling can be recast as a ``generalised'' 2D Liouville theory \cite{Solodukhin:1998tc}-\cite{Mertens:2020hbs} of quantum gravity with some complicated potential function.
  
The organisation and the summary of results of this chapter is as follows :

$\bullet$ In Section (\ref{S2}), following the standard procedure of dimensional reduction, we construct the most general theory of 2D Einstein-dilaton gravity coupled with U(1) gauge fields. Our theory contains all possible 2-derivative as well as the 4-derivative interaction terms allowed by the diffeomorphism invariance. 

$\bullet$ In Section (\ref{S3}), we obtain vacuum solutions of the 2D theory by treating the higher derivative interactions as ``perturbations''. We observe that the scalar curvature corresponding to the 2D theory diverges in the deep IR limit due to the presence of higher derivative interactions. On the other hand, the vacuum solution in the UV limit is dominated by Lifshitz$_2$ with dynamical exponent ($z=\frac{7}{3}$). On the other hand, if we switch off the 4-derivative interactions then, the space-time geometry becomes $AdS_2$ in IR limit and Lifshitz$_2$ (with dynamical exponent $z=\frac{3}{2}$) in UV limit which is consistent with  \cite{Lala:2020lge} . 

$\bullet$ In Section (\ref{S4}), we obtain the Gibbons-Hawking-York (GHY) boundary terms \cite{Gibbons:1976ue},\cite{Hawking:1995fd}-\cite{York:1972sj} for the 2D model that is needed for the successful implementation of the variational principle. Finally, we estimate counter terms which lead to the ``renormalised'' action.

$\bullet$ In Section (\ref{S5}), we use the renormalised action  to determine the boundary stress tensor and the central charge in the Fefferman Graham gauge \cite{fefferman},\cite{deHaro:2000vlm}. We observe that the central charge associated with the boundary theory goes as the inverse power of the quartic coupling ($\kappa$). This further implies that, a smooth $\kappa\rightarrow0$ limit of the central charge does not exist.

$\bullet$ In Section (\ref{SBH}), we obtain black hole solution for the 2D theory by treating the higher derivative interactions as a perturbative corrections over the pure JT gravity solutions.

$\bullet$ In Section (\ref{SBH1}), we explore thermal properties of 2D black holes in our model. In particular, we discuss the Wald entropy \cite{Wald:1993nt, Brustein:2007jj, Pedraza:2021cvx} for 2D black holes and observe that the Wald entropy diverges near the horizon due to the presence of higher derivative interactions. We interpret these divergences in terms of the density of states \cite{thooft, Brustein:2010ms}. 

$\bullet$ In Section (\ref{SS8}), we investigate the near horizon structure of 2D black holes in the presence of quartic interactions. We observe that the trace of the stress tensor vanishes in the near horizon limit which indicates the presence of a conformal field theory in the vicinity of the horizon. Finally, we transform the 2D theory into the ``generalised'' Liouville theory \cite{Solodukhin:1998tc}-\cite{Mertens:2020hbs} using the proper field re-definition and calculate the associated central charge. We observe that the central charge corresponding to the generalised Liouville theory diverges due to the presence of higher derivative interactions. 

$\bullet$ In Section (\ref{GLT}), we discuss the Weyl transformation \cite{Jackiw:2005su} properties of the generalised Liouville theory. We observe that the generalised Liouville theory is not invariant under the Weyl re-scaling. On top of it, the trace of its stress tensor does not vanish and comes out to be proportional to its central charge. We identify this as the Weyl (or trace) anomaly \cite{Castro:2019vog} for the generalised Liouville theory.

$\bullet$ Finally, we conclude the chapter in Section (\ref{sum}).

\section{Construction of the \texorpdfstring{$2D$}{2D} action }\label{S2}
The purpose of this section is to discuss the basic methodology that leads to the most general 2D action for Einstein-dilaton gravity coupled to U(1) gauge fields. We start with the most general theory of Einstein gravity coupled with U(1) gauge fields in five-dimensions \footnote{See Appendix \ref{Non-abelian} for a discussion on the Non-abelian sector.} \cite{Myers:2009ij}.
\begin{align}\label{action 5D}
    S_{(5D)}=&\int d^5x\sqrt{-g_{(5)}}\Big [(12+R)-\frac{\eta_1}{4}F^2+\eta_2[R_{MNOP}]^2+\eta_3 F^4+\eta_4F^{SP}F_{PR}\times\nonumber\\
&F^{RQ}F_{QS} +\eta_5\bigtriangledown_MF^{MN}\bigtriangledown^OF_{ON}+\epsilon^{MNOPQ}\Big(\eta_{6}F_{MN}F_{OP}\bigtriangledown^RF_{RQ}+\nonumber\\
&\eta_{7}F_{MN}F_{OR}\bigtriangledown^RF_{PQ}+\eta_{8}F_{MN}F_{OR}\bigtriangledown_P{F_Q}^R+\eta_{9}A_MR_{NOIJ}{R_{PQ}}^{IJ}\Big) \Big ]
\end{align}
where, $\eta_i$ $(i=1,..,9)$ are the respective coupling constants. The key feature of this model is that it contains the 4-derivative interaction terms along with the usual 2-derivative interactions. These higher derivative terms are the key contents of our model.

In principle, it is possible to add several other 4-derivative terms to the above action (\ref{action 5D}). However, all such terms can be eliminated using a proper redefinition of fields as demonstrated in Appendix \ref{gen action 5D}. Therefore, the action (\ref{action 5D}) is the most general theory of gravity (coupled to U(1) gauge fields) containing both 2-derivative and 4-derivative interaction terms.

We are interested in studying the JT gravity model with chemical potential in the context of $AdS_2/CFT_1$ correspondence. On that note, we will require to get rid of the extra dimensions present in the 5D theory (\ref{action 5D}).

Systematically, this can be achieved following a reduction ansatz for the metric as well as the gauge field \cite{Davison:2016ngz}
\begin{align}\label{ansatznewjh}
    ds^2_{(5)}&=ds^2_{(2)}+\phi(t,z)^{\frac{2}{3}}(dx^2+dy^2+dz^2),\hspace{2mm}A_Mdx^M=A_{\mu}dx^{\mu},\hspace{2mm} A_{\mu}\equiv A_{\mu}(x^{\nu})
\end{align}
where M is the 5 dimensional index and $\mu$ stands for the 2 dimensional space-time index.

Using the above ansatz (\ref{ansatznewjh}), one arrives at the required Einstein Hilbert action in 2 dimensions\footnote{See Appendix \ref{covariance} for a detailed discussion on the general covariance of the action.}
\begin{align}\label{action 2D}
S_{EH} =& \int d^2x\sqrt{-g_{(2)}}\phi \Bigg [\frac{12}{L^2}+R-\frac{\xi}{4}F^2+\kappa L^2\Big[(R_{\mu\nu\alpha\beta})^2+\frac{3}{4}\Big(\bigtriangledown_{\mu}\phi^{\frac{2}{3}}\Big)^4\nonumber+\\
&4\Big\{\frac{3}{4}\Big(\bigtriangledown_{\lambda}\phi^{\frac{2}{3}}\Big)\Big(\bigtriangledown_{\beta}\phi^{\frac{2}{3}}\Big)\Gamma^{\lambda}_{\alpha\mu}\Gamma^{\beta}_{\rho\sigma}g^{\alpha\rho}g^{\mu\sigma}+\frac{2}{3}\Gamma^{\lambda}_{\alpha\mu}\Big(\bigtriangledown_{\lambda}\phi^{\frac{2}{3}}\Big)(\bigtriangledown^{\alpha}\phi)(\bigtriangledown^{\mu}\phi)\phi^{\frac{-4}{3}}\nonumber\\
&-\Gamma^{\lambda}_{\alpha\mu}\Big(\bigtriangledown_{\lambda}\phi^{\frac{2}{3}}\Big)\{\partial_{\beta}(\bigtriangledown_{\sigma}\phi)\}g^{\alpha\beta}g^{\mu\sigma}\phi^{\frac{-1}{3}}-\frac{4}{9}\phi^{\frac{-5}{3}}(\bigtriangledown^{\alpha}\phi)(\bigtriangledown^{\mu}\phi)\{\partial_{\alpha}(\bigtriangledown_{\mu}\phi)\}+\nonumber\\
&\frac{4}{27}(\bigtriangledown_{\mu}\phi)^4\phi^{\frac{-8}{3}}-\frac{1}{3}\{\partial_{\alpha}(\bigtriangledown_{\mu}\phi)\}\{\partial_{\beta}(\bigtriangledown_{\rho}\phi)\}g^{\alpha\beta}g^{\mu\rho}\phi^{\frac{-2}{3}}\Big\}+F^4+F^{\mu\nu}F_{\nu\lambda}F^{\lambda\sigma}F_{\sigma\mu}\nonumber\\
&+\bigtriangledown_{\mu}F^{\mu\nu}\bigtriangledown^{\lambda}F_{\lambda\nu}\Big] \Bigg ],
\end{align}
where we introduce the length scale $L$ using dimensional analysis.

Notice that, in order to arrive (\ref{action 2D}), we make a special choice of coupling constants namely, $\eta_1=\xi$ and $\eta_2=\eta_3=\eta_4=\eta_5=\kappa $. Furthermore we treat these dimensionless coupling constants ($\xi$ and $\kappa$) to be small enough such that the 2-derivative and 4-derivative interaction terms can be treated as pertubations over pure JT gravity. On variation of (\ref{action 2D}) one arrives at the following structure  
\begin{eqnarray}
\delta S_{EH} &=&\int d^2x\sqrt{-g_{(2)}}[H_{\mu\nu}\delta g^{\mu\nu}+H_{\phi}\delta\phi+H^{\mu}\delta A_{\mu}].
\end{eqnarray}

Equations of motion for the metric, dilaton and the gauge field in bulk will be given by equating  the individual coefficients $\sqrt{-g}H_{\mu\nu}$, $\sqrt{-g}H_{\phi}$ and $\sqrt{-g}H_{\mu}$ to zero. Technically speaking, it will be easy to handle these equations using the static gauge given below
\begin{eqnarray}\label{sg}
ds^2= e^{2\omega(z)}(-dt^2+dz^2), \hspace{3mm}A_{\mu}=(A_{t}(z),0).
\end{eqnarray}
\section{Vacuum solutions}\label{S3}
Even in the static gauge, it is difficult to solve the bulk equations of motion exactly. Therefore, we will prefer to solve these equations perturbatively treating $\xi$ and $\kappa$ as an expansion parameter. 

Systematically, one can expand these fields in terms of the expansion parameters as shown in equation (\ref{e1})-(\ref{e3})
\begin{eqnarray}
 \omega&=&\omega_{(0)}+\xi\omega_{(1)}+\kappa\omega_{(2)},\label{e1}\\
\phi&=&\phi_{(0)}+\xi\phi_{(1)}+\kappa\phi_{(2)},\label{e2}\\
A_{t} &=& A_{t(0)}+\frac{\kappa}{\xi} A_{t(1)}\hspace{1mm},\hspace{3mm}\Big|\frac{\kappa}{\xi}\Big|<<1.\label{e3}
\end{eqnarray}

 In the above equation, the subscript (0) in ($\phi$, $\omega$) denotes the pure JT gravity fields whereas subscripts (1) and (2) denote the contributions coming from 2-derivative and 4-derivative interaction terms in (\ref{action 2D}). Notice that, the expansion of the gauge field ($A_t$) is different from $\phi$ and $\omega$ because it is absent in pure JT gravity theory. Gauge fields start appearing in the action as 2-derivative and 4-derivative interaction with coupling constants $\xi$ and $\kappa$ respectively. Therefore, the subscripts (0) and (1) in $A_t$ denote the contributions due to 2-derivative and 4-derivative interaction terms respectively. Finally, using equation (\ref{e1})-(\ref{e3}) we expand the coefficients $H_{\phi}$ , $H_{\mu}$ and $H_{\mu\nu}$ as follows
\begin{align}\label{ee}
H_{\phi}=H_{\phi}^{(0)}+\xi H_{\phi}^{(2)}+\kappa H_{\phi}^{(4)},\hspace{1mm} H_{\mu}=H_{\mu}^{(2)}+\frac{\kappa}{\xi}H_{\mu}^{(4)},\hspace{1mm}
H_{\mu\nu}=H_{\mu\nu}^{(0)}+\xi H_{\mu\nu}^{(2)}+\kappa H_{\mu\nu}^{(4)}.
\end{align}
Here, the superscript (0) denotes the contribution due to JT gravity. On the other hand, the superscripts (2) and (4) denote the contributions due to 2-derivative and 4-derivative interaction terms.

The action constructed in equation (\ref{action 2D}) exhibits both vacuum solution as well as black hole solution. In this section, we study vacuum solution in detail. The general plan is to solve equations (\ref{ee}) at different order in perturbation as discussed in the following subsections.
\subsection{Zeroth order solutions}
In order to find out the vacuum solutions $\omega_{(0)}^{vac}$ and $\phi_{(0)}^{vac}$ , we set $\xi=\kappa=0$ in $\sqrt{-g}H_{\mu\nu}$, $\sqrt{-g}H_{\phi}$ and $\sqrt{-g}H^{\mu}$. This yields the following set of equations
\begin{eqnarray}
\phi_{(0)}''-\omega_{(0)}'\phi_{(0)}'-\frac{6}{L^2}\phi_{(0)}e^{2\omega_{(0)}}=0,\label{v01}\\
 \omega_{(0)}'\phi_{(0)}'-\frac{6}{L^2}\phi_{(0)}e^{2\omega_{(0)}}=0,\label{v02}\\
 \frac{12}{L^2}- 2e^{-2\omega_{(0)}}\omega_{(0)}'' = 0.\label{v03}
\end{eqnarray}

On solving (\ref{v01}), (\ref{v02}) and (\ref{v03}) we get
\begin{eqnarray}\label{v0s}
e^{2\omega_{(0)}^{vac}} =\frac{1}{6z^2}, \hspace{3mm}\phi_{(0)}^{vac}=-\frac{C_1}{z}
\end{eqnarray}
where $C_is$ are the dimenionsionful integration constants, for instance $C_1$ have a dimension of length $L$. Equation (\ref{v0s}) stands for the vacuum solutions of pure JT gravity.  
\subsection{First order solutions in \texorpdfstring{$\xi$}{xi} }
Next, we note down leading order solutions (due to 2-derivative terms) by equating the coefficient of $\xi$ in $\sqrt{-g}H_{\mu\nu}$, $\sqrt{-g}H_{\phi}$ and $\sqrt{-g}H^{\mu}$ to zero
\begin{align}
2\Big(\omega_{(0)}'\phi_{(1)}'+\omega_{(1)}'\phi_{(0)}'+2\omega_{(1)}\phi_{(0)}'\omega_{(0)}'\Big)-(\phi_{(1)}''+2\omega_{(1)}\phi_{(0)}'')=0,\label{v21}\\
\frac{12}{L^2}e^{2\omega_{(0)}}\omega_{(1)}-\omega_{(1)}''+\frac{1}{4}e^{-2\omega_{(0)}} ({A_{t(0)}}')^2=0,\label{v22}\\
\partial_z\Big[\phi_{(0)}e^{-2\omega_{(0)}}{A_{t(0)}}'\Big]=0,\label{v23}
\end{align}
where (\ref{v21}) is corresponding to $\sqrt{-g}(H_{tt}+H_{zz})$. On solving equation (\ref{v21})-(\ref{v23}) we find, 
\begin{eqnarray}
&&A_{t(0)}^{vac}=-\frac{L^2C_3}{C_1}\log z+C_4,\label{vs21}\\
&&\omega_{(1)}^{vac}=C_5z^2+\frac{C_6}{z}+\frac{z^2C_3^2(-1+3\log z )L^2}{6C_1^2},\label{vs22}\\
&&\phi_{(1)}^{vac}= \frac{L^2C_3^2z(-4+3\log z)}{3C_1}+C_1\Big(2C_5z-\frac{C_6}{z^2}\Big)-\frac{C_7}{z} + C_8,\label{vs23}
\end{eqnarray}
 To summarise, (\ref{vs21})-(\ref{vs23}) are the first order corrections to the pure JT gravity solutions due to 2-derivative interactions present in (\ref{action 2D}).
\subsection{First order solutions in \texorpdfstring{$\kappa$ }{kappa}}
Next, we note down leading order contributions due to the presence of 4-derivative interactions in (\ref{action 2D}). This can be calculated by equating the coefficients of $\kappa$ in $\sqrt{-g}H_{\mu\nu}$, $\sqrt{-g}H_{\phi}$ and $\sqrt{-g}H^{\mu}$ to zero.
\begin{align} 
&\partial_z\Big [-e^{-2\omega_{(0)}}\phi_{(0)} A_{t(1)}'-24L^2e^{-6\omega_{(0)}}({A_{t(0)}}')^3\phi_{(0)}+ 2L^2\partial_z\big\{e^{-4\omega_{(0)}}\phi_{(0)}(2\omega_{(0)}'{A_{t(0)}}'\nonumber\\
&-{A_{t(0)}}'')\big\}\Big ]=0 ,\label{v41}\\
&\frac{24}{L^2}e^{2\omega_{(0)}}\omega_{(2)}-2\omega_{(2)}''+ \frac{24}{z^2}-\frac{5552}{27z^2}\Big(-\frac{C_1}{z}\Big)^{\frac{4}{3}}-\frac{40}{3z^2}\Big(-\frac{C_1}{z}\Big)^{\frac{8}{3}}
-\frac{3888z^2C_3^4L^4}{C_1^4}-\nonumber\\
&\frac{L^2C_3}{C_1^2}\Big(108C_3+C_9\Big)=0,\label{v42} \\
&\frac{6}{L^2}e^{2\omega_{(0)}}(\phi_{(2)}+4\omega_{(2)}\phi_{(0)})+(\phi_{(0)}'\omega_{(2)}'+\phi_{(2)}'\omega_{(0)}'+2\omega_{(2)}\phi_{(0)}'\omega_{(0)}')-(\phi_{(2)}''+2\omega_{(2)}\phi_{(0)}'')-\nonumber\\
&\frac{1}{18z^5C_1^3}\Bigg [216z^2C_1^4+976zC_1^5\Big(-\frac{C_1}{z}\Big)^{\frac{1}{3}}+8C_1^6\Big(-\frac{C_1}{z}\Big)^{\frac{2}{3}} +11664z^6C_3^4L^4+\nonumber\\
&9z^4C_1^2C_3C_9L^2\Bigg ]=0,\label{v43}\\
 &-\frac{6}{L^2}e^{2\omega_{(0)}}(\phi_{(2)}+4\omega_{(2)}\phi_{(0)})+(\phi_{(0)}'\omega_{(2)}'+\phi_{(2)}'\omega_{(0)}'+2\omega_{(2)}\phi_{(0)}'\omega_{(0)}')+\frac{1}{18z^3}\Bigg [216C_1+\nonumber\\
&24z\Big(-\frac{C_1}{z}\Big)^{\frac{11}{3}}+16z\Big(-\frac{C_1}{z}\Big)^{\frac{7}{3}}\Big\{-133+30\Big(-\frac{C_1}{z}\Big)^{\frac{1}{3}}\Big\}+11664z^4\frac{C_3^4L^4}{C_1^3}+\nonumber\\
&9z^2\frac{L^2C_3}{C_1}(72C_3+C_9)\Bigg]=0.\label{v44}
\end{align}

Notice that, (\ref{v44}) contains only single derivative terms which means that it is a constraint equation. We will use this constraint and equation (\ref{v0s}) in order to find $\phi_{(2)}^{vac}$ from equation (\ref{v43}). On the other hand, $A_{t(1)}^{vac}$ and $\omega_{(2)}^{vac}$ can be calculated using equations (\ref{v41}) and (\ref{v42}) respectively
\begin{align}
A_{t(1)}^{vac} = & \frac{L^4432 z^2 C_3^3}{C_1^3}+\frac{L^2\log z(72C_3+C_9)}{6C_1}+C_{10},\label{v4s1}\\
\omega_{(2)}^{vac} = & C_{11}z^2+\frac{C_{12}}{z}-6-\frac{1388}{15}\Big(-\frac{C_1}{z}\Big)^{\frac{4}{3}}-\frac{z^2}{18C_1^2}(-1+3\log z)C_3(108C_3+C_9)L^2\nonumber\\
&-\frac{6}{7}\Big(-\frac{C_1}{z}\Big)^{\frac{8}{3}}-\frac{972}{5C_1^4}z^4C_3^4L^4,\label{v4s2}\\
\phi_{(2)}^{vac} =& \frac{z^3C_{13}+C_{14}}{z}+\frac{2C_1^3}{7z^3}\Big(-\frac{C_1}{z}\Big)^{\frac{2}{3}}-\frac{12}{455}\Big(-\frac{C_1}{z}\Big)^{\frac{7}{3}}\Big\{2009+130\Big(-\frac{C_1}{z}\Big)^{\frac{1}{3}}\Big\}-\nonumber\\
&\frac{648z^3L^4C_3^4}{5C_1^3}+C_1\Big(2zC_{11}-\frac{C_{12}}{z^2}\Big)+\frac{L^2}{9C_1}\Big[zC_3\{108(1-3\log z)C_3+\nonumber\\
&(4-3\log (z)C_9)\}\Big].\label{v4s3}
\end{align}
Equations (\ref{v4s1})-(\ref{v4s3}) are the first order corrections to pure JT gravity due to 4-derivative interactions in (\ref{action 2D}).

Now, we have a complete set of solutions corresponding to metric, gauge fields and dilaton up to linear order in $\xi$ and $\kappa$. Collecting all these fields at different order, we can approximate the space-time metric (\ref{sg}) for vacuum solution as
\begin{align}\label{VM}
    ds_{vac}^2\approx e^{2\omega_{(0)}^{vac}}(1+2\xi\omega_{(1)}^{vac}+2\kappa\omega_{(2)}^{vac})(-dt^2+dz^2).
\end{align}

Below, we check the behaviour of space-time metric in two different limits -
\begin{itemize}
    \item  Case 1 : IR limit i.e. $z\rightarrow\infty$
\begin{align}
e^{2\omega^{vac}}=&\frac{L^2}{6z^2}+\xi L^2\Big\{\frac{-3C_3^2L^2+9\log (z) C_3^2L^2+18C_1^2C_5}{54C_1^2}+\frac{C_6}{3z^3}\Big\}+\kappa L^2\Big\{-\frac{2}{z^2}-\nonumber\\
&\frac{324z^2C_3^4 L^4}{5C_1^4}+\frac{1}{54C_1^2}\Big(108C_3^2L^2-324L^2\log (z)C_3^2+C_3C_9L^2\nonumber\\
&-3L^2\log(z)C_3C_9+18C_1^2C_{11}\Big)+\frac{1}{45z^3}\Big(1388C_1\Big(-\frac{C_1}{z}\Big)^{\frac{1}{3}}+15C_{12}\Big)\Big\},\label{vir}
\end{align}
\item Case 2 : UV limit i.e. $z\rightarrow0$
\begin{align}
e^{2\omega^{vac}}=&\frac{L^2}{6z^2}+\xi L^2\Big\{\frac{-3C_3^2L^2+9L^2\log(z)C_3^2+18C_1^2C_5}{54C_1^2}+\frac{C_6}{3z^3}\Big\}+\kappa L^2\Big\{-\frac{2}{z^2}-\nonumber\\
&\frac{2C_1^2(-C_1)^{\frac{2}{3}}}{7z^{\frac{14}{3}}}-\frac{324z^2C_3^4L^4}{5C_1^4}+\frac{1}{54C_1^2}\Big(108C_3^2L^2-324L^2\log(z)C_3^2+\nonumber\\
&C_3C_9L^2-3\log(z)C_3C_9L^2+18C_1^2C_{11}\Big)+\frac{1}{45z^3}\Big(1388C_1\Big(-\frac{C_1}{z}\Big)^{\frac{1}{3}}+\nonumber\\
&15C_{12}\Big)\Big\}.\label{vuv}
\end{align}
\end{itemize}

It is evident from (\ref{vir}) and (\ref{vuv}) that the 2-derivative and 4-derivative interaction terms present in our model alter the $AdS_2$ geometry of vacuum both in the UV and IR limits. In the UV limit (\ref{vuv}), the space-time geometry is dominated by the Lifshitz$_2$ with dynamical exponent $z=\frac{7}{3}$. On the other hand, the space-time metric exhibits a divergence as we move in the deep IR limit (\ref{vir}). 

In order to solidify our claim, we further compute the corresponding scalar curvature of the theory (\ref{action 2D}) which shows a divergence in the deep IR namely,
$$R\big|_{z\rightarrow \infty}\sim\kappa\Big(\frac{C_3}{C_1}\Big)^4z^4L^2,$$
where $\frac{C_3}{C_1}$ is precisely the coefficient that appears in the near boundary expansion of (\ref{VM}). This clearly reveals the fact that the space-time singularity is caused due to the presence of 4-derivative interactions in the original action (\ref{action 2D}). We identify this as the unique feature of higher derivative corrections in the theory (\ref{action 2D}).
\section{Boundary terms and renormalised action }\label{S4}
The boundary of space-time manifold in our theory (\ref{action 2D}) is located at $z=0$. Therefore, one must add  suitable boundary terms in action for a successful execution of variational principle \cite{Castro:2008ms}. 

The boundary term is given by standard Gibbons-Hawking-York term
\begin{eqnarray}\label{GHY}
S_{GHY}&=&D_1\int_0^{\beta} dt\sqrt{-\gamma}\phi K , \hspace{2mm} K=n^z\frac{\partial_z\sqrt{-\gamma}}{\sqrt{-\gamma}},\hspace{3mm}n^z=-\frac{1}{\sqrt{g_{zz}}},
\end{eqnarray} 
where $\gamma$ is the determinant of induced metric on boundary, $K$ is the trace of extrinsic curvature and $\beta$ is the inverse temperature \cite{Gibbons:1976ue}. We multiply the boundary term (\ref{GHY}) with an overall dimensionless constant $D_1$ which will prove to be useful in construction of counter terms.

On substituting equation (\ref{VM}) in (\ref{GHY}), we obtain  $S_{GHY}=-\beta(\phi \omega')D_1$. Using this expression, one can easily write down the on-shell Gibbons-Hawking-York boundary term as well as the on-shell Einstein-Hilbert action (\ref{action 2D}) as follows
\begin{align}
S_{GHY}^{on}=&-D_1 \beta \Bigg [\frac{C_1}{z^2}+\frac{103172\kappa  C_1^2}{585z^3}\Big(-\frac{C_1}{z}\Big)^{\frac{1}{3}}+\frac{24\kappa C_1^2}{7z^3}\Big(-\frac{C_1}{z}\Big)^{\frac{2}{3}}-\frac{18\kappa C_1^3}{7z^4}\Big(-\frac{C_1}{z}\Big)^{\frac{2}{3}}\nonumber\\
&-\frac{6\kappa L^2 C_3^2}{C_1}+\frac{7L^2\xi C_3^2}{6C_1}+\frac{72L^2\kappa \log(z)C_3^2}{C_1}-\frac{2L^2\xi \log(z)C_3^2}{C_1}+\frac{4536L^4z^2\kappa C_3^4}{5C_1^3}+\nonumber\\
&\frac{2\xi C_1C_{6}}{z^3}+\frac{\xi C_7}{z^2}-\frac{\xi C_8}{z}-\frac{7\kappa C_3C_9L^2}{18C_1}+\frac{2\kappa \log(z)C_3C_9L^2}{3C_1}-4\kappa C_1 C_{11}+\frac{2\kappa C_1 C_{12}}{z^3}\nonumber \label{SGHY}\\
&-4\xi C_1C_{5}-z\kappa C_{13}-\frac{\kappa C_{14}}{z^2}\Bigg],\\
S_{EH}^{on}=& -\frac{\beta \kappa}{27 C_1^3}\Bigg [\frac{6096C_1^5}{5z^3}\Big(-\frac{C_1}{z}\Big)^{\frac{1}{3}}+\frac{576C_1^5}{7z^3}\Big(-\frac{C_1}{z}\Big)^{\frac{5}{3}}+69984L^4z^2C_3^4 +27\log(z)\times\nonumber\\
&L^2C_1^2C_3(72C_3+C_9) \Bigg ],\label{SEH}
\end{align}
where we have truncated the above expressions (\ref{SGHY}) and (\ref{SEH}) up to linear order in $\xi$ and $\kappa$.

It should be noted that in the boundary limit i.e. $z\rightarrow 0$, both the equations (\ref{SGHY}) and (\ref{SEH}) diverge. Therefore, one requires to add counter terms in the action (\ref{action 2D}) to tame such UV divergences. These counter terms should be some function of the fields at boundary.

After a careful inspection, we come up with the following counter term
\begin{align}\label{SCT}
S_{CT}=& \int_{0}^{\beta}dt\sqrt{-\gamma}\Bigg[d_1\phi+d_2\sqrt{-\gamma}K^2+d_3\xi\sqrt{-\gamma^{\mu\nu}A_{\mu}A_{\nu}}+\xi \frac{d_4}{\sqrt{-\gamma}}\phi^3 \nonumber\\
&+\kappa \frac{d_5}{\sqrt{-\gamma}}\phi^3+\kappa\frac{d_6}{\sqrt{-\gamma}}\phi^2\Bigg ],
\end{align}
where $d_i$, $i=1,2..,6$ are some dimensionful constant coefficients.

Equation (\ref{SCT}) cures all the UV divergences of $S_{EH}^{on}+S_{GHY}^{on}$ (up to linear order in $\xi$ and $\kappa$) with a particular choice of coefficients\footnote{See Appendix \ref{const} for a detailed derivation of the coefficients. }
\begin{align}
    &d_1=\frac{6.3186}{L},\hspace{1mm}d_2=-C_10.1394,\hspace{1mm}d_3=-C_33.5904,\hspace{1mm}d_4=-\frac{C_6}{C_1^2}0.2788,\hspace{1mm}d_5=-\frac{C_{12}}{C_1^2}0.2788\nonumber\\
   & d_6=-\frac{0.16}{\kappa C_1}\left(-0.871+96.6\kappa\right),\hspace{1mm}D_1=-2.5795,\hspace{1mm}C_3=0.0283C_9,
\end{align}

where the constants $C_1$ have dimensions of length $L$, $C_3$ and $C_9$ have dimensions of length $L^{-1}$, and $C_6$ and $C_{12}$ have dimensions of length $L$.

With all these preliminaries, the complete renormalised  action can be schematically expressed as
\begin{align}\label{S2D}
    S_{2D}=S_{EH}+S_{GHY}+S_{CT}.
\end{align}

The variation of the full action (\ref{S2D}) is given by
\begin{align}\label{VS2D}
\delta S_{2D}=\int dt\sqrt{-\gamma}\Big[G^{ab}\delta \gamma_{ab}+G_{\phi}\delta{\phi}+G^a\delta A{_a}\Big ] +\hspace{1mm} bulk \hspace{2mm}terms
\end{align} 
where  (a,b) are the boundary indices. 

The bulk part of (\ref{VS2D}) is already discussed in Sections (\ref{S2}) and (\ref{S3}). On the other hand, the variation of boundary action yields
\begin{align}
G_{\phi}=&D_1K+Ld_1+3\xi\frac{d_4}{\sqrt{-\gamma}}\phi^2+3\kappa\frac{d_5}{\sqrt{-\gamma}}\phi^2 ,\\
G^t=&-\xi n_{\alpha} F^{\alpha t}\phi-\xi d_3\frac{\gamma^{tt}A_t}{\sqrt{-\gamma^{tt}A_tA_t}},\\
G^{tt}=&\frac{1}{\sqrt{g_{zz}}}\Big\{\partial_z\gamma^{tt}\phi-\partial_z\phi\gamma^{tt}+\frac{2}{z}\gamma^{tt}\phi\Big\}+\frac{D_1}{2}\frac{\phi}{\sqrt{g_{zz}}}\Big\{-\partial_z\gamma^{tt}-\frac{2}{z}\gamma^{tt}-\nonumber\\
&\frac{1}{\sqrt{-\gamma}}\gamma^{tt}\partial_z\sqrt{-\gamma}\Big\}+\frac{Ld_1}{2}\gamma^{tt}\phi+d_2\frac{K}{\sqrt{g_{zz}}}\Big\{-\partial_z\gamma^{tt}\sqrt{-\gamma}-\frac{2}{z}\gamma^{tt}\sqrt{-\gamma}-\nonumber\\
& \gamma^{tt}\partial_z\sqrt{-\gamma}\Big\}+d_3\frac{\xi}{2}\Bigg\{\gamma^{tt}\sqrt{-\gamma^{tt}A_t A_t}-\frac{A^t A^t}{\sqrt{-\gamma^{tt}A_tA_t}}\Bigg\}.\label{bgtt}
\end{align}

In arriving at equation (\ref{bgtt}), we have used (\ref{GHY}) and the dominating terms in the expansion of $\delta\gamma_{tt}$ near boundary. It is important to note that the above variation (\ref{VS2D}) makes sense only when the individual variations of the metric ($\delta\gamma_{ab}$), dilaton $(\delta\phi)$ and the gauge ($\delta A_a$) field vanishes at the boundary.

In order to check this explicitly we expand the variation of all fields near boundary which yield
\begin{align}
\delta\phi=&\frac{1}{9C_1}\Big[z\{3L^2(36\kappa-4\xi+3(-36\kappa+\xi)\log(z))C_3^2+\kappa L^2(4-3\log(z))C_3C_9+\nonumber \\
&18C_1^2(\xi C_5+\kappa C_{11})\}\Big ]+O[z]^2,\label{ve1}\\
\delta A_t=&432\frac{C_3^3}{C_1^3}\frac{\kappa}{\xi}z^2L^4+O[z]^3\hspace{2mm},\hspace{5mm}\delta\gamma_{tt}= \kappa \frac{324}{5}\frac{C_3^4}{C_1^4}z^2L^6+O[z]^3.\label{ve2}
\end{align}

From equations (\ref{ve1}) and (\ref{ve2}), it is quite evident that the individual variation of fields $\delta\phi$, $\delta A_t$ and $\delta \gamma_{tt}$ vanishes in the boundary limit $z\rightarrow 0$.  Therefore, the results derived above are all reliable and we will use them in deriving the boundary stress tensor in the next section.
\section{Stress tensor and central charge}\label{S5}
Having done the required background work, we now proceed towards computing the stress  tensor as well as the central charge for the boundary theory. Boundary stress tensor is defined as the variation of the action (\ref{S2D}) with respect to the induced metric ($\gamma_{ab}$)
\begin{eqnarray}\label{std}
T^{ab}&=&\frac{2}{\sqrt{-\gamma}}\frac{\delta S_{2D}}{\delta\gamma_{ab}}=2G^{ab},
\end{eqnarray}
where $G^{ab}$ is given by equation (\ref{bgtt}).

So far, our computations have been performed in the light cone gauge (\ref{sg}). However, it is not convenient to identify the central charge in this gauge. Therefore, we switch to so called Fefferman Graham gauge \cite{fefferman} in which it is quite straightforward to figure out the central charge.
\subsection{The Fefferman-Graham gauge }
In this section, we will demonstrate how to write down the background fields in the Fefferman-Graham gauge. In order to do that, we first make a coordinate transformation that takes us into the Fefferman-Graham gauge from the light cone gauge. This can be done as follows.

Consider the line element in light cone gauge   
\begin{eqnarray}\label{lm}
ds^2=-e^{2\omega(z)}dt^2+e^{2\omega(z)}dz^2.
\end{eqnarray}

Now, consider the following transformation
\begin{eqnarray}
d\eta &=&e^{\omega(z)}dz,
\end{eqnarray}
which by virtue of (\ref{e1}) and (\ref{v0s}) yields
\begin{eqnarray}\label{fgge}
\eta &=&\int\frac{1}{\sqrt{6}z}(1+\xi \omega_{(1)}+\kappa\omega_{(2)})dz.
\end{eqnarray}

 In principle, one can evaluate (\ref{fgge}) using equation (\ref{vs22}) and (\ref{v4s2}).  This will give us $\eta$ as a function of $z$ i.e. $\eta\equiv\eta(z)$. One can therefore revert (\ref{fgge}) to express $z$ as a function of $\eta$  and plug it back into equation (\ref{lm}). This yields the desired form of the line element in the Fefferman-Graham gauge\footnote{It is noteworthy to mention that various forms of the Fefferman-Graham metric exist in the literature \cite{Castro:2008ms}, \cite{deHaro:2000vlm}. For example, in \cite{deHaro:2000vlm}, the authors use the form $$ds^2=\frac{l^2}{r^2}\left(dr^2+g_{ij}(x,r)dx^idx^j\right).$$
However, both forms are equivalent, and one can derive the Fefferman-Graham form of (\ref{fgm}) in 2D from the above form using a suitable coordinate transformation, i.e., $r=le^{\eta/l}$, and identifying $h_{tt}=e^{-2\eta/l}g_{tt}$.} \cite{Castro:2008ms}
\begin{eqnarray}\label{fgm}
ds^2=h_{tt}(\eta)dt^2+d\eta^2.
\end{eqnarray}

In order to simplify our analysis further, we expand equation (\ref{fgge}) in the boundary limit ($z\rightarrow0$) and retain only dominating terms in the expansion. Notice that, the boundary in the  Fefferman-Graham gauge is located at $\eta=\infty$. 

Upon solving equation (\ref{fgge}) and expressing $z$ as a function of $\eta$ we get 
\begin{eqnarray}\label{zfg}
z= \frac{3^{\frac{9}{16}}\kappa^{\frac{3}{8}}\tilde{C_1}L^{\frac{3}{8}}}{2^{\frac{15}{16}}\times7^{\frac{3}{8}}\eta^{\frac{3}{8}}},
\end{eqnarray}
where $\tilde{C_1}=-C_1$. The above expression (\ref{zfg}) will be used while converting the light cone gauge into the Fefferman-Graham gauge and vice-versa. 

Using  (\ref{lm}), (\ref{zfg}) and (\ref{e1})-(\ref{e3}) we finally end up with the following expressions for the background fields as well as the stress tensor in the Fefferman-Graham gauge
\begin{eqnarray}
h_{tt}(\eta)\Big|_{\eta\rightarrow\infty}&=&\frac{32\times2^{\frac{3}{8}}\times 7^{\frac{3}{4}}\times\eta^{\frac{3}{2}}\Big(\frac{\eta^{\frac{3}{8}}}{L^{\frac{3}{8}}\kappa^{\frac{3}{8}}}\Big)^{\frac{2}{3}}\sqrt{L}}{9\times3^{\frac{5}{8}}\sqrt{\kappa}\tilde{C_1}^2}+...,\label{fgh}\\
\phi(\eta)\Big|_{\eta\rightarrow\infty}&=&-\frac{16\times2^{\frac{7}{16}}\times 7^{\frac{3}{8}}\times\eta^{\frac{9}{8}}\Big(\frac{\eta^{\frac{3}{8}}}{L^{\frac{3}{8}}\kappa^{\frac{3}{8}}}\Big)^{\frac{2}{3}}}{9\times3^{\frac{1}{16}}\kappa^{\frac{1}{8}}L^{\frac{9}{8}}}+...,\label{fgp}\\
A_t(\eta)\Big|_{\eta\rightarrow\infty}&=&-L^2\log\Bigg(\frac{3^{\frac{9}{16}}\kappa^{\frac{3}{8}}\tilde{C_1}L^{\frac{3}{8}}}{2^{\frac{15}{16}}\times7^{\frac{3}{8}}\eta^{\frac{3}{8}}}\Bigg)\Bigg(\frac{((72\kappa-6\xi)C_3+\kappa C_9)}{6\xi\tilde{C_1}}\Bigg)+..,\label{fga}\\
T_{tt}(\eta)\Big|_{\eta\rightarrow\infty}&\approx& -3175.934\frac{\eta^{3}}{L\kappa^{1.5}\tilde{C_1}^3}-294.245\frac{\eta^{3}}{L^2\kappa^{1.125}\tilde{C_1}^2}+...\label{fgt}
\end{eqnarray}
where (...) represents all the sub leading terms in an expansion near the boundary.
\subsection{Transformation properties of the stress tensor}
In the present Section, we study the transformation properties of the boundary stress tensor under diffeomorphism. Under diffeomorphism, $x^{\mu}\rightarrow x^{\mu}+\epsilon^{\mu}(x)$ the space time metric, gauge fields and the dilaton transform as follows
\begin{eqnarray}
\delta_{\epsilon}g_{\mu\nu}&=&\bigtriangledown_{\mu}\epsilon_{\nu}+\bigtriangledown_{\nu}\epsilon_{\mu},\label{df1}\\
\delta_{\epsilon}A_{\mu}&=&\epsilon^{\nu}\bigtriangledown_{\nu}A_{\mu}+A_{\nu}\bigtriangledown_{\mu}\epsilon^{\nu},\label{df2}\\
\delta_{\epsilon}\phi&=&\epsilon^{\mu}\bigtriangledown_{\mu} \phi.\label{df3}
\end{eqnarray}

Using (\ref{fgm}), (\ref{fgh}) and (\ref{df1}), one can find an expression for the parameter ($\epsilon_{\mu}$) of diffeomorphism which turns out to be
\begin{eqnarray}\label{dfp}
\epsilon_{\eta}= a\Xi'(t),\hspace{2mm}\epsilon_t=b\eta^{\frac{7}{4}}\Xi(t)+\frac{4}{3}a\eta\Xi''(t),
\end{eqnarray}
where $\Xi(t)$ is an arbitrary function of time while the constants (a, b) will be fixed latter on.

Recall, that we are working in a gauge in which $A_{\eta}$ is set to be zero. From equations (\ref{df2}) and (\ref{dfp}), it is easy to check that $\delta_{\epsilon}A_{\eta}\neq 0$ which means that the differomorphism destroys the gauge condition. Therefore, to retain the gauge condition, we make another gauge transformation i.e.  $A_{\mu}\rightarrow A_{\mu}+\partial_{\mu}\lambda$, where we choose $\lambda$ such that $(\delta_{\epsilon}+\delta_{\lambda})A_{\eta}=0$, which determines $\lambda$ at leading order as
\begin{align}\label{flam}
\lambda=&-\frac{1}{256\times2^{\frac{3}{8}}\times7^{\frac{3}{4}}\eta^{\frac{7}{8}}\xi}\Bigg[3^{\frac{5}{8}}aL^{\frac{7}{4}}\Big(\frac{\eta^{\frac{3}{8}}}{\kappa^{\frac{7}{8}}}\Big)^{\frac{1}{3}}\kappa^{\frac{7}{8}}\Big(8+15\log(2)-9\log(3)+6\log(7)\nonumber\\
&-16\log\Big(\frac{\kappa^{\frac{3}{8}}L^{\frac{3}{8}} \tilde{C_1}}{\eta^{\frac{3}{8}}}\Big)\Big)\tilde{C_1}((72\kappa-6\xi)C_3+\kappa C_9)\Xi''(t)\Bigg].
\end{align}

Using (\ref{dfp}) and (\ref{flam}), one can finally pin down the variations of the background fields under the diffeomorphism and the gauge transformation as
\begin{align}
\delta_{\epsilon}h_{tt}&=\frac{2}{27}\sqrt{\eta}\Big(27b\eta^{\frac{5}{4}}+\frac{28\times6^{\frac{3}{8}}\times7^{\frac{3}{4}}a}{\sqrt{\kappa}\tilde{C_1}^2}\sqrt{L}\Big(\frac{\eta^{\frac{3}{8}}}{L^{\frac{3}{8}}\kappa^{\frac{3}{8}}}\Big)^{\frac{2}{3}}\Big)\Xi'(t)+\frac{8}{3}a\eta\Xi'''(t),\\
(\delta_{\epsilon}+\delta_{\lambda})A_t&=\frac{L^{\frac{3}{2}}}{896\eta\Big(\frac{\eta^{\frac{3}{8}}}{L^{\frac{3}{8}}\kappa^{\frac{3}{8}}}\Big)^{\frac{2}{3}}\xi\tilde{C_1}}((72\kappa-6\xi)C_3+\kappa C_9)\Bigg[\Big(56a\Big(\frac{\eta^{\frac{3}{8}}}{L^{\frac{3}{8}}\kappa^{\frac{3}{8}}}\Big)^{\frac{2}{3}}-3\times6^{\frac{5}{8}}\times\nonumber\\
&7^{\frac{1}{4}}b\eta^{\frac{5}{4}}\sqrt{\kappa} \log\Big(\frac{3^{\frac{9}{16}}\kappa^{\frac{3}{8}}\tilde{C_1}L^{\frac{3}{8}}}{2^{\frac{15}{16}}\times7^{\frac{3}{8}}\eta^{\frac{3}{8}}}\Big)\tilde{C_1}^2\Big)\Xi'(t)-2\times6^{\frac{5}{8}}\times7^{\frac{1}{4}}a\sqrt{\eta}\sqrt{\kappa}\tilde{C_1}^2\Xi'''(t)\Bigg],\\
\delta_{\epsilon}\phi&=-\frac{22\times2^{\frac{7}{16}}\times7^{\frac{3}{8}}a\eta^{\frac{1}{8}}\Xi'(t)}{9\times3^{\frac{1}{16}}\kappa^{\frac{1}{8}}L^{\frac{9}{8}}}\Big(\frac{\eta^{\frac{3}{8}}}{L^{\frac{3}{8}}\kappa^{\frac{3}{8}}}\Big)^{\frac{2}{3}}.
\end{align}
 
 In order to proceed further, we first convert the stress tensor (see equation (\ref{std}) and (\ref{bgtt})) into  Fefferman-Graham coordinate and then explore its properties under diffeomorphism and gauge transformation. After doing all the calculations, we end up with the following expression
\begin{align}\label{dst1}
(\delta_{\epsilon}+\delta_{\lambda})T_{tt}\approx &\Big(-58.928 b\frac{\eta^{3}}{L^2\kappa^{0.375}}-636.04b\frac{\eta^{3}}{L\kappa^{0.75}\tilde{C_1}}\Big)\Xi'(t)\nonumber\\
&
-672.835a\frac{\eta^{3}}{\kappa^{0.75}\tilde{C_1}L^2}\Xi'''(t),
\end{align}
where we have retained only the dominant terms in the (boundary) limit $\eta\rightarrow \infty$. 

After a proper re-scaling, the boundary stress tensor and its variation under diffeomorphism and gauge transformation may be defined as,
\begin{eqnarray}\label{rst}
\tilde{T_{tt}}=\lim_{\eta\rightarrow\infty}\frac{L^2}{\eta^3}T_{tt}\hspace{2mm}and\hspace{2mm}(\delta_{\epsilon}+\delta_{\lambda})\tilde{T_{tt}}=\lim_{\eta\rightarrow\infty}\frac{L^2}{\eta^3}(\delta_{\epsilon}+\delta_{\lambda})T_{tt}.
\end{eqnarray}

With a proper choice of the constant $b =\frac{9.986}{\tilde{C_1}^2\kappa^{0.75}}$, one can express the equation (\ref{rst}) in a more elegant way. Using (\ref{fgt}), this finally leads to the transformation of the stress tensor\footnote{Since we are working in a static gauge therefore, $\Xi(t)\partial_t\tilde{T_{tt}}$ is trivially zero. } as follows
\begin{eqnarray}\label{cdm1}
(\delta_{\epsilon}+\delta_{\lambda})\tilde{T_{tt}}=2\tilde{T_{tt}}\Xi'(t)- \frac{ca}{\tilde{C_1}}\Xi'''(t).
\end{eqnarray}
  
  (\ref{cdm1}) is the standard form of variation of the boundary stress tensor (\ref{std}) under the action of  both diffeomorphism and gauge transformation. Finally, we have reached a stage where one can identify the central charge of the boundary theory. The constant $``c"$ appearing in (\ref{cdm1}) (as the coefficient of $\Xi'''(t)$) is the central charge associated to our boundary theory (\ref{S2D}) which is given by the following expression
 \begin{align}\label{cc}
     c=\frac{672.835}{\kappa^{0.75}}.
 \end{align}

Notice that (\ref{cc}) is purely a `dimensionless' large number as we are working in the small $\kappa$ regime. This also makes the entity in (\ref{cc}) highly non-perturbative in the sense that it diverges in the smooth $\kappa \rightarrow 0$ limit and does not connect to the pure JT gravity theory. Therefore, these theories are not smoothly connected to their conformal cousins, which are dual to pure JT gravity.
\section{Black hole solutions}\label{SBH}
We now explore black hole solutions\footnote{From now onwards we set the length scale $L=1$.} of the 2D gravity model (\ref{action 2D}). Like before, these solutions are expressed perturbatively with the gauge choice as discussed in Section (\ref{S3}).
\subsection{Zeroth order solution }
In order to calculate the zeroth order solution, we solve equations (\ref{v01}), (\ref{v02}) and (\ref{v03}) simultaneously which yields
\begin{eqnarray}\label{bh0}
e^{2\omega_{(0)}^{bh}}=\frac{8\mu}{12[\sinh(2\sqrt{\mu} z)]^2}\hspace{1mm},\hspace{2mm}\phi_{(0)}^{bh}=\frac{\sqrt{\mu}}{6}\coth(2z\sqrt{\mu}).
\end{eqnarray}
The above solutions (\ref{bh0}) correspond to black hole solutions in pure JT gravity \cite{Almheiri:2014cka}.
\subsection{First order corrections in \texorpdfstring{$\xi$}{xi}}
Leading order corrections to (\ref{bh0}) can be estimated by using equations (\ref{v21})-(\ref{v23}). These equations will be easy to handle if we change the coordinate as follows
\begin{eqnarray}
z=\frac{1}{2\sqrt{\mu}}\coth^{-1}\Big(\frac{\rho}{\sqrt{\mu}}\Big).
\end{eqnarray}\label{ze}

Using (\ref{bh0}) and (\ref{ze}), we can express the first order solution as 
\begin{align}
A_{t(0)}^{bh}=&-2Q\log(\rho)+d_1,\label{bh21}\\
\omega_{(1)}^{bh}=&\frac{1}{4\mu^{\frac{3}{2}}}\Bigg[3\Big\{-2\rho \tanh^{-1}\Big(\frac{\rho}{\sqrt{\mu}}\Big)\log(\rho)+2\sqrt{\mu}(1+\log(\rho))-\rho \text{PolyLog}\Big[2,-\frac{\rho}{\sqrt{\mu}}\Big]\nonumber \\ 
&+\rho \text{PolyLog}\Big[2,\frac{\rho}{\sqrt{\mu}}\Big]\Big\}Q^2+4\mu \Big\{\rho d_1-\sqrt{\mu}d_2+\rho \tanh^{-1}\Big(\frac{\rho}{\sqrt{\mu}}\Big)d_2\Big\}\Bigg], \label{bh22}
\end{align}
\begin{align}
\phi_{(1)}^{bh}=&\frac{1}{48\mu^{\frac{3}{2}}}\Bigg[3\Big\{2\mu+4\sqrt{\mu}\rho+4\sqrt{\mu}\rho \log(\rho)+\tanh^{-1}\Big(\frac{\rho}{\sqrt{\mu}}\Big)\Big\{6\rho^2-8\sqrt{\mu}\rho +\nonumber\\
&4(\mu-\rho^2)\log(\rho)\Big\} -\mu \log\Big(1-\frac{\rho}{\sqrt{\mu}}\Big)-8\sqrt{\mu}\rho \log\Big(1-\frac{\rho}{\sqrt{\mu}}\Big)+\nonumber\\
&6\rho^2\log\Big(1-\frac{\rho}{\sqrt{\mu}}\Big)-\mu \log\Big(1+\frac{\rho}{\sqrt{\mu}}\Big)+4\sqrt{\mu}\rho \log(-\mu+\rho^2)+(\mu-3\rho^2)\times\nonumber \\
&\log\Big(1-\frac{\rho^2}{\mu}\Big)-(\mu-\rho^2)\Big(4\text{PolyLog}\Big(2,\frac{\rho}{\sqrt{\mu}}\Big)-\text{PolyLog}\Big(2,\frac{\rho^2}{\mu}\Big)\Big)\Big\}Q^2\nonumber \\
&+4\mu\Big\{2\sqrt{\mu}\rho+2\rho^2\tanh^{-1}\Big(\frac{\rho}{\sqrt{\mu}}\Big)+\mu \log(-\sqrt{\mu}+\rho)-\mu \log(\sqrt{\mu}+\rho)\Big\}d_2+\nonumber\\
&48\mu^{\frac{3}{2}}(d_3+\rho d_4)+8\mu \rho^2d_1\Bigg]\label{bh23},
\end{align}
where $Q$ is the charge of the $U(1)$ gauge theory and $d_is$ are the constants where i takes the value 1, 2, 3, ... Equations (\ref{bh21}), (\ref{bh22}) and (\ref{bh23}) correspond to first order corrections to zeroth order (black hole) solutions due to 2-derivative interaction terms in (\ref{action 2D}).
\subsection{First order corrections in \texorpdfstring{$\kappa$}{kappa}}
Let us first estimate corrections to gauge fields due to 4-derivative interactions in (\ref{action 2D}). These can be estimated by comparing the coefficient of $\kappa$ in equation of motion for $A_{\mu}$ 
\begin{align}\label{bh41}
-e^{-2\omega_{(0)}}\phi_{(0)}A_{t(1)}'-24e^{-6\omega_{(0)}}\phi_{(0)}(A_{t(0)}')^3+2\partial_z[e^{-4\omega_{(0)}}\phi_{(0)}(2\omega_{(0)}'A_{t(0)}'-A_{t(0)}'')]=d_6.
\end{align}

Notice that the above equation (\ref{bh41}) is expressed in terms of $z$ and its derivatives. Upon solving equation (\ref{bh41}) in terms of $\rho$ and using equation (\ref{bh0}) we finally obtain
\begin{eqnarray}\label{bh4a}
A_{t(1)}^{bh}=\frac{12Q(\mu-72Q^2)}{\rho^2}+2\log(\rho)(12Q+d_5)+d_6.
\end{eqnarray}

Next, we collect the coefficient of $\kappa$ in equation of motion for $\phi$. After simplifying the expression we get,
\begin{align}\label{bh42}
&\frac{4}{243}(\mu-\rho^{2})\Bigg[-5832+\frac{6^{1 / 3}}{\rho^{\frac{8}{3}}}\Big\{-\mu^{2}(40 \times 6^{1 / 3}+\rho^{4 / 3})-2 \mu \rho^{2}(116 \times 6^{1 / 3}+7 \rho^{4 / 3})+\nonumber \\
&\rho^{4}(1388 \times 6^{1 / 3}+15 \rho^{4 / 3})\Big\}-972 \omega_{(2)}+\frac{486}{\rho^4}\Big\{54(-\mu+\rho^{2}) Q^{2}+1944 Q^{4}+3 \rho^{2} Q d_{5}+\nonumber\\
&\rho^{4}\Big(2 \rho \frac{\partial\omega_{(2)}}{\partial\rho}+(-\mu+\rho^{2}) \frac{\partial^2\omega_{(2)}}{\partial \rho^2}\Big)\Big\}\Bigg]=0.
\end{align}

In general, one can solve (\ref{bh42}) exactly for $\omega_{(2)}$. However, for the purpose of our present analysis, we are interested in the near boundary expression of this function. Therefore, we expand $\omega_{(2)}$ in the limit $\rho \rightarrow \infty$ and retain only leading order terms.

After simplification, one can express $\omega_{(2)}$ in the following form
\begin{align}
\omega_{(2)}^{bh}|_{\rho\rightarrow\infty}=-0.0072108\rho^{\frac{8}{3}}-8.48718\rho^{\frac{4}{3}} + F\big(\log(\mu),\log(\rho)\big)\rho- 0.0228342\mu\rho^{\frac{2}{3}}-6,  
\end{align}
where $F\big(\log(\mu),\log(\rho)\big)$ is given by
\begin{align}
    F=&\hspace{1mm}3.72651\mu^{\frac{1}{6}}-3.72651\Big(-\frac{1}{\sqrt{\mu}}\Big)^{\frac{2}{3}}\sqrt{\mu}+0.0239781\mu^{\frac{5}{6}}+0.0239781\Big(-\frac{1}{\sqrt{\mu}}\Big)^{\frac{1}{3}}\mu+\nonumber\\
    &\frac{3.375}{\mu^\frac{3}{2}}\Big\{4\log\Big(-\frac{1}{\sqrt{\mu}}\Big)^2-\log(\mu)^2+8\log\Big(-\frac{1}{\sqrt{\mu}}\Big)\log(\rho)+4\log(\mu)\log(\rho)-\nonumber\\
    &8\log\Big(-\frac{\rho}{\sqrt{\mu}}\Big)-8\log(\rho)\log\Big(-\frac{\rho}{\sqrt{\mu}}\Big)+8\log\Big(\frac{\rho}{\sqrt{\mu}}\Big)+8\log(\rho)\log\Big(\frac{\rho}{\sqrt{\mu}}\Big)\Big\}Q^2+\nonumber\\
    &\frac{972}{\mu^\frac{5}{2}}\Big(\log\Big(-\frac{\rho}{\sqrt{\mu}}\Big)Q^4-\log\Big(\frac{\rho}{\sqrt{\mu}}\Big)Q^4\Big)-\frac{0.1875Qd_5}{\mu^\frac{3}{2}}\Big\{-4\log\Big(-\frac{1}{\sqrt{\mu}}\Big)^2+\nonumber\\
    &\log(\mu)^2-8\log\Big(-\frac{1}{\sqrt{\mu}}\Big)\log(\rho)-4\log(\mu)\log(\rho)+8\log(\rho)\log\Big(-\frac{\rho}{\sqrt{\mu}}\Big)-\nonumber\\
    &8\log(\rho)\log\Big(\frac{\rho}{\sqrt{\mu}}\Big)\Big\}+\frac{d_7}{\sqrt{\mu}}+\frac{0.5}{\sqrt{\mu}}\Big(-\log\Big(\frac{-\rho}{\sqrt{\mu}}\Big)d_8+\nonumber\\
    &\log\Big(\frac{\rho}{\sqrt{\mu}}\Big)d_8\Big).
\end{align}

With all these expressions at hand, one can approximate the black hole metric (\ref{sg}) as 
\begin{eqnarray}\label{2dbh}
ds^2_{bh}=\frac{2}{3}(\rho^2-\mu)\Big(1+2(\xi \omega_{(1)}^{bh}+\kappa \omega_{(2)}^{bh})\Big)\Bigg(-dt^2+\frac{d\rho^2}{4(\mu-\rho^2)^2}\Bigg),
\end{eqnarray}
where the black hole horizon is located at $\rho=\sqrt{\mu}$.

In order to calculate $\phi_{(2)}$, we compare the coefficient of $\kappa$ in equations of motion of $g_{tt}$ and $g_{zz}$. On Subtracting $g_{zz}$ from $g_{tt}$ and after some simplification we find
\begin{align}\label{bh43}
12e^{2\omega_{(0)}}(\phi_{(2)}+4\phi_{(0)}\omega_{(2)})-\Big(\frac{\partial\rho}{\partial z}\frac{\partial}{\partial \rho}\Big\{\frac{\partial\rho}{\partial z}\Big(\frac{\partial \phi_{(2)}}{\partial \rho}\Big)\Big\}+2\omega_{(2)}\frac{\partial\rho}{\partial z}\frac{\partial}{\partial \rho}\Big\{\frac{\partial\rho}{\partial z}\Big(\frac{\partial \phi_{(0)}}{\partial \rho}\Big)\Big\}\Big)\nonumber\\+A(\rho)=0,
\end{align}
where $A(\rho)$ is given by 
\begin{align}
   A(\rho)&= \frac{2(\mu-\rho^2)}{729\rho^{\frac{1}{3}}}\Big\{6^{\frac{1}{3}}\mu^2-5832\rho^{\frac{4}{3}}-432\times6^{\frac{2}{3}}\rho^{\frac{8}{3}}+180\times6^{\frac{1}{3}}\rho^3+6^{\frac{1}{3}}\rho^4+\nonumber\\
    &2\mu(54\times6^{\frac{2}{3}}\rho^{\frac{2}{3}}-90\times6^{\frac{1}{3}}\rho-6^{\frac{1}{3}}\rho^2)\Big\}+\frac{4Q}{\rho^3}\Big\{6Q(4\mu^2-6\mu\rho^2+\rho^4+\nonumber\\
    &36(-\mu+\rho^2)Q^2)+\rho^2(-\mu+\rho^2)d_5\Big\}.
\end{align}

Technically speaking, it is very difficult to solve (\ref{bh43}) exactly. Therefore we will solve this equation in two different limits.

\begin{itemize}
    \item 
\underline{Case 1} : \textbf{Near boundary analysis} ($\rho\rightarrow\infty$) :\\
Using the near boundary expansion of $\omega_{(2)}$ and $A(\rho)$ in equation  (\ref{bh43}) we get
\begin{eqnarray}\label{BH44}
4\rho^2\frac{d^2\phi_{(2)}}{d\rho^2}+8\rho\frac{d\phi_{(2)}}{d\rho}-8\phi_{(2)}+0.02\rho^{\frac{11}{3}}=0,
\end{eqnarray}
where we consider $\rho^2>>\mu$ and retain only dominant terms in the above expression. Equation (\ref{BH44}) can be easily solved for $\phi_{(2)}$ 
\begin{eqnarray}\label{bh45}
\phi_{(2)}^{(bh)}|_{\rho\rightarrow\infty}=-\frac{9}{27200}\rho^{\frac{11}{3}}+\rho d_{9}+\frac{d_{10}}{\rho^2}.
\end{eqnarray}
\item \underline{Case 2} : \textbf{Near horizon analysis} ($\rho\rightarrow \sqrt{\mu}$) :\\
Converting equation (\ref{bh43}) in terms of $\rho$ and taking the limit $\rho\rightarrow\sqrt{\mu}$, we arrive at the following equation
\begin{eqnarray}
12e^{2\omega_0}\phi_2-\frac{\partial\rho}{\partial z}\frac{\partial}{\partial \rho}\Big\{\frac{\partial\rho}{\partial z}\Big(\frac{\partial \phi_2}{\partial \rho}\Big)\Big\}-24\sqrt{\mu}Q^2=0,
\end{eqnarray}
which can be solved for $\phi_{(2)}$ to yield,
\begin{align}\label{bh46}
\phi_{(2)}^{(bh)}|_{\rho\rightarrow\sqrt{\mu}}=&\frac{1}{\sqrt{\mu}}\Big(\rho d_{11}-\sqrt{\mu}d_{12}+\rho \tanh^{-1}\Big(\frac{\rho}{\sqrt{\mu}}\Big)d_{12}\Big)+\frac{3}{4\mu}\Big\{\rho\Big(4\log(-\sqrt{\mu}+\rho)\nonumber\\
&-4\log(\sqrt{\mu}+\rho)+\log\Big(1-\frac{\rho}{\sqrt{\mu}}\Big)^2+2\log\Big(1-\frac{\rho}{\sqrt{\mu}}\Big)\times\nonumber\\
&\log\Big[\frac{1}{4}\Big(1+\frac{\rho}{\sqrt{\mu}}\Big)\Big]-\log\Big(1+\frac{\rho}{\sqrt{\mu}}\Big)^2\Big)-4\Big\{\sqrt{\mu}-\rho \tanh^{-1}\Big(\frac{\rho}{\sqrt{\mu}}\Big)\Big\}\nonumber\\
&\times\log(\mu-\rho^2)+4\rho \text{PolyLog}\Big(2,\frac{1}{2}-\frac{\rho}{2\sqrt{\mu}}\Big)\Big\}Q^2.
\end{align}
\end{itemize}
The above set of solutions (\ref{bh4a})-(\ref{bh46}) are the first order corrections to pure JT gravity black hole solutions due to 4-derivative interaction terms in (\ref{action 2D}).

 Now, we have obtained a complete set of black hole as well as vacuum solutions for generalized JT gravity models with an abelian one form. Our next task would be to compare these solutions in the near boundary limit. Let us first expand the black hole solutions (\ref{bh0})-(\ref{bh46}) in the limit $z\rightarrow0$, which reveals the following leading order behaviour for the background fields and the metric 
 \begin{eqnarray}\label{bh47}
\phi^{(bh)}\Big|_{z\rightarrow 0}\sim\frac{1}{z^{\frac{11}{3}}}\hspace{1mm},\hspace{2mm}A_{t}^{(bh)}\Big|_{z\rightarrow 0}\sim \log(z)\hspace{1mm},\hspace{2mm}e^{2\omega ^{(bh)}}\Big|_{z\rightarrow 0}\sim \frac{1}{z^{\frac{14}{3}}}.
\end{eqnarray}

On the other hand, for vacuum solutions (\ref{v0s})-(\ref{v4s3}), we find the leading order behaviour for the background fields as well as the metric
\begin{eqnarray}\label{bh48}
\phi^{(vac)}\Big|_{z\rightarrow 0}\sim\frac{1}{z^{\frac{11}{3}}}\hspace{1mm},\hspace{2mm}A_{t}^{(vac)}\Big|_{z\rightarrow 0}\sim \log(z)\hspace{1mm},\hspace{2mm}e^{2\omega^{(vac)}}\Big|_{z\rightarrow 0}\sim \frac{1}{z^{\frac{14}{3}}}.
\end{eqnarray}

Comparing (\ref{bh47}) and (\ref{bh48}) we note that the leading order behaviour of both the black hole and the vacuum solution is identical near the boundary. Hence, the UV central charge (\ref{cc}) for black hole phase will be identical to that with the vacuum solution as mentioned previously section (\ref{S5}). 
\section{Thermodynamics of 2D black holes}\label{SBH1}
In the present section, we investigate the thermal properties of 2D black holes (\ref{2dbh}). In particular, we discuss the Wald entropy \cite{Wald:1993nt} of a black hole and interpret its divergences near the black hole horizon. 

To start with, we calculate the Hawking temperature \cite{Hawking:1974sw} for the 2D black hole (\ref{2dbh})
\begin{align}\label{ht}
   T_H=\frac{1}{2\pi}\sqrt{-\frac{1}{4}g^{tt}g^{\rho\rho}(\partial_\rho g_{tt})^2}\Bigg|_{\rho\rightarrow\sqrt{\mu}}=\hspace{2mm}\frac{\sqrt{\mu}}{\pi}\big(1+6\kappa-\kappa(36d_2+d_8)\big),
\end{align}
where $\sqrt{\mu}$ is the location of the horizon. Notice that, in arriving at (\ref{ht}), we set $Q=\sqrt{\frac{4\mu d_2}{3\log(\mu)}}$, $d_5=0$ and $\mu<<1$ such that hawking temperature reduces to \cite{Lala:2020lge} in the limit $\kappa\rightarrow 0$.
\subsection{Wald entropy }\label{divw}
The  Wald entropy\footnote{One can also compute the thermodynamic entropy of 2D black holes using the Cardy Formula \cite{Cardy:1986ie} as discussed by authors in \cite{Castro:2008ms} and \cite{Strominger:1997eq}. See Appendix \ref{cardy} for a brief discussion on the Cardy formula.    } \cite{Wald:1993nt, Brustein:2007jj, Pedraza:2021cvx} is defined as
\begin{eqnarray}\label{weed}
S_W=-2\pi Y^{abcd}\epsilon_{ab}\epsilon_{cd}\hspace{2mm},\hspace{2mm}Y^{abcd}=\frac{\partial \mathcal{L}}{\partial R_{abcd}},
\end{eqnarray}
where $\mathcal{L}$ is the Lagrangian density\footnote{we have used the notation  $S=\int d^2x\sqrt{-g}\mathcal{L}$ }, $R_{abcd}$ is the Riemann curvature tensor and $\epsilon_{ab}$ is the anti-symmetric tensor with the normalisation condition, $\epsilon^{ab}\epsilon_{ab}=-2$. 

Using (\ref{weed}), one can estimate the Wald entropy for the action (\ref{action 2D})
\begin{eqnarray}\label{we1}
S_W=4\pi\phi-16\kappa\pi\phi e^{-6\omega}\frac{\partial \rho}{\partial z}\frac{d}{d\rho}\Big\{\frac{\partial \rho}{\partial z}\Big(\frac{\partial \omega}{\partial \rho}\Big)\Big\}.
\end{eqnarray}

One can expand the above expression (\ref{we1}) explicitly using the equations (\ref{bh0}), (\ref{bh23}) and (\ref{bh46}) up to leading order in $\xi$ and $\kappa$ as

\begin{align}\label{wee}
S_W=&4\pi\Bigg(\frac{\rho}{6}-\frac{9\kappa\rho}{(-\rho^2+\mu)^2}+\frac{1}{48\mu^{\frac{3}{2}}}\xi\Big\{3\Big(4\rho\sqrt{\mu}+2\mu+(-3\rho^2+\mu)\log\Big(1-\frac{\rho^2}{\mu}\Big)+\nonumber\\
&6\rho^2\log\Big(1-\frac{\rho}{\sqrt{\mu}}\Big)-8\rho\sqrt{\mu}\log\Big(1-\frac{\rho}{\sqrt{\mu}}\Big)-\mu \log\Big(1-\frac{\rho}{\sqrt{\mu}}\Big)-\mu \log\Big(1+\frac{\rho}{\sqrt{\mu}}\Big)\nonumber\\
&+4\rho\sqrt{\mu}\log(\rho)+\tanh^{-1}\Big(\frac{\rho}{\sqrt{\mu}}\Big)(6\rho^2-8\rho\sqrt{\mu}+4(-\rho^2+\mu)\log(\rho))+4\rho\sqrt{\mu}\times\nonumber\\
&\log(\rho^2-\mu)-(-\rho^2+\mu)\Big(-\text{PolyLog}\Big[2,\frac{\rho^2}{\mu}\Big]+\text{4PolyLog}\Big[2,\frac{\rho}{\mu}\Big]\Big)\Big)Q^2+8\rho^2\mu d_1\nonumber\\
&+4\mu\Big(2\rho\sqrt{\mu}+2\rho^2\tanh^{-1}\Big[\frac{\rho}{\sqrt{\mu}}\Big]+\mu \log(-\sqrt{\mu}+\rho)-\mu \log(\rho+\sqrt{\mu})\Big)d_2\nonumber\\
&+48\mu^{\frac{3}{2}}(d_3+\rho d_4)\Big\}+\kappa\Big\{\frac{3}{4\pi}\Big(\rho\Big(4\log(-\sqrt{\mu}+\rho)+\log\Big(1-\frac{\rho}{\sqrt{\mu}}\Big)^2+\nonumber\\
&2\log\Big(1-\frac{\rho}{\sqrt{\mu}}\Big)\log\Big(\frac{1}{4}\Big(1+\frac{\rho}{\mu}\Big)\Big)-\log\Big(1+\frac{\rho}{\sqrt{\mu}}\Big)^2-4\log(\sqrt{\mu}+\rho)\Big)\nonumber\\
&-4\Big(\sqrt{\mu}-\rho\tanh^{-1}\Big(\frac{\rho}{\sqrt{\mu}}\Big)\Big)\log(-\rho^2+\mu)+4\rho\text{PolyLog}\Big(2,\frac{1}{2}-\frac{\rho}{2\sqrt{\mu}}\Big)\Big)Q^2\nonumber\\
&-d_{12}+\frac{\rho}{\sqrt{\mu}}\Big(d_{11}+\tanh^{-1}\Big[\frac{\rho}{\sqrt{\mu}}\Big]d_{12}\Big)\Big\}\Bigg).
\end{align}

It is evident from the above expression (\ref{wee}) that the Wald entropy  diverges in the near horizon limit i.e. $\rho\rightarrow\sqrt{\mu}$. As we explain below, these divergences are due to the short range correlations between quantum modes across the horizon. A careful inspection, further reveals that these divergences are sourced due to the presence of the higher derivative interaction terms in (\ref{action 2D}).

Below, we explain more about this with the help of a toy model calculation.

$\bullet$ \textbf{A toy model calculation :}

Consider a massive scalar field ($\Phi$) in the black hole background (\ref{2dbh}) that satisfies the Klein-Gordan equation 
\begin{eqnarray}\label{kg1}
(\bigtriangledown^2-m^2)\Phi=0.
\end{eqnarray}

We demand that $\Phi$ satisfies the ``brick wall'' boundary condition i.e. $\Phi=0$ at $z=z_*$, where $z_*$ is the location of black hole horizon. This calculation is analogous to the ’t Hooft’s brick wall model as discussed in \cite{thooft}-\cite{Brustein:2010ms}.

In the black hole background (\ref{2dbh}), equation (\ref{kg1}) takes the form
\begin{eqnarray}\label{kg2}
-\frac{1}{e^{2\omega}}\partial^2_t\Phi+\frac{1}{e^{2\omega}}\partial^2_z\Phi-m^2\Phi=0.
\end{eqnarray}

One can solve the above equation (\ref{kg2}) using method of seperation of variables. We consider $\Phi=e^{iEt}f(z)$, and plug it back into (\ref{kg2}) which yields
\begin{eqnarray}\label{kg3}
\frac{1}{e^{2\omega}}\big(E^2f(z)+\partial^2_zf(z)\big)-m^2f(z)=0.
\end{eqnarray}

In order to proceed further, we substitute $f(z)=\tilde{\rho}(z)e^{iS(z)}$, where $\tilde{\rho}(z)$ is a slowly varying function in $z$ and $S(z)$ is the wildly oscillating phase. On plugging $f(z)$ into (\ref{kg3}), we get
\begin{eqnarray}\label{kg4}
f(z)=\tilde{\rho}(z)e^{\pm i\int dz\sqrt{E^2-e^{2\omega}m^2}}.
\end{eqnarray}

Now we impose an additional boundary condition\footnote{This is called the  Dirichlet boundary condition and the coordinate $\overline{z}$ is located far away from the horizon \cite{thooft}.} on $\Phi$ i.e $\Phi=0$ at $z=\overline{z}$ such that the integral in (\ref{kg4}) become discrete
\begin{eqnarray}\label{ds1}
\int_{z_*}^{\overline{z}} dz\sqrt{E^2-e^{2\omega}m^2}=n(E),
\end{eqnarray}
where $n(E)$ is the density of states which measures the total number of states having energy $E$.

Using (\ref{ze}), we can express the density of states (\ref{ds1}) in terms of $\rho$ as
\begin{eqnarray}\label{ds2}
n(E)=\int_{\sqrt{\mu}}^{\overline{\rho}}d\rho\sqrt{\frac{E^2}{4(\mu-\rho^2)^2}-\frac{e^{2\omega}m^2}{4(\mu-\rho^2)^2}}.
\end{eqnarray}

In principle, one can use the above expression (\ref{ds2}), to estimate the corresponding free energy (F) and entropy (S) for the scalar field ($\Phi$)  as 
\begin{align}\label{FS}
    F=\int_0^{\infty}\frac{n(E)}{1-e^{\beta E}} dE\hspace{2mm}\text{ and }\hspace{2mm}S=\beta^2\partial_{\beta}F\Big|_{\beta=\beta_H},
\end{align}
where $\beta$ is the inverse temperature and $\beta_H$ is the inverse Hawking temperature. 

It is evident from (\ref{ds2}), that the integrand blows up at lower limit i.e. $\rho=\sqrt{\mu}$. This means that the density of states for $\Phi$ diverges near the horizon which leads to divergences in the free energy and entropy (\ref{FS}). In order to get rid of such divergences, we shift the horizon location by an infinitesimal amount $\delta$ i.e. $\rho\rightarrow\sqrt{\mu}+\delta$, where $\delta<<\sqrt{\mu}$. On plugging the shifted horizon back into (\ref{ds2}), we get a finite answer both for the density of states as well as for the entropy \cite{Solodukhin:2011gn,Brustein:2010ms}. This toy model calculation for $\Phi$ gives us an important clue about the interpretation of the above divergences in the Wald entropy (\ref{wee}).

 Recall that, we formulate the action (\ref{action 2D}) by adding matter field content\footnote{By matter field content, we means 2-derivative and 4-derivative interaction terms.} to the pure JT gravity model. Addition of matter field content introduces new degrees of freedom in our theory (\ref{action 2D}), which is analogous to the scalar field ($\Phi$) in the above calculation. Therefore, the divergence in the Wald entropy (\ref{wee}) (that arises due to the addition of the matter field content) is analogous to the divergence in the density of states (\ref{ds2}) for $\Phi$ near the horizon. Therefore, following the above discussion, one can get rid of divergences in the Wald entropy (\ref{wee}) by shifting the actual location of the horizon by an infinitesimal amount namely, $\rho\rightarrow\sqrt{\mu}+\delta$, where $\delta<<\sqrt{\mu}$.

\section{Near horizon CFT}\label{SS8}
We now explore the near horizon modes of the theory (\ref{action 2D}). In particular, we look the evidence of a CFT in the near horizon limit and calculate the central charge associated with it.

We start by computing the trace of the stress tensor in the near horizon limit
\begin{eqnarray}\label{NHT1}
g^{\mu\nu}T_{\mu\nu}=\frac{1}{\sqrt{-g}}g^{\mu\nu}\frac{\delta S_{EH}}{\delta g^{\mu\nu}}.
\end{eqnarray}

One can schematically express the above expression (\ref{NHT1}) as 
\begin{eqnarray}\label{NHTS}
g^{\mu\nu}T_{\mu\nu}=T_0+\xi T_1+\kappa T_2,
\end{eqnarray}
where $T_0$ is the trace of the stress tensor for the pure JT gravity theory. On the other hand, $T_1$ and $T_2$ are the correction terms due to the presence of 2-derivative and 4-derivative interactions in (\ref{action 2D}).  

The trace of the stress tensor in the JT gravity is given by
\begin{eqnarray}\label{NHT2}
T_0=e^{-2\omega_0}(\phi_0''-12\phi_0e^{2\omega_0}),
\end{eqnarray}
which turns out to be zero by virtue of equations of motion (\ref{v01}) and (\ref{v02}).

On the other hand, the first order correction in (\ref{NHTS}) due to the presence of 2-derivative interactions is given by
\begin{align}\label{NHT3}
T_1=&e^{-2\omega_0}\Big[-12(e^{2\omega_0}\phi_1+2\omega_1\phi_0e^{2\omega_0})+\phi_1''+\frac{\phi_0}{2}A_{t(0)}'^2e^{-2\omega_0}\Big]+\nonumber\\
&2\omega_1e^{-2\omega_0}[-\phi_0''+12\phi_0e^{2\omega_0}],
\end{align}
which  vanishes identically by virtue of (\ref{bh0}) and (\ref{bh23}). 

Finally, we calculate the correction due to the presence of 4-derivative interactions
\begin{align}\label{NHT5}
T_2=&-e^{-2\omega_0}\Big[12e^{2\omega_0}(\phi_2+4\phi_0\omega_2)-(\phi_2''+2\omega_2\phi_0'')+A(z)\Big]+\nonumber\\
&4\omega_2e^{-2\omega_0}\Big[-\phi_0''+12\phi_0e^{2\omega_0}\Big],
\end{align}
which also vanishes identically due to equations (\ref{bh0}), (\ref{ze}) and (\ref{bh43}).

Combining (\ref{NHT2})-(\ref{NHT5}), we conclude that the trace of the stress tensor (\ref{NHT1}) vanishes identically in the near horizon limit. These calculations suggest that there exists a conformal field theory in the near horizon limit. Our next step would be to compute the central charge corresponding to this conformal field theory. 

In order to simplify our analysis, we switch off 4-derivative interactions\footnote{See Appendix \ref{vp} for the correction due to 4-derivative interactions.} for the moment and transform the Einstein-Hilbert action (\ref{action 2D}) into the Liouville theory using the following field redefinition \cite{Alishahiha:2008tv, Solodukhin:1998tc}
\begin{align}\label{NHT6}
   \phi=\Phi^2=q\Phi_H\psi \hspace{2mm}\text{and}\hspace{2mm} g_{\mu\nu}\rightarrow e^{\frac{2\psi}{q\Phi_H}}g_{\mu\nu},
\end{align}
where q is a constant and $\Phi_H=\Phi|_{horizon}$\footnote{Notice that, we have taken out a common factor $\frac{1}{4}$ in (\ref{NHT7}) in order to be consistent with \cite{Solodukhin:1998tc}.}.

We plug (\ref{NHT6}) into (\ref{action 2D}) which yields
\begin{align}\label{NHT7}
    S=\int d^2x\sqrt{-g}\Big[\frac{1}{4}q\Phi_H\psi R+\frac{1}{2}(\bigtriangledown_{\mu}\psi)^2+3q\Phi_He^{\frac{2\psi}{q\Phi_H}}\psi-\frac{\xi}{16}q\Phi_H\psi e^{-\frac{2\psi}{q\Phi_H}}F^2\Big].
\end{align}

Next, we integrate out the gauge degrees of freedom in the action (\ref{NHT7}) which by virtue of the equation of motion (\ref{v23}), yields
\begin{align}\label{NHT8}
    S_L=\int d^2x\sqrt{-g}\Big[\frac{1}{2}(\bigtriangledown_{\mu}\psi)^2+\frac{1}{4}q\Phi_H\psi R+V(\psi)\Big],\hspace{1mm}
    V(\psi)=3q\Phi_He^{\frac{2\psi}{q\Phi_H}}\psi+\frac{\xi}{8}\frac{ e^{-\frac{2\psi}{q\Phi_H}}}{q\Phi_H\psi}b^2
\end{align}
where $V(\psi)$ is the potential\footnote{See Appendix \ref{vp} for the properties of the potential $V(\psi)$.} of the ``generalised" Liouville theory that contains the 2-derivative interaction term and $b$ is the integration constant. We discuss more about the generalised Liouville theory in the section [\ref{GLT}].

On varying (\ref{NHT8}) with respect to $g_{\mu\nu}$, we obtain the equation of motion for the metric as 
\begin{align}\label{NHT9}
    \frac{1}{2}(\partial_{\mu}\psi)(\partial_{\nu}\psi)-\frac{1}{4}g_{\mu\nu}(\bigtriangledown\psi)^2+\frac{q\Phi_H}{4}(g_{\mu\nu}\square\psi-\bigtriangledown_{\mu}\bigtriangledown_{\nu}\psi)-\frac{1}{2}g_{\mu\nu}V(\psi)=0.
\end{align}

We prefer to solve (\ref{NHT9}) in the following static gauge \cite{Solodukhin:1998tc}
\begin{align}\label{nhg}
    ds^2=-g(x)dt^2+\frac{dx^2}{g(x)}\hspace{2mm},\hspace{2mm}g(x)=\frac{2}{\beta_H}(x-x_H)+O(x-x_H)^2,
\end{align}
where the horizon is located at $x=x_H$.

In the near horizon limit, it is convenient to carry out an analysis in $(t,z)$ coordinate, where $z$ is given by 
\begin{align}
   z=\frac{\beta_H}{2}\log[x-x_H].
    \end{align}
    
In $(t,z)$ coordinates, (\ref{nhg}) reduces to
\begin{align}\label{nhgn}
    ds^2=-g(z)dt^2+g(z)dz^2\hspace{2mm},\hspace{2mm}g(z)=\frac{2}{\beta_H}e^{\frac{2z}{\beta_H}},
\end{align}
where the horizon is located at $z\rightarrow-\infty$.

Next, we note down the components of the stress tensor (\ref{NHT1}) of (\ref{NHT8}) in the gauge (\ref{nhgn}) 
\begin{align}
    T_{tt}&=\frac{1}{4}\Big[(\partial_t\psi)^2+(\partial_z\psi)^2\Big]-\frac{q\Phi_H}{4}\Big[\partial^2_z\psi-\frac{1}{\beta_H}\partial_z\psi\Big]+\frac{1}{2}g(z)V(\psi)\label{NHT10},\\
    T_{tz}&=\frac{1}{2}\partial_t\psi\partial_z\psi-\frac{q\Phi_H}{4}\Big[\partial_z\partial_t\psi-\frac{1}{\beta_H}\partial_t\psi\Big]\label{NHT11}.
\end{align}

Finally, we define the Virasoro generators \cite{Solodukhin:1998tc} in terms of the components of the stress tensor (\ref{NHT10})-(\ref{NHT11}) as 
\begin{align}\label{NHT12}
    L_n=\frac{L}{2\pi}\int _{-\frac{L}{2}}^{\frac{L}{2}}dze^{i\frac{2\pi}{L}nz}T_{++}(z),
\end{align}
where,  $T_{++}=T_{tt}+T_{tz}$ and the integration is on the circle of circumference $L$. At the end of the calculation, we stretch $L$ upto infinity.

Using (\ref{NHT10})-(\ref{NHT11}), in the near horizon limit i.e. $z\rightarrow-\infty$, we obtain
\begin{align}\label{NHT13}
    T_{++}=\frac{1}{4}\Big[(\partial_t+\partial_z)\psi\Big]^2-\frac{q\Phi_H}{4}\Big[\partial_z(\partial_z+\partial_t)\psi-\frac{1}{\beta_H}(\partial_z+\partial_t)\psi\Big].
\end{align}
Notice that, the expression of $T_{++}$ (\ref{NHT13}) does not depend on the form of the potential $V(\psi)$ in the near horizon limit. 

A straightforward calculation reveals that the Virasoro generators (\ref{NHT12}) along with (\ref{NHT13}) satisfy the following commutation relation
\begin{align}
    i\{L_k,L_n\}=(k-n)L_{n+k}+\frac{c_{H}}{12}k\Big(k^2+\Big(\frac{L}{2\pi\beta_H}\Big)^2\Big)\delta_{n+k},0,
\end{align}
where $c_H=3\pi q^2\Phi_H^2$ is the central charge associated with the conformal field theory near the horizon.

Using (\ref{bh0}) and (\ref{bh23}), one can further rewrite the central charge as
\begin{align}\label{cch}
    c_H=&\hspace{1mm}3\pi q^2\Bigg[\frac{\sqrt{\mu}}{6}+\xi\Bigg\{\Bigg\{\frac{3 }{8\sqrt{\mu}}-\frac{
    \tanh^{-1}\Big[\frac{\delta+\sqrt{\mu}}{\sqrt{\mu}}\Big]}{8\sqrt{\mu}}-\frac{\log[2]}{16\sqrt{\mu}}-\frac{\log\Big[1-\frac{(\delta+\sqrt{\mu})^2}{\mu}\Big]}{8\sqrt{\mu}}+\nonumber\\
    &\frac{\log[\sqrt{\mu}]}{4\sqrt{\mu}}-\frac{3\log\Big[1-\frac{\delta+\sqrt{\mu}}{\sqrt{\mu}}\Big]}{16\sqrt{\mu}}+\frac{\log[(\delta+\sqrt{\mu})^2-\mu]}{4\sqrt{\mu}}\Bigg\}Q^2+\Big\{\frac{1}{6}\sqrt{\mu}+\nonumber\\
    &\frac{1}{6}\sqrt{\mu}\tanh^{-1}\Big[\frac{\delta+\sqrt{\mu}}{\sqrt{\mu}}\Big]+\frac{1}{12}\sqrt{\mu}\log[\delta]-\frac{1}{12}\sqrt{\mu}\log[2\sqrt{\mu}]\Big\}\frac{3\log(\mu)}{4\mu}Q^2+\nonumber\\
    &\sqrt{\mu} d_4\Bigg\}\Bigg].
\end{align}
$\bullet$ Note : In arriving at (\ref{cch}), we write the full solution of the gauge field (\ref{e3}) and the dilaton (\ref{e2})
 $$
      A_t^{bh}=A_{t(0)}^{bh}+\frac{\kappa}{\xi}A_{t(1)}^{bh},\hspace{2mm}
      \phi^{bh}=\phi_{(0)}^{bh}+\xi\phi_{(1)}^{bh}+\kappa\phi_{(2)}^{bh}.
$$
  In the near horizon limit, we absorb the integration constant $d_3$ in $\phi_{(1)}^{bh}$ (\ref{bh23}) into the constant $d_{11}$ in $\phi_{(2)}^{bh}$ (\ref{bh46}) without any loss of generality. Similarly, we absorb the additive constant $d_1$ in $A_{t(0)}^{bh}$ (\ref{bh21}) into the constant $d_6$ in $A_{t(1)}^{bh}$ (\ref{bh4a}). Furthermore, we write the constant $d_2$ in terms of the charge $Q$ using (\ref{ht}).

Notice that, the expression (\ref{cch}) diverges near the horizon which is due to the divergences in the corresponding density of states as we have discussed in the Section (\ref{divw}). Therefore, in order to obtain a finite answer, we calculate the central charge in the limit $\rho\rightarrow\sqrt{\mu}+\delta$, where $\delta<<\sqrt{\mu}$.

\section{Generalised Liouville Theory and Weyl anomaly}\label{GLT}
In this section, we study the generalised Liouville theory in 2D that contains the 2-derivative interaction terms (\ref{NHT8}). In particular, we focus on the Weyl transformation properties of the generalised Liouville theory and the Weyl anomaly associated with it. 

Liouville theory is a conformal field theory in 2D which is dual to the Einstein gravity with negative cosmological constant in three-dimensions \cite{Grumiller:2007wb,Jackiw:2005su,Coussaert:1995zp,Li:2019mwb}. The action for the Liouville theory in two-dimensions is given by 
\begin{align}\label{sl1}
    S_L=\int d^2x\sqrt{-g}\Big[\frac{1}{2}g^{\mu\nu}\bigtriangledown_{\mu}\psi\bigtriangledown_{\nu}\psi+\frac{R}{\beta}\psi-\frac{m^2}{\beta^2}e^{\beta\psi}\Big],
\end{align}
where $R$ is the Ricci scalar, $\psi$ is the scalar field and ($\beta$, $m$) are the constants.

It is shown in \cite{Grumiller:2007wb} that one can construct the Liouville theory (\ref{sl1}) by consistent dimensional reduction of pure Einstein-Hilbert action in $D$ dimensions. The authors in \cite{Grumiller:2007wb} start with the following action
\begin{align}\label{sl2}
    S_D=\int d^Dx\sqrt{-g_{(D)}}R^{(D)}.
\end{align}

The dimensional reduction ansatz for space-time metric is given by
\begin{align}\label{sl3}
    ds^2=g_{\mu\nu}^{(D)}dx^{\mu}dx^{\nu}=g_{\alpha\beta}dx^{\alpha}dx^{\beta}+\frac{1}{\lambda}\phi^{\frac{2}{(D-2)}}d\Omega^2_{S_{D-2}},
    \end{align}
where ($\mu,\nu$) are $D$ dimensional indices, ($\alpha,\beta$) are 2 dimensional indices and $\lambda$ is the parameter having dimensions $[L]^{-2}$. Next, the authors parameterize the dimensions by $D=2+\epsilon$ and plug (\ref{sl3}) into (\ref{sl2}) to obtain the action for Liouville theory in the limit $\epsilon\rightarrow0$.

The stress energy tensor and the equation of motion for the field $\psi$ corresponding to the Liouville theory (\ref{sl1}) are  given by 
\begin{align}
T_{\mu\nu}&=\frac{1}{\sqrt{-g}}\frac{\partial S_{L}}{\partial g^{\mu\nu}}\nonumber\\
&=\frac{1}{2}\bigtriangledown_{\mu}\psi\bigtriangledown_{\nu}\psi-\frac{1}{4}g_{\mu\nu}(\bigtriangledown\psi)^2+\frac{1}{\beta}(g_{\mu\nu}\bigtriangledown^2\psi-\bigtriangledown_{\mu}\bigtriangledown_{\nu}\psi)+\frac{m^2}{2\beta^2}g_{\mu\nu}e^{\beta\psi},\label{sess1}\\
\bigtriangledown^2\psi&=\frac{R}{\beta}-\frac{m^2}{\beta}e^{\beta\psi}.\label{sess2}
\end{align}

Using (\ref{sess2}), one can compute the trace of stress tensor (\ref{sess1}) as
\begin{align}\label{trn}
   <T^{\mu}_{\mu}>= \frac{R}{\beta^2},
\end{align}
which does not vanish in curved space-time. This is what is known as the Weyl anomaly, where the coefficient $\frac{1}{\beta^2}$ is related to the central charge of the CFT.

Next, we look at the Weyl transformation properties of the Liouville theory (\ref{sl1}). In order to proceed, we consider the following field transformations \cite{Jackiw:2005su}
\begin{align}\label{sl4}
    g_{\mu\nu}\rightarrow e^{2\sigma}g_{\mu\nu}\hspace{2mm},\hspace{2mm}\psi\rightarrow\psi-\frac{2}{\beta}\sigma,
\end{align}
where $\sigma\equiv\sigma(t,z)$. Under the above transformation (\ref{sl4}), the action (\ref{sl1}) is transformed (up to boundary terms) as, $S_L\rightarrow S_L+\delta S_L$ where the difference is denoted as 
\begin{align}\label{sl5}
   \delta S_L=-\frac{2}{\beta^2}\int d^2x\sqrt{-g}\Big[R\sigma+g^{\mu\nu}\bigtriangledown_{\mu}\sigma\bigtriangledown_{\nu}\sigma\Big].
\end{align}

Equation (\ref{sl5}) suggests that the Liouville theory (\ref{sl1}) is not invariant under the transformation\footnote{See \cite{Jackiw:2005su} to obtain the Liouville theory in $D=2$ dimensions from Weyl invariant theories in $D>2$. } (\ref{sl4}).
However, the difference  $\delta S_L$ (\ref{sl5}) does not depend on the field $\psi$. As a result, the  equation of motion for $\psi$ (\ref{sess2})
remains invariant under the transformation (\ref{sl4}).

Notice that, the difference $\delta S_L$ (\ref{sl5}) can be set equal to zero (up to boundary terms) if we impose the equation of motion for $\sigma$
\begin{align}
    \bigtriangledown_{\mu}\bigtriangledown^{\mu}\sigma=R.\label{sric1}
\end{align}

One can solve (\ref{sric1}) for $\sigma$ in the  static light cone gauge (\ref{sg}) which yields a solution of the form
\begin{align}
 \sigma=-2\omega+z b_1+b_2,   
\end{align}
where $b_1$ and $b_2$ are the integration constants. Therefore, given the onshell condition (\ref{sric1}), the action (\ref{sl1}) is claimed to be invariant under the Weyl re-scaling (\ref{sl4}).

The Liouville theory (\ref{NHT8}) that we obtain is different from the standard Liouville theory (\ref{sl1}) in the sense that (\ref{NHT8}) does not reduce to (\ref{sl1}) in the limit $\xi\rightarrow0$. We are interested to look at the transformation properties of this generalised Liouville theory (\ref{NHT8}) under the following field redefinition 
\begin{align}\label{sft1}
     g_{\mu\nu}\rightarrow e^{2\sigma}g_{\mu\nu}\hspace{2mm},\hspace{2mm}\psi\rightarrow\psi-\tilde{c}_H\sigma\hspace{2mm},\hspace{2mm}\text{where}\hspace{2mm}\tilde{c}_H=\sqrt{\frac{c_H}{3\pi}}.
\end{align}

Under the above transformation (\ref{sft1}), the action (\ref{NHT8}) is transformed (upto boundary terms) as 
\begin{align}\label{stransact}
     \tilde{S_L}=&\int d^2x\sqrt{-g}\Big[\frac{1}{2}(\bigtriangledown_{\mu}\psi)^2+\frac{1}{4}\tilde{c}_H\psi R+3\tilde{c}_He^{\frac{2\psi}{\tilde{c}_H}}\psi+\frac{\xi b^2}{8}\frac{ e^{-\frac{2\psi}{\tilde{c}_H}}}{\tilde{c}_H(\psi-\sigma \tilde{c}_H)}e^{4
     \sigma}+\nonumber\\
     &\frac{1}{2}\tilde{c}_H\psi\bigtriangledown_{\mu}\bigtriangledown^{\mu}\sigma-\frac{\sigma}{4}\tilde{c}_H^2R-3\tilde{c}_H^2\sigma e^{\frac{2\psi}{\tilde{c}_H}}\Big],
\end{align}
 which is clearly not invariant. On top of that, even the dynamics of the scalar field $(\psi)$ is influenced deriving the transformation (\ref{sft1}). 
 
The stress energy tensor and the equation of motion for $\psi$ that follows from (\ref{NHT8}) are given by 
\begin{align}
T_{\mu\nu}&=\frac{1}{\sqrt{-g}}\frac{\delta S_{L}}{\delta g^{\mu\nu}}\nonumber\\
&=\frac{1}{2}\bigtriangledown_{\mu}\psi\bigtriangledown_{\nu}\psi-\frac{1}{4}g_{\mu\nu}(\bigtriangledown\psi)^2+\frac{1}{4}\tilde{c}_H(g_{\mu\nu}\bigtriangledown^2\psi-\bigtriangledown_{\mu}\bigtriangledown_{\nu}\psi)-\frac{1}{2}g_{\mu\nu}V(\psi),\label{sess3}\\
\bigtriangledown^2\psi&=\frac{1}{4}\tilde{c}_HR+V'(\psi).\label{sess4}
\end{align}

Using (\ref{sess4}), we compute the trace of stress tensor (\ref{sess3}), which yields
\begin{align}\label{sstn0}
   <T^{\mu}_{\mu}>\sim\frac{c_H}{48\pi}R+O\Big(\frac{1}{\sqrt{c_H}}\Big).
\end{align}
 
 Equation (\ref{sstn0}) confirms that the theory (\ref{NHT8}) is not Weyl Invariant. This is what we identify as the Weyl anomaly \cite{Castro:2019vog} for the generalised Liouville theory (\ref{stransact}). 
 
 After some algebra, one can express the transformed action (\ref{stransact}) as
 \begin{align}\label{svar1}
     S_L\rightarrow S_L+&\int d^2x\sqrt{-g}\Bigg[
     \frac{1}{2}\tilde{c}_H\psi\bigtriangledown_{\mu}\bigtriangledown^{\mu}\sigma-\frac{\sigma}{4}\tilde{c}_H^2R-3\tilde{c}_H^2\sigma e^{\frac{2\psi}{\tilde{c}_H}}+\nonumber\\
     &\frac{\xi b^2}{8}\frac{e^{-\frac{2\psi}{\tilde{c}_H}}}{\tilde{c}_H}\Big\{\frac{e^{4\sigma}}{(\psi-\sigma \tilde{c}_H)}-\frac{1}{\psi}\Big\}\Bigg].
 \end{align}

Notice that, following our previous arguments, the variation $\delta S_L$ (\ref{svar1}) can be set equal to zero if we impose the equation of motion for $\sigma$ 
\begin{align}\label{sem1}
    \frac{1}{2}\tilde{c}_H\psi\bigtriangledown_{\mu}\bigtriangledown^{\mu}\sigma-\frac{\sigma}{4}\tilde{c}_H^2R-3\tilde{c}_H^2\sigma e^{\frac{2\psi}{\tilde{c}_H}}+\frac{\xi b^2}{8}\frac{e^{-\frac{2\psi}{\tilde{c}_H}}}{\tilde{c}_H}\Big\{\frac{e^{4\sigma}}{(\psi-\sigma \tilde{c}_H)}-\frac{1}{\psi}\Big\}=0.
\end{align}

Next, we solve the equation of motion for $\sigma$  (\ref{sem1}) in the static light cone gauge (\ref{sg}). To start with, we perturbatively expand the fields $\psi$, $\omega$ and $\sigma$ treating $\xi$ as an expansion parameter
\begin{align}
    \psi&=\psi_0+\xi \psi_1,\label{sig2}\\
    \omega&=\omega_0+\xi \omega_1,\label{sig3}\\
    \sigma&=\sigma_0+\xi\sigma_1\label{sig4}.
\end{align}

The subscript (0) denotes  the zeroth order fields and the subscript (1) denotes the first order correction in the fields due to the presence of 2-derivative interactions. 

Using (\ref{sig2})-(\ref{sig4}), one can write the zeroth order equation of motion for $g_{\mu\nu}$ (\ref{sess3}), $\psi$ (\ref{sess4}) and $\sigma$ (\ref{sem1}) as
\begin{align}
    \psi_0'^2+\tilde{c}_H(\omega_0'\psi_0'-\psi_0'')+6\tilde{c}_He^{2\big(\frac{\psi_0}{\tilde{c}_H}+\omega_0\big)}\psi_0=0,\label{nee1}\\
       \psi_0'^2+\tilde{c}_H\omega_0'\psi_0'-6\tilde{c}_He^{2\big(\frac{\psi_0}{\tilde{c}_H}+\omega_0\big)}\psi_0=0,\label{nee2}\\
       \psi_0''+\frac{1}{2}\tilde{c}_H\omega_0''-3\psi_0'e^{2\big(\frac{\psi_0}{\tilde{c}_H}+\omega_0\big)}\Big(\tilde{c}_H+2\psi_0\Big)=0,\label{nee3}\\
       \psi_0\sigma_0''+\tilde{c}_H\Big[\omega_0''-6e^{2\big(\frac{\psi_0}{\tilde{c}_H}+\omega_0\big)}\Big]\sigma_0=0.\label{nee4}
\end{align}

Notice that, the above equations (\ref{nee1})-(\ref{nee4}) are the coupled non-linear differential equations and it is difficult to solve them exactly. Therefore, we solve these equations in the large $\tilde{c}_H$ limit and ignore all terms of the order $O(\frac{1}{\tilde{c}_H})$.

Using (\ref{nee1})-(\ref{nee3}), one can write the equation for $\psi_0$ in the large $\tilde{c}_H$ limit as
\begin{align}
    4\psi_0\psi_0'+\tilde{c}_H\psi_0'-\tilde{c}_H=0.\label{nee5}
\end{align}

On solving (\ref{nee5}) for $\psi_0$, we obtain
\begin{align}\label{nee6}
    \psi_0=\frac{1}{4}\Big(-\tilde{c}_H\pm\sqrt{8c_1+\tilde{c}_H\big(8z+\tilde{c}_H\big)}\Big),
\end{align}
where $c_1$ is the integration constant.

Finally, using (\ref{nee1})-(\ref{nee3}) and (\ref{nee6}) in the equation of motion for $\sigma$ (\ref{nee4}),  we obtain 
\begin{align}
   \sigma_0''+f(z)\sigma_0=0,\label{nee7}
\end{align}
where $f(z)$ is given by
\begin{align}\label{nee8}
    f(z)=\frac{32}{\tilde{c}_H^2\Big(\frac{8 c_1}{\tilde{c}_H^2}+\frac{8z}{\tilde{c}_H}+1\Big)^{\frac{3}{2}}\Big(1+\sqrt{\big(\frac{8 c_1}{\tilde{c}_H^2}+\frac{8z}{\tilde{c}_H}+1\big)}\Big)}=\frac{16}{\tilde{c}_H^2}+O\Big(\frac{1}{\tilde{c}_H^3}\Big).
\end{align}

On solving (\ref{nee7}) using (\ref{nee8}) in the large $\tilde{c_H}$ limit, we obtain
\begin{align}
    \sigma_0=c_2\cos{\Big(\frac{4z}{\tilde{c}_H}\Big)}+c_3\sin{\Big(\frac{4z}{\tilde{c}_H}\Big)},
\end{align}
where $c_2$ and $c_3$ are the integration constants.

Finally, we note down equations at leading order in $\xi$ which yield

\begin{align}
    &2\psi_1'\psi_0'+\tilde{c}_H(\omega_0'\psi_1'+\omega_1'\psi_0'-\psi_1'')+2e^{2\omega_0}\Bigg[3\tilde{c}_He^{2\frac{\psi_0}{\tilde{c}_H}}\Big(\frac{2}{\tilde{c}_H}\psi_0\psi_1+\psi_1\Big)+\frac{b^2}{8\psi_0}\frac{1}{\tilde{c}_H}e^{-2\frac{\psi_0}{\tilde{c}_H}}+\nonumber\\
    &6\omega_1\psi_0\tilde{c}_He^{2\frac{\psi_0}{\tilde{c}_H}}\Bigg]=0,\label{nde1}\\
    &2\psi_1'\psi_0'+\tilde{c}_H(\omega_0'\psi_1'+\omega_1'\psi_0')-2e^{2\omega_0}\Bigg[3\tilde{c}_He^{2\frac{\psi_0}{\tilde{c}_H}}\Big(\frac{2}{\tilde{c}_H}\psi_0\psi_1+\psi_1\Big)+\frac{b^2}{8\psi_0}\frac{1}{\tilde{c}_H}e^{-2\frac{\psi_0}{\tilde{c}_H}}+\nonumber\\
    &6\omega_1\psi_0\tilde{c}_He^{2\frac{\psi_0}{\tilde{c}_H}}\Bigg]=0,\label{nde2}\\
    &\psi_1''+\frac{1}{2}\tilde{c}_H\omega_1''-3e^{2\big(\frac{\psi_0}{\tilde{c}_H}+\omega_0\big)}\Bigg[\tilde{c}_H\Big\{\psi_1'+\frac{2}{\tilde{c}_H}\psi_1\psi_0'+2\omega_1\psi_0'\Big\}+2\Big\{\psi_0'\psi_1+\psi_1'\psi_0+\nonumber\\
    &\frac{2}{\tilde{c}_H}\psi_1\psi_0'\psi_0+2\omega_1\psi_0'\psi_0\Big\}\Bigg]+\frac{b^2}{4\psi_0}\frac{1}{\tilde{c}_H}\psi_0'e^{2\big(-\frac{\psi_0}{\tilde{c}_H}+\omega_0\big)}\Bigg(\frac{1}{2\psi_0}+\frac{1}{\tilde{c}_H}\Bigg)=0,\label{nde3}\\
    & \psi_1\sigma_0''+ \psi_0\sigma_1''+\tilde{c}_H\Big[\omega_0''\sigma_1+\omega_1''\sigma_0\Big]-6\tilde{c}_He^{2\big(\frac{\psi_0}{\tilde{c}_H}+\omega_0\big)}\Big(\sigma_1+2\omega_1\sigma_0+\frac{2}{\tilde{c}_H}\psi_1\sigma_0\Big)+\nonumber\\
    &\frac{ b^2}{4}\frac{e^{-2\big(\frac{\psi_0}{\tilde{c}_H}-\omega_0\big)}}{\tilde{c}_H^2}\Bigg[\frac{e^{4\sigma_0}}{(\psi_0-\sigma_0 \tilde{c}_H)}-\frac{1}{\psi_0}\Bigg]=0.\label{nde4}
\end{align}
Obtaining solutions for (\ref{nde1})-(\ref{nde4}) are quite involved which we therefore do not pursue here.

\section{Conclusion}\label{sum}
To summarise, in the present chapter we extend the notion of 2D Einstein-Maxwell-Dilaton gravity by incorporating the most general form of quartic (non-linear) interactions between the $U(1)$ gauge fields and the space-time metric, $g_{\mu\nu}$ allowed by the diffeomorphism invariance. We further explore the effects of adding such quartic interactions on the dual field theory observables at strong coupling.

It is important to notice that the non-linear $U(1)$ gauge interactions are generally not conformally invariant and also break the $SO(2)$ duality invariance of the theory. However, recently, the authors in \cite{Bandos:2020jsw}-\cite{Kosyakov:2020wxv} proposed a radical non-linear generalisation of the standard Maxwell electrodynamics that preserves both the conformal symmetry as well as the $SO(2)$ duality invariance of Maxwell electrodynamics and goes under the name of the ``ModMax'' Electrodynamics. In the next chapter, we explore the effects of these non-linear interactions on the observables pertinent to the 1D boundary theory.



\chapter{$AdS_2$ holography and ModMax}
\allowdisplaybreaks
\pagestyle{fancy}
\fancyhead[LE]{\emph{\leftmark}}
\fancyhead[LO]{\emph{\rightmark}}
\rhead{\thepage}

 Our starting point is the Einstein's gravity in four dimensions accompanied by the ModMax Lagrangian \cite{Bandos:2020jsw}-\cite{Kosyakov:2020wxv}. The 2D gravity action is obtained following a suitable dimensional reduction which contains a 2D image of the 4D ModMax Lagrangian. We carry out a perturbative analysis to find out the vacuum structure of the theory which asymptotes to $AdS_2$ in the absence of $U(1)$ gauge fields. We estimate the holographic central charge and obtain corrections perturbatively upto quadratic order in the ModMax and the $U(1)$ coupling. We also find out ModMax corrected 2D black hole solutions and discuss their extremal limits.

\section{Overview and motivation}

The non-linear generalisations of Maxwell electrodynamics \cite{Born:1934gh}-\cite{Heisenberg:1936nmg} in four dimensions play a pivotal role in understanding the dynamics of charged particles in the strong field regime. For example, the Born-Infeld (BI) theory \cite{Born:1934gh} was proposed in order to obtain the finite self energy corrections for a charged particle in an electromagnetic field. On the other hand, the Heisenberg-Euler-Kockel (HEK) model \cite{Heisenberg:1936nmg} describes the vacuum polarization effects of Quantum Electrodynamics\footnote{See a recent review \cite{Sorokin:2021tge} for different versions of the non-linear modified theories of Maxwell electrodynamics.}. However, both of these (non-linear) theories meet the standard Maxwell electrodynamics in the limit of ``weak'' field approximations.

Generally, the non-linear generalizations of Maxwell electrodynamics (NLE) is characterised by an action that contains a Lorentz scalar and a pseudo scalar which are quadratic in the field strength ($F^{\mu\nu}$) \cite{Sorokin:2021tge}-\cite{Peres:1961zz}
\begin{align}\label{introlinv}
    S= \frac{1}{2}F_{\mu\nu}F^{\mu\nu}\hspace{1mm},\hspace{2mm}P=\frac{1}{2}F_{\mu\nu}\Tilde{F}^{\mu\nu},
\end{align}
where $\Tilde{F}^{\mu\nu}$ is the Hodge dual of $F^{\mu\nu}$.

For instance, the BI electrodynamics is described by the following Lagrangian density \cite{Born:1934gh}
\begin{align}\label{introbi}
    \mathcal{L}_{BI}=T-\sqrt{T^2+\frac{T}{2}F_{\mu\nu}F^{\mu\nu}-\frac{1}{16}\big(F_{\mu\nu}\Tilde{F}^{\mu\nu}\big)^2},
\end{align}
where $T$ is the coupling parameter having the dimension of energy density.  Clearly, in the weak field limit ($T\rightarrow\infty$), the Lagrangian density (\ref{introbi}) reduces to the standard Maxwell electrodynamics. 

Unlike the standard Maxwell electrodynamics, its non-linear modifications are generally not invariant under the $SO(2)$ duality transformations and in fact break the conformal symmetry in four dimensions. For instance, the HEK theory \cite{Heisenberg:1936nmg} is not invariant under the electromagnetic duality and does not have a conformal symmetry. However, the BI electrodynamics is invariant under the $SO(2)$ duality \cite{Gibbons:1995cv} although it is not conformal invariant due to the presence of the dimensionful coupling ($T$) in the theory (\ref{introbi}).

Recently, there has been a radical proposal \cite{Bandos:2020jsw}-\cite{Kosyakov:2020wxv} to (non-linearly) generalize the Maxwell electrodynamics which retains its conformal invariance (in four dimensions) as well as preserves the $SO(2)$ duality symmetry. This goes under the name of the ``ModMax'' electrodynamics\footnote{For details, see the recent review \cite{Sorokin:2021tge}.}.  

The ModMax electrodynamics is a 1-parameter deformation of the Maxwell electrodynamics in four dimensions that is described by the following Lagrangian density\footnote{In the limit $\gamma\rightarrow0$, the ModMax electrodynamics  reduces to the standard Maxwell electrodynamics.} \cite{Bandos:2020jsw}-\cite{Kosyakov:2020wxv} 
\begin{align}\label{introld}
     \mathcal{L}_{MM}&=\frac{1}{2}\Big(S \cosh{\gamma}-\sqrt{S^2+P^2}\sinh{\gamma}\Big),
\end{align}
where $\gamma$ is the dimensionless coupling constant that measures the strength of the electromagnetic self interaction.

The physical requirements that the theory must be unitary and preserves the causality restrict the ModMax parameter $(\gamma)$ to take only positive values ($\gamma>0$)  \cite{Bandos:2020jsw}. The above restriction guarantees that the Lagrangian density (\ref{introld}) is a convex function of the electric field strength $E^i$.  

There have been some further modifications to the ModMax electrodynamics in the literature which include the 1-parameter generalisation of the BI theory\footnote{In the weak field limit, the ($\gamma$BI) theory reduces to the standard ModMax electrodynamics (\ref{introld}).} ($\gamma$BI) \cite{Bandos:2020hgy}  and $\mathcal{N}=1$ supersymmetric extension of the ModMax electrodynamics\footnote{See \cite{Kruglov:2021bhs}-\cite{ Cano:2021tfs} for further details. }  \cite{Bandos:2021rqy}. The supersymmetric version of the ModMax electrodynamics is invariant under the electromagnetic duality as well as posses the superconformal symmetry \cite{Bandos:2021rqy}.

The ModMax electrodynamics finds an extensive application in theories of gravity \cite{Barrientos:2022bzm}-\cite{Amirabi:2020mzv} as well. In fact, a large number of solutions have been obtained down the line. For instance, accelerated black holes \cite{Barrientos:2022bzm}, the Taub-NUT \cite{BallonBordo:2020jtw}-\cite{Flores-Alfonso:2020nnd} and Reissner-Nordstorm solutions \cite{Amirabi:2020mzv} in diverse spacetime dimensions have been constructed in the presence of ModMax interactions and the effects of non-linearity were explored on their thermal properties. Recently, the non-linear models of electrodynamics have also found their applications in the context of strongly correlated systems  \cite{Cai:2008in}-\cite{Hartnoll:2008vx} by means of the celebrated $AdS_{d+1}/CFT_d$ correspondence  \cite{Maldacena:1997re}-\cite{Gubser:1998bc}.

Despite of several notable applications those are alluded to the above, ModMax theories are least explored in $ AdS_2$ holography and in particular in the context of the JT/SYK correspondence \cite{Jackiw:1984je}-\cite{Lala:2019inz}, \cite{Davison:2016ngz}-\cite{Castro:2008ms}. In this chapter, we fill up some of these gaps in the literature and find out an interpretation for the projected ModMax interactions within the realm of 2D gravity theories.

In this chapter, we cook up a theory of JT gravity in the presence of 2D ``projected'' ModMax interactions and compute various physical entities associated with the boundary theory. For instance, we construct the holographic stress-energy tensor \cite{Castro:2008ms, Rathi:2021aaw}, \cite{Cadoni:2000gm}-\cite{Narayan:2020pyj} and compute the associated central charge  \cite{Castro:2008ms, Rathi:2021aaw}, \cite{Cadoni:2000gm}-\cite{Narayan:2020pyj} for the boundary theory. Finally, we construct black hole solutions in two dimensions and explore the effects of projected ModMax interactions on their thermal behaviour.

The organisation for the rest of the chapter is as follows :

$\bullet$ In Section \ref{secmd2d}, we follow suitable dimensional reduction procedure \cite{Davison:2016ngz, Lala:2020lge, Rathi:2021aaw} to construct a model for JT gravity in the presence of 2D projected ModMax interactions. We also clarify the meaning of projected ModMax interactions in 2D and in particular present a detail comparison with the 4D ModMax interactions.

$\bullet$ In Section \ref{secsol}, we calculate the conformal dimensions of different scalar operator in deep IR limit and make a comparative analysis between them. We further explore the vacuum structure of the theory using the Fefferman-Graham gauge \cite{Rathi:2021aaw,fefferman} by treating the non-linear $U(1)$ gauge interactions as ``perturbations'' over the pure JT gravity solutions. We estimate these solutions upto quadratic order in the gauge and ModMax couplings.

$\bullet$ In Section \ref{secstcc}, we construct the ``renormalised'' boundary stress tensor and investigate its transformation properties under the combined action of the diffeomorphism and the $U(1)$ gauge transformations  \cite{Castro:2008ms}. {We compute the central charge $(c_M)$ associated with the boundary theory  \cite{Castro:2008ms} up to quadratic order in the (ModMax and $U(1)$) couplings.

$\bullet$ In Section \ref{secbhwithmm}, we construct the black hole solutions upto quadratic order in the couplings. We observe that the non-linear interactions (or the projected ModMax interactions) play a crucial role in obtaining a finite value for the background fields at the horizon. 

Furthermore, we compute the Hawking temperature for 2D black holes \cite{Hawking:1975vcx} and calculate the associated Wald entropy \cite{Wald:1993nt, Brustein:2007jj, Pedraza:2021cvx}. We also investigate the ``extremal'' limit associated with these 2D black hole solutions and calculate the corresponding Wald entropy.

$\bullet$ Finally, we conclude this chapter in Section \ref{secconc}.

\section{JT gravity and 2D projected ModMax }\label{secmd2d}
The ModMax theory coupled to Einstein gravity in four dimensions is defined as \cite{Bandos:2020jsw, Kosyakov:2020wxv,Barrientos:2022bzm}
\begin{align}\label{lmd4d}
    I^{(4)}=\frac{1}{16\pi G_4}\int d^4x\sqrt{-g_{(4)}}\Big(R^{(4)}-2\Lambda -4\kappa\mathcal{L}^{(4)}_{\text{MM}}\Big),
\end{align}
where $R^{(4)}$ is the Ricci scalar in 4 dimensions, $\Lambda=-3$ is the cosmological constant\footnote{Here, we set the $AdS$ length $l=1$.}, $G_4$ is the Newton's constant in four dimensions,  $\kappa$ is the coupling constant and $\mathcal{L}^{(4)}_{\text{MM}}$ is the ModMax Lagrangian density in four dimensions \cite{Bandos:2020jsw, Kosyakov:2020wxv,Barrientos:2022bzm}

\begin{align}\label{momdmaxld}
    \mathcal{L}^{(4)}_{\text{MM}}&=\frac{1}{2}\Big(S \cosh{\gamma}-\sqrt{S^2+P^2}\sinh{\gamma}\Big),\nonumber\\
    \hspace{2mm}S&=\frac{1}{2}F_{MN}F^{MN}\hspace{1mm},\hspace{2mm}P=\frac{1}{2}F_{MN}\Tilde{F}^{MN}\hspace{1mm}, \hspace{2mm}\Tilde{F}^{MN}=\frac{1}{2}\epsilon^{MNUV}F_{UV}.
\end{align}

Here, $\gamma$ is the ModMax parameter and $(M,N)$ are the 4 dimensional space-time indices. Clearly, the standard Maxwell electrodynamics is recovered in the limit $\gamma\rightarrow 0$  \cite{Bandos:2020jsw, Kosyakov:2020wxv,Barrientos:2022bzm}.

The imprint of the ModMax theory (\ref{momdmaxld}) in two dimensions can be obtained via dimensional reduction \cite{Davison:2016ngz,Lala:2020lge,Rathi:2021aaw} of the following form
\begin{align}\label{ansatz}
    ds_{(4)}^2&=ds_{(2)}^2+\Phi(x^{\mu}) dx_i^2\hspace{1mm},\hspace{2mm}ds_{(2)}^2=g_{\mu\nu}(x^{\alpha})dx^{\mu}dx^{\nu},\nonumber\\
    A_{\mu}&\equiv A_{\mu}(x^{\nu}),\hspace{1mm}A_i\equiv A_{i}(x^{\mu}),
\end{align}
where $(\mu,\nu)$ are the two dimensional indices and $(i,j)$ are the indices of the compact dimensions.

Substituting (\ref{ansatz}) into (\ref{lmd4d}) and integrating over the compact directions, one finds\footnote{The Newton's constant in two and four dimensions are related by $G_2=\frac{G_4}{V_2}$, where $V_2$ is the volume of the compact space.}
\begin{align}\label{lmd2d}
  I_{\text{bulk}}=\frac{1}{16\pi G_2}\int d^2x\sqrt{-g_{(2)}}\Big(\Phi R^{(2)}-2\Lambda\Phi-4\kappa\Phi\mathcal{L}^{(2)}_{\text(MM)}\Big), 
\end{align}
where $R^{(2)}$ is the Ricci scalar in two dimensions, $G_2$ is the Newton's constant in two dimensions and 
\begin{align}\label{onlylmm2d}
    \mathcal{L}^{(2)}_{\text{MM}}&=\frac{1}{2}\Big(s\cosh{\gamma}-\sqrt{s^2+p^2}\sinh{\gamma}\Big),\nonumber\\
    s&=\frac{1}{2}F_{\mu\nu}F^{\mu\nu}+\Phi^{-1}\Big((\partial \chi)^2+(\partial \xi)^2\Big),\hspace{1mm}p=-2\Phi^{-1}\epsilon^{\mu\nu}\partial_{\mu}\chi\partial_{\nu}\xi
\end{align}
is what we define as the Lagrangian density of the projected ModMax theory in two dimensions. Here, we denote $A_{2}=\chi(x^{\mu}),\hspace{1mm} A_{3}=\xi(x^{\mu})$ and introduce $\epsilon^{\mu\nu}=\frac{\varepsilon^{\mu\nu}}{\sqrt{-g_{(2)}}}$ as the Levi-Civita tensor in two dimensions. 

Notice that, in the limit $\gamma\rightarrow0$, we do not recover the standard Maxwell electrodynamics in two dimensions \cite{Castro:2008ms, Lala:2020lge, Rathi:2021mla, Rathi:2021aaw}. On contrary, we do have additional contributions coming from non-vanishing scalar fields $\xi$ and $\chi$ which arise by virtue of the dimensional reduction procedure. This turns out to be the unique feature of the projected ModMax interactions in two dimensions. The $\gamma\rightarrow 0$ limit is what we refer as the 2D Maxwell interaction in this chapter.

$\bullet$ \textbf{A comparative study of 4D ModMax and the 2D projected ModMax:}

Below, we draw a comparative analysis between 4D ModMax \cite{Bandos:2020jsw}-\cite{ Kosyakov:2020wxv} and its 2D projection which plays the central role in what follows. 4D ModMax preserves the conformal invariance in its usual sense which is also evident from the generic structure of the associated stress-energy tensor 
 
\begin{align}\label{tr4}
    T^{(4)}_{MN}\sim f(\gamma)\Bigg(-\frac{1}{2}F^2g_{MN}+2g^{QP}F_{QM}F_{PN}\Bigg),
\end{align}
where we define the function
\begin{align}
   f(\gamma)= \Bigg(\cosh{\gamma}-\frac{F^2\sinh{\gamma}}{\sqrt{\big(F_{RS}F^{RS}\big)^2+\big(F_{RS}\Tilde{F}^{RS}\big)^2}}\Bigg).
\end{align}

Clearly, the trace $T^{M(4)}_M$ vanishes identically in four dimensions. On the other hand, the trace of the projected ModMax in two dimensions turns out to be
\begin{align}
   T^{\mu(2)}_{\mu}= g^{\mu\nu}T_{\mu\nu}^{(2)}= \frac{\Phi F^2}{2}\Bigg(\cosh{\gamma}-\frac{s\sinh{\gamma}}{\sqrt{s^2+p^2}}\Bigg),
\end{align}
which is a non-vanishing entity. 

This reflects to the fact that the projected theory losses its conformal invariance in two dimensions. Furthermore, the absence of the (Hodge) dual two form ($\Tilde{F}^{\mu\nu}$) in two dimensions spoils the electromagentic SO(2) duality invariance of the 2D projected theory in comparison to its 4D cousin. However, it is noteworthy to mention that the ModMax coupling $(\gamma)$ that appears in the 2D projected version is same as that of the 4D parent theory.

The equations of motion corresponding to different field contents can be obtained by varying the action (\ref{lmd2d}) 

\begin{align}\label{bulkvar}
        \delta I_{\text{bulk}}= \frac{1}{16\pi G_2}\int d^2x\sqrt{-g}\Big(\mathcal{H}_{\mu\nu}\delta g^{\mu\nu}+\mathcal{H}_{\Phi}\delta \Phi+\mathcal{H}_{\mu}\delta A^{\mu}+\mathcal{H}_{\chi}\delta\chi+\mathcal{H}_{\xi}\delta\xi\Big),
    \end{align}
where we define individual entities as 
\begin{align}
\mathcal{H}_{\Phi}=\hspace{1mm}&R-2\Lambda-4\kappa\mathcal{L}^{(2)}_{\text{MM}}+2\kappa\Phi^{-1}\Bigg[\Big((\partial \xi)^2+(\partial \chi)^2\Big)\cosh{\gamma}-\Bigg\{\frac{s\Big((\partial \xi)^2+(\partial \chi)^2\Big)}{\sqrt{s^2+p^2}}-\nonumber\\
    & \frac{2p\epsilon^{\mu\nu}\nabla_{\mu}\chi\nabla_{\nu}\xi}{\sqrt{s^2+p^2}}\Bigg\}\sinh{\gamma}\Bigg]=0,\label{geneomi}\\
\mathcal{H}_{\mu\nu}=\hspace{1mm}& \square \Phi g_{\mu\nu}-\nabla _{\mu}\nabla_{\nu}\Phi+\Lambda\Phi g_{\mu\nu} -2\kappa\Phi\Bigg[\mathcal{F_{\mu\nu}}\cosh{\gamma}-\frac{\sinh{\gamma}}{\sqrt{s^2+p^2}}\Bigg(s\mathcal{F_{\mu\nu}}-\frac{1}{2}s^2g_{\mu\nu}\Bigg)\nonumber\\&-\frac{s}{2}g_{\mu\nu}\cosh{\gamma}\Bigg]=0,\label{geneomijhi}\\
    \mathcal{H}_{\chi}=\hspace{1mm}&\kappa\nabla_{\mu}\Bigg[\nabla^{\mu}\chi \cosh{\gamma}-\frac{s\nabla^{\mu}\chi-p\epsilon^{\mu\nu}\nabla_{\nu}\xi}{\sqrt{s^2+p^2}}\sinh{\gamma}\Bigg]=0,\\
     \mathcal{H}_{\xi}=\hspace{1mm}&\kappa\nabla_{\mu}\Bigg[\nabla^{\mu}\xi \cosh{\gamma}-\frac{s\nabla^{\mu}\xi+p\epsilon^{\mu\nu}\nabla_{\nu}\chi}{\sqrt{s^2+p^2}}\sinh{\gamma}\Bigg]=0,\label{geneomijhf}\\
      \mathcal{H}_{\mu}=\hspace{1mm}&\kappa\nabla_{\mu}\Bigg[\Phi \Bigg(\cosh{\gamma}-\frac{s\sinh{\gamma}}{\sqrt{s^2+p^2}}\Bigg)F^{\mu\nu}\Bigg]=0,\label{geneomf}
\end{align}
along with the function
    \begin{align}
   \mathcal{F_{\mu\nu}}=&\hspace{1mm}F_{\mu\alpha}F_{\nu\beta}g^{\alpha\beta}+\Phi^{-1}\Big(\partial_{\mu}\xi\partial_{\nu}\xi+\partial_{\mu}\chi\partial_{\nu}\chi\Big).
   \end{align}
   
    
 \section{General solution with 2D projected ModMax}\label{secsol}
 The purpose of this Section is to obtain the most general solutions of (\ref{geneomi})-(\ref{geneomf}) in the Fefferman-Graham gauge\footnote{The explicit form of these equations (\ref{geneomi})-(\ref{geneomf}) have been provided in the Appendix \ref{geneomap}.} \cite{Rathi:2021aaw,fefferman}
 \begin{align}\label{fggauge}
       ds^2=\hspace{1mm}&d\eta^2+h_{tt}(t,\eta)dt^2,\hspace{1mm}A_{\mu}dx^{\mu}=\hspace{1mm}A_t(t,\eta)dt,\nonumber\\
       \Phi=\hspace{1mm}&\Phi(t,\eta),\hspace{1mm}\chi=\chi(t,\eta),\hspace{1mm}\xi=\xi(t,\eta).
 \end{align}

$\bullet$ \textbf{A note on conformal dimensions:}

 Here, we present a calculation on the conformal dimensions of the dual operators $\Delta_{\chi}$, $\Delta_{\xi}$ and $\Delta_{\Phi}$ corresponding to the bulk scalar fields $\chi$, $\xi$ and $\Phi$ respectively. This allows us to make a comparative study between various operator dimensions in the deep IR limit.

 The IR fixed point \cite{Castro:2018ffi}-\cite{Castro:2021wzn} is defined as the set of solutions to the equations of motion (\ref{geneomi})-(\ref{geneomf}) for constant values of the scalar fields
 \begin{align}\label{irjfp}
\chi(t,\eta)=\chi^{*}\hspace{1mm},\hspace{2mm}\xi(t,\eta)=\xi^{*}\hspace{1mm},\hspace{2mm} \Phi(t,\eta)=\Phi^{*},
 \end{align}
 where the superscript `*' denotes the values of the background scalars at the IR fixed point.

Using (\ref{irjfp}), one can solve the above set of equations (\ref{geneomi})-(\ref{geneomf}) in the Fefferman-Graham gauge (\ref{fggauge}) to obtain 
 \begin{align}
\omega^*=&\hspace{1mm}\alpha(t)e^{\sqrt{2}\eta\lambda}+\beta(t)e^{-\sqrt{2}\eta\lambda},\\
     A_{t}^*=&\hspace{1mm}\mu(t)+\frac{c}{\sqrt{2}\lambda}\Big(\alpha(t)e^{\sqrt{2}\eta\lambda}-\beta(t)e^{-\sqrt{2}\eta\lambda}\Big),
 \end{align}
where we define  $\lambda=\sqrt{-\Lambda}=\sqrt{3}$, $\omega=\sqrt{-h_{tt}}$ and $c$ is the integration constant. Here, $\alpha(t)$, $\beta(t)$ and $\mu(t)$ are some arbitrary functions of time.

 In order to compute the conformal dimensions of the dual operators, we expand the scalar fields ($\chi$, $\xi$ and $\Phi$) around the fixed point (\ref{irjfp}) and retain the equations of motion (\ref{geneomijhi})-(\ref{geneomijhf}) upto linear order in scalar fluctuations which yields 

 \begin{align}
     \Bigg[\partial^2_{\eta}+\frac{1}{\omega^*}\Big(\partial_{\eta}\omega^*\Big)\partial_{\eta}-\frac{1}{\omega^*}\partial_t\Bigg(\frac{1}{\omega^*}\partial_t\Bigg)-m^2\Bigg]\Tilde{\Phi}=&\hspace{1mm}0,\label{pertj1}\\
     \Bigg[\partial_{\eta}\Big(\omega^*\partial_{\eta}\Big)-\partial_t\Bigg(\frac{1}{\omega^*}\partial_t\Bigg)\Bigg]\Tilde{\chi}=&\hspace{1mm}0,\label{pertj2}\\
     \Bigg[\partial_{\eta}\Big(\omega^*\partial_{\eta}\Big)-\partial_t\Bigg(\frac{1}{\omega^*}\partial_t\Bigg)\Bigg]\Tilde{\xi}=&\hspace{1mm}0,\label{pertj3}
 \end{align}
 where we define $m^2=\big(6-2c^2\kappa e^{-\gamma}\big)$ and scalar fluctuations $\Tilde{\mathcal{Y}}=\mathcal{Y}-\mathcal{Y}^*$, where $\mathcal{Y}$ collectively denotes the scalar fields ($\Phi$, $\chi$ and $\xi$).
 
 It should be noted that, the mass-squared term ($m^2$) defined above must satisfy the Breitenlohner-Freedman (BF) bound\footnote{In ($d+1$) spacetime dimensions, the BF bound is defined as $m^2\geq-\Big(\frac{d^2}{2L}\Big)^2,$ where $L$ is the AdS length.} \cite{Ramallo:2013bua}, which for the present example sets a constraint of the form $c\leq\sqrt{\frac{25e^{\gamma}}{8\kappa}}$. Notice that, unlike (\ref{pertj1}), the equations of motion for scalar fluctuations $\Tilde{\chi}$ (\ref{pertj2}) and $\Tilde{\xi}$ (\ref{pertj3}) do not contain any mass-squared term. This indicates that these scalar fields ($\chi$ and $\xi$) are massless. This is consistent with the fact that these scalar fields ($\chi$ and $\xi$) carry only kinetic terms in the Lagrangian (\ref{onlylmm2d}).

 From the above set of equations (\ref{pertj1})-(\ref{pertj3}), one could finally decode the conformal dimensions\footnote{The conformal dimension of the dual operator ($\Delta$) is defined as $\Delta(\Delta-1)=m^2$ \cite{Ramallo:2013bua, Garcia-Garcia:2020ttf,Castro:2018ffi} where $m$ represents the mass of the scalar field.} of the dual operators as 
 \begin{align}\label{jhepsop}
     \Delta_{\Phi\pm}=\hspace{1mm}\frac{1}{2} \left(1\pm\sqrt{25-8 e^{-\gamma } c^2 \kappa }\right)\hspace{1mm},\hspace{2mm}\Delta_{\chi}=\hspace{1mm}\Delta_{\xi}=\hspace{1mm}1,
 \end{align}
where the subscript `$\pm$' denotes the two possible values of $\Delta_{\Phi}$.  
 
It is interesting to notice that the conformal dimension, $(\Delta_{\chi}=\Delta_{\xi})>\Delta_{\Phi+}$ for the range of constant, $\sqrt{\frac{3e^{\gamma}}{\kappa}}<c\leq\sqrt{\frac{25e^{\gamma}}{8\kappa}}$. This indicates that the dynamics of the dilaton fluctuation $\Tilde{\Phi}$ dominates \cite{Castro:2018ffi}--\cite{Castro:2021wzn} over the scalar fluctuations $\Tilde{\chi}$ and $\Tilde{\xi}$ in the deep IR. 
 
 On the other hand, one could set the conformal dimension, $(\Delta_{\chi}=\Delta_{\xi})<\Delta_{\Phi+}$ given the range $0\leq c<\sqrt{\frac{3e^{\gamma}}{\kappa}}$, which suggests that the IR dynamics is dominated by the scalar fluctuation $\Tilde{\chi}$ and $\Tilde{\xi}$. However,  for a particular choice of constant $c=\sqrt{\frac{3e^{\gamma}}{\kappa}}$, the  conformal dimensions, $\Delta_{\chi}=\Delta_{\xi}=\Delta_{\Phi+}=1$. In this case, the dynamics of all scalar fluctuations $\Tilde{\mathcal{Y}}$ are equally important in the deep IR.  

On a similar note, one finds that the maximum value of the conformal dimension\footnote{In this case, the constant $c$ is restricted to the range $\sqrt{\frac{3e^{\gamma}}{\kappa}}\leq c\leq\sqrt{\frac{25e^{\gamma}}{8\kappa}}$. If $c<\sqrt{\frac{3e^{\gamma}}{\kappa}}$, then the conformal dimension $\Delta_{\Phi-}$ become negative.} $\Delta_{\Phi-}$ is $1/2$. Therefore, in this case, the dilaton fluctuation $(\Tilde{\Phi})$ always dominates over the scalar fluctuations. Therefore, to summarise, one could conjecture that the dilaton fluctuation always dominates over scalar fluctuation if the constant falls in the range $\sqrt{\frac{3e^{\gamma}}{\kappa}}<c\leq\sqrt{\frac{25e^{\gamma}}{8\kappa}}$. 

Finally, it is noteworthy to compare our results with the existing literature \cite{Castro:2018ffi}-\cite{Castro:2021fhc}. The authors in \cite{Castro:2018ffi}, construct a 2D theory of gravity in the presence of a dilaton ($e^{-2\psi}$), scalar field ($\chi$) and a $U(1)$ gauge field following a consistent reduction of Einstein gravity in five dimensions. Unlike the present example, the authors in \cite{Castro:2018ffi} obtained a mass-squared term for the scalar field ($\chi$) which is thereby used to calculate the conformal dimension of the dual operator. Interestingly, they found that the dual operator is always irrelevant compared to the dilaton operator in the IR. In other words, the dilaton fluctuation  always dominates over the scalar fluctuations in the deep IR. \\

$\bullet$ \textbf{Remarks about perturbative solutions:}
 
 Now, we compute the most general solutions of (\ref{geneomi})-(\ref{geneomf}) in the Fefferman-Graham gauge. Clearly, these  equations (\ref{geneomi})-(\ref{geneomf}) are quite difficult to solve exactly in the ModMax coupling $\gamma$. Therefore, to proceed further, we simplify the fields as $\Phi\equiv\Phi(\eta),\hspace{1mm}h_{tt}\equiv h_{tt}(\eta),\hspace{1mm}A_t\equiv A_t(\eta),\hspace{1mm}\xi\equiv\xi(\eta)\hspace{1mm}\text{and}\hspace{1mm}\chi\equiv\chi(t) $ and solve them ``perturbatively'' treating the 2D Maxwell coupling ($\kappa$) and the 2D ModMax coupling $(\gamma)$ as expansion parameters.

One can systematically expand these fields using the expansion parameters ($\kappa$ and $\gamma$) as
\begin{align}
   & \mathcal{A}=\mathcal{A}_0+\kappa\mathcal{A}_1+\gamma\kappa\mathcal{A}_2+\kappa^2\mathcal{A}_3+...,\label{Afieldexp}\\
   & \mathcal{B}=\mathcal{B}_1+\gamma\mathcal{B}_2+\kappa\mathcal{B}_3+...,\hspace{2mm}|\kappa|<<1,\hspace{1.5mm}|\gamma|<<1,\label{Bfieldexp}
\end{align}
where $\mathcal{A}$ collectively denotes the fields ($\Phi$, $\omega$) and $\mathcal{B}$ denotes the remaining fields ($A_t$, $\chi$, $\xi$). Here, the subscript `0' denotes the pure JT gravity solution. On the other hand, subscripts `$1$' and `$2$' denote the leading order corrections due to the 
 2D Maxwell term and the 2D projected ModMax interaction respectively.  Furthermore, the subscript `$3$' stands for the quadratic order corrections due to the 2D Maxwell term alone.

Notice that, the $\mathcal{B}$ fields (\ref{Bfieldexp}) are expanded differently from that of the $\mathcal{A}$ fields (\ref{Afieldexp}). This is due to the fact that the $\mathcal{B}$ fields are coupled with an overall  2D Maxwell coefficient, $\kappa$ in the Lagrangian (\ref{lmd2d}). Therefore, one should think of the expansion (\ref{Bfieldexp}) to be multiplied with an overall factor of $\kappa$. On the other hand, the effects of the 2D projected ModMax comes into the picture at the quadratic level ($\gamma\kappa$). To summarise, we solve the equations of motion  (\ref{geneomi})-(\ref{geneomf}) up to quadratic order ($\gamma\kappa$ and $\kappa^2$) in the couplings and ignore all the higher order corrections.

\subsection{Zeroth order solution}
In order to obtain the pure JT gravity solutions, one has to take the limits $\kappa\rightarrow0$ and $\gamma\rightarrow0$ in the equations (\ref{geneomi})-(\ref{geneomf}), which yields
\begin{align}
\omega_0''+\Lambda\omega_0=&\hspace{1mm}0,\label{zeroeomi}\\
    \Phi_0''+\Lambda\Phi_0=&\hspace{1mm}0,\\
    \frac{\Phi_0'\omega_0'}{\omega_0}+\Lambda\Phi_0=&\hspace{1mm}0.\label{zeroeomf}
\end{align}

On solving (\ref{zeroeomi})-(\ref{zeroeomf}), one finds
\begin{align}
\omega_0=&\hspace{1mm}a_1 e^{\eta  \lambda }+a_2 e^{-\eta  \lambda },\label{zerosoli}\\
    \Phi_0=&\hspace{1mm}\frac{b_1 }{a_1 \lambda }e^{-\eta  \lambda } \left(a_1 e^{2 \eta  \lambda }-a_2\right),\label{zerosolf}
    \end{align}
    where $a_1,a_2$ and $b_1$ are the integration constants. 
    
    Equations (\ref{zerosoli})-(\ref{zerosolf}) are the zeroth order solutions of the theory (\ref{lmd2d}). In the following Sections, we will be using these solutions to obtain the next to leading order corrections for $\mathcal{A}$ and $\mathcal{B}$.
    
\subsection{ Order $\kappa$ solution}
The leading order corrections to the fields $\mathcal{A}$ and $\mathcal{B}$ are due to the presence of the Maxwell interactions in (\ref{lmd2d}),
\begin{align}\label{ONLYMAXWELLACT}
    \mathcal{L}_{\text{Maxwell}}=\frac{1}{4}F_{\mu\nu}F^{\mu\nu}+\frac{1}{2}\Phi^{-1}\Big((\partial \chi)^2+(\partial \xi)^2\Big).
\end{align}

On comparing the coefficients of $\kappa$ in the equations (\ref{geneomi})-(\ref{geneomf}), we obtain 
\begin{align}
 \omega_0\Phi_1''+\omega_1\Phi_0''-\Big(\Phi_0'\omega_1'+\Phi_1'\omega_0'\Big)+2\omega_0\Bigg(\xi_1'^2+\frac{\dot{\chi_1}^2}{\omega_0^2}\Bigg)=&\hspace{1mm}0,\label{oneeqi}\\
 \omega_1''+\Lambda\omega_1-\frac{A_{t1}'^2}{\omega_0}=&\hspace{1mm}0,\\
 \partial_{\eta}\Bigg(\frac{\Phi_0}{\omega_0}A_{t1}'\Bigg)=&\hspace{1mm}0,\\
    \partial_{\eta}\Big(\omega_0\xi_1'\Big)=&\hspace{1mm}0,\\
    \ddot{\chi}_1=&\hspace{1mm}0. \label{oneeqf}
    \end{align}
   
Using the zeroth order solutions (\ref{zerosoli})-(\ref{zerosolf}), one can solve the above set of equations to yield
    \begin{align}       
    \Phi_1=&\hspace{1mm}\frac{e^{-\eta  \lambda } }{4 \lambda ^2}\Bigg(\frac{4 \lambda }{a_1} \left(a_3 b_1 e^{2 \eta  \lambda }+a_2 \log \left(a_2-a_1 e^{2 \eta
    \lambda }\right)\right)+4 \lambda  e^{2 \eta  \lambda } \Big(2 \eta  \lambda -\nonumber\\
    &\log \left(a_2-a_1 e^{2
   \eta  \lambda }\right)\Big)+\tan ^{-1}\left(\frac{\sqrt{a_1} e^{\eta  \lambda
   }}{\sqrt{a_2}}\right)\Bigg(\frac{1}{a_1^{3/2} \sqrt{a_2}}-\frac{e^{2 \eta  \lambda } }{\sqrt{a_1} a_2^{3/2}}\Bigg)\Bigg),\label{onesoli}\\
   \omega_1=&\hspace{1mm}\frac{c_1^2 }{4 a_2}e^{-\eta  \lambda } \left(2 \eta  \lambda  e^{2 \eta  \lambda }-\frac{1}{a_1}\left(a_1 e^{2 \eta  \lambda
   }+a_2\right) \log \left(a_2-a_1 e^{2 \eta  \lambda }\right)\right)+a_3 e^{\eta  \lambda },\\
    A_{t1}=&\hspace{1mm}c_1 \Bigg[\log \left(a_2-a_1 e^{2 \eta  \lambda }\right)- \eta  \lambda\Bigg]+c_2, \\
   \xi_1=&\hspace{1mm}\frac{e_1 }{\lambda }\tan ^{-1}\left(\frac{\sqrt{a_1} e^{\eta  \lambda }}{\sqrt{a_2}}\right)+e_2,\\
    \chi_1=&\hspace{1mm}d_1t+d_2,\label{onesolf}
    \end{align}
    where $a_3,c_i,d_i$ and $e_i$, $(i=1,2)$ are the integration constants.
    
    Equations (\ref{onesoli})-(\ref{onesolf}) represent the leading order corrections to the fields $\mathcal{A}$ and $\mathcal{B}$ in the presence of the 2D Maxwell interactions (\ref{ONLYMAXWELLACT}).
    
    \subsection{ Order $\gamma\kappa$ solution}
    
    Next, we take into account the projected ModMax interactions and their imprint on the background fields $\mathcal{A}$ (\ref{Afieldexp}) and $\mathcal{B}$ (\ref{Bfieldexp}).
    
    A straight forward analysis reveals the following set of equations at order $\gamma\kappa$
    \begin{align}
      \omega_0\Phi_2''+\omega_2\Phi_0''-\Phi_0'\omega_2'-\Phi_2'\omega_0'+4\omega_0\xi_1'\xi_2'+\frac{4}{\omega_0}\dot{\chi_1}\dot{\chi_2}-f_0=&\hspace{1mm}0,\label{twoeqi}\\
       \omega_2''-\omega_2\lambda^2-\frac{2}{\omega_0}A_{t1}'A_{t2}'+\frac{s_0 A_{t1}'^2}{\omega_0\sqrt{s_0^2+p_0^2}}=&\hspace{1mm}0,\\
       \partial_{\eta}\Bigg[\frac{\Phi_0}{\omega_0}\Bigg(A_{t2}'-\frac{s_0}{\sqrt{s_0^2+p_0^2}}A_{t1}'\Bigg)\Bigg]=&\hspace{
      1mm}0,\\
      \partial_{\eta}\Bigg[\omega_0\xi_2'-\frac{\Big(s_0\xi_1'\omega_0-p_0\dot{\chi_1}\Big)}{\sqrt{s_0^2+p_0^2}}\Bigg]=&\hspace{1mm}0,\\
      \ddot{\chi_2}=&\hspace{1mm}0,\label{twoeqf}
    \end{align}
    where we identify the above functions as
    \begin{align}
        s_0=&\hspace{1mm}-\frac{1}{\omega_0^2}A_{t1}'^2+\frac{1}{\Phi_0}\Bigg(-\frac{\dot{\chi_1^2}}{\omega_0^2}+\xi_1'^2\Bigg)\hspace{1mm},\hspace{2mm}p_0=-\frac{2}{\Phi_0\omega_0}\dot{\chi_1}\xi_1',\\
       f_0=&\hspace{1mm} \frac{2\omega_0s_0}{\sqrt{s_0^2+p_0^2}}\Bigg(\xi_1'^2+\frac{\dot{\chi_1}^2}{\omega_0^2}\Bigg).
    \end{align}
    
   The above set of equations (\ref{twoeqi})-(\ref{twoeqf}) are difficult to solve for generic values of $\eta$. However, for our present purpose, it will be sufficient to solve them near the asymptotic limit ($\eta\rightarrow\infty$) of the space-time which yields 
    
    \begin{align}
   \Phi_2=&\hspace{1mm}\frac{1}{\lambda }\Big(b_2 e^{\eta  \lambda }-b_3 \lambda +e^{-\eta  \lambda }\Big)-\frac{b_1 }{a_1}\eta  e^{-\eta  \lambda },\label{soltwoi}\\
   \omega_2=&\hspace{1mm}e^{-\eta  \lambda } \left(a_4 e^{2 \eta  \lambda }+a_5+\eta  \lambda \right),\\
    \xi_2=&\hspace{1mm}\frac{e_3 }{\lambda }e^{-\eta  \lambda }+e_4,\\
    A_{t2}=&\hspace{1mm}c_3\eta\lambda+c_4,\\
     \chi_2=&\hspace{1mm}d_3t+d_4.\label{soltwof}
    \end{align}
where $a_i,b_j,c_k,d_k$ and $e_k$, $(i=4,5,\hspace{1mm}j=2,3,\hspace{1mm}k=3,4)$ are the integration constants.

As we show below, not all of these integration constants are actually important for our analysis. In fact, a few of them finally survive which can be fixed by making use of the residual gauge freedom \cite{Castro:2008ms} in the Fefferman-Graham gauge (\ref{fggauge}). In particular,  the re-scaling of the time coordinate $t\rightarrow a_1t$ preserves the gauge condition $g_{\eta t}=0$ and $g_{\eta\eta}=1$. Therefore, we can use this freedom to fix the constant\footnote{Interestingly, with this particular choice of the integration constant $a_1$ (\ref{constc1fix}), the final expression of the central charge (\ref{cfinalexp}) appears to be independent of all the remaining integration constants. }
\begin{align}\label{constc1fix}
    a_1=\frac{1}{a_2b_3}.
\end{align}

\subsection{Order $\kappa^2$  solution }\label{quadeomap}
Finally, we estimate the quadratic order ($\kappa^2$) corrections due to the Maxwell (\ref{ONLYMAXWELLACT}) term alone.

The resulting equations of motion (\ref{geneomi})-(\ref{geneomf}) can be expressed as
\begin{align}
    \omega_0\partial_{\eta}\Bigg(\frac{\Phi_0A_{t3}'}{\omega_0}-\frac{\Phi_0\omega_1A_{t1}'}{\omega_0^2}+\frac{\Phi_1A_{t1}'}{\omega_0}\Bigg)-\omega_1\partial_{\eta}\Bigg(\frac{\Phi_0A_{t1}'}{\omega_0}\Bigg)=&\hspace{1mm}0,\label{quadkappaapi}\\
    \omega_0\Phi_3''+\omega_1\Phi_1''+\omega_3\Phi_0''-\big(\Phi_0'\omega_3'+\Phi_1'\omega_1'+\Phi_3'\omega_0'\big)+f_2=&\hspace{1mm}0,\\
    \omega_3''-\lambda^2\omega_3-\frac{1}{\omega_0}\Bigg(2A_{t1}'A_{t3}'-\frac{\omega_1}{\omega_0}A_{t1}'^2\Bigg)=&\hspace{1mm}0,\\
    \partial_{\eta}\big(\omega_0\xi_3'+\omega_1\xi_1'\big)-\frac{\omega_1}{\omega_0}\partial_{\eta}\big(\omega_0\xi_1'\big)=&\hspace{1mm}0,\\
    \ddot{\chi}_3=&\hspace{1mm}0,\label{quadkappaapf}
\end{align}
where
\begin{align}
    f_2=2\big(2\omega_0\xi_1'\xi_3'+\omega_1\xi_1'^2\big)+
\frac{2}{\omega_0}\Bigg(2\dot{\chi}_1\dot{\chi}_3-\dot{\chi}_1^2\frac{\omega_1}{\omega_0}\Bigg).
\end{align}

The above set of equations (\ref{quadkappaapi})-(\ref{quadkappaapf}) could be solved near the asymptotics ($\eta\rightarrow\infty$) of the spacetime which yield
\begin{align}
    \Phi_3= &\hspace{1mm}-\frac{1}{2
   a_1^2 \lambda ^2}\Big(e^{-\eta  \lambda } (2 \eta  \lambda +3) \left(2 a_2 a_3 \lambda -b_1 c_1 c_5\right)\Big)+\frac{b_4 e^{\eta  \lambda }}{\lambda }+\frac{a_1 \left(a_1^2+1\right)+a_3}{6 a_1^2 \left(a_1^2+1\right) a_2},\\
   \omega_3=&\hspace{1mm}\frac{1}{4} e^{-\eta  \lambda } \left(\frac{c_1 (2 \eta  \lambda +1) \left(a_3 c_1 \lambda -2
   a_1 c_5\right)}{a_1^2 \lambda }+4 a_6 e^{2 \eta  \lambda }+4 a_7\right),\\
   \xi_3=&\hspace{1mm} \frac{1}{36 a_1^{5/2} \lambda }\Big(\sqrt{a_2} e_1 e^{-3 \eta  \lambda } \left(c_1^2 (1-6 \eta  \lambda )-12 a_2
   a_3\right)\Big)-\frac{e_5 e^{-\eta  \lambda }}{\lambda }+e_6,\\
   A_{t3}=&\hspace{1mm}\frac{c_1 e^{-\eta  \lambda }}{4 a_1 a_2 b_1 \lambda }+c_6 \eta +c_5,\\
   \chi_3=&\hspace{1mm}d_5 t+d_6,
\end{align}
where $a_i,b_j,c_k,d_k$ and $e_k$ ($i=6,7,\hspace{1mm}j=4,5,\hspace{1mm}k=5,6$) are the integration constants.
     

\section{Boundary stress tensor and central charge}\label{secstcc}
In this Section, we work out the ``renormalised'' boundary stress tensor \cite{Castro:2008ms, Hartman:2008dq,Rathi:2021aaw,Cadoni:2000gm,Narayan:2020pyj} and study its transformation properties under both the diffeomorphism and the $U(1)$ gauge transformations. In particular, we examine the effects of the projected ModMax interactions on the central charge of the boundary theory.  

To begin with, we workout the boundary terms\footnote{The boundary in the Ferrerman-Graham gauge is located near $\eta\rightarrow\infty$.} for the action (\ref{lmd2d}). This is required in order to implement a consistent variational principle \cite{Castro:2008ms,Rathi:2021aaw}. Systematically, one can decompose the boundary terms into following two pieces,
\begin{align}
    I_{\text{boundary}}&\hspace{1mm}=I_{\text{GHY}}+I_{\text{counter}},\label{fullba}
    \end{align}
    where $I_{\text{GHY}}$ is the standard Gibbons-Hawking-York boundary term and $I_{\text{counter}}$ represents the boundary counter terms.

    The Gibbons-Hawking-York boundary term \cite{Gibbons:1976ue,Castro:2008ms,Rathi:2021aaw} in 2D gravity is given by
    
    \begin{align}
        I_{\text{GHY}}=\frac{1}{8\pi G_2}\int_{0}^{\beta}dt\sqrt{-h}\Phi K\hspace{1mm},\hspace{2mm}K=\frac{1}{2}h^{tt}\partial_{\eta}h_{tt},
    \end{align}
    where $K$ is the trace of extrinsic curvature, $\beta$ is the inverse temperature  and $h_{tt}$  is the induced metric on the boundary. 

    On the other hand, the counter term that is required to absorb all the near boundary divergences of the on-shell action can be expressed as 
    
    \begin{align}\label{countertermdefi}
        I_{\text{counter}}=-\frac{1}{8\pi G_2}\int_{0}^{\beta}dt\sqrt{-h}\Bigg(\lambda\Phi+2\kappa\frac{b_1}{c_1} \sqrt{-h^{ab}A_aA_b}\Bigg),
    \end{align}
    where $(a,b)$ are the one dimensional boundary indices\footnote{Here, we set the constant $c_6=-\frac{\pi c_1^3}{16\sqrt{a_1a_2^3}b_1(a_1-c_1^2)}$ in order to cancel the boundary divergences up to quadratic order ($\gamma\kappa$ and $\kappa^2$) in the couplings.}. 
    
    Finally, the complete renormalised action is given by 
    \begin{align}
        I_{\text{renormalised}}=I_{\text{bulk}}+I_{\text{boundary}},\label{renormaction}
    \end{align}
    where $I_{\text{bulk}}$ and $I_{\text{boundary}}$ are given in (\ref{lmd2d}) and (\ref{fullba}) respectively.

    Notice that, the combination of the $U(1)$ gauge field in the $I_{\text{counter}}$ (\ref{countertermdefi}) seems to break the gauge invariance under the transformation 
    \begin{align}\label{u1gaugetrans}
        A_\alpha\rightarrow A_{\alpha}+\partial_{\alpha}\Sigma,
    \end{align}
    which yields the following extra piece under the $U(1)$ gauge (\ref{u1gaugetrans}) 
    \begin{align}\label{countergaugebreak}
        I_{\text{counter}}\sim \int_{0}^{\beta}dt\sqrt{-h}\Big(\sqrt{-h^{ab}A_aA_b}\Big)\rightarrow\int_{0}^{\beta}dt(A_t+\partial_t\Sigma).
    \end{align}

    However, one can preserve the gauge invariance by imposing the condition that $\partial_t\Sigma$ (see (\ref{sigmavalue})) must vanish near the boundary, $\eta\rightarrow\infty$ \cite{ Castro:2008ms}.

    Using the renormalised action (\ref{renormaction}), it is now straightforward to calculate the variation $\delta I_{\text{boundary}}$ under the combined action of the diffeomorphism and the $U(1)$ gauge, where $\delta I_{\text{boundary}}$ can be systematically expressed as\footnote{$\delta I_{\text{boundary}}$ already incorporates the bulk contributions $(\delta I_{\text{bulk}})$ near the asymptotic limit, $\eta\rightarrow\infty$. }
    \begin{align}
        \delta I_{\text{boundary}}= \frac{1}{16\pi G_2}\int dt\sqrt{-h}\Big(\mathcal{G}^{ab}\delta h_{ab}+\mathcal{G}_{\Phi}\delta \Phi+\mathcal{G}^a\delta A_a+\mathcal{G}_{\chi}\delta\chi+\mathcal{G}_{\xi}\delta\xi\Big).
    \end{align}
    
    Here, the boundary contributions can be expressed as
    \begin{align}
        \mathcal{G}^{ab}=&\hspace{1mm}n_{\mu}\nabla^{\mu}\Phi h^{ab}+n^{\mu}\frac{\Phi}{\sqrt{-h}}\Big(\partial_{\mu}\sqrt{-h}\Big)h^{ab}-\lambda\Phi h^{ab}-2\kappa \frac{b_1}{c_1} h^{ab}\sqrt{-h^{cd}A_cA_d}\nonumber\\
        &+2\kappa \frac{b_1}{c_1}\frac{A^aA^b}{\sqrt{-h^{cd}A_cA_d}},\label{stresstensordef}\\
        \mathcal{G}^a=&\hspace{1mm}-4\kappa n_{\mu}\Phi\Bigg(\cosh{\gamma}-\frac{s}{\sqrt{s^2+p^2}}\sinh{\gamma}\Bigg)F^{\mu a}+4\kappa \frac{b_1}{c_1} \frac{h^{ab}A_b}{\sqrt{-h^{cd}A_cA_d}},\\
         \mathcal{G}_{\chi}=&\hspace{1mm}-4\kappa n_{\mu}\Bigg(\nabla^{\mu}\chi\cosh{\gamma}-\frac{s\nabla^{\mu}\chi-p\epsilon^{\mu a}\nabla_{a}\xi}{\sqrt{s^2+p^2}}\sinh{\gamma}\Bigg),\\
        \mathcal{G}_{\xi}=&\hspace{1mm}-4\kappa n_{\mu}\Bigg(\nabla^{\mu}\xi\cosh{\gamma}-\frac{s\nabla^{\mu}\xi+p\epsilon^{\mu a}\nabla_{a}\chi}{\sqrt{s^2+p^2}}\sinh{\gamma}\Bigg),\\
         \mathcal{G}_{\Phi}=&\hspace{1mm}2K-2\lambda,
    \end{align}
     where $n^{\mu}=\delta^{\mu}_{\eta}$ is the unit normal vector at the boundary.

With all these preliminaries, we now introduce the boundary stress tensor \cite{Castro:2008ms,Rathi:2021aaw} corresponding to the action (\ref{renormaction})
\begin{align}\label{rstressdef}
    T^{ab}=\frac{2}{\sqrt{-h}}\frac{\delta I_{\text{boundary}}}{\delta h_{ab}}=\frac{\mathcal{G}^{ab}}{8\pi G_2},
\end{align}
where $\mathcal{G}^{ab}$ is given in (\ref{stresstensordef}).
     
Our next task is to explore the transformation properties of the background fields (\ref{Afieldexp})-(\ref{Bfieldexp}) and hence the boundary stress tensor (\ref{rstressdef}) under the combined effects of the diffeomorphism and the $U(1)$ gauge transformation.

Under the diffeomorphism,
\begin{align}\label{ctdef}
    x^{\mu}\rightarrow x^{\mu}+\epsilon^{\mu} (x),
\end{align}
the background fields  (\ref{Afieldexp})-(\ref{Bfieldexp}) transform as 
\begin{align}
    \delta_{\epsilon}A_{\mu}=&\hspace{1mm}\epsilon^{\nu}\nabla_{\nu}A_{\mu}+A_{\nu}\nabla_{\mu}\epsilon^{\nu},\label{diffeotranscoordi}\\
\delta_{\epsilon}g_{\mu\nu}=&\hspace{1mm}\nabla_{\mu}\epsilon_{\nu}+\nabla_{\nu}\epsilon_{\mu},\label{diffeotranscoordm}\\
\delta_{\epsilon}\mathcal{S}=&\hspace{1mm}\epsilon^{\mu}\nabla_{\mu}\mathcal{S},\label{diffeotranscoordf} 
\end{align}
where $\mathcal{S}$ collectively denotes the scalar fields $\Phi,\hspace{1mm}\xi$ and $\chi$.

The diffeomorphism parameter, $\epsilon^{\mu}(x)$ can be obtained using (\ref{diffeotranscoordm}) and the space-time metric (\ref{fggauge}), which yields the following
\begin{align}\label{parametersdiffeo}
   \epsilon_t=e^{2\eta\lambda}f(t)+\frac{1}{2\lambda}\Bigg(\frac{2 \left(a_1^2+1\right)}{3 a_1^2 \lambda }-\frac{4 a_3 \kappa
   }{3 a_1^3 \lambda }\Bigg)\partial^2_tf(t) \hspace{1mm},\hspace{2mm}\epsilon_{\eta}=\Bigg(\frac{2 \left(a_1^2+1\right)}{3 a_1^2 \lambda }-\frac{4 a_3 \kappa
   }{3 a_1^3 \lambda }\Bigg)\partial_tf(t),
\end{align}
where $f(t)$ is some function\footnote{In the Fefferman-Graham gauge \cite{Rathi:2021aaw,fefferman}, the variation of the space-time metric (under diffeomorphism (\ref{diffeotranscoordm})) yields a set of coupled differential equations that contain the  derivatives of the diffeomorphism parameters $\epsilon_t$ and $\epsilon_{\eta}$ with respect to the variable ``$\eta$''. Therefore, the function $f(t)$ in these equations appears as an integration constant. However, one can further compute the function $f(t)$ using suitable boundary conditions for the background fields $\mathcal{A}$ (\ref{Afieldexp}) and $\mathcal{B}$ (\ref{Bfieldexp}).} of time \cite{ Castro:2008ms}.

It should be noted that, we perform all the analysis in a gauge in which one of the components of the $U(1)$ gauge field, $A_{\eta}$ is set to be zero (\ref{fggauge}). On the other hand, under the diffeomorphism (\ref{ctdef}), $A_{\eta}$ transforms as 
\begin{align}\label{sigmagaugeeqn}
    \delta_{\epsilon}A_{\eta}=A_t\partial_{\eta}\Bigg(\frac{\epsilon_t}{h_{tt}}\Bigg)\neq 0,
\end{align}
which breaks the gauge condition $A_{\eta}=0$.

In order to restore this gauge condition, we employ the $U(1)$ gauge transformation, $A_\alpha\rightarrow A_{\alpha}+\partial_{\alpha}\Sigma$ and compute the $U(1)$ gauge parameter $\Sigma$ such that $(\delta_{\epsilon}+\delta_{\Sigma})A_{\eta}=0$, which yields the following 
\begin{align}\label{sigmaintermedstep}
   \Sigma= \hspace{1mm}-\int d\eta A_t\partial_{\eta}\Bigg(\frac{\epsilon_t}{h_{tt}}\Bigg),
\end{align}
where we have used the variation (\ref{sigmagaugeeqn}).

Now, one can perform the above integration (\ref{sigmaintermedstep}) using the background fields (\ref{Afieldexp})-(\ref{Bfieldexp}) and the diffeomorphism parameter (\ref{parametersdiffeo}), which yields
\begin{align}\label{sigmavalue}
    \Sigma=&\hspace{1mm}\frac{e^{-2 \eta  \lambda } }{12 a_1^5 a_2 \lambda ^3}\Bigg(f''(t) \big(2
   \left(a_1^2+1\right) a_2 a_1 \big(c_1 \lambda  \left(2 \log
   \left(a_1\right)-(\gamma -1) (2 \eta  \lambda +1)\right)+2 \gamma
    c_3 \lambda +\nonumber\\
    &\hspace{1mm}\kappa  \left(2 \lambda  \left(c_5 \eta
   +c_4\right)+c_5\right)\big)+a_1^2 c_1 \kappa  \lambda 
   \left(c_1^2 \log \left(a_1\right)-4 a_2 a_3\right) \left(2 \log
   \left(a_1\right)+2 \eta  \lambda +1\right)+\nonumber\\
   &\hspace{1mm}c_1 \kappa  \lambda 
   \left(c_1^2 \log \left(a_1\right)-8 a_2 a_3\right) \left(2 \log
   \left(a_1\right)+2 \eta  \lambda +1\right)\big)-3 a_1 a_2 c_1^3
   \kappa  \lambda ^3 f(t) \big(2 \log \left(a_1\right)+\nonumber\\
   &\hspace{1mm}2 \eta 
   \lambda +1\big)\Bigg).
\end{align}
It is interesting to notice that the $U(1)$ gauge parameter $\Sigma$ vanishes naturally in the asymptotic limit ($\eta\rightarrow\infty$), which is consistent with the gauge preserving condition (\ref{countergaugebreak}).

Finally, we note down the transformation of the boundary stress tensor (\ref{rstressdef}) under the combined action of the diffeomorphism (\ref{ctdef}) and the $U(1)$ gauge transformation which yields
\begin{align}
    (\delta_{\epsilon}+\delta_{\Sigma})T_{tt}=&\hspace{1mm}\frac{1}{8\pi G_2}\Bigg[\Bigg(\partial_{\eta}\Phi-\lambda\Phi-2\kappa\frac{b_1}{c_1}\frac{ A_t}{\omega}\Bigg)(\delta_{\epsilon} h_{tt})+4\kappa\frac{b_1\omega}{c_1} \big((\delta_{\epsilon}+\delta_{\Sigma}) A_t\big)\nonumber\\
    &+\frac{\Phi}{2}\partial_{\eta}(\delta_{\epsilon} h_{tt})-\frac{1}{2}\Big(\partial_{\eta}\omega^2-2\lambda\omega^2\Big)(\delta_{\epsilon}\Phi)-\partial_{\eta}(\delta_{\epsilon}\Phi)\omega^2\Bigg].\label{variationstresstensor}
\end{align}

 The variations of the background fields $h_{tt}, \hspace{1mm} A_t$ and $ \Phi$ can be obtained using (\ref{diffeotranscoordi})-(\ref{parametersdiffeo}) and (\ref{sigmavalue}), which yields the following

\begin{align}
(\delta_{\epsilon}+\delta_{\Sigma})A_t=&\hspace{1mm}\mathcal{H}_1(\eta)\partial_t f(t)+\mathcal{H}_2(\eta)\partial_t^3 f(t)\label{fv1},\\
   \delta_{\epsilon} h_{tt} =&\hspace{1mm} \mathcal{H}_3(\eta) \partial_tf(t) +\mathcal{H}_4(\eta)\partial_t^3f(t),\\
   \delta_{\epsilon} \Phi =&\hspace{1mm}\mathcal{H}_5(\eta)\partial_t f(t)\label{fv2},
\end{align}
where the explicit form of the functions $\mathcal{H}_i(\eta)$, $(i=1,2...5)$ are given in the Appendix \ref{h1h2fucdetailap}.

Using these variations (\ref{fv1})-(\ref{fv2}), the transformation of the boundary stress tensor (\ref{variationstresstensor}) can be expressed in a more elegant way 
\begin{align}
    (\delta_{\epsilon}+\delta_{\Sigma})\Tilde{T}_{tt}=2\Tilde{T}_{tt}\partial_tf(t)+f(t)\partial_t\Tilde{T}_{tt}- c_M\partial^3_tf(t).\label{stvariation}
\end{align}

Here, we define the re-scaled stress tensor as 
\begin{align}
    \Tilde{T}_{tt}= \frac{T_{tt}}{b_3(1+a_2^2b_3^2)},
\end{align}
 and identify the coefficient ``$c_M$'' (coefficient of $\partial_t^3f(t)$) as being the central charge \cite{Castro:2008ms,Rathi:2021aaw} of the boundary theory,
\begin{align}\label{cfinalexp}
    c_M=\frac{1}{144 \sqrt{3} \pi  G_2}\Big(\kappa-12 \gamma \kappa +2 \kappa^2\Big),
\end{align}
where we substitute $\lambda=\sqrt{3}$.

It should be noted that the above expression of the central charge (\ref{cfinalexp}) is a perturbative result up to quadratic order in the ModMax coupling ($\gamma$) and the $U(1)$ gauge coupling ($\kappa$). Furthermore, $c_M$ must be positive for the unitary theory \cite{Kiritsis:2019npv}, implying that the gauge couplings must lie within the range $\frac{1}{12} \leq \gamma < 1$ and $\frac{1}{2}(12\gamma - 1) < \kappa < 1$. Clearly, in the limit $\gamma\rightarrow 0$, the central charge (\ref{cfinalexp}) reduces to $\sim\frac{1}{G_2}$ which is consistent with the existing result in the literature \cite{Castro:2008ms}.
\section{Black holes and 2D projected ModMax  }\label{secbhwithmm}
We now construct the 2D black hole solutions and investigate their thermal properties in the presence of 2D projected ModMax interactions (\ref{lmd2d}). In particular, we emphasise on the role played by the ModMax parameter, that is required to set all the fields ``finite'' near the horizon. These solutions are further used to compute the Wald entropy \cite{Wald:1993nt, Brustein:2007jj, Pedraza:2021cvx} associated with these 2D black holes. Finally, we also comment on the possibilities for extremal black hole solutions in two dimensions. 

\subsection{Black hole solutions}
We estimate the 2D black hole solutions of (\ref{lmd2d}) by means of  perturbative techniques up to quadratic order in the ModMax parameter $(\gamma)$ and the Maxwell's coupling ($\kappa$). Technically speaking, it is not convenient to determine the black hole horizon in the Ferrferman-Graham gauge due to the presence of the non-trivial couplings in $U(1)$ gauge fields (\ref{lmd2d}). However, one can perform an elegant calculation using the light cone gauge. In this gauge, the space-time metric can be expressed as  
\begin{align}\label{bhlcgauge}
    ds^2=&\hspace{1mm}e^{2\omega(z)}\big(-dt^2+dz^2\big),\hspace{1mm}A_{\mu}dx^{\mu}=\hspace{1mm}A_t(z)dt,\nonumber\\
       \Phi=\hspace{1mm}&\Phi(z),\hspace{1mm}\chi=\chi(t),\hspace{1mm}\xi=\xi(z).
\end{align}

Like before as in (\ref{Afieldexp})-(\ref{Bfieldexp}), one can systematically expand the background fields in the couplings $\kappa$ and $\gamma$ as
\begin{align}
   & \mathcal{A}^{(bh)}=\mathcal{A}_0^{(bh)}+\kappa\mathcal{A}_1^{(bh)}+\gamma\kappa\mathcal{A}_2^{(bh)}+\kappa^2\mathcal{A}_3^{(bh)}...,\label{fieldexpansionblackhole1}\\
   & \mathcal{B}^{(bh)}=\mathcal{B}_1^{(bh)}+\gamma\mathcal{B}_2^{(bh)}+\kappa \mathcal{B}_3^{(bh)}...,\hspace{2mm}|\kappa|<<1,\hspace{1.5mm}|\gamma|<<1,\label{fieldexpansionblackhole2}
\end{align}
where $\mathcal{A}^{(bh)}$ collectively represents the fields ($\Phi$, $\omega$) and $\mathcal{B}^{(bh)}$ represents the remaining fields ($A_t$, $\chi$, $\xi$). Furthermore, the superscript ``${bh}$'' in $\mathcal{A}^{(bh)}$ and $\mathcal{B}^{(bh)}$ denote the black hole solution.
\subsection{Zeroth order solution}
In order to calculate black hole solutions at zeroth order, we switch off the $U(1)$ gauge couplings ($\kappa\rightarrow0$, $\gamma\rightarrow0$) in the equations of motion (\ref{geneomi})-(\ref{geneomf}), which yields the following set of equations
\begin{align}
    \Phi_0''-\omega_0'\Phi_0'+\Lambda e^{2\omega_0}\Phi_0=&\hspace{1mm}0,\label{bhzeroeomi}\\
    \omega_0'\Phi_0'+\Lambda e^{2\omega_0}\Phi_0=&\hspace{1mm}0,\\
     \omega_0''+e^{2\omega_0}\Lambda=&\hspace{1mm}0,\label{bhzeroeomf}
\end{align}
where $'$ denotes the derivative with respect to $z$.

On solving the equations (\ref{bhzeroeomi})-(\ref{bhzeroeomf}), one finds 
\begin{align}\label{bhzerosollc}
    e^{2\omega_{0}^{(bh)}}=-\frac{4\mu}{\Lambda \sinh^2{(2\sqrt{\mu}}z)}\hspace{1mm},\hspace{2mm}\Phi_0^{(bh)}=\phi_0,
\end{align}
where $\phi_0$ is a constant. 

It should be noted that we treat the dilaton ($\Phi$) as constant while taking the limits $\kappa\rightarrow0$ and $\gamma\rightarrow0$. However, it possesses a non-trivial profile in the presence of $U(1)$ gauge fields (see Section (\ref{subsecbhk}) and (\ref{subsecbhg})).

\subsection{Order $\kappa$ solution}\label{subsecbhk}
The leading order corrections to $\mathcal{A}^{(bh)}$ and $\mathcal{B}^{(bh)}$ could be estimated by solving the equations of motion (\ref{geneomi})-(\ref{geneomf}) at order $\kappa$

\begin{align}
\Phi_1''-2\big(\omega_0'\Phi_1'+\omega_1'\Phi_0'\big)+2\big(\dot{\chi}_1^2+\xi_1'^2\big)=&\hspace{1mm}0,\label{bhoneeomi}\\
\omega_1''+2\Lambda\omega_1e^{2\omega_0}-e^{-2\omega_0}A_{t1}'^2=&\hspace{1mm}0,\\
    \partial_{z}\Big(\Phi_0e^{-2\omega_0}A_{t1}'\Big)=&\hspace{1mm}0,\\
    \xi_1''=&\hspace{1mm}0,\\
    \ddot{\chi}_1=&\hspace{1mm}0,\label{bhoneeomf}
\end{align}
where . and $'$ denote the derivatives with respect to $t$ and $z$ respectively. 

In order to solve the above differential equations (\ref{bhoneeomi})-(\ref{bhoneeomf}), we adopt the following change in coordinates 
\begin{align}\label{bhctrhoz}
    \rho=\sqrt{\mu}\coth{(2\sqrt{\mu}z)}.
\end{align}

Using the zeroth order solutions (\ref{bhzerosollc}) together with (\ref{bhctrhoz}), one finds
\begin{align}
    \omega_1^{(bh)}=&\hspace{1mm}\frac{q_2}{\sqrt{\mu}}\rho\tanh^{-1}{\Bigg(\frac{\rho}{\sqrt{\mu}}\Bigg)}+\frac{q_1\rho}{\sqrt{\mu}}+\frac{m_1^2}{2\Lambda\phi_0^2}-q_2,\label{bhsolonei}\\
    \Phi_1^{(bh)}=&\hspace{1mm}-\frac{(n_1^2+l_1^2)\rho}{4\mu^{\frac{3}{2}}}\tanh^{-1}{\Bigg(\frac{\rho}{\sqrt{\mu}}\Bigg)}+g_1\rho,\label{phi1bh}\\
 \xi_1^{(bh)}=&\hspace{1mm}\frac{l_1}{2\sqrt{\mu}}\coth^{-1}
    {\Bigg(\frac{\rho}{\sqrt{\mu}}\Bigg)}+l_2\label{xi1bh},\\
      A_{t1}^{(bh)}=&\hspace{1mm}\frac{2m_1\rho}{\Lambda\phi_0}+m_2,\\
    \chi_1^{(bh)}=&\hspace{1mm}n_1t+n_2\label{bhsolonef},
\end{align}
where $m_i,n_i,l_i,q_i$ and $g_1$, $(i=1,2)$ are the integration constants.

\subsection{Order $\gamma\kappa$  solution}\label{subsecbhg}
The contributions due to the projected ModMax interactions could be estimated by solving the equations of motion (\ref{geneomi})-(\ref{geneomf}) at order $\gamma\kappa$

\begin{align}
\Phi_2''-2\big(\omega_0'\Phi_2'+\omega_2'\Phi_0'\big)+4\big(\dot{\chi}_1\dot{\chi}_2+\xi_1'\xi_2'\big)-\frac{2s_0}{\sqrt{s_0^2+p_0^2}}\big(\dot{\chi}_1^2+\xi_1'^2\big)=&\hspace{1mm}0,\label{bhtwoeomi}\\
\omega_2''+2\Lambda\omega_2e^{2\omega_0}-2e^{-2\omega_0}A_{t1}'A_{t2}'-e^{2\omega_0}\sqrt{s_0^2+p_0^2}+f_1=&\hspace{1mm}0,\\
    \partial_z\Bigg[\xi_2'-\frac{1}{\sqrt{s_0^2+p_0^2}}\Big(s_0\xi_1'-p_0\dot{\chi}_1\Big)\Bigg]=&\hspace{1mm}0,\\
    \partial_z\Bigg[e^{-2\omega_0}\Phi_0\Bigg(A_{t2}'-\frac{A_{t1}'s_0}{\sqrt{s_0^2+p_0^2}}\Bigg)\Bigg]=&\hspace{1mm}0,\\
    \ddot{\chi}_2=&\hspace{1mm}0,\label{bhtwoeomf}
\end{align}
where we define the above quantities as 
\begin{align}
   f_1=&\hspace{1mm}\frac{1}{\Phi_0\sqrt{s_0^2+p_0^2}}\Big(s_0\big(-\dot{\chi}_1^2+\xi_1'^2\big) -2p_0\dot{\chi}_1\xi_1'\Big)\hspace{1mm},\hspace{2mm} p_0=-2\Phi_0^{-1}e^{-2\omega_0}\dot{\chi}_1\xi_1',\nonumber\\ s_0=&\hspace{1mm}-e^{-4\omega_0}A_{t1}'^2+\Phi_0^{-1}e^{-2\omega_0}\big(-\dot{\chi}_1^2+\xi_1'^2\big).  
\end{align}

Clearly, the above differential equations (\ref{bhtwoeomi})-(\ref{bhtwoeomf}) are quite non trivial to solve exactly in the radial variable ($z$). However, for the purpose of our present analysis, it is sufficient to solve them near the black hole horizon.

 Using (\ref{bhctrhoz}), the location of the horizon $(\rho_H)$ can be determined by noting the spacetime metric (\ref{bhlcgauge})
\begin{align}\label{expbhmetric}
    ds^2_{(bh)}\approx\frac{4(\mu-\rho^2)}{\Lambda}\Big(1+2\kappa\omega_1^{(bh)}+2\gamma\kappa\omega_2^{(bh)}\Big)\Bigg(-dt^2+\frac{d\rho^2}{4(\mu-\rho^2)^2}\Bigg),
\end{align}
 which yields $\rho=\rho_H=\sqrt{\mu}$.

Finally, the near horizon solutions of the equations of motion (\ref{bhtwoeomi})-(\ref{bhtwoeomf}) could be listed as
\begin{align}
\Phi_2^{(bh)}=&\hspace{1mm}\frac{\rho  }{192 \mu ^{5/2}}\Bigg(64 \mu  \rho  \left(\rho -6 \sqrt{\mu }\right)+n_1^2 \rho  \Bigg(\frac{64 \mu ^{3/2} }{n_1^8l_1^2} \Bigg(8 \sqrt{\mu } (n_1^2-l_1^2)^2 \left(n_1^2+l_1^2\right) \left(\rho -9 \sqrt{\mu
   }\right)\nonumber\\
   &\hspace{1mm}-3n_1^4l_1^4\Bigg)-15 \sqrt{\mu
   }+2
   \rho \Bigg)+2 n_1 n_3 \rho  \left(2 \rho -15 \sqrt{\mu }\right)+\rho  l_1 \Big(l_1
   \left(\rho -12 \sqrt{\mu }\right)+\nonumber\\
   &3 l_3 \left(\rho -9 \sqrt{\mu }\right)\Big)+192 \mu
   ^{5/2} g_2\Bigg)+\frac{\rho  \left(n_1^2+2 n_3 n_1+l_1 \left(l_1+2
   l_3\right)\right) \log \left(\rho -\sqrt{\mu }\right)}{8 \mu ^{3/2}}\label{phi2bh},\\
   \xi_2^{(bh)}=&\hspace{1mm}\frac{1}{16 \mu ^{3/2}}\Bigg(-\frac{256 m_1^4 n_1^2 l_1 \mu ^2 \rho  \left(\rho -2 \sqrt{\mu }\right)}{\Lambda ^2 \text{$\phi_0 $}^2(n_1^2+l_1^2)^3}-\frac{32 \mu  \rho  \left(2 \sqrt{\mu }
   \left(n_1^2-2 l_1^2\right)+l_1^2 \rho \right)}{n_1^2l_1}+\nonumber\\
    &\hspace{1mm}(l_{3}+l_1)\Bigg(-\frac{1}{2} \rho\left(\rho -6
   \sqrt{\mu }\right)-4 \mu \log \left(\rho -\sqrt{\mu }\right)\Bigg)\Bigg)+l_{4},\label{xi2bh}\\
    A_{t2}^{(bh)}=&\hspace{1mm}\frac{32 \sqrt{\mu } m_1^3 n_1^2  l_1^2  \rho\left(\rho -2 \sqrt{\mu }\right)}{\Lambda ^2 \text{$\phi_0$}^2(n_1^2+l_1^2)^3
   }-m_3 \rho +m_4,\\
   \omega_2^{(bh)}=&\hspace{1mm}\frac{m_1 }{2
   \text{$\phi_0$}^2}\left(\frac{m_1 \left(n_1^2-l_1^2\right)}{\Lambda (n_1^2+l_1^2) }-m_3 \text{$\phi_0 $}\right)+q_3 I_0\left(\Tilde{\rho}\right)+q_4 K_0\left(\Tilde{\rho}\right),\\
   \chi_2^{(bh)}=&\hspace{1mm}n_3t+n_4,\label{chi2bh}
\end{align}
where we define $\Tilde{\rho}=2 \sqrt{\frac{\rho }{\sqrt{\mu }}-1}$ and $m_i,n_i,l_i,q_i$, $g_2$, $(i=3,4)$ are the integration constants. Furthermore, here $I_0\left(\Tilde{\rho}\right)$ and $ K_0\left(\Tilde{\rho}\right)$ are respectively the modified Bessel functions \cite{mpfunction} of the first $(I_{n}(\Tilde{\rho}))$ and the second kind $(K_{n}(\Tilde{\rho}))$.

\subsection{Order $\kappa^2$  solution }
The contribution due to the Maxwell (\ref{ONLYMAXWELLACT}) term alone at quadratic ($\kappa^2$) could be estimated by solving the equations (\ref{geneomi})-(\ref{geneomf}) at order $\kappa^2$ 
\begin{align}
-2\omega_1\partial_z\Big(\Phi_0e^{-2\omega_0}A_{t1}'\Big)+\partial_z\Big[e^{-2\omega_0}\Big(-2\omega_1\Phi_0A_{t1}'+\Phi_1A{t1}'+\Phi_0A_{t3}'\Big)\Big]&=0,\label{bhjhep1eq}\\
    \omega_3''+2\Lambda e^{2\omega_0}\big(\omega_1^2+\omega_3\big)-2e^{-2\omega_0}\big(-\omega_1 A_{t1}'^2+A_{t1}'A_{t3}'\big)&=0,\\
\Phi_3''-2\big(\omega_3'\Phi_0'+\omega_0'\Phi_3'+\omega_1'\Phi_1'\big)+4(\dot{\chi}_1\dot{\chi}_3+\xi_1'\xi_3')&=0,\\
    \xi_3''&=0,\\
    \ddot{\chi}_3&=0.\label{bhjhepfeq}
\end{align}

The solutions of the above equations (\ref{bhjhep1eq})-(\ref{bhjhepfeq}) are quite complicated, therefore we mention them in the Appendix \ref{jhepappendixk2}. Like before, one can further simplify these solutions (\ref{phi2bh})-(\ref{chi2bh}) and (\ref{jhepeopmkappa2i})-(\ref{jhepeopmkappa2f}) by making use of the residual gauge freedom in the light cone gauge (\ref{expbhmetric}). In particular, the re-scaling of the time coordinate, $t\rightarrow n_1t$  does not affect the gauge condition $g_{t\rho}=0$. Therefore, one can use this freedom to fix the constant $ n_1=\sqrt{1-l_1^2}$. 

It is evident from (\ref{phi1bh}), (\ref{xi1bh}), (\ref{jhepeopmkappa2i1}) and (\ref{jhepeopmkappa2i2})  that the leading order ($\kappa$) corrections as well as the quadratic order $(\kappa^2)$ corrections diverge as we move closer towards the black hole horizon ($\rho\sim\rho_H=\sqrt{\mu}$). Similar divergences persist even at quadratic order ($\gamma\kappa$) (see (\ref{phi2bh}) and (\ref{xi2bh})). However, for a particular choice of constants 
\begin{align}\label{n3l3}
 n_3=&\hspace{1mm}\frac{1 }{2\gamma\sqrt{1-l_1^2}}\Big((2l_1^2-1)(1+\gamma)-2\kappa n_5 \sqrt{1-l_1^2}-\kappa q_2\Big),\\
 l_3=&\hspace{1mm}\frac{1}{\gamma l_1}\Big(-l_1^2(1+\gamma)+\kappa n_5\sqrt{1-l_1^2}\Big),
\end{align}
the divergences at order $\kappa$ and $\kappa^2$ cancel with those at the quadratic order $(\gamma \kappa)$ thereby resulting in a finite expression for $\xi^{(bh)}$ and $\Phi^{(bh)}$ near the horizon $(\rho\sim\sqrt{\mu})$. This turns out to be a unique feature of projected ModMax interactions in two dimensions.

\subsection{2D Black hole thermodynamics}
With the above solutions at hand, we now explore the thermal properties of 2D black holes in the presence of projected ModMax interactions. In particular, we compute the Wald entropy \cite{Wald:1993nt, Brustein:2007jj, Pedraza:2021cvx} for 2D black holes. Finally, we also comment on the Wald entropy associated with the \emph{extremal} black holes in two dimensions. 

To begin with, we compute the Hawking temperature \cite{Hawking:1975vcx} for the 2D black holes which receives quadratic order corrections due to $U(1)$ gauge and ModMax couplings
\begin{align}\label{hawktdef}
    T_H=\frac{1}{2\pi}\sqrt{-\frac{1}{4}g^{tt}g^{\rho\rho}\big(\partial_{\rho}g_{tt}\big)^2}\Bigg|_{\rho\rightarrow\sqrt{\mu}}=\frac{\sqrt{\mu}}{\pi}\Bigg[1-\Big(\kappa q+ \gamma\kappa q+\kappa^2p\Big)\Bigg],
\end{align}
where we set the constants $q_4=q_{2}=q$ and $p$ is defined as
\begin{align}\label{jpcostbh}
    p=\frac{m_1^2}{12\Lambda\mu\phi_0^3}\Big(1+24q\mu\phi_0+\log(4)-8\mu^{\frac{3}{2}}g_1\Big).
\end{align}

The Wald entropy \cite{Wald:1993nt, Brustein:2007jj, Pedraza:2021cvx} is defined as
\begin{align}\label{walddefblackhole}
    S_W=-2\pi\frac{\delta \mathcal{L}}{\delta R_{\mu\nu\alpha\beta}}\epsilon_{\mu\nu}\epsilon_{\alpha\beta},
\end{align}
where $R_{\mu\nu\alpha\beta}$ is the Riemann curvature tensor, $\mathcal{L}$ is the Lagrangian density\footnote{Here we used the convention, $I=\int d^2x\sqrt{-g}\mathcal{L}$. } in two dimensions and  $\epsilon_{\mu\nu}$ is the anti-symmetric rank two tensor having the normalization condition, $\epsilon^{\mu\nu}\epsilon_{\mu\nu}=-2$.  

Using (\ref{renormaction}), the Wald entropy (\ref{walddefblackhole}) for 2D black holes turns out to be\footnote{Here, the entities $\phi_1$, $\phi_2$ and $\phi_3$ are respectively the values of  $\Phi_1^{(bh)}$ (\ref{phi1bh}), $\Phi_2^{(bh)}$ (\ref{phi2bh}) and $\Phi_3^{(bh)}$ (\ref{jhepeopmkappa2i1}) at the horizon $\rho=\rho_H=\sqrt{\mu}$.}
\begin{align}\label{waldentexp01}
 S_W=\frac{\Phi^{(bh)}}{4G_2}\Bigg|_{\rho\rightarrow\sqrt{\mu}}=\hspace{1mm}\frac{1}{4G_2}\Big(\phi_0+\kappa\phi_1+\gamma\kappa\phi_2+\kappa^2\phi_3\Big),
\end{align}
where we denote the above entities as
\begin{align}
    \phi_1=&\hspace{1mm}\sqrt{\mu } g_1-\frac{1}{192 \mu }\Big(12 \log (4 \mu )+2 l_1^2-13\Big),\label{phih}\\
    \phi_2=&\hspace{1mm}\frac{64 \mu  }{3 l_1^2}\frac{\left(1-2 l_1^2\right){}^2}{\left(l_1^2-1\right){}^3}+\frac{l_1^2}{l_1^2-1}+\sqrt{\mu } g_2-\frac{5}{3}\label{ph2h},\\
    \phi_3=&\hspace{1mm}\frac{1}{192 \mu }\Bigg(192 \mu ^{3/2} \left(g_1
   \left(q_1+q\right)+g_3\right)+2 \sqrt{1-l_1^2} n_5+q
   (36 \log (\mu )\nonumber\\
   &+13-24 \log (2))-48 q_1\Bigg),\label{jph3h}
\end{align}
and $\phi_0$ is the usual constant dilaton solution in the limit $\kappa\rightarrow0$ and $\gamma\rightarrow0$ (\ref{bhzerosollc}). 
 
 \subsection{A special case : Extremal 2D black holes}

As a special case, we study the extremal 2D black hole solutions and compute the associated Wald entropy. Extremal black holes correspond to the vanishing of the Hawking temperature (\ref{hawktdef}) 
\begin{align}\label{extremalconstraint}
    \kappa q+ \gamma\kappa q+\kappa^2p=1,
\end{align}
which for the present example stands as an extremality condition in two dimensions.

Using (\ref{extremalconstraint}) and (\ref{waldentexp01}), the Wald entropy for 2D extremal black holes $\big(S_{W}^{\text{(ext)}}\big)$ turns out to be  
  \begin{align}
      S_{W}^{\text{(ext)}}=\frac{1}{4G_2}\Bigg[\phi_0+\frac{\phi_2}{q}+\kappa\Big(\phi_1-\phi_2\Big)+\kappa^2\Bigg(\phi_3-\frac{p}{q}\phi_2\Bigg)\Bigg],
 \end{align}
 where the entities $p$, $\phi_1$, $\phi_2$ and $\phi_3$ are respectively given in (\ref{jpcostbh}), (\ref{phih}), (\ref{ph2h}) and (\ref{jph3h}).

\section{Concluding remarks}\label{secconc}
To summarise, in this chapter, we construct the 2D analogue of the four dimensional ModMax electrodynamics (coupled with Einstein gravity) using the notion of dimensional reduction. We investigate the effects of projected ModMax interactions on various physical entities associated with the boundary theory in one dimension. Finally, we construct the associated 2D black hole solutions and explore their thermal properties.

This completes the second part of the thesis, whose primary goal was to study the transformation properties of the 1D boundary stress-energy tensor under the combined action of both diffeomorphism and $U(1)$ gauge transformation and hence compute the central charge associated with the boundary theory at strong coupling.


\chapter{ Conclusion and Outlook}
\allowdisplaybreaks
\pagestyle{fancy}
\fancyhead[LE]{\emph{\leftmark}}
\fancyhead[LO]{\emph{\rightmark}}
We conclude the thesis first by summarising the key observations in Chapters 2-5 and then highlighting some of the future extensions down the line. As mentioned before, this thesis contains two parts: (i) in the first part, we investigated the phase transitions in 2D gravitational theories, and (ii) in the second part, we examined the transformation properties of the 1D boundary stress-energy tensor under the combined action of both diffeomorphism and $U(1)$ gauge transformation which eventually leads to the central charge associated with the 1D boundary theory at strong coupling.

Below, we summarize the key outcome of each of the individual chapters.

In Chapter 2, we constructed a 2D theory of gravity starting from a 5D gravity coupled with $U(1)$ and $SU(2)$ Yang-mills fields following a dimensional reduction. We explored the thermal properties of the vacuum and the black hole solutions for the 2D gravity model using a canonical ensemble. In particular, we computed the free energy and thermal entropy of the system and found that the thermal radiation beyond a certain critical temperature would collapse to form a globally stable black hole, indicating the analog of the Hawking-Page (HP) transition in 2D gravity.

In Chapter 3, we investigated the phase stability of $U(1)$ gauge charged Euclidean wormhole solutions in two dimensions. To be specific, we obtained the free energy (density) and the total charge of the system and explored their dependencies on the temperature at a fixed chemical potential. To our surprise, the free energy and the total charge of the system suffer from a discontinuity at a critical temperature ($T=T_0$), indicating the onset of a first-order phase transition. Furthermore, we found that beyond the critical temperature, the wormhole solutions transit into a pair of black holes at finite chemical potential via a first-order phase transition.

In Chapter 4, we build up a model of JT gravity that contains 2-derivative as well as 4-derivative interactions between the $U(1)$ gauge fields and the space-time metric. We computed the renormalized boundary stress-energy tensor, examined its transformation properties under the diffeomorphism and the $U(1)$ gauge transformation in the Fefferman-Graham gauge, and obtained the central charge associated with the 1D boundary theory. We notice that, the central charge goes inversely with the coupling constant associated with the 4-derivative interactions.

We further obtained the black hole solutions and investigated the effects of quartic interactions on their thermal properties. We found that the Wald entropy associated with the black hole diverges at the horizon due to the presence of higher derivative interactions in the theory. We interpreted the near-horizon divergences in terms of the density of states. Finally, we investigated the near-horizon modes of the theory and computed the central charge associated with the CFT$_{1}$ in the near-horizon limit. In addition, we found that the near-horizon CFT could be recast as 2D Liouville theory with generalized potential. We examined the Weyl invariance of the Liouville theory and computed the associated Weyl anomaly.  

In Chapter 5, following a dimension reduction, we constructed a 2D theory of gravity starting from a 4D gravity in the presence of ``ModMax'' interactions. We explored the effects of these non-trivial interactions on the central charge associated with the 1D boundary theory up to quadratic order in the ModMax and $U(1)$ gauge couplings. We further computed the associated black hole solutions in two dimensions and examined the effects of ModMax interactions on their thermal properties. Interestingly, we found that the ModMax interactions play a crucial role in obtaining the finite values of the background fields near the horizon.

After briefly summarizing the key results of the thesis, below we outline some interesting projects that can be persuaded in the future. 
\begin{itemize}
   
\item It would be indeed an interesting project to look for the dual 1D boundary theory pertinent to the modified models of JT gravity as discussed in Chapters 2-5 and identify the corresponding Schwarzian degrees of freedom. The Schwarzian  \cite{Maldacena:2016upp} action can be obtained following a standard procedure in which one integrates out the dilaton ($ \Phi $). In the process, one is left with the boundary term which is regarded as the low energy effective theory living on the boundary. Clearly, for the class of theories that we study, one would expect a modified Schwarzian in the presence of a chemical potential ($ \mu $).

As for example, below, we outline a simple toy model calculation for a particular case \cite{Rathi:2021mla}, in which we show schematically how does this extra term is generated due to the presence of the chemical potential ($ \mu $). To begin with, we consider an $ AdS_2 $ metric of the form
 \begin{equation}
     ds^2=\frac{1}{z^2}(dt^2+dz^2),
 \end{equation}
 where the boundary is located at, $z \sim \varepsilon \sim 0$.
 
Taking the first term on the R.H.S. of (\ref{action}) into account, one can show that the boundary action typically looks like (here we absorb the coupling constants of the bulk theory into $ \mu $)
\begin{eqnarray}
S_B \sim \int du~\Phi_r (u)  \text{Sch}\lbrace t(u), u\rbrace +\mu  \int du \frac{\Phi_r^2 (u)}{\varepsilon},
\end{eqnarray}
where we parameterize the coordinates ($t , z$) in terms of the boundary time coordinate ($u$) and $\Phi_r (u)$ is the coupling constant of the boundary field theory which is defined in terms of the boundary data of the form $\Phi|_{bdy}=\frac{\Phi_r}{\epsilon}$ \cite{Maldacena:2016upp}. It would be indeed an interesting project to explore such theories in detail an extract Green's functions etc. out of it. We leave all these issues for future investigations.

 \item It would be an interesting project to generalise the results of Chapter 2 in the presence of ModMax interactions (see Chapter 5) and investigate the possible deviations in the Hawking-Page transition in the presence of projected ModMax interactions. Moreover, one can further compute their imprints on various physical observables like the boundary stress-energy tensor and the central charge associated with the 1D boundary theory.

  \item  It would be nice to construct the 2D wormhole solutions \cite{Rathi:2021mla} and explore their thermal stability for the ModMax corrected JT gravity models \cite{Rathi:2023vhw} and/or the 2D gravity theory that contains the most generic quartic interactions \cite{Rathi:2021aaw}.

\item Recall that, in Section \ref{secwh} we compute the ``annealed'' Free energy \cite{Garcia-Garcia:2020ttf} of the wormhole solution and explore the associated phase stability (see Section \ref{thermoref2}). However, one can further refine the computation by including the corrections due to the replica wormholes \cite{Engelhardt:2020qpv,Almheiri:2019qdq} and compute the ``quenched'' Free energy \cite{Garcia-Garcia:2020ttf}. It would be nice to explore the properties of ``quenched'' Free energy and study the corresponding phase stability of the configuration at finite chemical potential ($\mu$).

   \item It is natural to further generalise the results of Chapter 4 in the presence of $SU(2)$ Yang-Mills fields and look for its imprints on the holographic stress tensor as well as central charge associated with the 1D boundary theory. It is noteworthy to mention that the $SU(2)$ Yang-Mills fields are responsible for the first order phase transition in 2D gravity \cite{Lala:2020lge}. Therefore, it would be an interesting project to explore the phase transition in the presence of such quartic couplings.

   \item In the literature, there exists an alternative way to derive the thermodynamic entropy of 2D black holes by noting the asymptotic growth of the physical states of a CFT by means of the Cardy formula ($S_C$)  \cite{Castro:2008ms, Cardy:1986ie, Strominger:1997eq}
 \begin{align}\label{cardyformula}
     S_C= 2\pi\sqrt{\frac{c_M \Delta}{6}},
 \end{align}
 where $\Delta$ is the  eigen value of the associated Virasoro generator $L_0$. 
 
 The authors in  \cite{Castro:2008ms} establish a 2D/3D dictionary which by virtue of the Cardy formula (\ref{cardyformula}) predicts the correct Bekenstein-Hawking entropy for 2D black holes. Therefore, it would be indeed an interesting project to uplift the 2D black hole solutions into three dimensions and establish a suitable 2D/3D mapping in the presence  most generic quartic interactions \cite{Rathi:2021aaw} and/or 2D projected ModMax interactions \cite{Rathi:2023vhw}.

  \item It would be an interesting project to explore the holographic renormalisation group flow and holographic c-theorem \cite{Myers:2010tj}-\cite{Suh:2020qnl} in the context of 2D gravity models that contain the most generic quartic interactions \cite{Rathi:2021aaw} and/or 2D projected ModMax interactions \cite{Rathi:2023vhw}.

   \item As mentioned in the introduction (see Section \ref{introthm}), the authors in \cite{Witten:2020ert} investigated the phase transition in the deformed JT gravity and the matrix integral model. It would be interesting to further generalize the matrix integral models associated with JT gravity in the presence of non-trivial gauge couplings (see Chapters 2-5) and explore the possible deviations in the Hawking-Page transition.

\end{itemize}

\appendix


\chapter{Dimensional reduction from $5D$ to $2D$}\label{dimred}

In order to analyze the dimensional reduction of the $5D$ theory (\ref{act:5D}) to $2D$, in the following we rewrite the ansatz
given in (\ref{anz:red}):
\begin{equation}\label{anz:red2}
ds^{2}=\Phi^{\alpha}\;d\tilde{s}^{2}+\Phi^{\beta}dx_{i}^{2},\quad A_{M}^{a}dx^{M}
=A_{\mu}^{a}dx^{\mu}, \quad A_{M}dx^{M}=A_{\mu}dx^{\mu}~;~\alpha,\beta  \in \mathbb{R}.
\end{equation}

With this choice of the metric we note that
\begin{equation}\label{g:measure}
 \sqrt{-g}=\sqrt{-\tilde{g}}\;\Phi^{\alpha+3\beta/2},
\end{equation} 
where $\tilde{g}$ is the determinant of the two-dimensional metric.

It is then easy to check that the first two terms in the action (\ref{act:5D}) can be written as
\begin{equation}\label{RL:5D}
\sqrt{-g}(\mathcal{R}-3\Lambda)=\sqrt{-\tilde{g}}\Phi^{3\beta/2}(\tilde{\mathcal{R}}
-3\Lambda\Phi^{\alpha}+\Phi^{\alpha}g^{ij}\mathcal{R}_{ij}). 
\end{equation}

In the next step, we rewrite the third term within the curly braces in (\ref{RL:5D}) as a total derivative term given by
\begin{equation}\label{totDv}
\Phi^{\alpha+3\beta/2}g^{ij}\mathcal{R}_{ij}\sim \nabla_{\mu}\Big[g^{\mu\nu}
g^{\lambda\sigma}g_{\lambda\sigma}\left(\partial_{\nu}\Phi^{\beta}\right)
\Phi^{\alpha+\beta/2}\Big],
\end{equation}
which therefore does not contribute to the bulk equations of motion (\ref{5a})-(\ref{eom:ab}). Notice that, the JT gravity
model (\ref{actsim:2D}) can be obtained once we set the parameters $\alpha=0$ and $\beta=2/3$. In other words, these
choices of parameters are quite pertinent to the JT gravity model discussed in this paper. The remaining terms in
(\ref{actsim:2D}) involving gauge fields can be obtained in a similar manner.

\chapter{Equations of motion at different order in the perturbation series}\label{apen:eom}
In the following, we note down the equations of motion upto leading order in perturbative expansion. For example, substituting 
(\ref{pert:exp}) into (\ref{nor:sclr1}) we obtain
\begin{subequations}
\setlength{\jot}{8pt}
\begin{align}
\mathcal{O}(0): \quad& 2\Phi_{(0)}''+6\Lambda\Phi_{(0)} e^{2\omega_{(0)}}=0,  \label{dil:zero}\\
\mathcal{O}(\xi): \quad& 2\left(\Phi_{(1)}^{ab}\right)''+6\Lambda e^{2\omega_{(0)}}\left(2
\omega_{(1)}^{ab}\Phi_{(0)}+\Phi_{(1)}^{ab}\right)+\frac{Q^{2}}{\Phi_{(0)}} e^{2\omega_{(0)}}=0, 
\label{dil:xi}\\
\mathcal{O}(\kappa): \quad& 2\left(\Phi_{(1)}^{na}\right)''+6\Lambda e^{2\omega_{(0)}}\left(2 \omega_{(1)}^{na}\Phi_{(0)}+\Phi_{(1)}^{na}\right)+\frac{1}{g_{s}^{2}}\Phi_{(0)}e^{-2\omega_{(0)}}
\left(\left(\chi_{(0)}'\right)^{2}+\chi_{(0)}^{2}\eta_{(0)}^{2}\right)\nonumber\\
&=0.  \label{dil:kapa}
\end{align}
\end{subequations}

Similarly (\ref{nor:guge}) could be arranged upto leading order in the perturbative expansion as
\begin{subequations}
\setlength{\jot}{8pt}
\begin{align}
\mathcal{O}(0): \quad& 4\omega_{(0)}''+6\Lambda e^{2\omega_{(0)}}=0,  \label{omga:zero}\\
\mathcal{O}(\xi): \quad& 4\left(\omega_{(1)}^{ab}\right)''+12\Lambda\omega_{(1)}^{ab} 
e^{2\omega_{(0)}}-\frac{Q^{2}}{\Phi_{(0)}^{2}}e^{2\omega_{(0)}}=0,  \label{omga:xi}\\
\mathcal{O}(\kappa): \quad& 4\left(\omega_{(1)}^{na}\right)''+12\Lambda\omega_{(1)}^{na} 
e^{2\omega_{(0)}}-\frac{e^{-2\omega_{(0)}}}{g_{s}^{2}}\left(\left(\chi_{(0)}'\right)^{2}
+\chi_{(0)}^{2}\eta_{(0)}^{2}\right)=0.  \label{omga:kapa}
\end{align}
\end{subequations}

Substituting (\ref{pert:exp}) into the constraint equation (\ref{cnst:nab}) we obtain
 \begin{subequations}
\setlength{\jot}{8pt}
\begin{align}
\mathcal{O}(0): &\quad \Phi_{(0)}e^{-2\omega_{(0)}}\chi_{(0)}^{2}\eta_{(0)}=0,  \label{cnst:0}\\
\mathcal{O}(\xi): &\quad  \Phi_{(0)}\left(\chi_{(0)}\eta_{(1)}^{ab}+2\eta_{(0)}\chi_{(1)}^{ab}\right)
+\chi_{(0)}\eta_{(0)}\left(\Phi_{(1)}^{ab}-2\Phi_{(0)}\omega_{(1)}^{ab}\right)=0,\hspace{1mm}
e^{-2\omega_{(0)}}\chi_{(0)}^{2}=0, \label{cnst:xi}\\
\mathcal{O}(\kappa): &\quad\Phi_{(0)}\left(\chi_{(0)}\eta_{(1)}^{na}+2\eta_{(0)}\chi_{(1)}^{na}\right)
+\chi_{(0)}\eta_{(0)}\left(\Phi_{(1)}^{na}-2\Phi_{(0)}\omega_{(1)}^{na}\right)=0,\hspace{1mm}
e^{-2\omega_{(0)}}\chi_{(0)}^{2}=0. \label{cnst:kapa}
\end{align}
\end{subequations}

Finally, from (\ref{nor:nab1}) and (\ref{nor:nab2}) the zeroth order equations could be recast as
\begin{subequations}
\setlength{\jot}{15pt}
\begin{align}
\partial_{z}\left(\Phi_{(0)}e^{-2\omega_{(0)}}\chi_{(0)}\eta_{(0)}\right)
+\Phi_{(0)}e^{-2\omega_{(0)}}\eta_{(0)}\left(\partial_{z}\chi_{(0)}\right)&=0,
\label{zero:nab1}\\
\partial_{z}\left(\Phi_{(0)}e^{-2\omega_{(0)}}\left(\partial_{z}\chi_{(0)}\right)
\right)-\Phi_{(0)}e^{-2\omega_{(0)}}\chi_{(0)}\eta_{(0)}^{2}&=0. \label{zero:nab2}
\end{align}
\end{subequations}

\chapter{Derivations of zeroth order solutions}\label{sol:zero}
In the conformal gauge \cite{Almheiri:2014cka}
\begin{equation}\label{conf:guge}
ds^{2}=-e^{2\omega(x^{+},x^{-})}dx^{+}dx^{-},\quad x^{\pm}=(t\pm z),
\end{equation}
the zeroth order equations of motion for the metric and the dilaton can be rewritten as
\begin{align}
4\partial_{+}\partial_{-}\Phi_{(0)}-3\Lambda\Phi_{(0)}e^{2\omega_{(0)}}&=0,\label{B1}\\ 
8\partial_{+}\partial_{-}\omega_{(0)}-3\Lambda e^{2\omega_{(0)}}&=0, \label{B2}\\
\partial_{+}\left(e^{2\omega_{(0)}}\partial_{+}\Phi_{(0)}\right)&=0, \label{B3}\\
\partial_{-}\left(e^{2\omega_{(0)}}\partial_{-}\Phi_{(0)}\right)&=0. \label{B4}
\end{align}

Clearly, the solution of (\ref{B2}) is given by
\begin{equation}\label{sol:B2}
e^{2\omega_{(0)}}=\left(-\frac{8}{3\Lambda}\right)\frac{1}{\left(x^{+}-x^{-}\right)^{2}}.
\end{equation}

Now integrating (\ref{B3}) and (\ref{B4}) we may write
\begin{align}
\partial_{+}\Phi_{(0)} &= \left(-\frac{8}{3\Lambda}\right)\frac{f(x^{-})}{\left(x^{+}
-x^{-}\right)^{2}},\label{sol:B3}\\
\partial_{-}\Phi_{(0)} &= \left(-\frac{8}{3\Lambda}\right)\frac{g(x^{+})}{\left(x^{+}
-x^{-}\right)^{2}}.\label{sol:B4}
\end{align}

Differentiating (\ref{sol:B3}) w.r.to $x^{-}$ and (\ref{sol:B4}) w.r.to $x^{+}$ and substituting them in (\ref{B1}) we obtain
\begin{align}\label{sol:phigen}
  \Phi_{(0)} =
    \begin{cases}
      \frac{2}{3\Lambda}\frac{\partial_{-}f(x^{-})\cdot (x^{+}-x^{-})+2f(x^{-})}
      {(x^{+}-x^{-})}  \\[10pt]
      \frac{2}{3\Lambda}\frac{\partial_{+}g(x^{+})\cdot (x^{+}-x^{-})-2g(x^{+})}
      {(x^{+}-x^{-})} .
    \end{cases}       
\end{align}

After a few easy algebraic steps, the general solution for the dilaton may be written as \cite{Almheiri:2014cka}
\begin{equation}\label{sol:phiF}
\Phi_{(0)}=\left(-\frac{2}{3\Lambda}\right)\frac{a+b(x^{+}+x^{-})+cx^{+}x^{-}}
{(x^{+}-x^{-})},
\end{equation}
where $a$, $b$, $c$ are real constants. For the vacuum we may choose $a=1$, $b=0$ and $c=0$ \cite{Almheiri:2014cka,
Kyono:2017jtc}.

In the next step, we find the corresponding solutions for the black hole by exploiting the $SL(2,\mathbb{R})$ invariance of the 
metric (\ref{conf:guge}). The general solutions can then be written as
\begin{align}
e^{2\omega_{(0)}}&=\left(-\frac{8}{3\Lambda}\right)\frac{\omega^{+}(x^{+})
\omega^{-}(x^{-})}{\left[\omega^{+}(x^{+})-\omega^{-}(x^{-})\right]^{2}}, 
\label{met:w}\\[10pt]
\Phi_{(0)}&=\left(-\frac{2}{3\Lambda}\right)\frac{1-\mu\;\omega^{+}(x^{+})
\omega^{-}(x^{-})}{\left[\omega^{+}(x^{+})-\omega^{-}(x^{-})\right]}, \label{dil:w}
\end{align}
where $\mu(>0)$ can be considered as the mass of the black hole, and $\omega^{\pm}(x^{\pm})$ are \emph{monotonic}
functions \cite{Almheiri:2014cka}. 

Finally, using the conformal transformations
\begin{equation}
\omega^{\pm}(x^{\pm})=\frac{1}{\sqrt{\mu}}\tanh\sqrt{\mu}x^{\pm}
\end{equation}
(\ref{met:w}) and (\ref{dil:w}) can be rewritten as
\begin{align}
e^{2\omega_{(0)}}&=\left(-\frac{8}{3\Lambda}\right)\frac{\mu}{\sinh^{2}\left(
\sqrt{\mu}(x^{+}-x^{-})\right)}, 
\label{met:H}\\[10pt]
\Phi_{(0)}&=\left(-\frac{2}{3\Lambda}\right)\sqrt{\mu}\coth\left(\sqrt{\mu}(x^{+}
-x^{-})\right). \label{dil:H}
\end{align}

\chapter{3D to 2D reduction}\label{dimreduction}
We begin by considering $ AdS_3 $ gravity coupled to Maxwell Chern-Simons term of topological gauge theories \cite{VanMechelen:2019ebr,Deser:1981wh}
  \begin{align}\label{3daction}
      S_{(3)}=\int d^3x\sqrt{-g_{(3)}}\Bigg(R_{(3)}+2+a_1F_{MN}F^{MN}+a_2\epsilon^{MNP}A_M\partial_N A_P+2(\partial_M f)^2\Bigg),
  \end{align}
 where $R_{(3)}$ is the 3D Ricci scalar, $f$ is the scalar field and $\epsilon^{MNP}=\frac{\varepsilon^{MNP}}{\sqrt{-g_{(3)}}}$ is the Levi-Civita tensor. Here $(M,N)$ represent the 3 dimensional indices.
 
In order to obtain the desired JT gravity model (\ref{action}), we propose a reduction ansatz of the following form 
 \begin{align}\label{3dansatz}
     ds_{(3)}^2&=\Phi(x^{\mu})^{-1}ds_{(2)}^2+\Phi(x^{\mu})^{2}dz^2,\hspace{2mm}ds_{(2)}^2=\Tilde{g}_{\mu\nu}(x^{\alpha})dx^{\mu}dx^{\nu},\nonumber\\
    A_{\mu}&\equiv A_{\mu}(x^{\nu}),\hspace{1mm}A_z\equiv \kappa(x^{\mu}),\hspace{1mm}f\equiv f(x^{\mu}),
 \end{align}
where $(\mu,\nu)$ are the 2 dimensional indices and $z$ is the compact dimension.

Substituting  (\ref{3dansatz}) into (\ref{3daction}) and integrating over compact dimension, we find
\begin{align}\label{2dactionreduce}
    S_{(2)}=&\int d^2x\sqrt{-g_{(2)}}\Bigg(\Phi R_{(2)}+2\big(1+(\partial_{\mu}f)^2\Phi\big)-\Phi^{-1}(\partial_{\mu}\Phi)^2+a_1\Phi^2F^2+\nonumber\\
    &2a_1\Phi^{-1}(\partial_{\mu}\kappa)^2 +a_2\epsilon^{\mu\nu}_{(2)}F_{\mu\nu}\kappa\Bigg).
\end{align}

Next, we redefine the fields as
\begin{align}\label{redefine}
 \kappa (\rho)=\frac{\Phi}{\sqrt{2a_1}},\hspace{2mm}f(\rho)=\int d\rho\sqrt{\frac{\Tilde{g}_{\rho\rho}(\Phi-1)}{\Phi}},
\end{align}
where $\rho$ is the radial direction of $ AdS_2 $.

Finally, substituting (\ref{redefine}) into (\ref{2dactionreduce}), we obtain the 2D gravity model
\begin{align}
    S_{(2)}=\int d^2x\sqrt{-g_{(2)}}\Bigg(\Phi \big(R_{(2)}+2\big)+a_1\Phi^2F^2+\Tilde{a}_2\epsilon^{\mu\nu}_{(2)}F_{\mu\nu}\Phi\Bigg),
\end{align}
where we define $\Tilde{a}_2=\frac{a_2}{\sqrt{2a_1}}$.

\chapter{Quantum stress-energy tensor for gauge fields }\label{QSTref2}
In this Appendix, we compute the expectation value for the quantum stress-energy tensor \cite{Garcia-Garcia:2020ttf} in the double trumpet background (\ref{gauge}). We use the point-splitting method of \cite{book}. As mentioned previously, the following derivation is technically different from that of the scalar field ($\chi$) \cite{Garcia-Garcia:2020ttf}. The reason for this stems from the fact that gauge fields in 2D are non-conformal. Therefore, one has to carry out a first principle derivation of the quantum stress-energy tensor considering the double trumpet geometry as the background.

The equation of motion for the gauge field, $A_{\mu}$ (\ref{eom3}) in the double trumpet background (\ref{gauge}) turns out to be 
\begin{align}\label{cylgaugeeom}
   \hat{\textbf{F}}A_{\tau}=J(\rho),
\end{align}
where $\hat{\textbf{F}}=\partial^2_{\rho}+(f(\rho)-2\tan(\rho))\partial_{\rho}$ is the corresponding differential operator. 

Furthermore, here we define
$$f(\rho)=2\Phi^{-1}\partial_{\rho}\Phi\hspace{1mm},\hspace{2mm} J(\rho)=\frac{ a_2}{2a_1}\frac{\partial_{\rho}\Phi}{\Phi^{2}\cos^2\rho}.$$

In order to proceed further, we segregate out gauge field components from (\ref{action})
\begin{align}\label{gaugeaction}
    S_{gauge}=a_1\int_{(I)} d^2x\sqrt{-g}\Phi^{2}F^2+a_2\int_{(II)} d^2x\sqrt{-g}\Phi\varepsilon^{\mu\nu}F_{\mu\nu}.
\end{align}

Notice that, the variation of the second integral $(II)$ with respect to the metric $g_{\mu\nu}$ vanishes. Therefore, only the first integral $(I)$ contributes to the expectation value of the quantum stress-energy tensor for gauge fields (\ref{breakref2}). 

In order to take the variation of (\ref{gaugeaction}), we decompose the (first) integral $(I)$ into the bulk and the boundary pieces as follows
\begin{align}\label{gaugeactionsplit}
    S_{gauge}&=2a_1\int d^2xA_{\nu}\partial_{\mu}\Big[\sqrt{-g}\Phi^{2}K^{\mu\alpha\nu\beta}\partial_{\beta}A_{\alpha}\Big]+a_2\int d^2x\sqrt{-g}\Phi\varepsilon^{\mu\nu}F_{\mu\nu}+S_{boundary},\nonumber\\
    K^{\mu\alpha\nu\beta}&=g^{\mu\alpha}g^{\nu\beta}-g^{\mu\beta}g^{\nu\alpha}.
\end{align}

The variation of (\ref{gaugeactionsplit}) with respect to the bulk metric $g^{\eta\kappa}$  yields the following expression
\begin{align}\label{gaugevariation1}
    &\frac{\delta S_{gauge}}{\delta g^{\eta\kappa}}=2a_1\int d^2x\Bigg[A_{\nu}\Big[\Phi^{2}\Big\{-\frac{1}{2}\Big(\partial_{\mu}(g_{\eta\kappa}\sqrt{-g})+\sqrt{-g}g_{\lambda\sigma}\frac{\delta(\partial_{\mu}g^{\lambda\sigma})}{\delta g^{\eta \kappa}}\Big)K^{\mu\alpha\nu\beta}-\nonumber\\
    &\frac{1}{2}\sqrt{-g}g_{\eta\kappa}(\partial_{\mu}K^{\mu\alpha\nu\beta})+\sqrt{-g}\Big(g^{\nu\beta}\frac{\delta(\partial_{\mu}g^{\mu\alpha})}{\delta g^{\eta\kappa}}+g^{\mu\alpha}\frac{\delta(\partial_{\mu}g^{\nu\beta})}{\delta g^{\eta\kappa}}-g^{\nu\alpha}\frac{\delta(\partial_{\mu}g^{\mu\beta})}{\delta g^{\eta\kappa}}-\nonumber\\
    &g^{\mu\beta}\frac{\delta(\partial_{\mu} g^{\nu\alpha})}{\delta g^{\eta\kappa}}\Big)\Big\}\Big]\dot(\partial_{\beta}A_{\alpha})+\Phi^{2}\Big\{A_{\nu}(\partial_{\eta}\sqrt{-g})g^{\nu\beta}\partial_{\beta}A_{\kappa}+A_{\eta}(\partial_{\mu}\sqrt{-g})g^{\mu\alpha}\partial_{\kappa}A_{\alpha}-\nonumber\\
    &A_{\nu}(\partial_{\eta}\sqrt{-g})g^{\nu\beta}\partial_{\kappa}A_{\beta}-A_{\eta}(\partial_{\mu}\sqrt{-g})g^{\mu\beta}\partial_{\beta}A_{\kappa}+\sqrt{-g}\Big(A_{\nu}\partial_{\eta}(g^{\nu\beta})\partial_{\beta}A_{\kappa}+\nonumber\\
    &A_{\eta}\partial_{\mu}(g^{\mu\alpha})\partial_{\kappa}A_{\alpha}-A_{\nu}\partial_{\eta}(g^{\nu\alpha})\partial_{\kappa}A_{\alpha}-A_{\eta}\partial_{\mu}(g^{\mu\beta})\partial_{\beta}A_{\kappa}\Big)\Big\}+\nonumber\\
    &\sqrt{-g}\Big\{A_{\nu}g^{\nu\beta}\partial_{\eta}(\Phi^{2}\partial_{\beta}A_{\kappa})-\frac{1}{2}A_{\nu}g_{\eta\kappa}K^{\mu\alpha\nu\beta}\partial_{\mu}(\Phi^{2}\partial_{\beta}A_{\alpha})+A_{\eta}g^{\mu\alpha}\partial_{\mu}(\Phi^{2}\partial_{\kappa}A_{\alpha})-\nonumber\\
    &A_{\nu}g^{\nu\alpha}\partial_{\eta}(\Phi^{2}\partial_{\kappa} A_{\alpha})-A_{\eta}g^{\mu\beta}\partial_{\mu}(\Phi^{2}\partial_{\beta}A_{\kappa})\Big\}\Bigg].
\end{align}

In the double trumpet background (\ref{gauge}), the above expression (\ref{gaugevariation1}) further simplifies as
\begin{align}\label{gaugevariation2}
    \frac{\delta S_{gauge}}{\delta g^{\tau\tau}}= \frac{\delta S_{gauge}}{\delta g^{\rho\rho}}=-a_1\int d^2x\sqrt{-g}A_{\tau}\hat{\textbf{L}} A_{\tau},
\end{align}
where we ignore all the derivatives of the metric variation i.e. $ \Big|\frac{\partial_{\mu}(\delta g^{\mu\alpha})}{\delta g^{\eta\kappa}}\Big|<<1$ and  $\hat{\textbf{L}}$ is the differential operator such that $\hat{\textbf{L}}A_{\tau}=\cos^2\rho\partial_{\rho}(\Phi^{2}\partial_{\rho}A_{\tau})$.  

Using (\ref{gaugevariation2}) as well as the point splitting method \cite{Garcia-Garcia:2020ttf}, the expectation value for the quantum stress-energy tensor\footnote{Notice that, there are two possible ways in which the differential operator $\hat{\textbf{L}}$ can act on the Green's function $G(\rho,\tau;\rho',\tau')$. However, both the possibilities give the same result. Therefore, we consider the average of both the possibilities in the definition of the expectation value of the stress tensor.} turns out to be 
\begin{align}\label{stcylgauge}
    <T_{\tau\tau }^{(gauge)}>_{qm}\hspace{1mm}=\hspace{1mm}<T_{\rho\rho}^{(gauge)}>_{qm}\hspace{1mm}= \frac{a_1}{2}\Big[&\lim_{x'\rightarrow x}\cos^2\rho\partial_{\rho}\Big(\Phi^{2}(\rho')\partial_{\rho'}G(\rho,\tau;\rho',\tau')\Big)+\nonumber\\
    &\lim_{x\rightarrow x'}\cos^2\rho'\partial_{\rho'}\Big(\Phi^{2}(\rho)\partial_{\rho}G(\rho,\tau;\rho',\tau')\Big)\Big],
\end{align}
where $x\equiv(\rho,\tau)$,  $x'\equiv(\rho',\tau')$.

Here $G(\rho,\tau;\rho',\tau')$ is the Green's function which satisfies the following equation
\begin{align}\label{green1}
   \hat{\textbf{F}}G(\rho,\tau;\rho',\tau')=-\delta(\rho-\rho')\delta(\tau-\tau'), 
\end{align}
where the operator $\hat{\textbf{F}}$ is given by (\ref{cylgaugeeom}).

In order to solve (\ref{green1}), we consider the following Fourier decomposition of $G(\rho,\tau;\rho',\tau')$ and $\delta(\tau-\tau')$
\begin{align}
    G(\rho,\tau;\rho',\tau')&=\sum_{m,m'\in \mathbb{Z}}\Tilde{G}(\rho,m;\rho',m')e^{2\pi i (m\tau+m'\tau')/b}\label{fd1},\\
    \delta(\tau-\tau')&=\frac{1}{b}\sum_{m,m'\in \mathbb{Z}}e^{2\pi i (m\tau+m'\tau')/b}\delta_{m+m'}.\label{fd2}
\end{align}

Plugging (\ref{fd1}) and (\ref{fd2}) into (\ref{green1}), we obtain
\begin{align}\label{green2}
    \Big[\partial^2_{\rho}+(f(\rho)-2\tan(\rho))\partial_{\rho}\Big]\Tilde{G}(\rho,m;\rho',m')=-\frac{1}{b}\delta(\rho-\rho')\delta_{m+m'}.
\end{align}

Now, we solve (\ref{green2}) using the following properties: \\

   $\bullet$ $\Tilde{G}$ is continuous in the limit $\rho\rightarrow \rho'$ : $\Tilde{G}(\rho)=\Tilde{G}(\rho')$.
   
   $\bullet$ The derivative of $\Tilde{G}$ is discontinuous in the limit\footnote{Here, we integrate (\ref{green2}) with respect to $\rho$ from $\rho=\rho'-\epsilon$ to $\rho=\rho'+\epsilon$ and take the limit $\epsilon\rightarrow 0.$} $\epsilon\rightarrow 0$ namely,
   \begin{align}\label{disc}
       \frac{d \Tilde{G}}{d\rho}\Big|_{\rho'+\epsilon}-\frac{d \Tilde{G}}{d\rho}\Big|_{\rho'-\epsilon}=-\frac{1}{b}.
   \end{align}

Notice that, $f(\rho)$ in (\ref{green2}) is an unknown function of $\rho$. Therefore one cannot solve (\ref{green2}) exactly for all values of $\rho$. However, we are interested in the near boundary analysis, therefore we solve (\ref{green2}) in the near boundary limits, $\rho\rightarrow\pm\frac{\pi}{2}$.\\

$\bullet$ \underline{Case 1} : Near  $\rho\sim\rho_R\sim\frac{\pi}{2} $ and $-\frac{\pi}{2}<\rho'<\rho<\frac{\pi}{2}$

Setting $m'=-m$ and upon solving equation (\ref{green2}) near $\rho\sim\frac{\pi}{2}$, we obtain 
 \begin{align}\label{npbgp1}
     \Tilde{G}_+(\rho,\rho',m)= &\hspace{1mm} \frac{1}{4}A(\rho',m)\Bigg(\frac{2\exp(p_0(\pi-2\rho)/2)}{\pi-2\rho}-p_0\text{Ei}(p_0(\pi-2\rho)/2)\Bigg)\nonumber \\
     &+B(\rho',m),
 \end{align}\\
 where $p_0=f\Big(\frac{\pi}{2}\Big)$ and the subscript $`+$' denotes the value of Green's function near the right boundary $\rho_R\sim+\frac{\pi}{2}$.\\
 
 $\bullet$ \underline{Case 2} : Near $\rho\sim\rho_L\sim-\frac{\pi}{2}$ and $-\frac{\pi}{2}<\rho<\rho'<\frac{\pi}{2}$

 On setting $m'=-m$ and solving equation (\ref{green2}) near $\rho\sim-\frac{\pi}{2}$, we obtain
 \begin{align}\label{npbgp2}
      \Tilde{G}_-(\rho,\rho',m)= &\hspace{1mm}  \frac{1}{4}C(\rho',m)\Bigg(-\frac{2\exp(-q_0(\pi+2\rho)/2)}{\pi+2\rho}-q_0\text{Ei}(-q_0(\pi+2\rho)/2)\Bigg)\nonumber \\
     &+D(\rho',m),
 \end{align}
  where $q_0=f\Big(-\frac{\pi}{2}\Big)$ and the subscript $`-$' denotes the value of Green's function near the left boundary $\rho_L\sim-\frac{\pi}{2}$.
  
Notice that, the differential equation (\ref{green2}) contains the derivative of\\ $\Tilde{G}(\rho,m;\rho',m')$ with respect to the quantity ``$\rho$''. Therefore, the entities $A(\rho',m)$, $B(\rho',m)$, $C(\rho',m)$ and $D(\rho',m)$ in (\ref{npbgp1}) and (\ref{npbgp2}) appears as an integration constants. However, one can compute the functions $A(\rho',m)$ and $C(\rho',m)$ using the properties of the Green's function those were mentioned in (\ref{disc}) but the functions $B(\rho',m)$ and $D(\rho',m)$ still remain undetermined. Therefore, we perform all the analysis keeping the general form of the functions $B(\rho',m)$ and $D(\rho',m)$ and finally impose a suitable condition on these functions (\ref{npbcond1}) using the fact that the left temperature of the wormhole near $\rho_L\sim-\frac{\pi}{2}$ must be identified with the right temperature near $\rho_R\sim\frac{\pi}{2}$.

Finally, (\ref{fd1}) can be systematically expressed as 
 \begin{align}\label{greenfunc}
     G(\rho,\tau;\rho',\tau')= \left\{
 \begin{array}{lr} 
      \sum\limits_{m\in \mathbb{Z}} \Tilde{G}_+(\rho,\rho',m)e^{2\pi i (\tau-\tau')m/b} & -\frac{\pi}{2}<\rho'<\rho<\frac{\pi}{2} \\
     \sum\limits_{m\in \mathbb{Z}} \Tilde{G}_-(\rho,\rho',m)e^{2\pi i (\tau-\tau')m/b}  & -\frac{\pi}{2}<\rho<\rho'<\frac{\pi}{2} 
      \end{array}
      \right.,
 \end{align}
 where the function $A(\rho',m)$ and $C(\rho',m)$ could be expressed as
 \begin{align}
    &A(\rho',m)=-\frac{1}{b}(\pi-2\rho')^2\Big(4b B(\rho',m)-4b D(\rho',m)+(\pi+2\rho')\big(2+\nonumber\\
     &\exp(q_0(\pi+2\rho')/2)q_0(\pi+2\rho')\text{Ei}(-q_0(\pi+2\rho')/2)\big)\Big)\Big(-p_0(\pi-2\rho')^2\times\nonumber\\
     &\text{Ei}(p_0(\pi-2\rho')/2)+\exp(p_0(\pi-2\rho')/2)\big(4\pi+\exp(q_0(\pi+2\rho')/2)q_0(\pi+2\rho')^2\times\nonumber\\
     &\text{Ei}(-q_0(\pi+2\rho')/2)\big)\Big)^{-1},\\
    &C(\rho',m)=\frac{1}{b}\exp(\pi q_0/2+(p_0+q_0)\rho')\Big(2\exp(p_0(\pi-2\rho')/2)(\pi+2\rho')^2(\pi-2\rho'-\nonumber\\
    &2b B(\rho',m)+2b D(\rho',m))-p_0(\pi^2-4\rho'^2)^2\text{Ei}(p_0(\pi-2\rho')/2)\Big)\Big(4\exp(p_0\pi/2)\pi-\nonumber\\
    &\exp(p_0\rho')p_0(\pi-2\rho')^2\text{Ei}(p_0(\pi-2\rho')/2)+\exp(\pi(p_0+q_0)/2+q_0\rho')q_0(\pi+2\rho')^2\times\nonumber\\
    &\text{Ei}(-q_0(\pi+2\rho')/2)\Big)^{-1}.
 \end{align}

Finally, plugging (\ref{greenfunc}) into (\ref{stcylgauge}), we obtain 
\begin{align}
      <T_{\tau\tau +}^{(gauge)}>_{qm}\hspace{1mm}=\hspace{1mm}<T_{\rho\rho +}^{(gauge)}>_{qm}\hspace{1mm}=&\hspace{1mm}\frac{1}{\mathcal{A}_4(\rho)}\Bigg[a_1\exp\big(p_0(\pi-2\rho)/2\big)\cos^2\rho\Phi^2\Big(4\mathcal{A}_1(\rho)+\nonumber\\
      &(\pi-2\rho)\big(\mathcal{A}_2(\rho)-\mathcal{A}_3(\rho)\big)\Big)\Bigg],\label{stgaugepref2new}\\
       <T_{\tau\tau -}^{(gauge)}>_{qm}\hspace{1mm}=\hspace{1mm}<T_{\rho\rho -}^{(gauge)}>_{qm}\hspace{1mm}=&\hspace{1mm}\frac{1}{\mathcal{B}_3(\rho)}\Bigg[a_1\exp(p_0\rho)\cos^2\rho\Phi^2\Big(\mathcal{B}_1(\rho)+\nonumber\\
       &\hspace{1mm}(\pi+2\rho)\mathcal{B}_2(\rho)\Big)\Bigg],\label{stgaugemref2new}
 \end{align}
where the functions $\mathcal{A}_1(\rho),.. \mathcal{A}_4(\rho), \mathcal{B}_1(\rho),.. \mathcal{B}_3(\rho)$ are given by
\begin{align}
    &\mathcal{A}_1(\rho)=\Bigg(4b B(\rho,m)-4b D(\rho,m)+(\pi+2\rho)\Big(2+\exp(q_0(\pi+2\rho)/2)q_0(\pi+2\rho)\times\nonumber\\
    &\text{Ei}(-q_0(\pi+2\rho)/2)\Big)\Bigg)\Bigg(-p_0(\pi-2\rho)^2\text{Ei}(p_0(\pi-2\rho)/2)+\exp(p_0(\pi-2\rho)/2)\times\nonumber\\
    &\Big(4\pi+\exp(q_0(\pi+2\rho)/2)q_0(\pi+2\rho)^2\text{Ei}(-q_0(\pi+2\rho)/2)\Big)\Bigg),\\
   & \mathcal{A}_2(\rho)=\Bigg(4b B(\rho,m)-4b D(\rho,m)+(\pi+2\rho)\Big(2+\exp(q_0(\pi+2\rho)/2)q_0(\pi+2\rho)\times\nonumber\\
    &\text{Ei}(-q_0(\pi+2\rho)/2)\Big)\Bigg)\Bigg(2\exp(p_0(\pi-2\rho)/2)p_0(\pi-2\rho)+4p_0(\pi-2\rho)\times\nonumber\\
   &\text{Ei}(p_0(\pi-2\rho)/2)-\exp(p_0(\pi-2\rho)/2)p_0\Big(4\pi+\exp(q_0(\pi+2\rho)/2)q_0(\pi+2\rho)^2\times\nonumber\\
   &\text{Ei}(-q_0(\pi+2\rho)/2)\Big)+\exp(p_0(\pi-2\rho)/2)q_0(\pi+2\rho)\Big(2+\exp(q_0(\pi+2\rho)/2)\times\nonumber\\
   &(4+\pi q_0+2q_0\rho)\text{Ei}(-q_0(\pi+2\rho)/2)\Big)\Bigg),\\
   &\mathcal{A}_3(\rho)= \Bigg(-p_0(\pi-2\rho)^2\text{Ei}(p_0(\pi-2\rho)/2)+\exp(p_0(\pi-2\rho)/2)\Big(4\pi+\nonumber\\
   &\exp(q_0(\pi+2\rho)/2)q_0(\pi+2\rho)^2\text{Ei}(-q_0(\pi+2\rho)/2)\Big)\Bigg)\Bigg(4+2\exp(q_0(\pi+2\rho)/2)\times\nonumber\\
   &q_0(\pi+2\rho)\text{Ei}(-q_0(\pi+2\rho)/2)+q_0(\pi+2\rho)\Big(2+\exp(q_0(\pi+2\rho)/2)(2+\pi q_0+\nonumber\\
   &2q_0\rho)\text{Ei}(-q_0(\pi+2\rho)/2)\Big)+4b\partial_{\rho}B(\rho,m)-4b\partial_{\rho}D(\rho,m)\Bigg),\\
  & \mathcal{A}_4(\rho)=b(\pi-2\rho)\Bigg(p_0(\pi-2\rho)^2\text{Ei}(p_0(\pi-2\rho)/2)-\exp(p_0(\pi-2\rho)/2)\Big(4\pi+\nonumber\\
  &\exp(q_0(\pi+2\rho)/2)q_0(\pi+2\rho)^2\text{Ei}(-q_0(\pi+2\rho)/2)\Big)\Bigg)^2,
   \end{align}
    \begin{align}
    &\mathcal{B}_1(\rho)=(p_0+q_0)\Bigg(2\exp(p_0(\pi-2\rho)/2)(\pi+2\rho)^2(\pi-2\rho-2bB(\rho,m)+2bD(\rho,m))\nonumber\\
  &-p_0(\pi^2-4\rho^2)^2\text{Ei}(p_0(\pi-2\rho)/2)\Bigg)\Bigg[\Bigg(4\exp(p_0\pi/2)\pi-\exp(p_0\rho)p_0(\pi-2\rho)^2\times\nonumber\\
  &\text{Ei}(p_0(\pi-2\rho)/2)+\exp\big(\pi(p_0+q_0)/2+q_0\rho\big)q_0(\pi+2\rho)^2\text{Ei}(-q_0(\pi+2\rho)/2)\Bigg)\nonumber\\
  &-\frac{1}{(p_0+q_0)}\Bigg(2\exp(p_0\pi/2)p_0(\pi-2\rho)+2\exp(p_0\pi/2)q_0(\pi+2\rho)+4\exp(p_0\rho)p_0\times\nonumber\\
  &(\pi-2\rho)\text{Ei}(p_0(\pi-2\rho)/2)-\exp(p_0\rho)p_0^2(\pi-2\rho)^2\text{Ei}(p_0(\pi-2\rho)/2)+q_0(\pi+2\rho)\times\nonumber\\
  &4\exp\big((p_0+q_0)\pi/2+q_0\rho\big)\text{Ei}(-q_0(\pi+2\rho)/2)+\exp\big((p_0+q_0)\pi/2+q_0\rho\big)q_0^2(\pi+2\rho)^2\nonumber\\
  &\times\text{Ei}(-q_0(\pi+2\rho)/2)\Bigg)\Bigg],\\
  &\mathcal{B}_2(\rho)=2\Bigg(4\exp(p_0\pi/2)\pi-\exp(p_0\rho)p_0(\pi-2\rho)^2\text{Ei}(p_0(\pi-2\rho)/2)+ q_0(\pi+2\rho)^2\times\nonumber\\
  &\exp\big((p_0+q_0)\pi/2+q_0\rho\big)\text{Ei}(-q_0(\pi+2\rho)/2)\Bigg)\Bigg(\exp(p_0(\pi-2\rho)/2)p_0(\pi-2\rho)(\pi+2\rho)\nonumber\\
  &+4\exp(p_0(\pi-2\rho)/2)(\pi-2\rho-2bB(\rho,m)+2bD(\rho,m))-\exp(p_0(\pi-2\rho)/2)p_0\times\nonumber\\
  &(\pi+2\rho)(\pi-2\rho-2bB(\rho,m)+2bD(\rho,m))+8p_0\rho(\pi-2\rho)\text{Ei}(p_0(\pi-2\rho)/2)-\nonumber\\
  &2\exp(p_0(\pi-2\rho)/2)(\pi+2\rho)(1+b\partial_{\rho}B(\rho,m)-b\partial_{\rho}D(\rho,m))\Bigg),\\
   & \mathcal{B}_3(\rho)=b(\pi+2\rho)^2\Bigg(4\exp(p_0\pi/2)\pi-\exp(p_0\rho)p_0(\pi-2\rho)^2 \text{Ei}(p_0(\pi-2\rho)/2)+\nonumber\\
  &\exp\big((p_0+q_0)\pi/2+q_0\rho\big)q_0(\pi+2\rho)^2  \text{Ei}(-q_0(\pi+2\rho)/2)\Bigg)^2.
\end{align}
Here the subscript $`\pm$' denotes the  expectation value of the quantum stress-energy tensor near the boundary  $\rho=\pm\frac{\pi}{2}$.

The leading order terms in the quantum stress-energy tensor could be expressed as follows
 \begin{align}
      &<T_{\tau\tau +}^{(gauge)}>_{qm}\Big|_{\rho\rightarrow\frac{\pi}{2}}\hspace{1mm}=\hspace{1mm}<T_{\rho\rho +}^{(gauge)}>_{qm}\Big|_{\rho\rightarrow\frac{\pi}{2}}\hspace{1mm}=\hspace{1mm}\frac{a_1}{b \pi}(\pi-2\rho)\Big(\pi+b B(\pi/2,m)-\nonumber\\
       &b D(\pi/2,m)+\exp(\pi q_0)\pi^2q_0\text{Ei}(-\pi q_0)\Big)\Phi\Big(\frac{\pi}{2}\Big)^{2}\Big(1+\exp(\pi q_0)\pi q_0\text{Ei}(-\pi q_0)\Big)^{-1},\label{stcylgaugep}
       \end{align}
       \begin{align}
       &<T_{\tau\tau -}^{(gauge)}>_{qm}\Big|_{\rho\rightarrow-\frac{\pi}{2}}\hspace{1mm}=\hspace{1mm}<T_{\rho\rho -}^{(gauge)}>_{qm}\Big|_{\rho\rightarrow-\frac{\pi}{2}}\hspace{1mm}=\hspace{1mm}\frac{a_1}{b \pi}(\pi+2\rho)\Big(b\exp(p_0\pi)\times \nonumber\\
       &B(-\pi/2,m)-b \exp(p_0\pi)D(-\pi/2,m)+\pi\big(-\exp(p_0\pi)+p_0\pi\text{Ei}(\pi p_0)\big)\Big)\Phi\Big(-\frac{\pi}{2}\Big)^{2}\nonumber\\
       &\times\Big(-\exp(p_0\pi)+\pi p_0\text{Ei}(\pi p_0)\Big)^{-1}.\label{stcylgaugem}
 \end{align}
 
Before we conclude, it is noteworthy to mention that one can also evaluate the expectation value of the quantum stress-energy tensor for the scalar field $\chi$ following the method as described above. In the flat space-time limit ($\rho\rightarrow0$), these results boil down into an expression in the cylindrical coordinates \cite{Garcia-Garcia:2020ttf} as given below 
\begin{align}
     <T_{\rho\rho}^{(\chi)}>_{qm}&\approx\sum_{m\in \mathbb{Z}}-\frac{m\pi}{b^2\tanh(2\pi^2m/b)}+\Bigg(-\frac{m\pi\coth(2\pi^2m/b)}{b^2}-\frac{\tanh(m\pi^2/b)}{4m\pi}\Bigg)\rho^2\nonumber\\
     &=T_{cyl}+O(\rho^2).
\end{align}

\chapter{Detailed expressions of $F_{\pm}$}\label{expref2}
The functions $F_{\pm}$, are given by 
  \begin{align}
      F_+=&\hspace{1mm}\Bigg(-C_3+\frac{1}{48}\big(\pi(-4r_0+12r_2+\pi(r_1-r_3))-8\big((-1+6\gamma)r_1+r_3\big)\big)\alpha+\nonumber\\
      &r_1\alpha \log\Big(\frac{2}{\pi}\Big)+\frac{\pi}{2}X_{\tau\tau}(b)+\frac{1}{36}\Bigg(-1-9C_1^2\pi+\frac{9a_1(8+\pi^2)}{b \pi}\Big(\pi+b B(\pi/2,m)\nonumber\\
      &-b D(\pi/2,m)+\exp(\pi q_0)\pi^2q_0\text{Ei}(-\pi q_0)\Big)\Phi\Big(\frac{\pi}{2}\Big)^{2}\Big(1+\exp(\pi q_0)\pi q_0\times\nonumber\\
      &\text{Ei}(-\pi q_0)\Big)^{-1}\Bigg)\Bigg),\label{fphi1}\\
        F_-=&\hspace{1mm}\Bigg(\frac{1}{36}+C_4+\frac{C_1^2\pi}{4}+\frac{1}{48}\big(8\big((-1+6\gamma)s_1+s_3\big)+\pi(-4s_0+12s_2+\pi(-s_1\nonumber\\
      &+s_3))\big)\beta+s_1\beta \log\Big(\frac{\pi}{2}\Big)-\frac{\pi}{2}X_{\tau\tau}(b)+\frac{a_1}{4b \pi}(8+\pi^2)\Big(b\exp(p_0\pi) B(-\pi/2,m)\nonumber\\
       &-b \exp(p_0\pi)D(-\pi/2,m)+\pi\big(-\exp(p_0\pi)+p_0\pi\text{Ei}(\pi p_0)\big)\Big)\Phi\Big(-\frac{\pi}{2}\Big)^{2}\times\nonumber\\
       &\Big(\exp(p_0\pi)-\pi p_0\text{Ei}(\pi p_0)\Big)^{-1}\Bigg),\label{fphi2}
       \end{align}
       where we denote $r_0=\hspace{1mm}\Phi\Big |_{\rho=\frac{\pi}{2}},\hspace{1mm}r_1=\Phi'\Big|_{\rho=\frac{\pi}{2}},\hspace{1mm}r_2=\Phi''\Big|_{\rho=\frac{\pi}{2}},\hspace{1mm}r_3=\Phi'''\Big|_{\rho=\frac{\pi}{2}},$ 
       
       \begin{align}
       \hspace{1mm}s_0=\Phi\Big|_{\rho=-\frac{\pi}{2}},\hspace{1mm}s_1=\Phi'\Big|_{\rho=-\frac{\pi}{2}},\hspace{1mm}s_2=\Phi''\Big|_{\rho=-\frac{\pi}{2}},\hspace{1mm}s_3=\Phi'''\Big|_{\rho=-\frac{\pi}{2}}. 
         \end{align}
 Here $\gamma$ is the Euler's constant and $C_3,C_4$ are the integration constants.

 \chapter{Black hole solution}\label{bhsolref2}
In the following section, we evaluate the black hole solution by substituting (\ref{fexp}) into the equations of motion (\ref{eom1})-(\ref{eom4}) and solve them at different orders in the coupling.
\section{Zeroth order solutions}
Zeroth order solutions are obtained by setting the expansion parameters as $a_1=a_2=0$. The corresponding equations of motion (\ref{eom1})-(\ref{eom4}) turn out to be
\begin{align}
    \Phi_0''-2e^{2\omega_0}\Phi_0&=0,\label{bh01}\\
    \omega_0''-e^{2\omega_0}&=0,\label{bh02}\\
    \chi_0''&=0.\label{bh03}
\end{align}

On solving (\ref{bh01})-(\ref{bh03}), we find the zeroth order solutions as
\begin{align}\label{bh0sol}
    \omega_0=\frac{1}{2}\log\Bigg(\frac{4r_H}{\sinh^2(2\sqrt{r_H}z)}\Bigg),\hspace{1mm}\Phi_0=\sqrt{r_H}\coth(2\sqrt{r_H}z),\hspace{1mm}\chi_0=b_1z+b_2,
\end{align}
where $r_H,b_1$ and $b_2$ are integration constants. 
\section{First order solutions}
We now estimate the leading order contributions due to $U(1)$ gauge fields. The corresponding equations of motion (\ref{eom1})-(\ref{eom4}) turn out to be
\begin{align}
    \Phi_1''-2(\omega_1'\Phi_0'+\omega_0'\Phi_1')-2\chi_0'\chi_1'&=0,\label{bh11}\\
    \omega_1''-2\omega_0''\omega_1-2\Phi_0e^{-2\omega_0}A_{\tau1}'^2&=0,\label{bh12}\\
    \partial_z(\Phi_0^2e^{-2\omega_0}A_{\tau1}')&=0,\label{bh13}\\
    \chi_1''-2\chi_0''\omega_1&=0.\label{bh14}
\end{align}

In order to solve (\ref{bh11})-(\ref{bh14}), we adopt the following change in coordinates 
\begin{align}
    z=\frac{1}{2\sqrt{r_H}}\coth^{-1}\Bigg(\frac{r}{\sqrt{r_H}}\Bigg),
\end{align}
where $r_H$ denotes the location of the black hole horizon.

Upon solving (\ref{bh11})-(\ref{bh14}) we find first order corrections to the background fields as
\begin{align}
A_{\tau1}=&\hspace{1mm}\frac{2b_3}{r}+\mu_{bh},\label{bhsol11}\\
\omega_1=&\hspace{1mm}-\frac{b_3^2}{r_H^2r}\Big(r_H+r^2(-2\log(r)+\log(-\sqrt{r_H}+r)+\log(\sqrt{r_H}+r))\Big)-b_5+\nonumber\\
&\frac{r}{\sqrt{r_H}}\Bigg(b_4+\tanh^{-1}\Bigg(\frac{r}{\sqrt{r_H}}\Bigg)b_5\Bigg),\label{bhsol12}\\
\chi_1=&\hspace{1mm}\frac{b_6}{\sqrt{r_H}}\tanh^{-1}\Bigg(\frac{r}{\sqrt{r_H}}\Bigg)+b_7,\label{bhsol13}\\
    \Phi_1=&\hspace{1mm}\frac{b_3^2}{r_H^2}\Big((-r_H+r^2)(2\log r-\log(-\sqrt{r_H}+r)-\log(\sqrt{r_H}+r))\Big)+\frac{1}{8r_H^{\frac{3}{2}}}\Bigg(8r_H r^2b_4\nonumber\\
    &+4r_H\Bigg(2\sqrt{r_H}r+2r^2\tanh^{-1}\Bigg(\frac{r}{\sqrt{r_H}}\Bigg)+r_H\log(-\sqrt{r_H}+r)-\nonumber\\
    &r_H\log(\sqrt{r_H}+r)\Bigg)b_5+r\Big(-\log(-\sqrt{r_H}+r)+\log(\sqrt{r_H}+r)\Big)b_1b_6\Bigg)\nonumber\\
    &+b_8+\rho b_{9}\label{bhsol14},
\end{align}
where $b_3,b_4,.,b_{9}$ are the integration constants and $\mu_{bh}$ is the chemical potential for the black hole phase.

The black hole solution up to leading order in $a_1$ can be summarised as 
\begin{align}
     \omega_{bh}=&\hspace{1mm}\frac{1}{2}\log(-4r_H+4r^2)+a_1\Bigg(
-\frac{b_3^2}{r_H^2r}\Big(r_H+r^2(-2\log(r)+\log(-\sqrt{r_H}+r)+\nonumber\\
&\log(\sqrt{r_H}+r))\Big)-b_5+
\frac{r}{\sqrt{r_H}}\Bigg(b_4+\tanh^{-1}\Bigg(\frac{r}{\sqrt{r_H}}\Bigg)b_5\Bigg)\Bigg),\\
\Phi_{bh}=&\hspace{1mm}r+a_1\Bigg(\frac{b_3^2}{r_H^2}\Big((-r_H+r^2)(2\log r-\log(-\sqrt{r_H}+r)-\log(\sqrt{r_H}+r))\Big)+\nonumber\\
    &\frac{1}{8r_H^{\frac{3}{2}}}\Bigg(8r_H r^2b_4+4r_H\Bigg(2\sqrt{r_H}r+2r^2\tanh^{-1}\Bigg(\frac{r}{\sqrt{r_H}}\Bigg)+r_H\log(-\sqrt{r_H}+r)-\nonumber\\
    &r_H\log(\sqrt{r_H}+r)\Bigg)b_5+r\Big(-\log(-\sqrt{r_H}+r)+\log(\sqrt{r_H}+r)\Big)b_1b_6\Bigg)+b_8+\nonumber\\
    &\rho b_{9}\Bigg),\\
    \chi_{bh}=&\hspace{1mm}\frac{b_1\coth^{-1}(r/\sqrt{r_H})}{2\sqrt{r_H}}+b_2+a_1\Bigg(\frac{b_6}{\sqrt{r_H}}\tanh^{-1}\Bigg(\frac{r}{\sqrt{r_H}}\Bigg)+b_7\Bigg),\\
    A_{\tau}^{bh}=&\frac{2b_3}{r}+\mu_{bh}.
\end{align}

\chapter{Non-abelian generalization of JT gravity}\label{Non-abelian}
In this Appendix, we will derive the most general 2D action for Einstein-dilaton gravity coupled with U(1) gauge and SU(2) Yang-Mills fields that contains the 4-derivative interaction terms. 

We start with the most general 5D action that contains the 2-derivative interaction terms \begin{align}\label{Na 5D}
S^{(2)}_{5D}=& \int d^5x\sqrt{-g_{(5)}}\Big[\alpha_1(\lambda + R)-\frac{\alpha_2}{4}F^2+\frac{\alpha_3}{3}\epsilon^{MNOPQ}A_MF_{NO}F_{PQ}-\frac{\alpha_4}{4}\big(F^{(a)}\big)^2+\nonumber\\
&\frac{\alpha_5}{4}\epsilon^{MNOPQ}A_MF^{(a)}_{NO}F^{(a)}_{PQ}\Big]
\end{align}
 where $\alpha_i$ $(i=1,..,5)$ are the respective coupling constants. 
 
 Next, we will add the following 4-derivative gauge invariant interaction terms to the above action (\ref{Na 5D})
 
\begin{align}\label{Na4 5D}
S^{(4)}_{5D}=&\int d^5x\sqrt{-g_{(5)}}\Big[\beta_1 R^2+\beta_2 (R_{MN})^2 + \beta_3 (R_{MNOP})^2+\beta_4 RF^2+ \beta_5R^{MN}F_{MO}{F_{N}}^O +\nonumber\\
&\beta_6 R_{MNOP}F^{MN}F^{OP}+\beta_7 R_{MNOP}F^{MO}F^{NP}+\beta_8 (F^2)^2+ \beta_9 F^{MN}F_{NO}F^{OP}F_{PM}+\nonumber\\
&\beta_{10}\bigtriangledown^MF_{MN}\bigtriangledown^O{F_{O}}^N+\beta_{11}\bigtriangledown_MF_{NO}\bigtriangledown^MF^{NO}+\beta_{12}\bigtriangledown_MF_{NO}\bigtriangledown^NF^{MO}+\nonumber\\
&\beta_{13}\bigtriangledown^2F^2+\beta_{14}\bigtriangledown_M\bigtriangledown^NF_{NO}F^{MO}+\beta_{15}\bigtriangledown^N\bigtriangledown_MF_{NO}F^{MO}+\epsilon^{MNOPQ}\big\{F_{MN}\times\nonumber\\
&(\beta_{16}F_{OP}\bigtriangledown^RF_{RQ}+\beta_{17}F_{OR}\bigtriangledown^RF_{PQ}+\beta_{18}F_{OR}\bigtriangledown_PF_{QS}g^{RS})+\beta_{19}A_M R_{NOKL}\times\nonumber\\
&{R_{PQ}}^{KL}\big\}+\beta_{20}\epsilon_{NOPQR}\epsilon^{NIJKL}F^{OP}F^{QR}F_{IJ}F_{KL}+\delta_1R(F^{(a)})^2+\delta_2R^{MN}F^{(a)}_{MO}{F^{(a)}_{N}}^O\nonumber\\
&+\delta_3 R_{MNOP}F^{(a)MN}F^{(a)OP}+\delta_4 R_{MNOP}F^{(a)MO}F^{(a)NP}+\delta_5((F^{(a)})^2)^2+\nonumber\\
&\delta_6F^{(a)}_{MN}F^{(b)MN}F^{(a)}_{OP}F^{(b)OP}+\delta_7F^{(a)MN}F^{(a)}_{NO}F^{(b)OP}F^{(b)}_{PM}+\delta_8F^{(a)MN}F^{(b)}_{NO}F^{(a)OP}F^{(b)}_{PM}\nonumber\\
&+\delta_9F^{(a)MN}F^{(b)}_{NO}F^{(b)OP}F^{(a)}_{PM}+\delta_{10}\bigtriangledown^2(F^{(a)})^2+\delta_{11}\bigtriangledown_M\bigtriangledown^NF^{(a)}_{NO}F^{(a)MO}+\nonumber\\
&\delta_{12}\bigtriangledown^N\bigtriangledown_MF^{(a)}_{NO}F^{(a)MO}+\epsilon_{NOPQR}\epsilon^{NIJKL}\big\{\delta_{13}F^{(a)OP}F^{(a)QR}F^{(b)}_{IJ}F^{(b)}_{KL}+\nonumber\\
&\delta_{14}F^{(a)OP}F^{(b)QR}F^{(a)}_{IJ}F^{(b)}_{KL}\big\}+\delta_{15}F^2(F^{(a)})^2+\delta_{16}\epsilon_{MNORQ}(\bigtriangledown_PF^{PM})F^{(a)NO}F^{(a)RQ}\nonumber\\
&+\delta_{17}F_{MN}F_{OP}F^{(a)MN}F^{(a)OP}+\epsilon^{MOPQR}\epsilon_{MIJKL}\big\{\delta_{18}F_{OP}F_{QR}F^{(a)IJ}F^{(a)KL}+\nonumber\\
&\delta_{19}F_{OP}F^{(a)}_{QR}F^{IJ}F^{(a)KL}\big\}+\delta_{20}\epsilon_{MNORQ}F^{PM}\bigtriangledown_P(F^{(a)NO}F^{(a)RQ})+\nonumber\\
&\delta_{21}\bigtriangledown_P(\epsilon_{MNORQ}F^{PM}F^{(a)NO}F^{(a)RQ})+\delta_{22}F^{(a)RP}F^{(a)}_{SP}F^{SQ}F_{RQ}+\nonumber\\
&\epsilon^{MNOPQ}\big\{\delta_{23}(\bigtriangledown_M{F_N}^R)F^{(a)}_{OR}F^{(a)}_{PQ}+\delta_{24}\bigtriangledown_M(F^{(a)}_{OR}F^{(a)}_{PQ}){F_N}^R+\delta_{25}(\bigtriangledown^RF_{MN})F^{(a)}_{OR}F^{(a)}_{PQ}\nonumber\\
&+\delta_{26}\bigtriangledown^R(F^{(a)}_{OR}F^{(a)}_{PQ})F_{MN}\big\}+\delta_{27}\bigtriangledown_M(\epsilon^{MNOPQ}{F_N}^RF^{(a)}_{OR}F^{(a)}_{PQ})+\nonumber\\
&\delta_{28}\bigtriangledown^R(\epsilon^{MNOPQ}F_{MN}F^{(a)}_{OR}F^{(a)}_{PQ})\Big]
\end{align}
where $\beta_i$ and $\delta_j$ $(i=1,..,20$ and $j=1,..,28)$ are the respective coupling constants.
 
 On adding (\ref{Na 5D}) and (\ref{Na4 5D}), we get the required action that contains all 2-derivative as well as 4-derivative interactions. However, we can eliminate the various interaction terms in (\ref{Na4 5D}) using a proper redefinition of fields.
 
 Consider the following redefinition of fields
 \begin{align}
     g^{RS}\rightarrow g^{RS}+\delta g^{RS},\hspace{2mm}A_N\rightarrow A_N+\delta A_N,\hspace{2mm}A_N^{(a)}\rightarrow A_N^{(a)}+\delta A_N^{(a)}
 \end{align}
 such that the action transform as $S\rightarrow S'=S+\delta S$ where $S=S_{5D}^{(2)}+S_{5D}^{(4)}$ and $S'$ is called the transformed action.
 
 The most general variation of fields that contains 2-derivative interactions takes the form
 \begin{align}
 \delta g^{RS} = & \mu_1R^{RS}+\mu_2F^{RP}{F^S}_P+\mu_3Rg^{RS}+\mu_4F^2g^{RS}+\mu_5g^{RS}+\mu_6(F^{(a)})^2g^{RS}+\nonumber\\
&\mu_7F^{(a)RP}{F^{(a)S}}_P,\label{naa1}\\
\delta A_N= &\lambda_1A_N+\lambda_2\bigtriangledown^MF_{MN}+\lambda_3\epsilon_{NOPQR}F^{OP}F^{QR}+\lambda_4\epsilon_{NOPQR}F^{(a)OP}F^{(a)QR},\label{naa2}\\
\delta A^{(a)}_N= &\sigma_1A^{(a)}_N+\sigma_2\bigtriangledown^MF^{(a)}_{MN}+\sigma_3\epsilon_{NOPQR}F^{(a)OP}F^{QR},\label{naa3}
 \end{align}
 where $\mu_i$, $\lambda_j$ and $\sigma_k$ $(i=1,..,7,j=1,..,4$ and $ k=1,..,3)$ are the respective constants. 
  
  In order to get rid of various interaction terms in (\ref{Na 5D}) and (\ref{Na4 5D}), we transformed the action $S$ (where $S=S_{5D}^{(2)}+S_{5D}^{(4)}$) into $S'$ using (\ref{naa1})-(\ref{naa3}) with the following particular choice of constants 
  \begin{align}
      &\mu_1=-\frac{4\beta_2}{3\alpha_1},\hspace{2mm}\mu_2 = \frac{-2\alpha_2\beta_2}{3\alpha_1^2} ,\hspace{2mm}\mu_3=\frac{4}{9\alpha_1}(2\beta_1+\beta_2), \hspace{2mm}\mu_4=\frac{2\alpha_2}{27\alpha_1^2}(2\beta_2+\beta_1),\hspace{2mm}\mu_5 = -\frac{2}{3},\nonumber\\
&\mu_6 = \frac{2}{27\alpha_1}[9\delta_2+6\delta_1+\frac{\alpha_4}{\alpha_1}(2\beta_2+\beta_1)],\hspace{2mm} \mu_7=\frac{-2}{3\alpha_1}[\delta_2+\frac{\alpha_4}{\alpha_1}\beta_2],\hspace{2mm}\lambda_1=\frac{-1}{3},\nonumber\\
&\lambda_2=\frac{-4\alpha_2}{3\alpha_3\alpha_5}[2\delta_{18}-\frac{\alpha_5}{\alpha_3}\beta_{20}],\hspace{3mm}\lambda_3=\frac{4\beta_{20}}{3\alpha_3},\hspace{3mm} \lambda_4=\frac{1}{3\alpha_3}[2\delta_{18}-\frac{\alpha_5}{\alpha_3}\beta_{20}]
  \end{align}
  and $\sigma_k=0$ \footnote{We choose $\sigma_k=0$ to preserve the gauge invariance.} which yields
  \begin{align}\label{Na4f 5D}
      S'=&\int d^5x\sqrt{-g_{(5)}}\Big[\eta_1\lambda +\eta_2R-\frac{\eta_3}{4}F^2-\frac{\eta_4}{4}\big(F^{(a)}\big)^2+
\frac{\eta_5}{4}\epsilon^{MNOPQ}A_MF^{(a)}_{NO}F^{(a)}_{PQ}+\nonumber\\
&\eta_6[R_{MNOP}]^2+\eta_7(F^2)^2+\eta_8F^{SP}F_{PR}F^{RQ}F_{QS}+\eta_9\bigtriangledown_MF^{MN}\bigtriangledown^OF_{ON}+\nonumber\\
&+\epsilon^{MNOPQ}\big\{\eta_{10}F_{MN}F_{OP}\bigtriangledown^RF_{RQ}+\eta_{11}F_{MN}F_{OR}\bigtriangledown^RF_{PQ}+\eta_{12}F_{MN}F_{OR}\bigtriangledown_P{F_Q}^R+\nonumber\\
&\eta_{13}A_MR_{NOIJ}{R_{PQ}}^{IJ}\big\}+\eta_{14}\big(\big(F^{(a)}\big)^2\big)^2+\eta_{15}F^2\big(F^{(a)}\big)^2+\eta_{16}F^{(a)RP}F^{(a)}_{SP}F^{SQ}F_{RQ}+\nonumber\\
&\eta_{17}F^{(a)RP}F^{(a)}_{PS}F^{(b)SQ}F^{(b)}_{QR}+\eta_{18}\epsilon_{NOPQR}\epsilon^{NIJKL}F^{(a)OP}F^{(a)QR}F^{(b)}_{IJ}F^{(b)}_{KL}+\nonumber\\
&\eta_{19}R_{MNOP}F^{(a)MN}F^{(a)OP}+\eta_{20}R_{MNOP}F^{(a)MO}F^{(a)NP}+\eta_{21}F_{MN}F_{PQ}F^{(a)MN}F^{(a)PQ}+\nonumber\\
&\eta_{22}\epsilon^{MOPQR}\epsilon_{MIJKL}F_{OP}F^{(a)}_{QR}F^{IJ}F^{(a)KL}\Big]
  \end{align}
  where $\eta_i$ $(i=1,..,22)$ are the new coupling constants respectively. 
  
  We can express the new coupling constants $\eta_i$ in terms of the old coupling constants $\alpha_j$, $\beta_k$ and $\delta_l$ as follows 
  \begin{align}
      &\eta_1 =\frac{8}{3}\alpha_1, \hspace {2mm}\eta_2=2\alpha_1-\frac{4\lambda}{9}(5\beta_1+\beta_2),\hspace{2mm}\eta_3=\frac{\lambda\alpha_2}{27\alpha_1}(4\beta_2+20\beta_1)+\frac{2\alpha_2}{3},\nonumber\\
&\eta_4=\frac{8\lambda}{3}\Big(2\delta+\frac{5}{3}\delta_1\Big)+\frac{4\lambda\alpha_4}{27\alpha_1}(5\beta_1+\beta_2)+\frac{4\alpha_4}{3},\hspace{2mm}\eta_5=\frac{2\alpha_5}{3},\hspace{2mm}\eta_6=\frac{4\beta_3}{3},\nonumber\\
&\eta_7=\frac{\alpha_2^2}{108\alpha_1^2}(\beta_1-7\beta_2)-\frac{4\beta_8}{3},\hspace{2mm}\eta_8=\frac{1}{3}\Big(\frac{\alpha_2^2}{\alpha_1^2}\beta_2-4\beta_9\Big),\hspace{2mm}\eta_9=\frac{-4\alpha_2^2}{3\alpha_3\alpha_5}\Big(2\delta_{18}-\frac{\alpha_5}{\alpha_3}\beta_{20}\Big),\nonumber
\end{align}
\begin{align}
&\eta_{10}=\frac{8\alpha_2}{3}\Big(\frac{\beta_{20}}{\alpha_3}-\frac{\delta_{18}}{\alpha_5}\Big),\hspace{2mm}\eta_{11}=\frac{-2\beta_{17}}{3},\hspace{2mm}\eta_{12}=\frac{-2\beta_{18}}{3},\hspace{2mm}\eta_{13}=\frac{2\beta_{19}}{3},\nonumber\\
&\eta_{14}=\frac{\alpha_4}{108\alpha_1}\Big[6\delta_1+\frac{\alpha_4}{\alpha_1}(\beta_1-7\beta_2)\Big], \hspace{2mm}\eta_{15}=\frac{\alpha_4\alpha_1}{54\alpha_1^2}(\beta_1-7\beta_2)+\frac{\alpha_2\delta_1}{\alpha_118}-\frac{2\delta_{15}}{3},\nonumber\\
&\eta_{16}=\frac{1}{3}\Big[\frac{\alpha_2}{\alpha_1}(\delta_2+\frac{2\alpha_4\beta_2}{\alpha_1})-2\delta_{22}\Big],\hspace{2mm}\eta_{17}=\frac{\alpha_4}{3\alpha_1}\Big(\delta_2+\frac{\alpha_4}{\alpha_1}\beta_2\Big),\hspace{2mm}\eta_{18}=\frac{\alpha_{5}}{12\alpha_3}\Big(2\delta_{18}-\frac{\alpha_5\beta_{20}}{\alpha_3}\Big)\nonumber\\
&\eta_{19}=\frac{2\delta_3}{3},\hspace{2mm}\eta_{20}=\frac{2\delta_4}{3},\hspace{2mm}\eta_{21}=\frac{-2\delta_{17}}{3},\hspace{2mm}\eta_{22}=\frac{-2\delta_{19}}{3}.
\end{align}
  
With all these preliminaries, equation (\ref{Na4f 5D}) is the most general 5D action of gravity coupled with U(1) gauge and SU(2) Yang-Mills fields that contains all the 4-derivative interaction terms. 

Our next step is to truncate the 5D action to 2D using the following ansatz
\begin{align}
      &ds^2_{(5)}=ds^2_{(2)}+\phi(t,z)^{\frac{2}{3}}(dx^2+dy^2+dz^2),\label{AMA}\\
      &A_Mdx^M=A_{\mu}dx^{\mu},\hspace{2mm} A_{\mu}\equiv A_{\mu}(x^{\nu}),\label{AAA}\\
       &A_M^{(a)}dx^M=A_{\mu}^{(a)}dx^{\mu},\hspace{2mm} A_{\mu}^{(a)}\equiv A_{\mu}^{(a)}(x^{\nu}).\label{ANA}
\end{align}
Using the above ansatz (\ref{AMA})-(\ref{ANA}) in (\ref{Na4f 5D}) we finally obtain
\begin{align}\label{NA 2D}
    S_{(2D)} =& \int d^2x\sqrt{-g_{(2)}}\phi\Bigg[\eta_1\lambda+\eta_2R-\frac{\eta_3}{4}F^2-\frac{\eta_4}{4} \big(F^{(a)}\big )^2+\eta_6\Big[(R_{\mu\nu\alpha\beta})^2+\nonumber\\
&\frac{3}{4}\big(\bigtriangledown_{\mu}\phi^{\frac{2}{3}}\big)^4+4\Big\{\frac{3}{4}(\bigtriangledown_{\lambda}\phi^{\frac{2}{3}})(\bigtriangledown_{\beta}\phi^{\frac{2}{3}})\Gamma^{\lambda}_{\alpha\mu}\Gamma^{\beta}_{\rho\sigma}g^{\alpha\rho}g^{\mu\sigma}+\frac{2}{3}\Gamma^{\lambda}_{\alpha\mu}(\bigtriangledown_{\lambda}\phi^{\frac{2}{3}})\times\nonumber\\
&(\bigtriangledown^{\alpha}\phi)(\bigtriangledown^{\mu}\phi)\phi^{\frac{-4}{3}}-\Gamma^{\lambda}_{\alpha\mu}(\bigtriangledown_{\lambda}\phi^{\frac{2}{3}})\{\partial_{\beta}(\bigtriangledown_{\sigma}\phi)\}g^{\alpha\beta}g^{\mu\sigma}\phi^{\frac{-1}{3}}+\frac{4}{27}(\bigtriangledown_{\mu}\phi)^4\phi^{\frac{-8}{3}}\nonumber\\
&-\frac{4}{9}\phi^{\frac{-5}{3}}(\bigtriangledown^{\alpha}\phi)(\bigtriangledown^{\mu}\phi)\{\partial_{\alpha}(\bigtriangledown_{\mu}\phi)\}-\frac{1}{3}\{\partial_{\alpha}(\bigtriangledown_{\mu}\phi)\}\{\partial_{\beta}(\bigtriangledown_{\rho}\phi)\}g^{\alpha\beta}g^{\mu\rho}\phi^{\frac{-2}{3}}\Big\}\Big]\nonumber\\
&+\eta_7F^4+\eta_8F^{\mu\nu}F_{\nu\lambda}F^{\lambda\sigma}F_{\sigma\mu}+\eta_9\bigtriangledown_{\mu}F^{\mu\nu}\bigtriangledown^{\lambda}F_{\lambda\nu}+\eta_{14}\big(\big(F^{(a)}\big)^2\big)^2+\nonumber\\
&\eta_{15}F^2\big(F^{(a)}\big)^2+\eta_{16}F^{(a)\mu\nu}F^{(a)}_{\lambda\nu}F^{\lambda\sigma}F_{\mu\sigma}+\eta_{17}F^{(a)\mu\nu}F^{(a)}_{\lambda\nu}F^{(b)\lambda\sigma}F^{(b)}_{\mu\sigma}+\nonumber\\
&\eta_{19}R_{\mu\nu\sigma\lambda}F^{(a)\mu\nu}F^{(a)\sigma\lambda}+\eta_{20}R_{\mu\nu\lambda\sigma}F^{(a)\mu\lambda}F^{(a)\nu\sigma}+\eta_{21}F_{\mu\nu}F_{\lambda\sigma}F^{(a)\mu\nu}F^{(a)\lambda\sigma}\Bigg],
\end{align}
where we identify  (\ref{NA 2D}) as the most general 2D action of gravity coupled with U(1) gauge and SU(2) Yang-Mills fields that contains all the 4-derivative interaction terms.

\chapter{Most general 4-derivative action in 5D}\label{gen action 5D}
The purpose of this Appendix is to discuss the most general 5D action of gravity coupled with U(1) gauge fields that contains the 4-derivative interaction terms\footnote{This can be achieved by simply removing the SU(2) Yang-Mills fields in Appendix \ref{Non-abelian}.}. 

In the first place, we consider the most general 2-derivative action of the following form
\begin{eqnarray}\label{A2d}
S^{(2)}=&& \int d^5x\sqrt{-g}\Bigg[12 + R-\frac{\alpha_1}{4}F^2+\frac{\alpha_2}{3}\epsilon^{MNOPQ}A_MF_{NO}F_{PQ}\Bigg],
\end{eqnarray}
where $\alpha_1$ and $\alpha_2$ are the respective coupling constants.

Next, we note down the most general 4-derivative gauge invariant interaction terms as follows
\begin{align}\label{A4d}
S^{(4)}=&\int d^5x\sqrt{-g} \Big [\beta_1 R^2+\beta_2 [R_{MN}]^2 + \beta_3 [R_{MNOP}]^2+\beta_4 RF^2+ \beta_5R^{MN}F_{MO}{F_{N}}^O +\nonumber\\
&\beta_6 R_{MNOP}F^{MN}F^{OP}+\beta_7 R_{MNOP}F^{MO}F^{NP}+\beta_8 F^4+ \beta_9 F^{MN}F_{NO}F^{OP}F_{PM}+\nonumber\\
&\beta_{10}\bigtriangledown^MF_{MN}\bigtriangledown^O{F_{O}}^N+\beta_{11}\bigtriangledown_MF_{NO}\bigtriangledown^MF^{NO}+\beta_{12}\bigtriangledown_MF_{NO}\bigtriangledown^NF^{MO}+\nonumber\\
&\beta_{13}\bigtriangledown^2F^2+\beta_{14}\bigtriangledown_M\bigtriangledown^NF_{NO}F^{MO}+\beta_{15}\bigtriangledown^N\bigtriangledown_MF_{NO}F^{MO}+\nonumber\\
&\epsilon^{MNOPQ}\big\{F_{MN}(\beta_{16}F_{OP}\bigtriangledown^RF_{RQ}+\beta_{17}F_{OR}\bigtriangledown^RF_{PQ}+\beta_{18}F_{OR}\bigtriangledown_PF_{QS}g^{RS})+\nonumber\\
&\beta_{19}A_M R_{NOKL}{R_{PQ}}^{KL}\big\}+\beta_{20}\epsilon_{NOPQR}\epsilon^{NIJKL}F^{OP}F^{QR}F_{IJ}F_{KL}\Big],
\end{align}
where $\beta_i$ $(i=1,..,20)$ are the respective coupling constants. On adding (\ref{A4d}) in (\ref{A2d}), we obtain the most general action for gravity (coupled to U(1) gauge fields) that contains 4-derivative interaction terms. Moreover, it is also possible to eliminate the various interaction terms in (\ref{A4d}) using a proper redefinition of fields as discussed below.

Consider the following redefinition of fields  $$g^{RS}\rightarrow g^{RS}+\delta g^{RS},\hspace{2mm}A_N\rightarrow A_N+\delta A_N$$ such that the action transform as $S\rightarrow S'=S+\delta S$. 

The most general 2-derivative variation of fields are given by
\begin{eqnarray}
\delta g^{RS} &=& \mu_1R^{RS}+\mu_2F^{RP}{F^S}_P+\mu_3Rg^{RS}+\mu_4F^2g^{RS}+\mu_5g^{RS}\label{aa1}\\
\delta A_N&=&\lambda_1A_N+\lambda_2\bigtriangledown^MF_{MN}+\lambda_3\epsilon_{NOPQR}F^{OP}F^{QR}.\label{aa2}
\end{eqnarray}
where $\mu_i$ and $\lambda_j$ ($i=1,..,4$ and $j=1,..,3$) are the respective coupling constants. 

 In order to get rid of various interaction terms in (\ref{A2d}) and (\ref{A4d}), we transformed the action $S$ (where $S=S_{5D}^{(2)}+S_{5D}^{(4)}$) into $S'$ using (\ref{aa1}) and (\ref{aa2}) with the following particular choice of constants 
\begin{eqnarray}
&&\mu_1=-\frac{4\beta_2}{3},\hspace{3mm}\mu_2=-\frac{2}{3}\alpha_1\beta_2 ,\hspace{3mm}\mu_3=\frac{4}{9}(2\beta_1+\beta_2), \hspace{3mm}\mu_4=\frac{2\alpha_1}{27}(2\beta_2+\beta_1),\nonumber\\
&&\mu_5 = -\frac{2}{3},\hspace{3mm}\lambda_1=\frac{-1}{3},\hspace{3mm}\lambda_2=\frac{4\alpha_1}{3\alpha_2^2}\beta_{20},\hspace{3mm}\lambda_3=\frac{4\beta_{20}}{3\alpha_2},
\end{eqnarray}
which yields
\begin{align}\label{FA4d}
    S'=&\int d^5x\sqrt{-g}\Big [(12+R)-\frac{\eta_1}{4}F^2+\eta_2[R_{MNOP}]^2+\eta_3 F^4+\eta_4F^{SP}F_{PR}F^{RQ}F_{QS}+ \nonumber\\
&\eta_5\bigtriangledown_MF^{MN}\bigtriangledown^OF_{ON}+\epsilon^{MNOPQ}(\eta_{6}F_{MN}F_{OP}\bigtriangledown^RF_{RQ}+\eta_{7}F_{MN}F_{OR}\bigtriangledown^RF_{PQ}+\nonumber\\
&\eta_{8}F_{MN}F_{OR}\bigtriangledown_P{F_Q}^R+\eta_{9}A_MR_{NOIJ}{R_{PQ}}^{IJ}) \Big ]
\end{align}
where $\eta_i$ ($i=1,..,9$) are the new coupling constants respectively. 

We can express the new coupling constants, $\eta_i$ in terms of old coupling constants, $\alpha_j$ and $\beta_k$ as follows
\begin{align}
    &\eta_1=\frac{\alpha_1}{6}(4\beta_2+20\beta_1)+\frac{\alpha_1}{4},\hspace{2mm}\eta_2=\frac{\beta_3}{2},\hspace{2mm}\eta_3=\frac{\alpha_1^2}{288}(\beta_1-7\beta_2)-\frac{\beta_8}{2},\hspace{2mm}\eta_4=\frac{1}{8}(\alpha_1^2\beta_2\nonumber\\ &-4\beta_9),\eta_5=\frac{\alpha_1^2}{2\alpha_2^2}\beta_{20},\hspace{2mm}
    \eta_{6}=\alpha_1\Big(\frac{\beta_{20}}{\alpha_2}\Big),\hspace{3mm}\eta_{7}=\frac{-\beta_{17}}{4},\hspace{3mm}\eta_{8}=\frac{-\beta_{18}}{4},\hspace{3mm}\eta_{9}=\frac{\beta_{19}}{4}.
\end{align}

Notice that, we have taken out a common factor $\frac{8}{3}$ in (\ref{FA4d}) and impose the condition on $\beta_{1,2}$ such that $5\beta_1+\beta_2=-\frac{1}{8}$. With all these preliminaries, we end up with a most general 5D action (\ref{FA4d}) of gravity coupled with U(1) gauge fields that contains all the 4-derivative interaction terms. 

\chapter{Covariance of the 2D action (\ref{action 2D})}\label{covariance}
In this section, we will discuss the general covariance of the action (\ref{action 2D}). After a careful observation,  we realize that the $\Gamma^{\alpha}_{\beta\lambda}$s appear in such a combination that the entire expression transform as a ``scalar'' under general coordinate transformation (GCT). 

As an illustration, let's look at a particular term in (\ref{action 2D}) as mentioned below:
\begin{align}
\Gamma^{\sigma}_{\alpha\mu}(\bigtriangledown_{\sigma}\phi)(\bigtriangledown^{\alpha}\phi)(\bigtriangledown^{\mu}\phi)=\Gamma^{\sigma}_{\alpha\mu}A_{\sigma}B^{\alpha}C^{\mu},\label{C1}
\end{align}
where $A_{\sigma},B^{\alpha}$ and $C^{\mu}$ are the tensors of rank 1. 

A straightforward computation reveals that under GCT 
\begin{align}\label{C3}
\Gamma^{\sigma}_{\alpha\mu}A_{\sigma}B^{\alpha}C^{\mu}\rightarrow\Gamma^{'\sigma}_{\alpha\mu}A'_{\sigma}B^{'\alpha}C^{'\mu}
=&\frac{\partial x^{'\sigma}}{\partial x^{\beta}}\frac{\partial x^{\rho}}{\partial x^{'\mu}}\frac{\partial x^{\tau}}{\partial x^{'\alpha}}\Gamma^{\beta}_{\tau\rho}\frac{\partial x^{\tilde{\beta}}}{\partial x^{'\sigma}}\frac{\partial x^{'\alpha}}{\partial x^{\tilde{\tau}}}\frac{\partial x^{'\mu}}{\partial x^{\tilde{\rho}}}A_{\tilde{\beta}}B^{\tilde{\tau}}C^{\tilde{\rho}}\nonumber\\
&-\frac{\partial x^{\tilde{\tau}}}{\partial x^{'\alpha}}\frac{\partial^2x^{'\sigma}}{\partial x^{\tilde{\tau}}\partial x^{\tilde{\beta}}}\frac{\partial x^{\tilde{\beta}}}{\partial x^{'\mu}}\frac{\partial x^{\beta}}{\partial x^{'\sigma}}\frac{\partial x^{'\alpha}}{\partial x^{\tau}}\frac{\partial x^{'\mu}}{\partial x^{\rho}}A_{\beta}B^{\tau}C^{\rho} \nonumber\\\nonumber 
=&\Gamma^{\beta}_{\tau\rho}A_{\tilde{\beta}}B^{\tilde{\tau}}C^{\tilde{\rho}}\delta^{\tilde{\beta}}_{\beta}\delta^{\rho}_{\tilde{\rho}}\delta^{\tau}_{\tilde{\tau}}-\delta^{\tilde{\tau}}_{\tau}\delta^{\tilde{\beta}}_{\rho}\frac{\partial}{\partial x^{\tilde{\tau}}}\Big(\frac{\partial x^{'\sigma}}{\partial x_{\tilde{\beta}}}\Big)\frac{\partial x^{\beta}}{\partial x^{'\sigma}}A_{\beta}B^{\tau}C^{\rho}\nonumber\\
=&\Gamma^{\beta}_{\tau\rho}A_{\beta}B^{\tau}C^{\rho}-\frac{\partial}{\partial x^{\tau}}\Big(\frac{\partial x^{'\sigma}}{\partial x_{\tilde{\beta}}}\Big)\frac{\partial x^{\beta}}{\partial x^{'\sigma}}A_{\beta}B^{\tau}C^{\tilde{\beta}}.
\end{align}

 Notice that, the last term in (\ref{C3}) vanishes identically as shown below
\begin{align}\label{C4}
\Bigg[\frac{\partial}{\partial x^{\tau}}\Big(\frac{\partial x^{'\sigma}}{\partial x_{\tilde{\beta}}}\Big)\frac{\partial x^{\beta}}{\partial x^{'\sigma}}\Bigg]A_{\beta}B^{\tau}C^{\tilde{\beta}}=&\Bigg[\frac{\partial}{\partial x^{\tau}}\Big(\frac{\partial x^{'\sigma}}{\partial x_{\tilde{\beta}}}\frac{\partial x^{\beta}}{\partial x^{'\sigma}}\Big)-\frac{\partial x^{'\sigma}}{\partial x_{\tilde{\beta}}}\frac{\partial^2x^{\beta}}{\partial x^{\tau}\partial x^{'\sigma}}\Bigg]A_{\beta}B^{\tau}C^{\tilde{\beta}}\nonumber\\
=&\Bigg[\frac{\partial}{\partial x^{\tau}}\Big(\delta^{\beta}_{\tilde{\beta}}\Big)-\frac{\partial x^{'\sigma}}{\partial x_{\tilde{\beta}}}\frac{\partial}{\partial x^{'\sigma}}\Big(\frac{\partial x^{\beta}}{\partial x^{\tau}}\Big)\Bigg]A_{\beta}B^{\tau}C^{\tilde{\beta}}\nonumber\\
=& 0.
\end{align}

\chapter{Solving the constants \texorpdfstring{$D_{i}$s}{Di}}\label{const}
 In order to cure the divergences in the Gibbons-Hawking-York term and Einstein-Hilbert action (\ref{SGHY})-(\ref{SEH}), we set the constants $d_{i}$s (\ref{SCT}) such that the coefficients of each divergent term vanish identically. 
 
 Below, we note  down the coefficients of each divergent terms as
 \begin{align}
     &\text{coefficient of $\frac{1}{z^\frac{10}{3}}$}\hspace{2mm}:\hspace{2mm}\frac{-2032}{45}-\frac{103172}{585}D_1-\frac{14188}{195}\sqrt{\frac{2}{3}}d_1L+\frac{11104}{45}d_2/C_1=0,\label{CONST1}\\
     &\text{coefficient of $\frac{1}{z^\frac{11}{3}}$}\hspace{2mm}:\hspace{2mm}-\frac{24}{7}D_1-\frac{4}{7}\sqrt{6}Ld_1=0,\label{CONST2}\\
     &\text{coefficient of $\frac{1}{z^\frac{14}{3}}$}\hspace{2mm}:\hspace{2mm}\frac{64}{21}+\frac{18}{7}D_1+\frac{4}{7}\sqrt{\frac{2}{3}}Ld_1-\frac{32}{7}d_2/C_1=0,\label{CONST3}\\
      &\text{coefficient of $\xi\log(z)$}\hspace{2mm}:\hspace{2mm}d_3/C_3+\frac{11213}{3123}=0,\label{CONST4}\\
            &\text{coefficient of $\kappa\log(z)$}\hspace{2mm}:\hspace{2mm}12d_3+\frac{19868C_3}{347}+\frac{d_3}{6C_3}C_9+\frac{1844}{9369}C_9=0,\label{CONST5}\\
            &\text{coefficient of $\frac{1}{z^2}$}\hspace{2mm}:\hspace{2mm}d_6\kappa C_1+\frac{16112}{1041}\kappa-\frac{871}{6246}=0,\label{CONST6}\\
            &\text{coefficient of $\frac{\xi}{z^3}$}\hspace{2mm}:\hspace{2mm}d_4C_1^2/C_6+\frac{871}{3123}=0,\label{CONST7}\\
            &\text{coefficient of $\frac{\kappa}{z^3}$}\hspace{2mm}:\hspace{2mm}d_5C_1^2/C_6+\frac{871}{3123}=0.\label{CONST8}
 \end{align}
 
 On solving (\ref{CONST1})-(\ref{CONST8}), we get the following values as solutions
 \begin{align}
     &d_1=\frac{6.3186}{L},\hspace{1mm}d_2=-C_10.1394,\hspace{1mm}d_3=-C_33.5904,\hspace{1mm}d_4=-\frac{C_6}{C_1^2}0.2788,\hspace{1mm}d_5=-\frac{C_{12}}{C_1^2}0.2788\nonumber\\
   & d_6=-\frac{0.16}{\kappa C_1}\left(-0.871+96.6\kappa\right),\hspace{1mm}D_1=-2.5795,\hspace{1mm}C_3=0.0283C_9.
\end{align}

\chapter{Cardy formula}\label{cardy}
Cardy formula \cite{Cardy:1986ie} measures the degrees of freedom (and hence the entropy) of a two-dimensional conformal field theory (CFT$_{2}$)\footnote{For the  generalization to d-dimensional CFT, see \cite{Verlinde:2000wg}} using the central charge (c) and the conformal weight for the ground state ($\Delta$). In the present Section, we derive the expression for the Cardy formula. We start by computing the partition function of CFT$_{2}$ on torus (T$^{2}$) and then use this partition function to estimate the entropy. 

The partition function of CFT$_{2}$ on torus (T$^{2}$) of modular parameter $\tau=x^1+ix^0$ is given by \cite{Cardy:1986ie}
\begin{align}\label{cd1}
    Z=\text{Tr}\Big[e^{-2\pi(Im\tau)H}e^{i2\pi(Re\tau)P}\Big],
\end{align}
where $H$ and $P$ are the time and space translation on the cylinder respectively and the factor of $2\pi$ in the definition of $Z$ is merely a convention.

One can derive $H$ and $P$ from the stress tensor $T_{\mu\nu}$ as 
\begin{align}\label{cd2}
    H=\frac{1}{2\pi}\int dx^1T_{00}\hspace{2mm}\text{and}\hspace{2mm}  P=\frac{1}{2\pi}\int dx^1T_{01},
\end{align}
where $T_{00}$ and $T_{01}$ can be expressed in terms of the stress tensor on the cylinder ($T_{cyl}$) as 
\begin{align}\label{cd3}
    T_{00}=-\big(T_{cyl}(z)+T_{cyl}(\overline{z})\big)\hspace{2mm}\text{and}\hspace{2mm}T_{01}=-\big(T_{cyl}(z)-T_{cyl}(\overline{z})\big).
\end{align}

The stress tensor on the cylinder, $ T_{cyl}(z)$ is given by \cite{Tong:2009np}-\cite{icts}
\begin{align}\label{cd4}
    T_{cyl}(z)=-\sum e^{-inz}L_n+\frac{c}{24},
\end{align}
where $L_n$ are the virasoro generators that satisfy the following commutation relations
\begin{align}\label{cd5}
    &[L_n,L_m]=(n-m)+\frac{c}{12}n(n^2-1)\delta_{n+m,0}\\
     &[\overline{L}_n,\overline{L}_m]=(n-m)+\frac{c}{12}n(n^2-1)\delta_{n+m,0}\\
     &[L_n,\overline{L}_m]=0.
\end{align}

Using equations (\ref{cd2})-(\ref{cd4}) in (\ref{cd1}), we get\footnote{Since central charge is the real number, therefore we use $\overline{c}=c$ in (\ref{cd6}).}
\begin{align}\label{cd6}
    Z(\tau,\overline{\tau})=\text{Tr}\Big[e^{2\pi i\tau\big(L_0-\frac{c}{24}\big)}e^{2\pi i\overline{\tau}\big(\overline{L}_0-\frac{c}{24}\big)}\Big].
\end{align}

Next, we trace (\ref{cd6}) over the energy eigen state $|E\rangle$ that satisfy the following eigen value equation
\begin{align}
    L_0|E\rangle=\Delta|E\rangle,
\end{align}
which yields\footnote{For simplicity, we have consider only first part of the trace.}
\begin{align}\label{cd7}
    Z(\tau)=\int_0^{\infty}d\Delta\rho(\Delta)e^{2\pi i\tau\big(\Delta-\frac{c}{24}\big)},
    \end{align}
where $\rho(\Delta)$ is the density of states correspond to the energy $\Delta$.

One can evaluate $\rho(\Delta)$ by taking the inverse Laplace transformation of (\ref{cd7}) as follows
\begin{align}\label{cd8}
    \rho(\Delta)=\oint_c d\tau Z(\tau)e^{-2\pi i\tau\big(\Delta-\frac{c}{24}\big)}.
\end{align}

We are interested in computing the entropy of CFT$_2$ in the high temperature limit (or $\Delta>>1$). Therefore, in this limit, the integral (\ref{cd8}) is dominated by $Z(\tau\rightarrow0)$. In order to find $Z(\tau)$ in the limit $\tau\rightarrow0$, we utilize the important fact that the partition function (\ref{cd6}) is modular invariant i.e. $Z(\tau)=Z(-\frac{1}{\tau})$. This can be understood in terms of the geometry of torus as described below. 

Consider a torus (T$^2$) which is parametrize by the two coordinates i.e. $\sigma_1$ and $\sigma_2$ in the range $0\leq\sigma_1\leq2\pi$ and $0\leq\sigma_2\leq2\pi$. The general metric of the torus is given by
\begin{align}\label{t1}
   ds^2=|d\sigma_1+\tau d\sigma_2|^2 \hspace{2mm},
\end{align}
where $\tau\in\mathbb{C}$ is the modular parameter. It is easy to check that (\ref{t1}) is invariant (up to conformal factor) under the following $SL(2,\mathbb{Z})$ transformation  
\begin{align}
\begin{pmatrix}
 \sigma_1\\
 \sigma_2
\end{pmatrix}=\begin{pmatrix}
 d & b\\
 c & a
\end{pmatrix}\begin{pmatrix}
 \sigma'_1\\
 \sigma'_2
\end{pmatrix}\hspace{2mm},\hspace{2mm}\tau'=\frac{a\tau+b}{c\tau+d}\hspace{2mm},
\end{align}
where $a,b,c,d\in\mathbb{Z}$ and $ad-bc=1$.

Therefore we conclude that the modular parameter $\tau'$ is equivalent to $\tau$ and for a particular choice of constants i.e. $a=0,b=-1,c=1,d=0$, we get $\tau'=-\frac{1}{\tau}$. 

Using this property in (\ref{cd6}), we obtain 
\begin{align}\label{cd9}
    Z(\tau)=e^{2\pi i\frac{1}{\tau}\frac{c}{24}}\overline{Z}(-\frac{1}{\tau}),\hspace{2mm}\text{where}\hspace{2mm}\overline{Z}(-\frac{1}{\tau})\equiv \text{Tr}e^{-2\pi i \frac{1}{\tau}L_0}.
\end{align}

In the limit $\tau\rightarrow 0$, the dominant contribution in $\overline{Z}(-\frac{1}{\tau})$ comes from the lowest eigen value of the operator $L_0$ which is set to be zero without any loss of generality. Therefore, (\ref{cd9}) reduces to $ Z(\tau)=e^{2\pi i\frac{1}{\tau}\frac{c}{24}}$. 

Using this expression in (\ref{cd8}), we finally obtain 
\begin{align}\label{cd10}
    \rho(\Delta)=\oint_c d\tau e^{-2\pi i \big(\tau\big(\Delta-\frac{c}{24}\big)-\frac{1}{\tau}\frac{c}{24}\big)}.
\end{align}

One can approximate the above integral (\ref{cd10}) using the saddle point approximation as 
\begin{align}\label{cd11}
    \rho(\Delta)\approx e^{-2\pi i f(\tau*)},\hspace{2mm}\text{where}\hspace{2mm}f(\tau)=\tau\big(\Delta-\frac{c}{24}\big)-\frac{1}{\tau}\frac{c}{24}
\end{align}
 and $\tau*$ is calculated using the following condition
 \begin{align}\label{cd12}
     \frac{d f(\tau)}{d \tau}\Big|_{\tau=\tau*}=0,\hspace{2mm}\text{which yields}\hspace{2mm}\tau*=i\sqrt{\frac{c}{24\big(\Delta-\frac{c}{24}\big)}}.
 \end{align}
 
On plugging the above value of $\tau*$ (\ref{cd12}) in (\ref{cd11}), we obtain the density of states $\rho(\Delta)$ in the limit $\Delta>>1$. 

The Cardy formula for the black hole is defined in terms of density of states $\rho(\Delta)$ \cite{icts} as 
 \begin{align}\label{CDF}
     S_{Cardy}=\log[\rho(\Delta)].
 \end{align}
 Using (\ref{cd12}) and (\ref{cd11}) in (\ref{CDF}), we obtain the Cardy formula for a 2D black hole in the limit $\Delta>>1$ as
  \begin{align}\label{CDF1}
    S_{Cardy}=2\pi\sqrt{\frac{c\Delta}{6}}.
      \end{align}
 Equation (\ref{CDF1}) is the entropy of a two-dimensional CFT.

It has been found in \cite{Strominger:1997eq} that the entropy computed using the Cardy formula (\ref{CDF1}) matches with the black hole entropy in the bulk. In particular, the author of \cite{Strominger:1997eq} considered the three-dimensional theory of gravity coupled with matter fields 
\begin{align}\label{ath1}
    S=\frac{1}{16\pi G}\int d^3x\sqrt{-g}\Big(R+\frac{2}{l^2}\Big)+S_m,
\end{align}
where $S_m$ contains the matter field. 

Next, they computed the central charge for the boundary theory corresponding to (\ref{ath1}) and estimated the boundary degrees of freedom using the Cardy formula (\ref{CDF1}). Remarkably, these boundary degrees of freedom precisely matched with the Bekenstein-Hawking entropy of the black holes\footnote{In literature, these black holes are called the BTZ black holes \cite{Banados:1992gq}} corresponding to (\ref{ath1}). Finally, they claimed that this result held for any consistent theory of quantum gravity.

A similar analysis has been performed by authors in \cite{Castro:2008ms}. In particular, they considered the 2D Einstein-dilaton gravity in the presence of $U(1)$ gauge fields and computed the central charge associated with the corresponding boundary theory. They established the 2D/3D dictionary and determined the boundary degrees of freedom using the Cardy formula, showing that it precisely matched with the corresponding Bekenstein-Hawking entropy formula for 2D black holes.

Following a similar spirit, one can also construct a suitable 2D/3D dictionary for 2D black holes in the presence of quartic interactions and compute their entropy using the Cardy formula (\ref{CDF1}), as discussed by the authors in \cite{Castro:2008ms}. Alternatively, one can also determine the eigenvalue of the dilatation operator $L_0$ by comparing the Wald entropy (\ref{weed}) with the Cardy formula, provided that there exists a suitable 2D/3D dictionary.

Below, we compute the $\Delta$ for the quartic interactions by assuming that the boundary degrees of freedom for the ground state are equivalent to the black hole entropy, or that there exists a suitable 2D/3D dictionary that establishes a connection between them. 

On comparing the Wald entropy of a 2D black hole (\ref{we1}) with the Cardy formula  (\ref{CDF1}), we obtain
\begin{align}
\Delta=\frac{3}{2}\frac{S_W^2}{c\pi^2}.
\end{align}

As our analysis reveals, $\Delta$  receives corrections both due to the presence of 2-derivative and 4-derivative interaction terms in (\ref{action 2D}). We re-scale  $\Delta$ $\rightarrow$ $\tilde{\Delta}=\frac{\Delta}{w}$ ( with $w=3.53\times10^{-5}\mu \kappa^{1.5}$) which finally yields the eigen value for the UV CFT as
\begin{align}\label{s3}
\Delta=\tilde{\Delta}\big|_{UV}=&\frac{c}{24}+\xi F_1+\kappa F_2,
\end{align}
where $F_1$ and $F_2$ are the corrections due to 2-derivative and 4-derivative interactions present in (\ref{action 2D})
\begin{align}\label{s4}
F_1=&\frac{ \kappa^{0.75} }{w}\Big\{\Big(0.0044587-0.0029725\log\Big(-\frac{\delta}{\sqrt{\mu}}\Big)+0.0029725\log(\delta\sqrt{\mu})+\nonumber\\
&0.00148625\log(\mu)\Big)Q^2+\Big(0.00198166+0.000990832\log(\delta)\nonumber\\
&-0.000990832\log\Big(-\frac{\delta}{\sqrt{\mu}}\Big)-0.000495416\log(\mu)\Big)\frac{3\log(\mu)}{4}Q^2+0.01189\mu d_4\Big\},\nonumber\\
F_2=&\frac{\kappa^{0.75}}{w}\Big\{-\frac{0.0267525}{\delta^2}-\frac{0.0535049}{\delta\sqrt{\mu}}+\Big(-0.00428444+0.03567\log(\delta)-\nonumber\\
&0.0123623\log\Big(-\frac{\delta}{\sqrt{\mu}}\Big)+0.00891749\log\Big(-\frac{\delta}{\sqrt{\mu}}\Big)^2-0.03567\log(2\sqrt{\mu})-\nonumber\\
&0.03567\log(-2\delta\sqrt{\mu})+0.03567\tanh^{-1}\Big(1+\frac{\delta}{\sqrt{\mu}}\Big)\log(-2\delta\sqrt{\mu})+\nonumber\\
&0.03567\text{PolyLog}\Big[2,-\frac{\delta}{2\sqrt{\mu}}\Big]\Big)Q^2+0.01189\sqrt{\mu}d_{11}+\Big(-0.01189\sqrt{\mu}+\nonumber\\
&0.01189\sqrt{\mu}\tanh^{-1}\Big(1+\frac{\delta}{\sqrt{\mu}}\Big)\Big)d_{12}\Big\}.
\end{align}

$\bullet$ Note : In arriving at (\ref{s4}), we write the full solution of the gauge field (\ref{e3}) and the dilaton (\ref{e2})
 $$
      A_t^{bh}=A_{t(0)}^{bh}+\frac{\kappa}{\xi}A_{t(1)}^{bh},\hspace{2mm}
      \phi^{bh}=\phi_{(0)}^{bh}+\xi\phi_{(1)}^{bh}+\kappa\phi_{(2)}^{bh}.
$$
  In the near horizon limit, we absorb the integration constant $d_3$ in $\phi_{(1)}^{bh}$ (\ref{bh23}) into the constant $d_{11}$ in $\phi_{(2)}^{bh}$ (\ref{bh46}) without any loss of generality. Similarly, we absorb the additive constant $d_1$ in $A_{t(0)}^{bh}$ (\ref{bh21}) into the constant $d_6$ in $A_{t(1)}^{bh}$ (\ref{bh4a}). Furthermore, we write the constant $d_2$ in terms of the charge $Q$ using (\ref{ht}).

Finally, we have used the fact that $\delta<<\sqrt{\mu}$ and retain terms up to leading order in the couplings $\xi$ and $\kappa$.

\chapter{Properties of the potential \texorpdfstring{$V(\psi)$}{Vpsi}}\label{vp}
In this Appendix, we discuss the stability of the potential function $V(\psi)$ for the generalised Liouville theory as constructed in (\ref{NHT8}). Finally, we generalise this theory (\ref{NHT8}) by considering 4-derivative interactions.

In order to study the stability of the potential, we first note down the extrema of $V(\psi)$ (\ref{NHT8}) by setting
\begin{align}\label{fa3}
    \frac{dV(\psi)}{d\psi}\Bigg|_{\psi=\psi_{i}}=0,\hspace{2mm}(i=1,2,3)
\end{align}
which reveals
\begin{align}
    \psi_1&=-\frac{1}{2}\sqrt{\frac{c_H}{3\pi}}\label{fa4},\\
   \psi_2&= \frac{1}{2}\sqrt{\frac{c_H}{3\pi}}\text{Productlog}\Bigg[-\pi\sqrt{\frac{3b^2\xi}{2c_H^2}}\Bigg]\label{fa5},\\
    \psi_3&= \frac{1}{2}\sqrt{\frac{c_H}{3\pi}}\text{Productlog}\Bigg[\pi\sqrt{\frac{3b^2\xi}{2c_H^2}}\Bigg]\label{fa6},
\end{align}
where we express $\Phi_H$ in terms of the central charge $c_H$ (\ref{cch}) associated with the theory (\ref{NHT8}).

Using (\ref{fa4})-(\ref{fa6}), we find that the potential $V(\psi)$ exhibits local minima at $\psi_1$ and $\psi_3$ (if $2c_H^2>3\xi\pi^2 b^2 e$) which we identify as the possible vacuua of the theory (\ref{NHT8}). However, notice that in the limit of large central charge $\psi_1\rightarrow-\infty$ which therefore corresponds to the most stable vacuua of the theory (\ref{NHT8}).

Finally, we write down the most general action  (\ref{NHT8}) by considering 4-derivative interactions. Under the following field redefinition
\begin{align}
    \phi=q\Phi_H\psi\hspace{2mm},\hspace{2mm}g_{\mu\nu}\rightarrow e^{2\sigma}g_{\mu\nu}\hspace{2mm},\text{where}\hspace{2mm}\sigma=\frac{\psi}{q\Phi_H},
\end{align}
4-derivative interactions in (\ref{action 2D}) transform as \begin{align}
    S^{(4)}=&q\Phi_H\kappa\int d^2x\sqrt{-g}\psi\Bigg[e^{-2\sigma}\Big\{\big[R_{\mu\nu\alpha\beta}-\bigtriangledown^2\sigma(g_{\mu\alpha}g_{\nu\beta}-g_{\mu\beta}g_{\nu\alpha})\big]\big[R^{\mu\nu\alpha\beta}-\bigtriangledown^2\sigma(g^{\mu\alpha}g^{\nu\beta}\nonumber\\
    &-g^{\mu\beta}g^{\nu\alpha})\big]\Big\}+\frac{3}{4}(\bigtriangledown_{\mu}(q\Phi_H\psi)^{\frac{2}{3}})^4e^{-2\sigma}+4e^{-2\sigma}\Big\{\frac{3}{4}(\bigtriangledown_{\lambda}(q\Phi_H\psi)^{\frac{2}{3}})(\bigtriangledown_{\beta}(q\Phi_H\psi)^{\frac{2}{3}})(\Gamma^{\lambda}_{\alpha\mu}+\nonumber\\
    &\frac{e^{-2\sigma}}{2}(\delta^{\lambda}_{\alpha}\bigtriangledown_{\mu}e^{2\sigma}+\delta^{\lambda}_{\mu}\bigtriangledown_{\alpha}e^{2\sigma}-g_{\alpha\mu}g^{\lambda\bar{\sigma}}\bigtriangledown_{\bar{\sigma}}e^{2\sigma}))(\Gamma^{\beta}_{\rho\bar{\sigma}}+\frac{e^{-2\sigma}}{2}(\delta^{\beta}_{\rho}\bigtriangledown_{\bar{\sigma}}e^{2\sigma}+\delta^{\beta}_{\bar{\sigma}}\bigtriangledown_{\rho}e^{2\sigma}-\nonumber\\
    &g_{\rho\bar{\sigma}}g^{\beta\lambda}\bigtriangledown_{\lambda}e^{2\sigma}))g^{\alpha\rho}g^{\mu\bar{\sigma}}+\frac{2}{3}    (\Gamma^{\lambda}_{\alpha\mu}+\frac{e^{-2\sigma}}{2}(\delta^{\lambda}_{\alpha}\bigtriangledown_{\mu}e^{2\sigma}+\delta^{\lambda}_{\mu}\bigtriangledown_{\alpha}e^{2\sigma}-g_{\alpha\mu}g^{\lambda\bar{\sigma}}\bigtriangledown_{\bar{\sigma}}e^{2\sigma}))\times\nonumber\\
    &\Big(\bigtriangledown_{\lambda}(q\Phi_H\psi)^{\frac{2}{3}}\Big)(\bigtriangledown^{\alpha}(q\Phi_H\psi))(\bigtriangledown^{\mu}(q\Phi_H\psi))(q\Phi_H\psi)^{\frac{-4}{3}}-(\Gamma^{\lambda}_{\alpha\mu}+\frac{e^{-2\sigma}}{2}(\delta^{\lambda}_{\alpha}\bigtriangledown_{\mu}e^{2\sigma}+\nonumber\\
    &\delta^{\lambda}_{\mu}\bigtriangledown_{\alpha}e^{2\sigma}-g_{\alpha\mu}g^{\lambda\bar{\sigma}}\bigtriangledown_{\bar{\sigma}}e^{2\sigma}))\Big(\bigtriangledown_{\lambda}(q\Phi_H\psi)^{\frac{2}{3}}\Big)\{\partial_{\beta}(\bigtriangledown_{\tilde{\sigma}}(q\Phi_H\psi))\}g^{\alpha\beta}g^{\mu\tilde{\sigma}}(q\Phi_H\psi)^{\frac{-1}{3}}-\nonumber\\
    &\frac{4}{9}(q\Phi_H\psi)^{\frac{-5}{3}}(\bigtriangledown^{\alpha}(q\Phi_H\psi))(\bigtriangledown^{\mu}(q\Phi_H\psi))\{\partial_{\alpha}(\bigtriangledown_{\mu}(q\Phi_H\psi))\}+\frac{4}{27}(\bigtriangledown_{\mu}(q\Phi_H\psi))^4\times\nonumber\\
    &(q\Phi_H\psi)^{\frac{-8}{3}}-\frac{1}{3}\{\partial_{\alpha}(\bigtriangledown_{\mu}(q\Phi_H\psi))\}\{\partial_{\beta}(\bigtriangledown_{\rho}(q\Phi_H\psi))\}g^{\alpha\beta}g^{\mu\rho}(q\Phi_H\psi)^{\frac{-2}{3}}\Big\}+e^{-6\sigma}F^4+\nonumber\\
    &e^{-6\sigma}F^{\mu\nu}F_{\nu\lambda}F^{\lambda\bar{\sigma}}F_{\bar{\sigma}\mu}+e^{-4\sigma}g^{\mu\Bar{\mu}}g^{\nu\Bar{\nu}}g^{\lambda\Bar{\lambda}}(\bigtriangledown_{\mu}F_{\Bar{\mu}\Bar{\nu}}-\frac{e^{-2\sigma}}{2}(\delta^{\rho}_{\mu}\bigtriangledown_{\Bar{\mu}}e^{2\sigma}+\delta^{\rho}_{\bar{\mu}}\bigtriangledown_{\mu}e^{2\sigma}-\nonumber\\
    &g_{\mu\bar{\mu}}g^{\rho\bar{\sigma}}\bigtriangledown_{\bar{\sigma}}e^{2\sigma})F_{\rho\bar{\nu}}-\frac{e^{-2\sigma}}{2}(\delta^{\rho}_{\mu}\bigtriangledown_{\Bar{\nu}}e^{2\sigma}+\delta^{\rho}_{\bar{\nu}}\bigtriangledown_{\mu}e^{2\sigma}-g_{\mu\bar{\nu}}g^{\rho\bar{\sigma}}\bigtriangledown_{\bar{\sigma}}e^{2\sigma})F_{\bar{\mu}\rho})\times(\bigtriangledown_{\bar{\lambda}}F_{\lambda\nu}-\nonumber\\
    &\frac{e^{-2\sigma}}{2}(\delta^{\rho}_{\lambda}\bigtriangledown_{\Bar{\lambda}}e^{2\sigma}+\delta^{\rho}_{\bar{\lambda}}\bigtriangledown_{\lambda}e^{2\sigma}-g_{\lambda\bar{\lambda}}g^{\rho\bar{\sigma}}\bigtriangledown_{\bar{\sigma}}e^{2\sigma})F_{\rho\nu}-\frac{e^{-2\sigma}}{2}(\delta^{\rho}_{\nu}\bigtriangledown_{\Bar{\lambda}}e^{2\sigma}+\delta^{\rho}_{\bar{\lambda}}\bigtriangledown_{\nu}e^{2\sigma}\nonumber\\
    &-g_{\nu\bar{\lambda}}g^{\rho\bar{\sigma}}\bigtriangledown_{\bar{\sigma}}e^{2\sigma})F_{\lambda\rho})\Bigg]\label{4v}.
\end{align}

Notice that, in arriving at (\ref{4v}), we express the Riemann tensor in 2D as $$R_{\mu\nu\alpha\beta}=\frac{R}{2}(g_{\mu\alpha}g_{\nu\beta}-g_{\mu\beta}g_{\nu\alpha}),$$ where $R$ is the Ricci scalar in 2D. Equation (\ref{4v}) represents corrections to the Liouville theory (\ref{NHT8}) due to the presence of 4-derivative interactions in (\ref{action 2D}).

\chapter{Equations of motion}\label{geneomap}
In this Appendix, we note down the most general form of the equations of motion (\ref{geneomi})-(\ref{geneomf}) in the Fefferman-Graham gauge (\ref{fggauge}),

 \begin{align}
    A_{t}\hspace{1mm}:\hspace{1mm}&\frac{\kappa}{\sqrt{-h_{tt}}}\partial_{\eta}\Bigg[\frac{\Phi A_t'}{\sqrt{-h_{tt}}}\Bigg(\cosh{\gamma}-\frac{s\sinh{\gamma}}{\sqrt{s^2+p^2}}\Bigg)\Bigg]=0,\label{gaugeeomiap}\\
    \chi\hspace{1mm}:\hspace{1mm}&\frac{\kappa}{\sqrt{-h_{tt}}}\partial_t\Bigg(-\frac{\dot{\chi}}{\sqrt{-h_{tt}}}\cosh{\gamma}+\frac{s\dot{\chi}+p\xi'\sqrt{-h_{tt}}}{\sqrt{-h_{tt}}\sqrt{s^2+p^2}}\sinh{\gamma}\Bigg)+\frac{\kappa}{\sqrt{-h_{tt}}}\partial_{\eta}\Bigg(\sqrt{-h_{tt}}\times\nonumber\\
    &\chi'\cosh{\gamma}-\frac{s\chi'\sqrt{-h_{tt}}+p\dot{\xi}}{\sqrt{s^2+p^2}}\sinh{\gamma}\Bigg)=0,\\
     \xi\hspace{1mm}:\hspace{1mm}&\frac{\kappa}{\sqrt{-h_{tt}}}\partial_t\Bigg(-\frac{\dot{\xi}}{\sqrt{-h_{tt}}}\cosh{\gamma}-\frac{-s\dot{\xi}+p\chi'\sqrt{-h_{tt}}}{\sqrt{-h_{tt}}\sqrt{s^2+p^2}}\sinh{\gamma}\Bigg)+\frac{\kappa}{\sqrt{-h_{tt}}}\partial_{\eta}\Bigg(\sqrt{-h_{tt}}\times\nonumber\\
    &\xi'\cosh{\gamma}-\frac{s\xi'\sqrt{-h_{tt}}-p\dot{\chi}}{\sqrt{s^2+p^2}}\sinh{\gamma}\Bigg)=0,
\end{align}

 \begin{align}
    \Phi\hspace{1mm}:\hspace{1mm}&\sqrt{-h_{tt}}(\sqrt{-h_{tt}})''+(\sqrt{-h_{tt}})^2\Lambda-\kappa A_t'^2\cosh{\gamma}+\frac{\kappa s A_t'^2}{\sqrt{s^2+p^2}}\sinh{\gamma}=0,\\
    g_{tt}\hspace{1mm}:\hspace{1mm}&\Phi''+\Lambda\Phi
+2\kappa\Phi\Bigg(\cosh{\gamma}-\frac{s\sinh{\gamma}}{\sqrt{s^2+p^2}}\Bigg)\Bigg(\Phi^{-1}(\chi'^2+\xi'^2)-\frac{s}{2}\Bigg)=0,\\
g_{\eta\eta}\hspace{1mm}:\hspace{1mm}&-\frac{1}{\sqrt{-h_{tt}}}\partial_{t}\Bigg(\frac{\dot{\Phi}}{\sqrt{-h_{tt}}}\Bigg)+\frac{(\sqrt{-h_{tt}})'\Phi'}{\sqrt{-h_{tt}}}+\Lambda\Phi-2\kappa\Phi\Bigg(\cosh{\gamma}-\frac{s\sinh{\gamma}}{\sqrt{s^2+p^2}}\Bigg)\times\nonumber\\
&\Bigg(\frac{\Phi^{-1}(\dot{\chi}^2+\dot{\xi}^2)}{(\sqrt{-h_{tt}})^2}+\frac{s}{2}\Bigg)=0,\\
g_{\eta t}\hspace{1mm}:\hspace{1mm}&-(\dot{\Phi})'+\frac{\dot{\Phi}(\sqrt{-h_{tt}})'}{\sqrt{-h_{tt}}}-2\kappa \Bigg(\cosh{\gamma}-\frac{s\sinh{\gamma}}{\sqrt{s^2+p^2}}\Bigg)(\dot{\xi}\xi'+\dot{\chi}\chi')=0,\label{gaugeeomfap}
\end{align}
along with the functions
\begin{align}
    s=&-\frac{A_t'^2}{(\sqrt{-h_{tt}})^2}+\Phi^{-1}\Bigg(-\frac{1}{(\sqrt{-h_{tt}})^2}(\dot{\chi}^2+\dot{\xi}^2)+\chi'^2+\xi'^2\Bigg),\\
    p=&-2\frac{\Phi^{-1}}{\sqrt{-h_{tt}}}\big(\dot{\chi}\xi'-\chi'\dot{\xi}\big),
\end{align}
where . and $'$ denote the derivatives with respect to $t$ and $\eta$ respectively. 

\chapter{Details of the functions $\mathcal{H}_i$'s}\label{h1h2fucdetailap}
In this Appendix, we present the explicit details of the functions $\mathcal{H}_i$, $(i=1,2,..5)$
\begin{align}
    \mathcal{H}_1(\eta)=&\hspace{1mm}\frac{1}{4} \Bigg[c_1 \Bigg(-\frac{a \kappa  e^{-\eta  \lambda }}{a_1 a_2 b_1}+\frac{\kappa 
   }{a_1^3}\left(a_3 \left(8 \log \left(a_2-a_1 e^{2 \eta  \lambda }\right)-8 \eta  \lambda
   \right)-\frac{e^{-\eta  \lambda }}{a_2 b_1 \lambda }\right)+\nonumber\\
   &\hspace{1mm}\frac{1}{a_1^2}\Big(4 (\gamma +1) \eta 
   \lambda -4 \log \left(a_2-a_1 e^{2 \eta  \lambda }\right)\Big)-4 a (\gamma +1) \lambda
   +\frac{8 a a_1 \lambda  e^{2 \eta  \lambda }}{a_1 e^{2 \eta  \lambda }-a_2}\Bigg)-\nonumber\\
   &\hspace{1mm}\frac{4
   }{a_1^2}\left(\kappa  \left(c_5 \left(\eta -a a_1^2\right)+c_4\right)+\gamma 
   c_3\right)-\frac{1}{a_1^4 a_2}\Big(c_1^3 \kappa  e^{-2 \eta  \lambda } \Big(2 a_1 e^{2 \eta  \lambda }\Big(2 \eta ^2 \lambda ^2+
   \nonumber\\
   &\hspace{1mm}\log \left(a_2-a_1 e^{2 \eta  \lambda }\right)^2-3 \eta  \lambda  \log \left(a_2-a_1
   e^{2 \eta  \lambda }\right)\Big)+a_2 \left(2 \log \left(a_1\right)+2
   \eta  \lambda +1\right)\Big)\Big)\Bigg],
\end{align}
\begin{align}
 \mathcal{H}_2(\eta)=&\hspace{1mm}-\frac{1}{8 a_1^3
   a_2 b_1 \lambda ^2}\Bigg(a e^{-3 \eta  \lambda } \Big(c_1 \Big(-2 a_1 a_2 b_1 \lambda  e^{\eta  \lambda }
   \Big(-2 \log \left(a_2-a_1 e^{2 \eta  \lambda }\right)+2 \log \left(a_1\right)\nonumber\\
   &\hspace{1mm}-\gamma +4
   \eta  \lambda +1\Big)+4 a_2 a_3 b_1 \kappa  \lambda  e^{\eta  \lambda } \Big(-2 \log
   \left(a_2-a_1 e^{2 \eta  \lambda }\right)+2 \log \left(a_1\right)+4 \eta  \lambda\nonumber\\
   &\hspace{1mm}
   +1\Big)+\kappa \Big)+b_1 c_1^3 \kappa  \lambda  e^{\eta  \lambda } \Big(2 \Big(\log
   ^2\left(a_2-a_1 e^{2 \eta  \lambda }\right)-3 \eta  \lambda  \log \left(a_2-a_1 e^{2 \eta 
   \lambda }\right)+\nonumber\\
   &\hspace{1mm}2 \eta ^2 \lambda ^2\Big)-\log \left(a_1\right) (2 \eta  \lambda +1)-2
   \log ^2\left(a_1\right)\Big)-2 a_1 a_2 b_1 c_5 \kappa  e^{\eta  \lambda }\Big)\Bigg),
   \end{align}
   
\begin{align}
   \mathcal{H}_3(\eta)=&\hspace{1mm}2 e^{2 \eta  \lambda }-\frac{1}{8 a_1 a_2^2 \left(a_1 e^{2 \eta  \lambda }-a_2\right)}a \Bigg[8 a_2 a_1^3 \lambda  e^{2 \eta  \lambda } \Big(4 a_2 \kappa  e^{2 \eta 
   \lambda } \left(a_4 \gamma +a_6 \kappa +a_3\right)-2 a_2^2+\nonumber\\
   &\hspace{1mm}c_1^2 \kappa  e^{2 \eta  \lambda } \left(2 \eta  \lambda -\log
   \left(a_2-a_1 e^{2 \eta  \lambda }\right)\right)\Big)+a_1^2 \kappa  e^{2 \eta  \lambda } \Big(c_1^4
   \kappa  \lambda  e^{2 \eta  \lambda } \Big(\log \left(a_2-a_1 e^{2 \eta  \lambda }\right)\nonumber\\
   &\hspace{1mm}-2 \eta  \lambda
   \Big){}^2+8 a_2 a_3 c_1^2 \kappa  \lambda  e^{2 \eta  \lambda } \left(2 \eta  \lambda -\log \left(a_2-a_1 e^{2
   \eta  \lambda }\right)\right)+8 a_2^2 \Big(2 a_3^2 \kappa  \lambda  e^{2 \eta  \lambda }-\nonumber\\
   &\hspace{1mm}c_1 \left(c_1 \lambda 
   \left(-\log \left(a_2-a_1 e^{2 \eta  \lambda }\right)+2 \eta  \lambda +1\right)+2 c_5 \kappa \right)\Big)-32
   a_2^3 \lambda  \big(a_4 \gamma +a_6 \kappa +\nonumber\\
   &\hspace{1mm} a_3\big)\Big)-a_2 a_1 \kappa ^2 \Big(16 a_2^2 \left(a_3^2
   \lambda  e^{2 \eta  \lambda }-c_1 c_5\right)+c_1^4 \lambda  e^{2 \eta  \lambda } \Big(\log \left(a_2-a_1 e^{2
   \eta  \lambda }\right)^2-\nonumber\\
   &\hspace{1mm}2 (2 \eta  \lambda +1) \log \left(a_2-a_1 e^{2 \eta  \lambda }\right)+4 \eta  \lambda 
   (\eta  \lambda +1)\Big)+8 a_2 a_3 c_1^2 \lambda  e^{2 \eta  \lambda } \Big(2 \eta  \lambda -\nonumber\\
   &\hspace{1mm}\log \left(a_2-a_1
   e^{2 \eta  \lambda }\right)\Big)\Big)-8 a_2^3 a_3 c_1^2 \kappa ^2 \lambda +16 a_2^2 a_1^4 \lambda  e^{4 \eta 
   \lambda }\Bigg],\\
   \mathcal{H}_4(\eta)=&\hspace{1mm}\frac{a}{\lambda},
    \end{align}
   
\begin{align}
   \mathcal{H}_5(\eta)=&\hspace{1mm} a \Bigg[\frac{1}{4} \kappa  e^{\eta  \lambda } \Bigg(\frac{2 b_1 c_1^2 e^{2 \eta  \lambda }}{a_2 \left(a_2-a_1
   e^{2 \eta  \lambda }\right)}+\frac{b_1 }{a_1 a_2}\Big(c_1^2 \left(-\log \left(a_2-a_1 e^{2 \eta  \lambda }\right)+2 \eta 
   \lambda +2\right)+\nonumber\\
   &\hspace{1mm}4 a_2 a_3\Big)-\frac{d_1^2 e^{\eta  \lambda }}{a_1 a_2 \lambda  e^{2 \eta  \lambda
   }+a_2^2 \lambda }-\frac{d_1^2 \tan ^{-1}\left(\frac{\sqrt{a_1} e^{\eta  \lambda }}{\sqrt{a_2}}\right)}{\sqrt{a_1}
   a_2^{3/2} \lambda }-\frac{a_1 e_1^2 e^{\eta  \lambda }}{a_1 \lambda  e^{2 \eta  \lambda }+a_2 \lambda
   }-\nonumber\\
   &\hspace{1mm}\frac{\sqrt{a_1}}{\sqrt{a_2}
   \lambda } e_1^2 \tan ^{-1}\left(\frac{\sqrt{a_1} e^{\eta  \lambda }}{\sqrt{a_2}}\right)\Bigg)+\frac{\kappa ^2 e^{-\eta  \lambda } }{2 a_1^2 \lambda }\Big(2 a_1^2 b_4 \lambda  e^{2 \eta  \lambda }+2 a_2 a_3
   \lambda  (2 \eta  \lambda +1)-\nonumber\\
   &\hspace{1mm}b_1 c_1 c_5 (2 \eta  \lambda +1)\Big)+b_2 \gamma  \kappa 
   e^{\eta  \lambda }+b_1 e^{\eta  \lambda }\Bigg],
\end{align}
where we denote the constant $a= \frac{2 \left(a_1^2+1\right)}{3 a_1^2 \lambda }-\frac{4 a_3 \kappa
   }{3 a_1^3 \lambda }$.

   \chapter{Order $\kappa^2$ solutions}\label{jhepappendixk2}
In this Appendix, we note down the solution of the equations (\ref{bhjhep1eq})-(\ref{bhjhepfeq}),
\begin{align}
    \omega_3^{(bh)}=&\hspace{1mm}\frac{1}{192 \Lambda ^2 \mu ^{3/2} \text{$\phi_0$}^4}\Bigg(64 g_1 \Lambda  \mu ^{3/2} m_1^2 \rho  \text{$\phi_0$}
   \left(\log \left(1-\frac{\rho }{\sqrt{\mu }}\right)+\log
   \left(\frac{\rho }{\sqrt{\mu }}+1\right)\right)-\nonumber\\
   &\hspace{1mm}16 \Lambda
    l_1^2 m_1^2 \text{$\phi_0 $} \Big(\sqrt{\mu } \log
   \left(\frac{\rho }{\sqrt{\mu }}+1\right)+\log
   \left(1-\frac{\rho }{\sqrt{\mu }}\right) \Big(\rho
   \left(-\log \left(\frac{\rho }{\sqrt{\mu
   }}+1\right)\right)+\nonumber\\
   &\hspace{1mm}\sqrt{\mu }+\rho  \log (4)\Big)+\rho
   \tanh ^{-1}\left(\frac{\rho }{\sqrt{\mu }}\right)-2 \rho
   \text{Li}_2\left(\frac{1}{2}-\frac{\rho }{2 \sqrt{\mu
   }}\right)+\rho \Big)+16 m_1 \times\nonumber\\
   &\hspace{1mm}\Big(3 \mu  \rho  \tanh
   ^{-1}\left(\frac{\rho }{\sqrt{\mu }}\right) \left(2
   \Lambda ^2 m_6 \text{$\phi_0$}^3-3 m_1^3\right)-\Lambda
   m_1 n_1^2 \text{$\phi_0$} \Big(\sqrt{\mu } \log
   \left(\frac{\rho }{\sqrt{\mu }}+1\right)\nonumber\\
   &\hspace{1mm}+\log
   \left(1-\frac{\rho }{\sqrt{\mu }}\right) \left(\rho
   \left(-\log \left(\frac{\rho }{\sqrt{\mu
   }}+1\right)\right)+\sqrt{\mu }+\rho  \log (4)\right)+\nonumber\\
   &\hspace{1mm}\rho
   \tanh ^{-1}\left(\frac{\rho }{\sqrt{\mu }}\right)-2 \rho
   \text{Li}_2\left(\frac{1}{2}-\frac{\rho }{2 \sqrt{\mu
   }}\right)+\rho \Big)\Big)-96 \Lambda  \sqrt{\mu } q_2
   \text{$\phi_0$}^2 \Big(\Lambda  q_1 \text{$\phi_0$}^2\times
   \nonumber\\
   &\hspace{1mm} \Big(-\sqrt{\mu }+2 \left(\mu -\rho ^2\right)
   \rho\tanh
   ^{-1}\left(\frac{\rho }{\sqrt{\mu }}\right)\Big)-4 \sqrt{\mu } m_1^2 \rho  \tanh
   ^{-1}\left(\frac{\rho }{\sqrt{\mu }}\right)\Big)\nonumber\\
   &\hspace{1mm}-3
   \Lambda ^2 \sqrt{\mu } q_2^2 \text{$\phi_0$}^4 \Big(
   \tanh ^{-1}\left(\frac{\rho }{\sqrt{\mu }}\right)32\left(\left(\mu -\rho ^2\right) \tanh
   ^{-1}\left(\frac{\rho }{\sqrt{\mu }}\right)+\sqrt{\mu }
   \rho \right)\nonumber\\
   &\hspace{1mm}+45 \sqrt{\mu } \rho \Big)+96 \Lambda ^2
   \sqrt{\mu } \rho  q_1^2 \text{$\phi_0$}^4 \Big(\rho
   -\sqrt{\mu } \tanh ^{-1}\left(\frac{\rho }{\sqrt{\mu
   }}\right)\Big)+192 \Lambda ^2 \mu  \text{$\phi_0$}^4\times\nonumber\\
   &\Big(q_6 \Big(\rho  \tanh ^{-1}\left(\frac{\rho
   }{\sqrt{\mu }}\right)-\sqrt{\mu }\Big)+\rho
   q_5\Big)\Bigg),\label{jhepeopmkappa2i}\\
   \Phi_3^{(bh)}=&\hspace{1mm}g_3\rho-\frac{1}{8 \mu ^2}\Bigg(2 \rho ^2 \tanh ^{-1}\left(\frac{\rho }{\sqrt{\mu
   }}\right) \left(q_1 \left(l_1^2+n_1^2\right)-4 g_1 \mu
   ^{3/2} q_2\right)+\mu  \log \left(\rho -\sqrt{\mu }\right)\times\nonumber\\
   &\hspace{1mm}\left(q_1 \left(l_1^2+n_1^2\right)-4 g_1 \mu ^{3/2}
   q_2\right)-\mu  \log \left(\sqrt{\mu }+\rho \right)
   \left(q_1 \left(l_1^2+n_1^2\right)-4 g_1 \mu ^{3/2}
   q_2\right)-\nonumber\\
   &\hspace{1mm}8 g_1 \mu ^{3/2} \rho ^2 q_1-8 g_1 \mu ^2 \rho
   q_2+2 \mu  q_2 \left(l_1^2+n_1^2\right) \log
   \left(1-\frac{\rho ^2}{\mu }\right)-\mu  \big(q_2
   \left(l_1^2+n_1^2\right)+\nonumber\\
   &\hspace{1mm}4 l_1 l_5+2 n_1 n_5\big) \log
   \left(\mu -\rho ^2\right)+2 q_2 \left(l_1^2+n_1^2\right)
   \left(\rho ^2-\mu \right) \tanh ^{-1}\left(\frac{\rho
   }{\sqrt{\mu }}\right)^2+\nonumber\\
   &\hspace{1mm}2 \sqrt{\mu } \rho  q_1
   \left(l_1^2+n_1^2\right)-\mu  \left(q_2
   \left(l_1^2+n_1^2\right)-4 l_1 l_5-2 n_1 n_5\right) \log
   \left(\sqrt{\mu }+\rho \right)+\nonumber\\
   &\hspace{1mm}\sqrt{\mu } \left(\rho
   -\sqrt{\mu }\right) \left(q_2 \left(l_1^2+n_1^2\right)-4
   l_1 l_5-2 n_1 n_5\right) \log \left(\rho -\sqrt{\mu
   }\right)-\sqrt{\mu } \rho  \big(q_2
   \left(l_1^2+n_1^2\right)\nonumber\\
   &\hspace{1mm}-4 l_1 l_5-2 n_1 n_5\big) \log
   \left(\sqrt{\mu }+\rho \right)+4 \sqrt{\mu } \rho  q_2
   \left(l_1^2+n_1^2\right) \tanh ^{-1}\left(\frac{\rho
   }{\sqrt{\mu }}\right)\Bigg)+g_4,\label{jhepeopmkappa2i1}
   \end{align}
   \begin{align}
   A_{t3}^{(bh)}=&\hspace{1mm}\frac{1}{8 \Lambda  \mu ^{3/2}
   \text{$\phi_0$}^2}\Bigg(2 \sqrt{\mu } \Big(m_1 \rho  \left(-4 g_1 \mu  \rho
   +l_1^2+n_1^2+8 \sqrt{\mu } \rho  q_1 \text{$\phi_0$}+8 \mu
    q_2 \text{$\phi_0$}\right)+\nonumber\\
    &\hspace{1mm}4 \Lambda  \mu  \text{$\phi
_0$}^2 \left(m_6 \rho +m_5\right)\Big)+2 m_1 \rho ^2
   \tanh ^{-1}\left(\frac{\rho }{\sqrt{\mu }}\right)
   \left(l_1^2+n_1^2+8 \mu  q_2 \text{$\phi_0$}\right)+\nonumber\\
   &\hspace{1mm}\mu
   m_1 \left(\log \left(\rho -\sqrt{\mu }\right)-\log
   \left(\sqrt{\mu }+\rho \right)\right) \left(l_1^2+n_1^2+8
   \mu  q_2 \text{$\phi_0$}\right)\Bigg),\\
   \xi_3^{(bh)}=&\hspace{1mm}\frac{l_5}{\sqrt{\mu}}\tanh ^{-1}\left(\frac{\rho }{\sqrt{\mu }}\right)+l_6,\label{jhepeopmkappa2i2}\\
   \chi_3^{(bh)}=&\hspace{1mm}n_5t+n_6,\label{jhepeopmkappa2f}
\end{align}
where $m_i,n_i,l_i,q_i$, $g_j$, $(i=5,6$, $j=3,4)$ are the integration constants and we define $\text{Li}_2(x)=\text{PolyLog}(2,x)$.



\clearpage
\phantomsection 
\addcontentsline{toc}{chapter}{\textbf{Bibliography}}
\label{Bibliography}
\lhead{\emph{Bibliography}}




\end{document}